\documentclass[aps,rmp,groupedaddress,twocolumn,floatfix,longbibliography]{revtex4-1}
\usepackage{graphicx,amssymb,amsmath,amsbsy,latexsym,braket} 
\usepackage{wasysym}
\usepackage[T1]{fontenc}
\pdfoutput=1
\usepackage{color}
\usepackage[colorlinks=false,linkcolor=red]{hyperref}%
\usepackage{pxfonts}           
\def\mb#1{\mathbf{#1}}            
\def\bs#1{\boldsymbol{#1}}
\def\m#1{\mathrm{#1}}
\def\mi#1{\mathit{#1}}  

\def\u0#1#2{u^{\scriptscriptstyle{\mathrm{\;#1}}}_{\scriptscriptstyle{0#2}}}

\def\slantfrac#1#2{\kern.1em^{#1}\kern-.3em/\kern-.1em_{#2}}
\def\ts#1{\textstyle{#1}}
\newcommand{\greeksym}[1]{{\usefont{U}{psy}{m}{n}#1}}

\newcommand{\inc}{\mbox{\small\greeksym{d}\hskip 0.05ex}}%

\newcommand{\umu}{\mbox{\greeksym{m}}}
\newcommand{\udelta}{\mbox{\greeksym{d}}}
\newcommand{\ulambda}{\mbox{\footnotesize\greeksym{l}}}

\newcommand{\urho}{\mbox{\greeksym{r}}}
\newcommand{\upsi}{\mbox{\greeksym{y}}}
\def\slantfrac#1#2{\kern.1em^{#1}\kern-.3em/\kern-.1em_{#2}}

\begin{document}
\definecolor{AliceBlue}{rgb}{0.94,0.97,1.00}
\definecolor{AntiqueWhite1}{rgb}{1.00,0.94,0.86}
\definecolor{AntiqueWhite2}{rgb}{0.93,0.87,0.80}
\definecolor{AntiqueWhite3}{rgb}{0.80,0.75,0.69}
\definecolor{AntiqueWhite4}{rgb}{0.55,0.51,0.47}
\definecolor{AntiqueWhite}{rgb}{0.98,0.92,0.84}
\definecolor{BlanchedAlmond}{rgb}{1.00,0.92,0.80}
\definecolor{BlueViolet}{rgb}{0.54,0.17,0.89}
\definecolor{CadetBlue1}{rgb}{0.60,0.96,1.00}
\definecolor{CadetBlue2}{rgb}{0.56,0.90,0.93}
\definecolor{CadetBlue3}{rgb}{0.48,0.77,0.80}
\definecolor{CadetBlue4}{rgb}{0.33,0.53,0.55}
\definecolor{CadetBlue}{rgb}{0.37,0.62,0.63}
\definecolor{CornflowerBlue}{rgb}{0.39,0.58,0.93}
\definecolor{DarkBlue}{rgb}{0.00,0.00,0.55}
\definecolor{DarkCyan}{rgb}{0.00,0.55,0.55}
\definecolor{DarkGoldenrod1}{rgb}{1.00,0.73,0.06}
\definecolor{DarkGoldenrod2}{rgb}{0.93,0.68,0.05}
\definecolor{DarkGoldenrod3}{rgb}{0.80,0.58,0.05}
\definecolor{DarkGoldenrod4}{rgb}{0.55,0.40,0.03}
\definecolor{DarkGoldenrod}{rgb}{0.72,0.53,0.04}
\definecolor{DarkGray}{rgb}{0.66,0.66,0.66}
\definecolor{DarkGreen}{rgb}{0.00,0.39,0.00}
\definecolor{DarkGrey}{rgb}{0.66,0.66,0.66}
\definecolor{DarkKhaki}{rgb}{0.74,0.72,0.42}
\definecolor{DarkMagenta}{rgb}{0.55,0.00,0.55}
\definecolor{DarkOliveGreen1}{rgb}{0.79,1.00,0.44}
\definecolor{DarkOliveGreen2}{rgb}{0.74,0.93,0.41}
\definecolor{DarkOliveGreen3}{rgb}{0.64,0.80,0.35}
\definecolor{DarkOliveGreen4}{rgb}{0.43,0.55,0.24}
\definecolor{DarkOliveGreen}{rgb}{0.33,0.42,0.18}
\definecolor{DarkOrange1}{rgb}{1.00,0.50,0.00}
\definecolor{DarkOrange2}{rgb}{0.93,0.46,0.00}
\definecolor{DarkOrange3}{rgb}{0.80,0.40,0.00}
\definecolor{DarkOrange4}{rgb}{0.55,0.27,0.00}
\definecolor{DarkOrange}{rgb}{1.00,0.55,0.00}
\definecolor{DarkOrchid1}{rgb}{0.75,0.24,1.00}
\definecolor{DarkOrchid2}{rgb}{0.70,0.23,0.93}
\definecolor{DarkOrchid3}{rgb}{0.60,0.20,0.80}
\definecolor{DarkOrchid4}{rgb}{0.41,0.13,0.55}
\definecolor{DarkOrchid}{rgb}{0.60,0.20,0.80}
\definecolor{DarkRed}{rgb}{0.55,0.00,0.00}
\definecolor{DarkSalmon}{rgb}{0.91,0.59,0.48}
\definecolor{DarkSeaGreen1}{rgb}{0.76,1.00,0.76}
\definecolor{DarkSeaGreen2}{rgb}{0.71,0.93,0.71}
\definecolor{DarkSeaGreen3}{rgb}{0.61,0.80,0.61}
\definecolor{DarkSeaGreen4}{rgb}{0.41,0.55,0.41}
\definecolor{DarkSeaGreen}{rgb}{0.56,0.74,0.56}
\definecolor{DarkSlateBlue}{rgb}{0.28,0.24,0.55}
\definecolor{DarkSlateGray1}{rgb}{0.59,1.00,1.00}
\definecolor{DarkSlateGray2}{rgb}{0.55,0.93,0.93}
\definecolor{DarkSlateGray3}{rgb}{0.47,0.80,0.80}
\definecolor{DarkSlateGray4}{rgb}{0.32,0.55,0.55}
\definecolor{DarkSlateGray}{rgb}{0.18,0.31,0.31}
\definecolor{DarkSlateGrey}{rgb}{0.18,0.31,0.31}
\definecolor{DarkTurquoise}{rgb}{0.00,0.81,0.82}
\definecolor{DarkViolet}{rgb}{0.58,0.00,0.83}
\definecolor{DeepPink1}{rgb}{1.00,0.08,0.58}
\definecolor{DeepPink2}{rgb}{0.93,0.07,0.54}
\definecolor{DeepPink3}{rgb}{0.80,0.06,0.46}
\definecolor{DeepPink4}{rgb}{0.55,0.04,0.31}
\definecolor{DeepPink}{rgb}{1.00,0.08,0.58}
\definecolor{DeepSkyBlue1}{rgb}{0.00,0.75,1.00}
\definecolor{DeepSkyBlue2}{rgb}{0.00,0.70,0.93}
\definecolor{DeepSkyBlue3}{rgb}{0.00,0.60,0.80}
\definecolor{DeepSkyBlue4}{rgb}{0.00,0.41,0.55}
\definecolor{DeepSkyBlue}{rgb}{0.00,0.75,1.00}
\definecolor{DimGray}{rgb}{0.41,0.41,0.41}
\definecolor{DimGrey}{rgb}{0.41,0.41,0.41}
\definecolor{DodgerBlue1}{rgb}{0.12,0.56,1.00}
\definecolor{DodgerBlue2}{rgb}{0.11,0.53,0.93}
\definecolor{DodgerBlue3}{rgb}{0.09,0.45,0.80}
\definecolor{DodgerBlue4}{rgb}{0.06,0.31,0.55}
\definecolor{DodgerBlue}{rgb}{0.12,0.56,1.00}
\definecolor{FloralWhite}{rgb}{1.00,0.98,0.94}
\definecolor{ForestGreen}{rgb}{0.13,0.55,0.13}
\definecolor{GhostWhite}{rgb}{0.97,0.97,1.00}
\definecolor{GreenYellow}{rgb}{0.68,1.00,0.18}
\definecolor{HotPink1}{rgb}{1.00,0.43,0.71}
\definecolor{HotPink2}{rgb}{0.93,0.42,0.65}
\definecolor{HotPink3}{rgb}{0.80,0.38,0.56}
\definecolor{HotPink4}{rgb}{0.55,0.23,0.38}
\definecolor{HotPink}{rgb}{1.00,0.41,0.71}
\definecolor{IndianRed1}{rgb}{1.00,0.42,0.42}
\definecolor{IndianRed2}{rgb}{0.93,0.39,0.39}
\definecolor{IndianRed3}{rgb}{0.80,0.33,0.33}
\definecolor{IndianRed4}{rgb}{0.55,0.23,0.23}
\definecolor{IndianRed}{rgb}{0.80,0.36,0.36}
\definecolor{LavenderBlush1}{rgb}{1.00,0.94,0.96}
\definecolor{LavenderBlush2}{rgb}{0.93,0.88,0.90}
\definecolor{LavenderBlush3}{rgb}{0.80,0.76,0.77}
\definecolor{LavenderBlush4}{rgb}{0.55,0.51,0.53}
\definecolor{LavenderBlush}{rgb}{1.00,0.94,0.96}
\definecolor{LawnGreen}{rgb}{0.49,0.99,0.00}
\definecolor{LemonChiffon1}{rgb}{1.00,0.98,0.80}
\definecolor{LemonChiffon2}{rgb}{0.93,0.91,0.75}
\definecolor{LemonChiffon3}{rgb}{0.80,0.79,0.65}
\definecolor{LemonChiffon4}{rgb}{0.55,0.54,0.44}
\definecolor{LemonChiffon}{rgb}{1.00,0.98,0.80}
\definecolor{LightBlue1}{rgb}{0.75,0.94,1.00}
\definecolor{LightBlue2}{rgb}{0.70,0.87,0.93}
\definecolor{LightBlue3}{rgb}{0.60,0.75,0.80}
\definecolor{LightBlue4}{rgb}{0.41,0.51,0.55}
\definecolor{LightBlue}{rgb}{0.68,0.85,0.90}
\definecolor{LightCoral}{rgb}{0.94,0.50,0.50}
\definecolor{LightCyan1}{rgb}{0.88,1.00,1.00}
\definecolor{LightCyan2}{rgb}{0.82,0.93,0.93}
\definecolor{LightCyan3}{rgb}{0.71,0.80,0.80}
\definecolor{LightCyan4}{rgb}{0.48,0.55,0.55}
\definecolor{LightCyan}{rgb}{0.88,1.00,1.00}
\definecolor{LightGoldenrod1}{rgb}{1.00,0.93,0.55}
\definecolor{LightGoldenrod2}{rgb}{0.93,0.86,0.51}
\definecolor{LightGoldenrod3}{rgb}{0.80,0.75,0.44}
\definecolor{LightGoldenrod4}{rgb}{0.55,0.51,0.30}
\definecolor{LightGoldenrodYellow}{rgb}{0.98,0.98,0.82}
\definecolor{LightGoldenrod}{rgb}{0.93,0.87,0.51}
\definecolor{LightGray}{rgb}{0.83,0.83,0.83}
\definecolor{LightGreen}{rgb}{0.56,0.93,0.56}
\definecolor{LightGrey}{rgb}{0.83,0.83,0.83}
\definecolor{LightPink1}{rgb}{1.00,0.68,0.73}
\definecolor{LightPink2}{rgb}{0.93,0.64,0.68}
\definecolor{LightPink3}{rgb}{0.80,0.55,0.58}
\definecolor{LightPink4}{rgb}{0.55,0.37,0.40}
\definecolor{LightPink}{rgb}{1.00,0.71,0.76}
\definecolor{LightSalmon1}{rgb}{1.00,0.63,0.48}
\definecolor{LightSalmon2}{rgb}{0.93,0.58,0.45}
\definecolor{LightSalmon3}{rgb}{0.80,0.51,0.38}
\definecolor{LightSalmon4}{rgb}{0.55,0.34,0.26}
\definecolor{LightSalmon}{rgb}{1.00,0.63,0.48}
\definecolor{LightSeaGreen}{rgb}{0.13,0.70,0.67}
\definecolor{LightSkyBlue1}{rgb}{0.69,0.89,1.00}
\definecolor{LightSkyBlue2}{rgb}{0.64,0.83,0.93}
\definecolor{LightSkyBlue3}{rgb}{0.55,0.71,0.80}
\definecolor{LightSkyBlue4}{rgb}{0.38,0.48,0.55}
\definecolor{LightSkyBlue}{rgb}{0.53,0.81,0.98}
\definecolor{LightSlateBlue}{rgb}{0.52,0.44,1.00}
\definecolor{LightSlateGray}{rgb}{0.47,0.53,0.60}
\definecolor{LightSlateGrey}{rgb}{0.47,0.53,0.60}
\definecolor{LightSteelBlue1}{rgb}{0.79,0.88,1.00}
\definecolor{LightSteelBlue2}{rgb}{0.74,0.82,0.93}
\definecolor{LightSteelBlue3}{rgb}{0.64,0.71,0.80}
\definecolor{LightSteelBlue4}{rgb}{0.43,0.48,0.55}
\definecolor{LightSteelBlue}{rgb}{0.69,0.77,0.87}
\definecolor{LightYellow1}{rgb}{1.00,1.00,0.88}
\definecolor{LightYellow2}{rgb}{0.93,0.93,0.82}
\definecolor{LightYellow3}{rgb}{0.80,0.80,0.71}
\definecolor{LightYellow4}{rgb}{0.55,0.55,0.48}
\definecolor{LightYellow}{rgb}{1.00,1.00,0.88}
\definecolor{LimeGreen}{rgb}{0.20,0.80,0.20}
\definecolor{MediumAquamarine}{rgb}{0.40,0.80,0.67}
\definecolor{MediumBlue}{rgb}{0.00,0.00,0.80}
\definecolor{MediumOrchid1}{rgb}{0.88,0.40,1.00}
\definecolor{MediumOrchid2}{rgb}{0.82,0.37,0.93}
\definecolor{MediumOrchid3}{rgb}{0.71,0.32,0.80}
\definecolor{MediumOrchid4}{rgb}{0.48,0.22,0.55}
\definecolor{MediumOrchid}{rgb}{0.73,0.33,0.83}
\definecolor{MediumPurple1}{rgb}{0.67,0.51,1.00}
\definecolor{MediumPurple2}{rgb}{0.62,0.47,0.93}
\definecolor{MediumPurple3}{rgb}{0.54,0.41,0.80}
\definecolor{MediumPurple4}{rgb}{0.36,0.28,0.55}
\definecolor{MediumPurple}{rgb}{0.58,0.44,0.86}
\definecolor{MediumSeaGreen}{rgb}{0.24,0.70,0.44}
\definecolor{MediumSlateBlue}{rgb}{0.48,0.41,0.93}
\definecolor{MediumSpringGreen}{rgb}{0.00,0.98,0.60}
\definecolor{MediumTurquoise}{rgb}{0.28,0.82,0.80}
\definecolor{MediumVioletRed}{rgb}{0.78,0.08,0.52}
\definecolor{MidnightBlue}{rgb}{0.10,0.10,0.44}
\definecolor{MintCream}{rgb}{0.96,1.00,0.98}
\definecolor{MistyRose1}{rgb}{1.00,0.89,0.88}
\definecolor{MistyRose2}{rgb}{0.93,0.84,0.82}
\definecolor{MistyRose3}{rgb}{0.80,0.72,0.71}
\definecolor{MistyRose4}{rgb}{0.55,0.49,0.48}
\definecolor{MistyRose}{rgb}{1.00,0.89,0.88}
\definecolor{NavajoWhite1}{rgb}{1.00,0.87,0.68}
\definecolor{NavajoWhite2}{rgb}{0.93,0.81,0.63}
\definecolor{NavajoWhite3}{rgb}{0.80,0.70,0.55}
\definecolor{NavajoWhite4}{rgb}{0.55,0.47,0.37}
\definecolor{NavajoWhite}{rgb}{1.00,0.87,0.68}
\definecolor{NavyBlue}{rgb}{0.00,0.00,0.50}
\definecolor{OldLace}{rgb}{0.99,0.96,0.90}
\definecolor{OliveDrab1}{rgb}{0.75,1.00,0.24}
\definecolor{OliveDrab2}{rgb}{0.70,0.93,0.23}
\definecolor{OliveDrab3}{rgb}{0.60,0.80,0.20}
\definecolor{OliveDrab4}{rgb}{0.41,0.55,0.13}
\definecolor{OliveDrab}{rgb}{0.42,0.56,0.14}
\definecolor{OrangeRed1}{rgb}{1.00,0.27,0.00}
\definecolor{OrangeRed2}{rgb}{0.93,0.25,0.00}
\definecolor{OrangeRed3}{rgb}{0.80,0.22,0.00}
\definecolor{OrangeRed4}{rgb}{0.55,0.15,0.00}
\definecolor{OrangeRed}{rgb}{1.00,0.27,0.00}
\definecolor{PaleGoldenrod}{rgb}{0.93,0.91,0.67}
\definecolor{PaleGreen1}{rgb}{0.60,1.00,0.60}
\definecolor{PaleGreen2}{rgb}{0.56,0.93,0.56}
\definecolor{PaleGreen3}{rgb}{0.49,0.80,0.49}
\definecolor{PaleGreen4}{rgb}{0.33,0.55,0.33}
\definecolor{PaleGreen}{rgb}{0.60,0.98,0.60}
\definecolor{PaleTurquoise1}{rgb}{0.73,1.00,1.00}
\definecolor{PaleTurquoise2}{rgb}{0.68,0.93,0.93}
\definecolor{PaleTurquoise3}{rgb}{0.59,0.80,0.80}
\definecolor{PaleTurquoise4}{rgb}{0.40,0.55,0.55}
\definecolor{PaleTurquoise}{rgb}{0.69,0.93,0.93}
\definecolor{PaleVioletRed1}{rgb}{1.00,0.51,0.67}
\definecolor{PaleVioletRed2}{rgb}{0.93,0.47,0.62}
\definecolor{PaleVioletRed3}{rgb}{0.80,0.41,0.54}
\definecolor{PaleVioletRed4}{rgb}{0.55,0.28,0.36}
\definecolor{PaleVioletRed}{rgb}{0.86,0.44,0.58}
\definecolor{PapayaWhip}{rgb}{1.00,0.94,0.84}
\definecolor{PeachPuff1}{rgb}{1.00,0.85,0.73}
\definecolor{PeachPuff2}{rgb}{0.93,0.80,0.68}
\definecolor{PeachPuff3}{rgb}{0.80,0.69,0.58}
\definecolor{PeachPuff4}{rgb}{0.55,0.47,0.40}
\definecolor{PeachPuff}{rgb}{1.00,0.85,0.73}
\definecolor{PowderBlue}{rgb}{0.69,0.88,0.90}
\definecolor{RosyBrown1}{rgb}{1.00,0.76,0.76}
\definecolor{RosyBrown2}{rgb}{0.93,0.71,0.71}
\definecolor{RosyBrown3}{rgb}{0.80,0.61,0.61}
\definecolor{RosyBrown4}{rgb}{0.55,0.41,0.41}
\definecolor{RosyBrown}{rgb}{0.74,0.56,0.56}
\definecolor{RoyalBlue1}{rgb}{0.28,0.46,1.00}
\definecolor{RoyalBlue2}{rgb}{0.26,0.43,0.93}
\definecolor{RoyalBlue3}{rgb}{0.23,0.37,0.80}
\definecolor{RoyalBlue4}{rgb}{0.15,0.25,0.55}
\definecolor{RoyalBlue}{rgb}{0.25,0.41,0.88}
\definecolor{SaddleBrown}{rgb}{0.55,0.27,0.07}
\definecolor{SandyBrown}{rgb}{0.96,0.64,0.38}
\definecolor{SeaGreen1}{rgb}{0.33,1.00,0.62}
\definecolor{SeaGreen2}{rgb}{0.31,0.93,0.58}
\definecolor{SeaGreen3}{rgb}{0.26,0.80,0.50}
\definecolor{SeaGreen4}{rgb}{0.18,0.55,0.34}
\definecolor{SeaGreen}{rgb}{0.18,0.55,0.34}
\definecolor{SkyBlue1}{rgb}{0.53,0.81,1.00}
\definecolor{SkyBlue2}{rgb}{0.49,0.75,0.93}
\definecolor{SkyBlue3}{rgb}{0.42,0.65,0.80}
\definecolor{SkyBlue4}{rgb}{0.29,0.44,0.55}
\definecolor{SkyBlue}{rgb}{0.53,0.81,0.92}
\definecolor{SlateBlue1}{rgb}{0.51,0.44,1.00}
\definecolor{SlateBlue2}{rgb}{0.48,0.40,0.93}
\definecolor{SlateBlue3}{rgb}{0.41,0.35,0.80}
\definecolor{SlateBlue4}{rgb}{0.28,0.24,0.55}
\definecolor{SlateBlue}{rgb}{0.42,0.35,0.80}
\definecolor{SlateGray1}{rgb}{0.78,0.89,1.00}
\definecolor{SlateGray2}{rgb}{0.73,0.83,0.93}
\definecolor{SlateGray3}{rgb}{0.62,0.71,0.80}
\definecolor{SlateGray4}{rgb}{0.42,0.48,0.55}
\definecolor{SlateGray}{rgb}{0.44,0.50,0.56}
\definecolor{SlateGrey}{rgb}{0.44,0.50,0.56}
\definecolor{SpringGreen1}{rgb}{0.00,1.00,0.50}
\definecolor{SpringGreen2}{rgb}{0.00,0.93,0.46}
\definecolor{SpringGreen3}{rgb}{0.00,0.80,0.40}
\definecolor{SpringGreen4}{rgb}{0.00,0.55,0.27}
\definecolor{SpringGreen}{rgb}{0.00,1.00,0.50}
\definecolor{SteelBlue1}{rgb}{0.39,0.72,1.00}
\definecolor{SteelBlue2}{rgb}{0.36,0.67,0.93}
\definecolor{SteelBlue3}{rgb}{0.31,0.58,0.80}
\definecolor{SteelBlue4}{rgb}{0.21,0.39,0.55}
\definecolor{SteelBlue}{rgb}{0.27,0.51,0.71}
\definecolor{VioletRed1}{rgb}{1.00,0.24,0.59}
\definecolor{VioletRed2}{rgb}{0.93,0.23,0.55}
\definecolor{VioletRed3}{rgb}{0.80,0.20,0.47}
\definecolor{VioletRed4}{rgb}{0.55,0.13,0.32}
\definecolor{VioletRed}{rgb}{0.82,0.13,0.56}
\definecolor{WhiteSmoke}{rgb}{0.96,0.96,0.96}
\definecolor{YellowGreen}{rgb}{0.60,0.80,0.20}
\definecolor{aliceblue}{rgb}{0.94,0.97,1.00}
\definecolor{antiquewhite}{rgb}{0.98,0.92,0.84}
\definecolor{aquamarine1}{rgb}{0.50,1.00,0.83}
\definecolor{aquamarine2}{rgb}{0.46,0.93,0.78}
\definecolor{aquamarine3}{rgb}{0.40,0.80,0.67}
\definecolor{aquamarine4}{rgb}{0.27,0.55,0.45}
\definecolor{aquamarine}{rgb}{0.50,1.00,0.83}
\definecolor{azure1}{rgb}{0.94,1.00,1.00}
\definecolor{azure2}{rgb}{0.88,0.93,0.93}
\definecolor{azure3}{rgb}{0.76,0.80,0.80}
\definecolor{azure4}{rgb}{0.51,0.55,0.55}
\definecolor{azure}{rgb}{0.94,1.00,1.00}
\definecolor{beige}{rgb}{0.96,0.96,0.86}
\definecolor{bisque1}{rgb}{1.00,0.89,0.77}
\definecolor{bisque2}{rgb}{0.93,0.84,0.72}
\definecolor{bisque3}{rgb}{0.80,0.72,0.62}
\definecolor{bisque4}{rgb}{0.55,0.49,0.42}
\definecolor{bisque}{rgb}{1.00,0.89,0.77}
\definecolor{black}{rgb}{0.00,0.00,0.00}
\definecolor{blanchedalmond}{rgb}{1.00,0.92,0.80}
\definecolor{blue1}{rgb}{0.00,0.00,1.00}
\definecolor{blue2}{rgb}{0.00,0.00,0.93}
\definecolor{blue3}{rgb}{0.00,0.00,0.80}
\definecolor{blue4}{rgb}{0.00,0.00,0.55}
\definecolor{blueviolet}{rgb}{0.54,0.17,0.89}
\definecolor{blue}{rgb}{0.00,0.00,1.00}
\definecolor{brown1}{rgb}{1.00,0.25,0.25}
\definecolor{brown2}{rgb}{0.93,0.23,0.23}
\definecolor{brown3}{rgb}{0.80,0.20,0.20}
\definecolor{brown4}{rgb}{0.55,0.14,0.14}
\definecolor{brown}{rgb}{0.65,0.16,0.16}
\definecolor{burlywood1}{rgb}{1.00,0.83,0.61}
\definecolor{burlywood2}{rgb}{0.93,0.77,0.57}
\definecolor{burlywood3}{rgb}{0.80,0.67,0.49}
\definecolor{burlywood4}{rgb}{0.55,0.45,0.33}
\definecolor{burlywood}{rgb}{0.87,0.72,0.53}
\definecolor{cadetblue}{rgb}{0.37,0.62,0.63}
\definecolor{chartreuse1}{rgb}{0.50,1.00,0.00}
\definecolor{chartreuse2}{rgb}{0.46,0.93,0.00}
\definecolor{chartreuse3}{rgb}{0.40,0.80,0.00}
\definecolor{chartreuse4}{rgb}{0.27,0.55,0.00}
\definecolor{chartreuse}{rgb}{0.50,1.00,0.00}
\definecolor{chocolate1}{rgb}{1.00,0.50,0.14}
\definecolor{chocolate2}{rgb}{0.93,0.46,0.13}
\definecolor{chocolate3}{rgb}{0.80,0.40,0.11}
\definecolor{chocolate4}{rgb}{0.55,0.27,0.07}
\definecolor{chocolate}{rgb}{0.82,0.41,0.12}
\definecolor{coral1}{rgb}{1.00,0.45,0.34}
\definecolor{coral2}{rgb}{0.93,0.42,0.31}
\definecolor{coral3}{rgb}{0.80,0.36,0.27}
\definecolor{coral4}{rgb}{0.55,0.24,0.18}
\definecolor{coral}{rgb}{1.00,0.50,0.31}
\definecolor{cornflowerblue}{rgb}{0.39,0.58,0.93}
\definecolor{cornsilk1}{rgb}{1.00,0.97,0.86}
\definecolor{cornsilk2}{rgb}{0.93,0.91,0.80}
\definecolor{cornsilk3}{rgb}{0.80,0.78,0.69}
\definecolor{cornsilk4}{rgb}{0.55,0.53,0.47}
\definecolor{cornsilk}{rgb}{1.00,0.97,0.86}
\definecolor{cyan1}{rgb}{0.00,1.00,1.00}
\definecolor{cyan2}{rgb}{0.00,0.93,0.93}
\definecolor{cyan3}{rgb}{0.00,0.80,0.80}
\definecolor{cyan4}{rgb}{0.00,0.55,0.55}
\definecolor{cyan}{rgb}{0.00,1.00,1.00}
\definecolor{darkblue}{rgb}{0.00,0.00,0.55}
\definecolor{darkcyan}{rgb}{0.00,0.55,0.55}
\definecolor{darkgoldenrod}{rgb}{0.72,0.53,0.04}
\definecolor{darkgray}{rgb}{0.66,0.66,0.66}
\definecolor{darkgreen}{rgb}{0.00,0.39,0.00}
\definecolor{darkgrey}{rgb}{0.66,0.66,0.66}
\definecolor{darkkhaki}{rgb}{0.74,0.72,0.42}
\definecolor{darkmagenta}{rgb}{0.55,0.00,0.55}
\definecolor{darkolive}{rgb}{0.33,0.42,0.18}
\definecolor{darkorange}{rgb}{1.00,0.55,0.00}
\definecolor{darkorchid}{rgb}{0.60,0.20,0.80}
\definecolor{darkred}{rgb}{0.55,0.00,0.00}
\definecolor{darksalmon}{rgb}{0.91,0.59,0.48}
\definecolor{darksea}{rgb}{0.56,0.74,0.56}
\definecolor{darkslate}{rgb}{0.18,0.31,0.31}
\definecolor{darkslate}{rgb}{0.18,0.31,0.31}
\definecolor{darkslate}{rgb}{0.28,0.24,0.55}
\definecolor{darkturquoise}{rgb}{0.00,0.81,0.82}
\definecolor{darkviolet}{rgb}{0.58,0.00,0.83}
\definecolor{deeppink}{rgb}{1.00,0.08,0.58}
\definecolor{deepsky}{rgb}{0.00,0.75,1.00}
\definecolor{dimgray}{rgb}{0.41,0.41,0.41}
\definecolor{dimgrey}{rgb}{0.41,0.41,0.41}
\definecolor{dodgerblue}{rgb}{0.12,0.56,1.00}
\definecolor{firebrick1}{rgb}{1.00,0.19,0.19}
\definecolor{firebrick2}{rgb}{0.93,0.17,0.17}
\definecolor{firebrick3}{rgb}{0.80,0.15,0.15}
\definecolor{firebrick4}{rgb}{0.55,0.10,0.10}
\definecolor{firebrick}{rgb}{0.70,0.13,0.13}
\definecolor{floralwhite}{rgb}{1.00,0.98,0.94}
\definecolor{forestgreen}{rgb}{0.13,0.55,0.13}
\definecolor{gainsboro}{rgb}{0.86,0.86,0.86}
\definecolor{ghostwhite}{rgb}{0.97,0.97,1.00}
\definecolor{gold1}{rgb}{1.00,0.84,0.00}
\definecolor{gold2}{rgb}{0.93,0.79,0.00}
\definecolor{gold3}{rgb}{0.80,0.68,0.00}
\definecolor{gold4}{rgb}{0.55,0.46,0.00}
\definecolor{goldenrod1}{rgb}{1.00,0.76,0.15}
\definecolor{goldenrod2}{rgb}{0.93,0.71,0.13}
\definecolor{goldenrod3}{rgb}{0.80,0.61,0.11}
\definecolor{goldenrod4}{rgb}{0.55,0.41,0.08}
\definecolor{goldenrod}{rgb}{0.85,0.65,0.13}
\definecolor{gold}{rgb}{1.00,0.84,0.00}
\definecolor{gray0}{rgb}{0.00,0.00,0.00}
\definecolor{gray100}{rgb}{1.00,1.00,1.00}
\definecolor{gray10}{rgb}{0.10,0.10,0.10}
\definecolor{gray11}{rgb}{0.11,0.11,0.11}
\definecolor{gray12}{rgb}{0.12,0.12,0.12}
\definecolor{gray13}{rgb}{0.13,0.13,0.13}
\definecolor{gray14}{rgb}{0.14,0.14,0.14}
\definecolor{gray15}{rgb}{0.15,0.15,0.15}
\definecolor{gray16}{rgb}{0.16,0.16,0.16}
\definecolor{gray17}{rgb}{0.17,0.17,0.17}
\definecolor{gray18}{rgb}{0.18,0.18,0.18}
\definecolor{gray19}{rgb}{0.19,0.19,0.19}
\definecolor{gray1}{rgb}{0.01,0.01,0.01}
\definecolor{gray20}{rgb}{0.20,0.20,0.20}
\definecolor{gray21}{rgb}{0.21,0.21,0.21}
\definecolor{gray22}{rgb}{0.22,0.22,0.22}
\definecolor{gray23}{rgb}{0.23,0.23,0.23}
\definecolor{gray24}{rgb}{0.24,0.24,0.24}
\definecolor{gray25}{rgb}{0.25,0.25,0.25}
\definecolor{gray26}{rgb}{0.26,0.26,0.26}
\definecolor{gray27}{rgb}{0.27,0.27,0.27}
\definecolor{gray28}{rgb}{0.28,0.28,0.28}
\definecolor{gray29}{rgb}{0.29,0.29,0.29}
\definecolor{gray2}{rgb}{0.02,0.02,0.02}
\definecolor{gray30}{rgb}{0.30,0.30,0.30}
\definecolor{gray31}{rgb}{0.31,0.31,0.31}
\definecolor{gray32}{rgb}{0.32,0.32,0.32}
\definecolor{gray33}{rgb}{0.33,0.33,0.33}
\definecolor{gray34}{rgb}{0.34,0.34,0.34}
\definecolor{gray35}{rgb}{0.35,0.35,0.35}
\definecolor{gray36}{rgb}{0.36,0.36,0.36}
\definecolor{gray37}{rgb}{0.37,0.37,0.37}
\definecolor{gray38}{rgb}{0.38,0.38,0.38}
\definecolor{gray39}{rgb}{0.39,0.39,0.39}
\definecolor{gray3}{rgb}{0.03,0.03,0.03}
\definecolor{gray40}{rgb}{0.40,0.40,0.40}
\definecolor{gray41}{rgb}{0.41,0.41,0.41}
\definecolor{gray42}{rgb}{0.42,0.42,0.42}
\definecolor{gray43}{rgb}{0.43,0.43,0.43}
\definecolor{gray44}{rgb}{0.44,0.44,0.44}
\definecolor{gray45}{rgb}{0.45,0.45,0.45}
\definecolor{gray46}{rgb}{0.46,0.46,0.46}
\definecolor{gray47}{rgb}{0.47,0.47,0.47}
\definecolor{gray48}{rgb}{0.48,0.48,0.48}
\definecolor{gray49}{rgb}{0.49,0.49,0.49}
\definecolor{gray4}{rgb}{0.04,0.04,0.04}
\definecolor{gray50}{rgb}{0.50,0.50,0.50}
\definecolor{gray51}{rgb}{0.51,0.51,0.51}
\definecolor{gray52}{rgb}{0.52,0.52,0.52}
\definecolor{gray53}{rgb}{0.53,0.53,0.53}
\definecolor{gray54}{rgb}{0.54,0.54,0.54}
\definecolor{gray55}{rgb}{0.55,0.55,0.55}
\definecolor{gray56}{rgb}{0.56,0.56,0.56}
\definecolor{gray57}{rgb}{0.57,0.57,0.57}
\definecolor{gray58}{rgb}{0.58,0.58,0.58}
\definecolor{gray59}{rgb}{0.59,0.59,0.59}
\definecolor{gray5}{rgb}{0.05,0.05,0.05}
\definecolor{gray60}{rgb}{0.60,0.60,0.60}
\definecolor{gray61}{rgb}{0.61,0.61,0.61}
\definecolor{gray62}{rgb}{0.62,0.62,0.62}
\definecolor{gray63}{rgb}{0.63,0.63,0.63}
\definecolor{gray64}{rgb}{0.64,0.64,0.64}
\definecolor{gray65}{rgb}{0.65,0.65,0.65}
\definecolor{gray66}{rgb}{0.66,0.66,0.66}
\definecolor{gray67}{rgb}{0.67,0.67,0.67}
\definecolor{gray68}{rgb}{0.68,0.68,0.68}
\definecolor{gray69}{rgb}{0.69,0.69,0.69}
\definecolor{gray6}{rgb}{0.06,0.06,0.06}
\definecolor{gray70}{rgb}{0.70,0.70,0.70}
\definecolor{gray71}{rgb}{0.71,0.71,0.71}
\definecolor{gray72}{rgb}{0.72,0.72,0.72}
\definecolor{gray73}{rgb}{0.73,0.73,0.73}
\definecolor{gray74}{rgb}{0.74,0.74,0.74}
\definecolor{gray75}{rgb}{0.75,0.75,0.75}
\definecolor{gray76}{rgb}{0.76,0.76,0.76}
\definecolor{gray77}{rgb}{0.77,0.77,0.77}
\definecolor{gray78}{rgb}{0.78,0.78,0.78}
\definecolor{gray79}{rgb}{0.79,0.79,0.79}
\definecolor{gray7}{rgb}{0.07,0.07,0.07}
\definecolor{gray80}{rgb}{0.80,0.80,0.80}
\definecolor{gray81}{rgb}{0.81,0.81,0.81}
\definecolor{gray82}{rgb}{0.82,0.82,0.82}
\definecolor{gray83}{rgb}{0.83,0.83,0.83}
\definecolor{gray84}{rgb}{0.84,0.84,0.84}
\definecolor{gray85}{rgb}{0.85,0.85,0.85}
\definecolor{gray86}{rgb}{0.86,0.86,0.86}
\definecolor{gray87}{rgb}{0.87,0.87,0.87}
\definecolor{gray88}{rgb}{0.88,0.88,0.88}
\definecolor{gray89}{rgb}{0.89,0.89,0.89}
\definecolor{gray8}{rgb}{0.08,0.08,0.08}
\definecolor{gray90}{rgb}{0.90,0.90,0.90}
\definecolor{gray91}{rgb}{0.91,0.91,0.91}
\definecolor{gray92}{rgb}{0.92,0.92,0.92}
\definecolor{gray93}{rgb}{0.93,0.93,0.93}
\definecolor{gray94}{rgb}{0.94,0.94,0.94}
\definecolor{gray95}{rgb}{0.95,0.95,0.95}
\definecolor{gray96}{rgb}{0.96,0.96,0.96}
\definecolor{gray97}{rgb}{0.97,0.97,0.97}
\definecolor{gray98}{rgb}{0.98,0.98,0.98}
\definecolor{gray99}{rgb}{0.99,0.99,0.99}
\definecolor{gray9}{rgb}{0.09,0.09,0.09}
\definecolor{gray}{rgb}{0.75,0.75,0.75}
\definecolor{green1}{rgb}{0.00,1.00,0.00}
\definecolor{green2}{rgb}{0.00,0.93,0.00}
\definecolor{green3}{rgb}{0.00,0.80,0.00}
\definecolor{green4}{rgb}{0.00,0.55,0.00}
\definecolor{greenyellow}{rgb}{0.68,1.00,0.18}
\definecolor{green}{rgb}{0.00,1.00,0.00}
\definecolor{grey0}{rgb}{0.00,0.00,0.00}
\definecolor{grey100}{rgb}{1.00,1.00,1.00}
\definecolor{grey10}{rgb}{0.10,0.10,0.10}
\definecolor{grey11}{rgb}{0.11,0.11,0.11}
\definecolor{grey12}{rgb}{0.12,0.12,0.12}
\definecolor{grey13}{rgb}{0.13,0.13,0.13}
\definecolor{grey14}{rgb}{0.14,0.14,0.14}
\definecolor{grey15}{rgb}{0.15,0.15,0.15}
\definecolor{grey16}{rgb}{0.16,0.16,0.16}
\definecolor{grey17}{rgb}{0.17,0.17,0.17}
\definecolor{grey18}{rgb}{0.18,0.18,0.18}
\definecolor{grey19}{rgb}{0.19,0.19,0.19}
\definecolor{grey1}{rgb}{0.01,0.01,0.01}
\definecolor{grey20}{rgb}{0.20,0.20,0.20}
\definecolor{grey21}{rgb}{0.21,0.21,0.21}
\definecolor{grey22}{rgb}{0.22,0.22,0.22}
\definecolor{grey23}{rgb}{0.23,0.23,0.23}
\definecolor{grey24}{rgb}{0.24,0.24,0.24}
\definecolor{grey25}{rgb}{0.25,0.25,0.25}
\definecolor{grey26}{rgb}{0.26,0.26,0.26}
\definecolor{grey27}{rgb}{0.27,0.27,0.27}
\definecolor{grey28}{rgb}{0.28,0.28,0.28}
\definecolor{grey29}{rgb}{0.29,0.29,0.29}
\definecolor{grey2}{rgb}{0.02,0.02,0.02}
\definecolor{grey30}{rgb}{0.30,0.30,0.30}
\definecolor{grey31}{rgb}{0.31,0.31,0.31}
\definecolor{grey32}{rgb}{0.32,0.32,0.32}
\definecolor{grey33}{rgb}{0.33,0.33,0.33}
\definecolor{grey34}{rgb}{0.34,0.34,0.34}
\definecolor{grey35}{rgb}{0.35,0.35,0.35}
\definecolor{grey36}{rgb}{0.36,0.36,0.36}
\definecolor{grey37}{rgb}{0.37,0.37,0.37}
\definecolor{grey38}{rgb}{0.38,0.38,0.38}
\definecolor{grey39}{rgb}{0.39,0.39,0.39}
\definecolor{grey3}{rgb}{0.03,0.03,0.03}
\definecolor{grey40}{rgb}{0.40,0.40,0.40}
\definecolor{grey41}{rgb}{0.41,0.41,0.41}
\definecolor{grey42}{rgb}{0.42,0.42,0.42}
\definecolor{grey43}{rgb}{0.43,0.43,0.43}
\definecolor{grey44}{rgb}{0.44,0.44,0.44}
\definecolor{grey45}{rgb}{0.45,0.45,0.45}
\definecolor{grey46}{rgb}{0.46,0.46,0.46}
\definecolor{grey47}{rgb}{0.47,0.47,0.47}
\definecolor{grey48}{rgb}{0.48,0.48,0.48}
\definecolor{grey49}{rgb}{0.49,0.49,0.49}
\definecolor{grey4}{rgb}{0.04,0.04,0.04}
\definecolor{grey50}{rgb}{0.50,0.50,0.50}
\definecolor{grey51}{rgb}{0.51,0.51,0.51}
\definecolor{grey52}{rgb}{0.52,0.52,0.52}
\definecolor{grey53}{rgb}{0.53,0.53,0.53}
\definecolor{grey54}{rgb}{0.54,0.54,0.54}
\definecolor{grey55}{rgb}{0.55,0.55,0.55}
\definecolor{grey56}{rgb}{0.56,0.56,0.56}
\definecolor{grey57}{rgb}{0.57,0.57,0.57}
\definecolor{grey58}{rgb}{0.58,0.58,0.58}
\definecolor{grey59}{rgb}{0.59,0.59,0.59}
\definecolor{grey5}{rgb}{0.05,0.05,0.05}
\definecolor{grey60}{rgb}{0.60,0.60,0.60}
\definecolor{grey61}{rgb}{0.61,0.61,0.61}
\definecolor{grey62}{rgb}{0.62,0.62,0.62}
\definecolor{grey63}{rgb}{0.63,0.63,0.63}
\definecolor{grey64}{rgb}{0.64,0.64,0.64}
\definecolor{grey65}{rgb}{0.65,0.65,0.65}
\definecolor{grey66}{rgb}{0.66,0.66,0.66}
\definecolor{grey67}{rgb}{0.67,0.67,0.67}
\definecolor{grey68}{rgb}{0.68,0.68,0.68}
\definecolor{grey69}{rgb}{0.69,0.69,0.69}
\definecolor{grey6}{rgb}{0.06,0.06,0.06}
\definecolor{grey70}{rgb}{0.70,0.70,0.70}
\definecolor{grey71}{rgb}{0.71,0.71,0.71}
\definecolor{grey72}{rgb}{0.72,0.72,0.72}
\definecolor{grey73}{rgb}{0.73,0.73,0.73}
\definecolor{grey74}{rgb}{0.74,0.74,0.74}
\definecolor{grey75}{rgb}{0.75,0.75,0.75}
\definecolor{grey76}{rgb}{0.76,0.76,0.76}
\definecolor{grey77}{rgb}{0.77,0.77,0.77}
\definecolor{grey78}{rgb}{0.78,0.78,0.78}
\definecolor{grey79}{rgb}{0.79,0.79,0.79}
\definecolor{grey7}{rgb}{0.07,0.07,0.07}
\definecolor{grey80}{rgb}{0.80,0.80,0.80}
\definecolor{grey81}{rgb}{0.81,0.81,0.81}
\definecolor{grey82}{rgb}{0.82,0.82,0.82}
\definecolor{grey83}{rgb}{0.83,0.83,0.83}
\definecolor{grey84}{rgb}{0.84,0.84,0.84}
\definecolor{grey85}{rgb}{0.85,0.85,0.85}
\definecolor{grey86}{rgb}{0.86,0.86,0.86}
\definecolor{grey87}{rgb}{0.87,0.87,0.87}
\definecolor{grey88}{rgb}{0.88,0.88,0.88}
\definecolor{grey89}{rgb}{0.89,0.89,0.89}
\definecolor{grey8}{rgb}{0.08,0.08,0.08}
\definecolor{grey90}{rgb}{0.90,0.90,0.90}
\definecolor{grey91}{rgb}{0.91,0.91,0.91}
\definecolor{grey92}{rgb}{0.92,0.92,0.92}
\definecolor{grey93}{rgb}{0.93,0.93,0.93}
\definecolor{grey94}{rgb}{0.94,0.94,0.94}
\definecolor{grey95}{rgb}{0.95,0.95,0.95}
\definecolor{grey96}{rgb}{0.96,0.96,0.96}
\definecolor{grey97}{rgb}{0.97,0.97,0.97}
\definecolor{grey98}{rgb}{0.98,0.98,0.98}
\definecolor{grey99}{rgb}{0.99,0.99,0.99}
\definecolor{grey9}{rgb}{0.09,0.09,0.09}
\definecolor{grey}{rgb}{0.75,0.75,0.75}
\definecolor{honeydew1}{rgb}{0.94,1.00,0.94}
\definecolor{honeydew2}{rgb}{0.88,0.93,0.88}
\definecolor{honeydew3}{rgb}{0.76,0.80,0.76}
\definecolor{honeydew4}{rgb}{0.51,0.55,0.51}
\definecolor{honeydew}{rgb}{0.94,1.00,0.94}
\definecolor{hotpink}{rgb}{1.00,0.41,0.71}
\definecolor{indianred}{rgb}{0.80,0.36,0.36}
\definecolor{ivory1}{rgb}{1.00,1.00,0.94}
\definecolor{ivory2}{rgb}{0.93,0.93,0.88}
\definecolor{ivory3}{rgb}{0.80,0.80,0.76}
\definecolor{ivory4}{rgb}{0.55,0.55,0.51}
\definecolor{ivory}{rgb}{1.00,1.00,0.94}
\definecolor{khaki1}{rgb}{1.00,0.96,0.56}
\definecolor{khaki2}{rgb}{0.93,0.90,0.52}
\definecolor{khaki3}{rgb}{0.80,0.78,0.45}
\definecolor{khaki4}{rgb}{0.55,0.53,0.31}
\definecolor{khaki}{rgb}{0.94,0.90,0.55}
\definecolor{lavenderblush}{rgb}{1.00,0.94,0.96}
\definecolor{lavender}{rgb}{0.90,0.90,0.98}
\definecolor{lawngreen}{rgb}{0.49,0.99,0.00}
\definecolor{lemonchiffon}{rgb}{1.00,0.98,0.80}
\definecolor{lightblue}{rgb}{0.68,0.85,0.90}
\definecolor{lightcoral}{rgb}{0.94,0.50,0.50}
\definecolor{lightcyan}{rgb}{0.88,1.00,1.00}
\definecolor{lightgoldenrod}{rgb}{0.93,0.87,0.51}
\definecolor{lightgoldenrod}{rgb}{0.98,0.98,0.82}
\definecolor{lightgray}{rgb}{0.83,0.83,0.83}
\definecolor{lightgreen}{rgb}{0.56,0.93,0.56}
\definecolor{lightgrey}{rgb}{0.83,0.83,0.83}
\definecolor{lightpink}{rgb}{1.00,0.71,0.76}
\definecolor{lightsalmon}{rgb}{1.00,0.63,0.48}
\definecolor{lightsea}{rgb}{0.13,0.70,0.67}
\definecolor{lightsky}{rgb}{0.53,0.81,0.98}
\definecolor{lightslate}{rgb}{0.47,0.53,0.60}
\definecolor{lightslate}{rgb}{0.47,0.53,0.60}
\definecolor{lightslate}{rgb}{0.52,0.44,1.00}
\definecolor{lightsteel}{rgb}{0.69,0.77,0.87}
\definecolor{lightyellow}{rgb}{1.00,1.00,0.88}
\definecolor{limegreen}{rgb}{0.20,0.80,0.20}
\definecolor{linen}{rgb}{0.98,0.94,0.90}
\definecolor{magenta1}{rgb}{1.00,0.00,1.00}
\definecolor{magenta2}{rgb}{0.93,0.00,0.93}
\definecolor{magenta3}{rgb}{0.80,0.00,0.80}
\definecolor{magenta4}{rgb}{0.55,0.00,0.55}
\definecolor{magenta}{rgb}{1.00,0.00,1.00}
\definecolor{maroon1}{rgb}{1.00,0.20,0.70}
\definecolor{maroon2}{rgb}{0.93,0.19,0.65}
\definecolor{maroon3}{rgb}{0.80,0.16,0.56}
\definecolor{maroon4}{rgb}{0.55,0.11,0.38}
\definecolor{maroon}{rgb}{0.69,0.19,0.38}
\definecolor{mediumaquamarine}{rgb}{0.40,0.80,0.67}
\definecolor{mediumblue}{rgb}{0.00,0.00,0.80}
\definecolor{mediumorchid}{rgb}{0.73,0.33,0.83}
\definecolor{mediumpurple}{rgb}{0.58,0.44,0.86}
\definecolor{mediumsea}{rgb}{0.24,0.70,0.44}
\definecolor{mediumslate}{rgb}{0.48,0.41,0.93}
\definecolor{mediumspring}{rgb}{0.00,0.98,0.60}
\definecolor{mediumturquoise}{rgb}{0.28,0.82,0.80}
\definecolor{mediumviolet}{rgb}{0.78,0.08,0.52}
\definecolor{midnightblue}{rgb}{0.10,0.10,0.44}
\definecolor{mintcream}{rgb}{0.96,1.00,0.98}
\definecolor{mistyrose}{rgb}{1.00,0.89,0.88}
\definecolor{moccasin}{rgb}{1.00,0.89,0.71}
\definecolor{navajowhite}{rgb}{1.00,0.87,0.68}
\definecolor{navyblue}{rgb}{0.00,0.00,0.50}
\definecolor{navy}{rgb}{0.00,0.00,0.50}
\definecolor{oldlace}{rgb}{0.99,0.96,0.90}
\definecolor{olivedrab}{rgb}{0.42,0.56,0.14}
\definecolor{orange1}{rgb}{1.00,0.65,0.00}
\definecolor{orange2}{rgb}{0.93,0.60,0.00}
\definecolor{orange3}{rgb}{0.80,0.52,0.00}
\definecolor{orange4}{rgb}{0.55,0.35,0.00}
\definecolor{orangered}{rgb}{1.00,0.27,0.00}
\definecolor{orange}{rgb}{1.00,0.65,0.00}
\definecolor{orchid1}{rgb}{1.00,0.51,0.98}
\definecolor{orchid2}{rgb}{0.93,0.48,0.91}
\definecolor{orchid3}{rgb}{0.80,0.41,0.79}
\definecolor{orchid4}{rgb}{0.55,0.28,0.54}
\definecolor{orchid}{rgb}{0.85,0.44,0.84}
\definecolor{palegoldenrod}{rgb}{0.93,0.91,0.67}
\definecolor{palegreen}{rgb}{0.60,0.98,0.60}
\definecolor{paleturquoise}{rgb}{0.69,0.93,0.93}
\definecolor{paleviolet}{rgb}{0.86,0.44,0.58}
\definecolor{papayawhip}{rgb}{1.00,0.94,0.84}
\definecolor{peachpuff}{rgb}{1.00,0.85,0.73}
\definecolor{peru}{rgb}{0.80,0.52,0.25}
\definecolor{pink1}{rgb}{1.00,0.71,0.77}
\definecolor{pink2}{rgb}{0.93,0.66,0.72}
\definecolor{pink3}{rgb}{0.80,0.57,0.62}
\definecolor{pink4}{rgb}{0.55,0.39,0.42}
\definecolor{pink}{rgb}{1.00,0.75,0.80}
\definecolor{plum1}{rgb}{1.00,0.73,1.00}
\definecolor{plum2}{rgb}{0.93,0.68,0.93}
\definecolor{plum3}{rgb}{0.80,0.59,0.80}
\definecolor{plum4}{rgb}{0.55,0.40,0.55}
\definecolor{plum}{rgb}{0.87,0.63,0.87}
\definecolor{powderblue}{rgb}{0.69,0.88,0.90}
\definecolor{purple1}{rgb}{0.61,0.19,1.00}
\definecolor{purple2}{rgb}{0.57,0.17,0.93}
\definecolor{purple3}{rgb}{0.49,0.15,0.80}
\definecolor{purple4}{rgb}{0.33,0.10,0.55}
\definecolor{purple}{rgb}{0.63,0.13,0.94}
\definecolor{red1}{rgb}{1.00,0.00,0.00}
\definecolor{red2}{rgb}{0.93,0.00,0.00}
\definecolor{red3}{rgb}{0.80,0.00,0.00}
\definecolor{red4}{rgb}{0.55,0.00,0.00}
\definecolor{red}{rgb}{1.00,0.00,0.00}
\definecolor{rosybrown}{rgb}{0.74,0.56,0.56}
\definecolor{royalblue}{rgb}{0.25,0.41,0.88}
\definecolor{saddlebrown}{rgb}{0.55,0.27,0.07}
\definecolor{salmon1}{rgb}{1.00,0.55,0.41}
\definecolor{salmon2}{rgb}{0.93,0.51,0.38}
\definecolor{salmon3}{rgb}{0.80,0.44,0.33}
\definecolor{salmon4}{rgb}{0.55,0.30,0.22}
\definecolor{salmon}{rgb}{0.98,0.50,0.45}
\definecolor{sandybrown}{rgb}{0.96,0.64,0.38}
\definecolor{seagreen}{rgb}{0.18,0.55,0.34}
\definecolor{seashell1}{rgb}{1.00,0.96,0.93}
\definecolor{seashell2}{rgb}{0.93,0.90,0.87}
\definecolor{seashell3}{rgb}{0.80,0.77,0.75}
\definecolor{seashell4}{rgb}{0.55,0.53,0.51}
\definecolor{seashell}{rgb}{1.00,0.96,0.93}
\definecolor{sienna1}{rgb}{1.00,0.51,0.28}
\definecolor{sienna2}{rgb}{0.93,0.47,0.26}
\definecolor{sienna3}{rgb}{0.80,0.41,0.22}
\definecolor{sienna4}{rgb}{0.55,0.28,0.15}
\definecolor{sienna}{rgb}{0.63,0.32,0.18}
\definecolor{skyblue}{rgb}{0.53,0.81,0.92}
\definecolor{slateblue}{rgb}{0.42,0.35,0.80}
\definecolor{slategray}{rgb}{0.44,0.50,0.56}
\definecolor{slategrey}{rgb}{0.44,0.50,0.56}
\definecolor{snow1}{rgb}{1.00,0.98,0.98}
\definecolor{snow2}{rgb}{0.93,0.91,0.91}
\definecolor{snow3}{rgb}{0.80,0.79,0.79}
\definecolor{snow4}{rgb}{0.55,0.54,0.54}
\definecolor{snow}{rgb}{1.00,0.98,0.98}
\definecolor{springgreen}{rgb}{0.00,1.00,0.50}
\definecolor{steelblue}{rgb}{0.27,0.51,0.71}
\definecolor{tan1}{rgb}{1.00,0.65,0.31}
\definecolor{tan2}{rgb}{0.93,0.60,0.29}
\definecolor{tan3}{rgb}{0.80,0.52,0.25}
\definecolor{tan4}{rgb}{0.55,0.35,0.17}
\definecolor{tan}{rgb}{0.82,0.71,0.55}
\definecolor{thistle1}{rgb}{1.00,0.88,1.00}
\definecolor{thistle2}{rgb}{0.93,0.82,0.93}
\definecolor{thistle3}{rgb}{0.80,0.71,0.80}
\definecolor{thistle4}{rgb}{0.55,0.48,0.55}
\definecolor{thistle}{rgb}{0.85,0.75,0.85}
\definecolor{tomato1}{rgb}{1.00,0.39,0.28}
\definecolor{tomato2}{rgb}{0.93,0.36,0.26}
\definecolor{tomato3}{rgb}{0.80,0.31,0.22}
\definecolor{tomato4}{rgb}{0.55,0.21,0.15}
\definecolor{tomato}{rgb}{1.00,0.39,0.28}
\definecolor{turquoise1}{rgb}{0.00,0.96,1.00}
\definecolor{turquoise2}{rgb}{0.00,0.90,0.93}
\definecolor{turquoise3}{rgb}{0.00,0.77,0.80}
\definecolor{turquoise4}{rgb}{0.00,0.53,0.55}
\definecolor{turquoise}{rgb}{0.25,0.88,0.82}
\definecolor{violetred}{rgb}{0.82,0.13,0.56}
\definecolor{violet}{rgb}{0.93,0.51,0.93}
\definecolor{wheat1}{rgb}{1.00,0.91,0.73}
\definecolor{wheat2}{rgb}{0.93,0.85,0.68}
\definecolor{wheat3}{rgb}{0.80,0.73,0.59}
\definecolor{wheat4}{rgb}{0.55,0.49,0.40}
\definecolor{wheat}{rgb}{0.96,0.87,0.70}
\definecolor{whitesmoke}{rgb}{0.96,0.96,0.96}
\definecolor{white}{rgb}{1.00,1.00,1.00}
\definecolor{yellow1}{rgb}{1.00,1.00,0.00}
\definecolor{yellow2}{rgb}{0.93,0.93,0.00}
\definecolor{yellow3}{rgb}{0.80,0.80,0.00}
\definecolor{yellow4}{rgb}{0.55,0.55,0.00}
\definecolor{yellowgreen}{rgb}{0.60,0.80,0.20}
\definecolor{yellow}{rgb}{1.00,1.00,0.00}

\title{Anderson's considerations on the flow of superfluid helium:\\
some offshoots}%
\author{Eric Varoquaux}%
\email{eric.varoquaux@cea.fr}
\affiliation{CNRS-URM 2464 and CEA-IRAMIS, Service de Physique de l'\'Etat Condens\'e, \\
  Centre d'\'Etudes de Saclay, 91191 Gif-sur-Yvette Cedex (France)}
\date{\today}%
\begin{abstract}
  Nearly five decades have elapsed since the seminal 1966 paper of
  P.W. Anderson on the flow of superfluid helium, $^4$He at that time. Some of
  his ``Considerations'' -- the role of the quantum phase as a dynamical
  variable, the interplay between the motion of quantised vortices and
  potential superflow, its incidence on dissipation in the superfluid and the
  appearance of critical velocities, the quest for the hydrodynamic analogues
  of the Josephson effects in helium -- and the way they have evolved over the
  past half-century are recounted below.  But it is due to key advances on the
  experimental front that phase slippage could be harnessed in the laboratory,
  leading to a deeper understanding of superflow, vortex nucleation, the
  various intrinsic and extrinsic dissipation mechanisms in superfluids,
  macroscopic quantum effects and the superfluid analogue of both {\it ac} and
  {\it dc} Josephson effects -- pivotal concepts in superfluid physics -- have
  been performed. Some of the experiments that have shed light on the more
  intimate effect of quantum mechanics on the hydrodynamics of the dense
  heliums are surveyed, including the nucleation of quantised vortices both by
  Arrhenius processes and by macroscopic quantum tunnelling, the setting up of
  vortex mills, and superfluid interferometry.
\end{abstract}
\pacs{} 
\maketitle
\tableofcontents


Superfluids display quantum properties over large distance. Superfluid
currents may persist indefinitely unlike those of ordinary fluids
\citep{Reppy:65}; the circulation of flow velocity has been found quantised
over meter-size paths \citep{Verbeek:74}. These manifestations of macroscopic
quantum phenomena have constituted one of the early hallmarks of experimental
condensed matter physics  as reviewed over the years by a number of
authors.\footnote{Among these reviews, see in particular  
  \citet{Vinen:63},~\citet{Khalatnikov:65},~\citet{Vinen:66},~\citet{Andronikashvili:66},~\citet{Vinen:68},~\citet{Putterman:71},~\citet{Nozieres:90},~\citet{Vollhardt:90},~\citet{Volovik:03},~\citet{Sonin:15}.}
They are viewed as supporting the concept of a macroscopic wavefunction
extending over the whole superfluid sample as put forward by
\citet{London:54}. Together with Landau's views on the irrotational nature of
superfluid motion \cite{Landau:41}, these ideas form the basis of the
two-fluid model, summarised in Sec.\ref{basics}, which not only describes
the hydrodynamics of superfluids \citep{Khalatnikov:65} but has been extended
to other fields of physics like superconductivity, the Bose-Einstein condensed
gases$\ldots$ \citep{Enz:74}

Hydrodynamics, and superfluid hydrodynamics in its wake, is expected to break
down at small scale when the typical lengths of the problem at hand are no
longer much larger than the interatomic separation or other microscopic lengths
characteristic of the internal structure of the fluid, the ``size'' of Cooper
pairs in superfluid $^3$He for instance.  It has however been known for
a long time, notably from the ion propagation measurements of
\citet{Rayfield:64} that the relevant scale in superfluid $^4$He is
surprisingly small, of the order of \aa ngstr\"oms. In these experiments, the
velocity of vortex rings could be measured in a direct way and compared to the
usual outcome of classical hydrodynamics (see
Sec.\ref{VortexDynamics}). Rayfield and Reif found that, at first sight at
least, hydrodynamics remains valid down to atomic sizes. Their
result holds for $^4$He, which is a dense fluid of bosons. The relevant scale
is much larger in superfluid $^3$He, a
BCS-type p-wave superfluid in which the characteristic length is fixed by the
``size'' of the Cooper pair. This article aims at reviewing how the
hydrodynamics of superfluid $^4$He and $^3$He evolves from large to small
scale and ultimately breaks down at close distance,
revealing the more intimate quantum properties of these fluids. This is no
mean feat, as noted long ago by Uhlenbeck, who is quoted to have said ``{\it One
must watch like a hawk to see Planck's constant appear in hydrodynamics}''
\cite{Putterman:74}.

The main object of study in the following is the time and space evolution of
the phase of the macroscopic wavefunction, often simply referred to as ``the
phase'', in so-called aperture flow. This concept of ``phase'' with
wave-mechanical properties governing the evolution of macroscopic quantities
has become so well-spread that its meaning is, wrongly perhaps, taken for
granted. It was put forward by P. W. Anderson in 1966, following the lead of
\citet{Feynman:55}, mainly by the recognition of the phase as the quantity
commandeering in superfluids both the putative Josephson-type effects and
dissipation caused by vortex motion. The dynamics of quantised vortices,
central to Anderson's ideas, is outlined in Sec.\ref{VortexDynamics}. A
detailed understanding of these phenomena is of fundamental importance as they
govern the breakdown of viscousless flow -- the most noteworthy feature of
superfluidity -- and the appearance of an entirely new class of phenomena --
the hydrodynamic analogues of the Josephson effects -- that underpin the sort
of interferometry that can be performed with the superfluid wavefunction.

These ideas were agitated in the mid-sixties, in particular at the Sussex
Meeting in 1965, by \citet{Anderson:66b}, and  by a number of prominent
physicists, notably Nozi\`eres and Vinen. Reliable experimental observations
were performed twenty years later only, as recalled in Sec.\ref{slippage},
giving a host of new results and insights on superfluid hydrodynamics, notably
an improved understanding of critical velocities and of the nucleation of
vortices, topics discussed in Secs.\ref{CriticalVelocities} and
\ref{PhaseSlipCriticalVelocity}, of possible mechanisms for formation of
vortex tangles, described in Sec.\ref{AllThat}, of the appearance of the
Josephson regime of superflow through tiny apertures, described in
Sec.\ref{Josephson}.

Presented below is a coverage of some of the ramifications of Anderson's ideas
on phase slippage in superfluids. It is intended to provide a gangway between
the many excellent monographs\footnote{See, for example, \citet{Nozieres:90},
  \citet{Tilley:90},\citet{Vollhardt:90}.} that provide the background
material on this subject and the more specialised research publications in the
literature that give the full, raw, sometimes arcane, coverage. As such, it
does not constitute a comprehensive review - space and time constraining - but
touches on a few selected issues that provide the backbone of this subfield of
superfluid hydrodynamics. Reviews with different flavours span over a quarter
of a century and show how this field has evolved.\footnote{The reviews include
  the work of 
  \citet{Varoquaux:87}, \citet{Avenel:87},
  \citet{Avenel:89}, \citet{Varoquaux:91}, \citet{Varoquaux:92},
  \citet{Bowley:92}, \citet{Avenel:93}, \citet{Varoquaux:94},
  \citet{Zimmermann:96}, \citet{Packard:98}, \citet{Varoquaux:99},
  \citet{Varoquaux:01a}, \citet{Varoquaux:01b},
  \citet{Davis:02},\citet{Packard:04} and \citet{Sato:12}.}

Particularly worthy of notice are the reviews of closely related subjects,
Sonin's description of vortex dynamics (\citep{Sonin:87,Sonin:15}, the Landau critical
velocity in superfluid $^4$He by \citet{McClintock:95} and in superfluid
$^3$He-B by \citet{Dobbs:00}, and vortex formation and dynamics
in superfluid $^3$He by \citet{Eltsov:05,Salomaa:87,Volovik:03}. The later
references also cover the exciting field of exotic topological defects in
superfluid $^3$He under rotation, not considered here.


\section{The basic superfluid: He-4}

\label{basics}

Helium-4 undergoes an ordering transition toward a superfluid state at $T\sim
2.17$ K under its saturated vapour pressure (SVP), which is now commonly viewed as a
form of Bose-Einstein condensation. A similar transition occurs in helium-3 at
$T\lesssim 2.7$ mK when Cooper pairing in a state with parallel spin $S=1$ and
relative orbital momentum $l=1$ occurs.

\subsection{The two-fluid hydrodynamics}
\label{TwoFluidModel}

The flowing superfluid helium must obey some form of hydrodynamic equations
given by the general conservation laws, Galilean invariance, and the
thermodynamic equation of state that should also include its superfluid
properties. These equations were written down for $^4$He by \citet{Landau:41}
\citep{Landau:Hydro,Khalatnikov:65} who made the key assumption that in order
to describe the viscous-less fluid flow the independent hydrodynamical
variables must include a velocity field $\mb v_\m s$, to which is associated a
fraction $\rho_\m s/\rho$ of the total density of the liquid, $\rho=m_4/v_4$
being the $^4$He atomic mass divided by the volume occupied by one atom.  This
ideal inviscid fluid velocity field conforms to the Euler equation for ideal fluid
flow. As a consequence, it also obeys the Kelvin-Helmholtz theorem, which
states that the vorticity $\nabla \times \mb v_\m s$ remains constant along
the fluid flowlines.\footnote{See \ \citet{Landau:Hydro}, \S 8.} In addition
to vorticity conservation, a more stringent condition was assumed by
\citet{Landau:41}, namely that the superfluid velocity field be at any instant
irrotational at all points in the superfluid:\footnote{This fundamental
  assumption, which sets an important difference between superfluids and ideal
  Euler fluids, is discussed below in this Section. It is best justified by
  its consequences, the subject matter of the principal part of this review.}

\begin{equation}        \label{rotvs0}
    \nabla \times \mb v_\m s = 0 \; .
\end{equation} 

The superfluid fraction velocity therefore derives from a velocity potential. This
property will be shown below to be intimately related to the microscopic
description of the superfluid and to have far-reaching consequences.

The remainder of the fluid, the normal fraction $\rho_\m n = \rho - \rho_\m
s$, to which is associated a ``normal'' velocity $\mb v_\m n$, obeys
an equation similar to the Navier-Stokes equation of viscous flow. The total
momentum density of the helium liquid is the sum of the contributions of these
two fluids:
\begin{equation}
     \mb j = \rho_\m s\mb v_\m s + \rho_\m n\mb v_\m n \; .
\end{equation}

\begin{figure}[t]       
  \begin{center}
    \includegraphics[height=50mm]{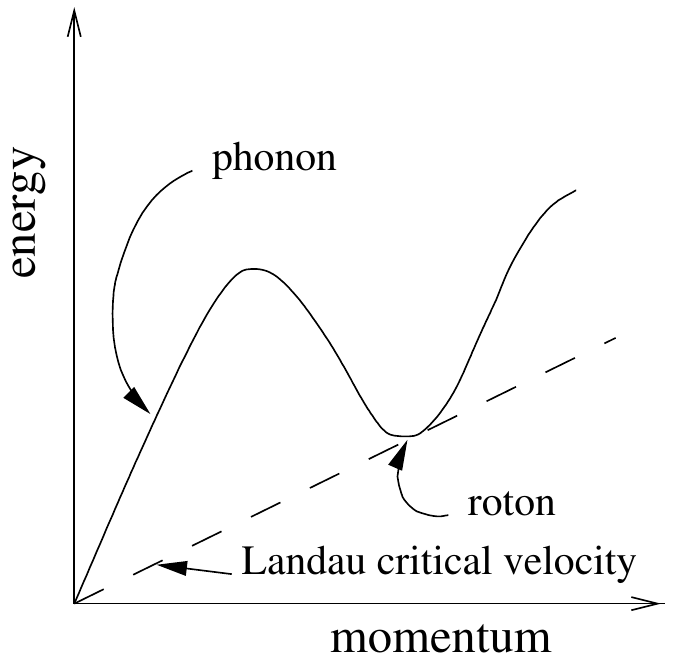}
    \caption{\label{DispersionCurve} Dispersion curve of
      the elementary excitations in superfluid $^4$He as devised by
      \citet{Landau:47}. }
  \end{center}
\end{figure}

The superfluid part of the flow being assumed ideal, it carries no
entropy. The fluid entropy is transported by the normal fluid described by
Landau as a gas of thermally-excited elementary excitations, the phonons -- or
sound quanta -- at long wavelengths and the rotons at wavelengths of the order of
interatomic spacing, as sketched in Fig.\ref{DispersionCurve}.
%
%
Landau attributed the non-viscous property of $^4$He flow to the intrinsic
shape of this spectrum. There exists no elementary excitation with a finite
momentum $\bf p =\hbar\,\bf k$ and a small enough energy $\epsilon(\bf p)$ to
couple to a solid obstacle or a wall at rest while conserving energy and
momentum. At least for small enough superflow velocities. This requirement
sets the Landau criterion for dissipationless flow: $v_\m s<v_\m L$. The
intricate problem of critical velocities above which dissipation arises in
superfluid flow is dealt with in Sec.\ref{CriticalVelocities}.

When $\rho_\m n$ and $\rho_\m s$ can be assumed incompressible -- {\it i.e.},
for small flow velocities, the separation between potential flow for $\mb v_\m
s$ and a Poiseuille flow for $\mb v_\m n$ becomes exact.\footnote{See
  \citet{Landau:Hydro}, ch. XVI.} In this approximation, superflow is
effectively decoupled from normal fluid motion. It has become customary to
talk somewhat loosely of the motion of the superfluid as fully distinct from
that of the normal fluid. This simplified view is adopted here but,
occasionally, it fails \citep{Idowu:00a,Idowu:00b}.

Landau's two-fluid hydrodynamics, based on early experiments on superfluids
and preceded by the original suggestion of \citet{Tisza:38}, accounts
remarkably well for a whole class of thermodynamical and hydrodynamical
properties, notably for the existence of second sound and for the
non-classical rotational inertia in Andronikashvili oscillating-disks
experiments \citep{Andronikashvili:66}. It features full internal consistency;
it assumes that the motion of $\rho_\m s$ is pure potential (irrotational)
flow and carries no entropy, and it reaches the conclusion that, below $v_\m
L$, this motion is indeed fully inviscid. It has been universally adopted.

The two-fluid hydrodynamics has been extended to the superfluid phases of
$^3$He. A number of new features appear owing to the anisotropy introduced
by the Cooper pairing in a $l=1$ state of orbital momentum. Some of these
features are discussed in Sect.\ref{CPR}, but the separation of the hydrodynamics between a
superfluid component and a ``normal'' component still holds as well as the
existence of a Landau critical velocity.

However, this model has a number of shortcomings pointed out, among others, by
\citet{London:54}.  In particular, for the purpose of this review, it does not
discuss the roots of the irrotationality condition, Eq.\,(\ref{rotvs0}), which
are to be found in the existence of a complex scalar order parameter arising
from the transition to a Bose-condensed state, as recognised at a very early
stage of the study of superfluids by
\citet{London:38}.\footnote{\citet{Landau:Hydro} --\S 128 -- trace its roots to
  the property of the energy spectrum for the low energy excitations of the
  superfluid, constituted by sound quanta, or phonons. The Landau school
  (\citet{Khalatnikov:65} -- Ch 8 -- and \citet{Abrikosov:61}) -- \S 1.3)
  merely states the irrotationality condition as an assumption justified by
  its experimental implications. \citet{London:54} -- \S 19 -- actually derives
  this condition from a peculiar variation principle devised by
  \citet{Eckart:38} but the actual significance of this principle has not been
  clarified, nor, as it seems, that of Eq.\,(\ref{rotvs0}) (see, for example, the
  pedagogical review of \citet{Essen:12} and the references therein on the
  similar problem of the Meissner effect in superconductivity.}  It gives no
clue as to when the hydrodynamics of the superfluid fails at close distance as
any macroscopic approach to hydrodynamic is bound to. Namely, it provides no
way of estimating the superfluid coherence length - or healing length of the
macroscopic wavefunction.  It also completely disregards the existence of
quantised vorticity,\footnote{This question was nonetheless treated by the
  Landau school \citep{Khalatnikov:65}.} which, as recognised by
\citet{Feynman:55} and \citet{Anderson:65}, is responsible for dissipation of
the superfluid motion and for different, and more commonly met in practice,
critical velocity mechanisms than that of $v_\m L$.

As already noted, the Landau criterion for critical velocities rests on the
fact that the elementary excitation energy spectrum is sharply defined, that
is, it rises linearly as $\epsilon = \m c p$ with no spread in energy: there
are {\it no} excitations with low energy and small moment capable of
exchanging momentum with the superfluid in slow motion, thus causing
dissipation. This sharpness of the energy spectrum $\epsilon(p)$ has been
checked directly by neutron scattering measurements of the dynamic structure
factor as discussed in particular by \citet{Glyde:93,Glyde:13}.

It may be pointed out here that $^3$He in the normal Fermi liquid state does
display an elementary excitation spectrum with a phonon branch and a roton
minimum in neutron diffraction experiments
\citep{Stirling:76,Griffin:87}. However, that spectrum is broad; $^3$He does
not exhibit superfluid properties until Cooper pairs of fermions form and
Bose-condense.  As stressed by \citet{Feynman:72}, it really is the lack -- the
``scarcity'' in Feynman's words -- of low-lying energy levels at finite
momenta, a property of the $N$-boson groundstate with a macroscopic number of
particles in it and the existence of few excited states well-separated in
energy, that results in superfluidity.


\subsection{The superfluid order parameter} \label{OrderParameter}


A different approach to superfluidity in which a central role was attributed
to the phase of the order parameter (assumed to describe this superfluidity)
was sketchily framed by Onsager in 1948, as reported by
\citet{London:54}.\footnote{As implied in a footnote of a paper by
  \citet{Onsager:49} and mentioned in the footnote in page 151 of London's
  book.} London himself did not make much use of this concept of phase,
although, on the one hand, he was the first to propose that superfluidity
arises from Bose-Einstein condensation and the appearance of a ``macroscopic
wave function'', and on the other hand, he had earlier realised the important
significance of the phase factor in quantum mechanical wavefunctions.

Indeed, as soon as wavefunctions are considered, the concept of quantum phase
becomes relevant. Its early origin can be found in a formalisation of
electrodynamics by Weyl \citep{Yang:03}, in which a gauge transformation
explicitly introduces a factor $\m e^{\theta}$ in the theory. A change of
gauge, $\mb A' \longrightarrow \mb A +\nabla\theta$, combined with a change of
the wavefunction $\psi' \longrightarrow \psi\, \exp\{\m i\,e\theta/\hbar
c\} $
leaves the Schr\"odinger equation unchanged. The application of Weyl's
prescription to quantum-mechanical systems led F. London to turn the exponent
$\theta$ into a purely imaginary quantity $\m i\varphi$, $\varphi$ then having
the significance of an actual phase \citep{Yang:03}. These historical developments
explain the somewhat inadequate terminology that refers to changes of the
phase as gauge transformations \cite{Greiter:05}.\footnote{In the words of
  C.N. Yang (2003) ``{\it Weyl in 1929 came back with an important paper that
  really launched what was called, and is still called, gauge theory of
  electromagnetism, a misnomer.  (It should have been called phase theory of
  electromagnetism.)'}'. }

But the unifying power of quantum field gauge theories ultimately carried the
day. The fact that a droplet of superfluid randomly picks up a (well-defined)
quantum phase when it nucleates out of vapour or out of normal fluid in a
confined geometry, is referred to as the breaking of gauge symmetry. The term
``Bose broken-symmetry'' promoted in particular by \citet{Griffin:87} to
describe the appearance of a macroscopic number of particles in the
groundstate of a Bose system with the same one-particle wavefunction
exhibiting the same phase factor only gained limited
acceptance.

The ground state wavefunction of a homogeneous system of structureless bosons
such as superfluid $^4$He at rest can be shown quite generally to have no
node;\footnote{See \citet{Feynman:55}, \citet{Penrose:56} or
  \citet{Landau:Quantum} \S 61.} it reduces to a complex scalar with a
constant phase and a modulus that remains finite at every point in the sample.
Atomic motion results in small scale, small amplitude fluctuations of this
complex scalar.  Averaging these fluctuations over finite, but still small,
volume elements leaves a ``coarse-grained'' average wavefunction. If the system is
inhomogeneous on a scale much larger than the coarse-graining volume, the
modulus and phase are slowly varying functions of the position $\mb r$,
\begin{equation}        \label{MacroscopicWaveFunction}
  \mi\Psi(\mb r) = f(\mb r)\, \m e^{\ts{\m i\varphi(\mb r)}}   \; . 
\end{equation}

This in essence is the macroscopic wavefunction considered by
\citet{London:38}, \citet{Onsager:49}, \citet{London:54} and
\citet{Feynman:55,Feynman:72}. The information on the localisation of the
bosons at $\mb r_1,\mb r_2,\ldots,\mb r_\m N$ and on their short-range
correlations has been lost in Eq.(\ref{MacroscopicWaveFunction}), which is not
an exact many-body ground state wavefunction anymore.\footnote{A useful
  discussion of this topic can be found in \citet{Nozieres:90}, Ch. 5 and also
  in \citet{Feynman:72}. The more rigorous approach discussed in
  \S{\ref{MicroscopicApproach}}, based on the density matrix formalism, shows
  how such a macroscopic wavefunction can be constructed.} However, considered
as a ``macroscopic matter field'' in Anderson's own words, it has provided a
lot of mileage in describing the properties of superfluids.

\subsection{The superfluid velocity}
%
The particle density $n(\mb r)$ at point $\mb r$ of the $N$-boson system is
given in terms of this macroscopic wavefunction,
Eq.(\ref{MacroscopicWaveFunction}), by
\begin{equation}        \label{NumberDensity}
  n(\mb r)  = \int \m d^3\mb r_1 \ldots \m d^3\mb r_\m N\,  
    \mi\Psi^\star(\mb r)\mi\Psi(\mb r)\,
    \sum_{i=1}^{\m N}\delta(\mb r -\mb r_i) \; ,
\end{equation}
and the particle current density by 
\begin{equation}        \label{Current}
\begin{split}      
    \mb j(\mb r) &= \int \m d^3\mb r_1 \ldots \m d^3\mb r_\m N 
    \\
    &\sum_{i=1}^{\m N}  \frac{\hbar}{2 \m i m_4}
    \bigg[ \mi\Psi^\star(\mb r)\delta(\mb r -\mb r_i)\mb \nabla_\mb r \mi\Psi(\mb r) +
    \mi\Psi(\mb r) \mb \nabla_\mb r \delta(\mb r -\mb r_i) \mi\Psi^\star(\mb r) \bigg]
    \\
    &= \int \m d^3\mb r_1 \ldots \m d^3\mb r_\m N
    \\
    &\sum_{i=1}^{\m N}  \frac{\hbar}{2 \m i m_4}\, \delta(\mb r -\mb r_i)
    \bigg[ \mi\Psi^\star(\mb r) \nabla_\mb r \mi\Psi(\mb r) - \mi\Psi(\mb r) 
    \nabla_\mb r \mi\Psi^\star(\mb r) \bigg]  
    \\
    &= n(\mb r) \frac{\hbar}{m_4} \nabla \varphi(\mb r) \; . 
\end{split}
\end{equation}

Equation (\ref{Current}) leads as a matter of course to the definition of the
local mean velocity of the bosons as
\begin{equation}        \label{vs}
  \mb v_\m s =  \frac{\hbar}{m_4} \nabla \varphi(\mb r) \; .
\end{equation}
According to this definition, the quantity $\mb v_\m s$ derives from the
velocity potential $ (\hbar/m_4) \varphi(\mb r)$ and is identified to the
quantity introduced in the two-fluid hydrodynamics under the same notation.
This identification implies that the quantity $ n(\mb r)$ given by
Eq.(\ref{NumberDensity}) stands for the superfluid number density
$\rho_\m s/m_4$. The strong correlations between bosons in the dense system --
in particular, the hard core interactions -- are averaged out in the
coarse-graining procedure.

Definition (\ref{vs}) of the superfluid velocity and the identification of
$f(\mb r)$ in Eq.(\ref{MacroscopicWaveFunction}) with $(\rho_\m
s\,/\,m_4)^{1/2}$ thus appears as a necessary formal construction that
reproduces, in the classical limit, the quantity postulated by Landau to set
up the two-fluid hydrodynamic model. At finite temperature, the number of
atoms involved in Eqs.(\ref{NumberDensity}) and (\ref{Current}) is simply
proportional to $\rho_\m s(T)/\rho$.  The macroscopic wavefunction thus takes
the following form:
\begin{equation}     \label{MacroscopicWavefunction2}
  \mi\Psi(\mb r) = n_\m s(\mb r)^{1/2} \m e^{\ts{\m i \varphi(\mb r)}} \; .
\end{equation}
%

\subsection{A more microscopic approach}
\label{MicroscopicApproach}
So far, the discussion has been based on the general properties of the
groundstate wavefunction of $N$-boson systems, turned into a ``macroscopic
wavefunction'' by coarse-grained averaging. No precise prescription on how
this averaging can be carried out in practice, no clue as to the suspected
relationship between Bose-Einstein condensation and superfluidity at the
microscopic level have been given.  Off-diagonal long range order (ODLRO)
represents the commonly acknowledged fundamental concept that achieves this
connection, underlying both superconductivity and superfluidity. It defines
the kind of order that prevails in a superfluid or a superconductor as put
forward by \citet{Yang:62} extending earlier work by \citet{Penrose:56} while
a parallel route was taken by Bogolyubov and other representatives of the
Russian school,\footnote{See \citet{Abrikosov:61}.} and in particular
\citet{Beliaev:58} for the system of interacting bosons.\footnote{See
  \citet{Kadanoff:13} for an insightful account of the historical genesis of
  the idea of ODLRO and a discussion of the role of the condensate in
  superfluidity and superconductivity.}

ODLRO stands for the correlation that exists between atoms in Bose-Einstein
condensates. In its simplest form, for a gas of $N$ non-interacting Bose
particles in a box of volume $V$, it is expressed by the single-particle
density matrix
\begin{eqnarray}        \label{DensityMatrix}
 \urho_1(\mb r,\mb r') &=& (N/V)\,\int \m d\mb r_2 \ldots \m d\mb r_N  
 \nonumber \\
 && \mi\Psi_{\m N}^\star(\mb r,\mb r_2,\ldots,\mb r_N)  \mi\Psi_{\m N}(\mb r',\mb r_2,\ldots,\mb r_N)
  \; ,
\end{eqnarray}
where $\mi\Psi_{\m N}$ is the eigenfunction of the groundstate of the
$N$-boson system at $T=0$ satisfying the boundary conditions at the box wall
(rigid walls, or periodic). As the particles of an ideal gas do not interact,
the many-body wavefunction $\mi\Psi_{\m N}$ is simply the product of $N$
identical single-particle wavefunctions $\psi(\mb r)$ evaluated at $\mb r =\mb
r_i$, the particle locations, suitably normalised and symmetrised. Upon
integration over the $N-1$ particles $\mb r_2 \ldots \mb r_N$, all what is
left is the product
\begin{equation}        \label{IdealBEDensityMatrix}
 \urho_1(\mb r,\mb r') = (N/V)  \psi^\star(\mb r)\,\psi(\mb r') \; ,
\end{equation}
of single-particle wavefunctions, which is quite simple but highly anomalous
in that it does not vanish when the two locations $\mb r$ and $\mb r'$ become
far apart as it would do for a classical ideal gas. This simple remark has
startling consequences. Even though the boson particles are assumed not to
interact in the ideal gas, they still show a large degree of
correlation. These correlations of statistical origin\footnote{More is said
  on these correlations in Sec.\ref{Conclusion}} preclude the use of the
grand canonical ensemble because two widely separated parts of the system
cannot be assumed to behave independently \citep{Ziff:77}.

The extension of the anomalous result Eq.(\ref{IdealBEDensityMatrix}) to
non-ideal Bose gases is non-trivial -- one may remember for instance that a
minute attractive interaction between bosons destabilises the gas. And yet a
further extension to non-equilibrium situations is mandatory to describe
superflow.

Such an extension to the weakly--repulsive Bose gas is implicit in the
pioneering work of \citet{Bogolyubov:47} who showed how second quantisation
techniques could be used to derive the property of linearity of the energy
spectrum at long wavelengths, $\epsilon = cp$, as asserted for the dense
superfluid helium by Landau. Further progress was carried out by
\citet{Beliaev:58} using field-theoretical techniques to express the
relationship between the particle number density $n$, the chemical potential
$\mu$, and the particle number density in the condensate $n_0$, which differs
from $n$ because the interaction between particles prevent all of them to fall
into the lowest energy state.

Various refinements have led to what now constitutes the conventional
way\footnote{See, for example, the monograph of \citet{Griffin:93} and his
  historical note \citep{Griffin:99}} to describe the nearly-ideal BEC gas
with a number density of atoms $n$, the Bose order parameter being written as
a complex number $\sqrt{n_0}\exp{(\m{i}\varphi)}$ involving the number
density of atoms in the ground state $n_0$, as expounded, for instance, by
\citet{Dalfovo:99}. This description is only well-grounded for small
depletion of the condensate, {\it i.e.}, for $n_0/n$ not too far from unity.  
For a dense, strongly interacting, Bose system such as liquid $^4$He, this
condition is not fulfilled. The zero-momentum ground state is strongly
depleted. 

The spirit of the definition of the order parameter for helium was given by
\citet{Penrose:56}, based on an analysis of the large-scale correlations in the
various terms of the single-particle density matrix.

In a usual fluid, the on-diagonal elements $\urho_1(\mb r,\mb r)$ are of order
of the particle number density $n(\mb r)$. Particle correlations decrease
rapidly as {\bf r-r}$'$ increases and so do the off-diagonal terms with $\mb r
\neq \mb r'$. By contrast, a superfluid can sustain a persistent current:
large scale correlations should be strong so that, when a particle is
deflected at {\bf r} by an obstacle and kicked out of the condensate, a
twin-sister particle is immediately relocated in the condensate at {\bf r}$'$
with no loss of order in momentum space. Such correlations should be described in
the density matrix by a term embodying the condensate of the same
``structure'' as the product in Eq.(\ref{IdealBEDensityMatrix}), supplemented
by other terms for the part of the system that cannot be accommodated in the
ground state because of interparticle collisions. Thus, following Penrose and Onsager,
the criterion for Bose-Einstein condensation must be traced to the
existence of one element of the form (\ref{IdealBEDensityMatrix}), namely a
product $ \Phi^*(\mb r)\Phi(\mb r')$ with a macroscopic size relative to other
elements, so that the density matrix takes the form: 
\begin{equation}        \label{ODLRODensityMatrix}
   \urho_1(\mb r,\mb r')= \Phi^*(\mb r)\Phi(\mb r') + 
      \sum{\mbox{other matrix elements}}
\; . 
\end{equation}
The sum in the right hand side of Eq.(\ref{ODLRODensityMatrix}) is a mixed bag
of terms describing the correlations between particles outside the condensate
as well as terms involving both condensate and non-condensate particles. The
function $\Phi(\mb r)$ can be viewed as playing the role of a single-particle
wavefunction of the interacting particles in the condensate, $n_0 = (1/V)
\int |\Phi(\mb r)|^2 \m d\mb r$ being the mean number density of those
particles.\footnote{\citet{Nozieres:90} give in Ch. 10 a very transparent account
  of ODLRO using the notation of field theory, in which the density matrix
  reads
\[
\rho_1({\mb r},{\mb r}') = \sum_n
\langle\phi|\upsi^\dagger({\mb
    r})|\phi_n\rangle\langle\phi_n|\upsi({\mb r}')|\phi\rangle \; ,
\]
$\upsi^\dagger({\mb r})$ and $\upsi({\mb r}')$ being the boson
creation and annihilation field operators, $|\phi_n\rangle$ a complete set of
eigenstates of the system and $|\phi\rangle$ the state in which the average is
expressed, which is taken as the ground state $|\phi_0\rangle$. Among the
intermediate states $|\phi_n\rangle$, those of special relevance to the
kicking-out and relocation processes discussed here  connect the
ground state with $N$ bosons to the ground state with $N-1$ bosons. So
attention must be focused on the following matrix element
\[
\Phi({\mb r}) = \langle\phi_0(N-1)|\upsi(\mb r)|\phi_0(N)\rangle
\] 
that is taken to represent the condensate wavefunction.  }
This number density $n_0$ can be orders of magnitude smaller than the total
density $n$, but is assumed to still remain macroscopic.

As the ground state wavefunction for a boson system is everywhere non-zero,
its absolute value for a homogeneous system, $|\Phi(\mb r)|$, is equal to
$\sqrt{n_0}$ to the extent that $n_0$ is constant in space. This reasoning can
be extended to situations that are slightly non-uniform in space. The term
$\Phi^*(\mb r)\Phi(\mb r')$ in Eq.(\ref{ODLRODensityMatrix}) does not decay as
the particle locations $\mb r$ and $\mb r'$ become far apart compared to
interatomic distances: it describes the long-range correlations in the
condensate, or ODLRO. The excited states with $\mb k \neq 0$ and distribution
$n_{\mb k}$ are not macroscopically populated and only have short range
coherence. The summation over all these remaining contributions in
Eq.(\ref{ODLRODensityMatrix}) may also amount to a macroscopic term $\sum_\mb
k n_\mb k$, of order $N$. Each of these terms decays as $|\mb r - \mb r'|$
becomes large but there are a large number of them: the whole of the excited
states is also macroscopically populated.

Needless to say, the single particle wavefunction $\Phi(\mb r)$ in
Eq.(\ref{ODLRODensityMatrix}) bears no relationship to that for free particles
in Eq.(\ref{IdealBEDensityMatrix}) for the ideal gas.  Neither the $\Phi^*(\mb
r)\Phi(\mb r')$ term in Eq.(\ref{ODLRODensityMatrix}) nor the incoherent terms
have been expressed in full for the dense helium 4,\footnote{As stated in
  the monograph of \citet{Nozieres:90}, \S 9.4} contrarily to near-ideal BEC
gases and to the BCS theory (for Cooper pairs and superfluid $^3$He). But in
superfluid $^4$He as in these other situations, ODLRO is found to be present
and to constitute a unifying feature sufficient to ensure flux quantisation
(BCS superconductors) or velocity circulation quantisation (dense superfluid
helium). That the simple factorisation of the coherent part of the density
matrix ensures superfluidity is a remarkable result. It has been established
on general grounds by \citet{Yang:62}.

\citet{Penrose:56} used various approximate forms for the groundstate
wavefunction of dense helium-4 at $T=0$ to illustrate the splitting of the
density matrix, Eq.(\ref{ODLRODensityMatrix}), and to evaluate the depletion
of the condensate, {\it i.e.}, the value of $n_0$. They have used in
particular Feynman's simple ansatz for the superfluid wavefunction
\citep{Feynman:55}, which assumes strong hard-core repulsion and weak 2-body
attraction with a minor role in interparticle correlations. Only the former
can be kept for an approximate evaluation of $n_0$. 

Building on this remark, Penrose and Onsager noticed that the depletion under
scrutiny can conveniently be derived from the (known) pair distribution for a
classical gas of hard spheres such as the one pictured in Fig.\,2. They found
that collisions between hard spheres with diameter 2.6 \AA~, 3.6 \AA~ apart,
leave only about 8 \% of the helium atoms in the zero-momentum state. This
value of $n_0/n$ in $^4$He has been confirmed by more elaborate theories and
by experiment.\footnote{For a recent review, see \citet{Glyde:13}.}  While the
depletion of the condensate is a small effect in low density atomic gases
\citep{Dalfovo:99}, it is considerable in liquid helium.
%
\begin{figure}[t]       
  \begin{center}
    \includegraphics[width=60mm]{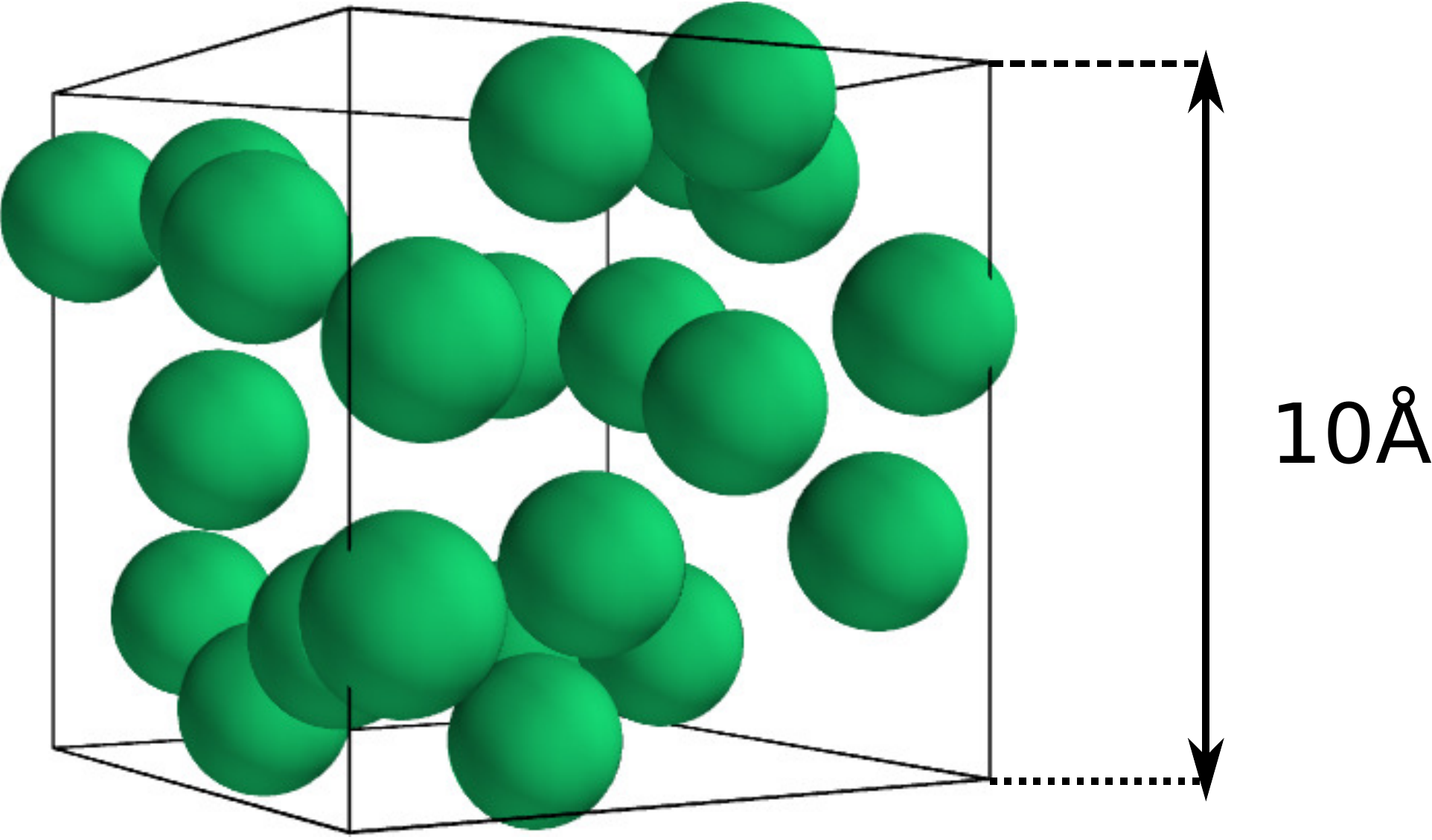}
    \caption{\label{HardSpheres}(Color online) 3D view of 2.6 \AA\ diameter
      hard spheres randomly distributed in a 10x10x10 \AA\ cube with a 3.6
      \AA\ mean spacing, as for liquid helium-4 at SVP. In helium, atoms are
      both strongly confined by hard core repulsion and dynamically
      delocalised by the zero point energy motion. Courtesy Nelle Varoquaux. }
  \end{center}
\end{figure}

This large depletion raises the following question: how is it that the
condensate fraction is only 8 \% at $T$ = 0 while the superfluid fraction in
the two-fluid model is 100 \% ? The answer is simple: these are not the same
quantities.\footnote{This point has been discussed  by
  \citet{Griffin:87,Griffin:93}.}  The superfluid density stands for the
inertia of the superfluid fraction, as measured for instance by a gyroscopic
device sensitive to trapped superfluid currents \citep{Reppy:65} or, less
directly, by the decoupling of the superfluid component in an oscillating disk
experiment \citep{Andronikashvili:66}.  In these experiments, only the
elementary quasiparticles couple to the transverse oscillations of the cell
walls; the remaining of the superfluid is not set into motion.

The superfluid fraction superfluid $\rho_\m s/\rho$ is not directly related to
the probability of finding bosons in the $\mb k =0$ quantum state. The
occupation of the condensate is seen experimentally as a hump at zero energy
transfer in the dynamic structure factor measured by neutron inelastic
scattering \citep{Glyde:13}, a quantity rather well-hidden from
experimentalists'sight in helium.  When the superfluid is set into motion, the
condensate enforces long-range order and drags the excited states along
through the short-range correlations; there is entrainment of the atoms in the
fluid by the condensate.  Microscopic theory is needed to describe this
process in detail.

Deferring to the end, \S\ref{Conclusion}, further discussion on the merits of,
and differences between, the microscopic approach -- the ODLRO concept of
\citet{Penrose:56} and \citet{Yang:62} -- and the macroscopic quantum field
point of view -- a discussion found in Appendix A1 of
\citet{Anderson:66a} -- the superfluid order parameter in $^4$He will be taken
in the following as the macroscopic wavefunction
Eq.(\ref{MacroscopicWavefunction2}), namely $\mi\Psi(\mb r) = f(\mb r) \m
e^{\m i\varphi(\mb r)}$ with $f = n_\m s^{{1/2}}$, $n_\m s$ being the
superfluid number density. This choice conforms to that of
\S\ref{OrderParameter} and leads straightforwardly to the two-fluid model. The
alternate choice to set $f = n_0^{{1/2}}$ would be less productive: the
condensate density $n_0$ remains half-buried in the formalism and is difficult
to access experimentally. This situation differs markedly from that in cold
atomic gases. There, the condensate can be imaged directly in momentum space;
it comprises most of the particles in the gas so that $n_0 \simeq n $; the
corresponding order parameter emerges seamlessly from perturbation theory
applied to the weakly-interacting Bose gas following Bogolyubov's
prescriptions.

For weakly interacting bosons, the macroscopic wavefunction
dynamics is governed by the Gross-Pitaevskii equation.\footnote{In the context
  of superfluids, see \citet{Gross:61,Pitaevskii:61,Langer:68,Nozieres:90}.}
The relevance of the Gross-Pitaevskii equation to the Bose-condensed systems
has been put in a bright new perspective for ultra-cold atoms in a trap, for
which it constitutes an excellent description
\citep{Dalfovo:99,Cohen-Tannoudji:01}.  However, the approach based on this
equation turns out to be less adequate for the dense superfluid, sometimes not
even qualitatively correct, as found for instance by \citet{Pomeau:93} in the
context of the breakdown of superflow. It will not be pursued here, except to
mention that it does provide an estimate of the distance over which the
two-fluid hydrodynamics needs to be supplemented by quantum corrections, as
discussed in Sect.\ref{Josephson}.

In all cases, the phase, a variable of lesser relevance in the older quantum
mechanics, is now assigned the role of governing superfluid dynamics. A
fundamental result, established early by \citet{Beliaev:58}, relates the phase
of the condensate particles to the chemical potential, $\varphi = \mu \,
t/\hbar $, where $t$ is time. The deep significance of this odd-looking relation
only became apparent later, when the Josephson effects came on stage.

The role of the phase as a dynamical variable was extended to superfluid
helium by \citet{Anderson:64} \citep{Anderson:65,Anderson:66b,Anderson:84b}
who noted that phase and particle number are canonically conjugate
variables. This property is well-known in quantum electrodynamics for
photons in a cavity. The number of photons in a given mode and their phase,
defined for coherent electromagnetic fields in the cavity, are non-commuting
operators. As such, they obey an uncertainty relation \citep{Heitler:54} that
reads:
\begin{equation}        \label{NumberPhaseUn}     
  \inc N \,\inc\varphi \sim 1 \; . 
\end{equation}
As remarked by Heitler, ``{\it if the number of quanta of a wave are
given it follows from eq.(\ref{NumberPhaseUn}) that the phase of this
wave is entirely undetermined and  vice versa. If for two waves the phase
difference is given (but not the absolute phase) the total number of light
quanta may be determined, but it is uncertain to which wave they belong}''. 
This remark will bear implications throughout this review.
 
Superfluidity is more than simply the absence of viscosity supplemented by the
condition that vortices have quantised circulation. The urge to observe the
role of the phase in a Josephson-type effect -- and the failure to do so for a
long period of time -- became quite pressing to confirm the picture drawn by
Anderson of helium as obeying quantum mechanics in a more profound way than
simply as an ideal inviscid fluid with quantised velocity circulation.

\subsection{Anderson{'}s phase slippage}
\label{PhaseSlippage}

Anderson's famed ``Considerations'' on the flow of superfluid $^4$He
\citep{Anderson:66b} provided the conceptual basis for this experimental
search for Josephson-type effects in neutral matter. Their underlying aim was
to convey a physical, laboratory-oriented, meaning to order parameter
Eq.(\ref{MacroscopicWavefunction2}) and, in particular, to its phase. These
Considerations provided the groundwork for phase slippage experiments in
$^4$He ; they were gradually fostered in a series of Lectures Notes
\citep{Anderson:64,Anderson:65,Anderson:84b} and built upon the ideas of
\citet{London:54}, \citet{Feynman:55}, and \citet{Penrose:56}, and also on the
quantum field theoretic approach of \citet{Beliaev:58}. 

In the absence of a fully-established microscopic theory of dense boson
systems, these considerations rest on the following set of well-argumented
conjectures:

\begin{enumerate}
\item{~By extrapolation of the properties of the coherent photon fields in quantum
    electrodynamics recalled above, $N$ and $\varphi$ are taken in dense
    liquid helium as canonically conjugate dynamical (quantum) variables in
    the sense that $N \leftrightarrow \m i(\partial/\partial\varphi)$ and
    $\varphi \leftrightarrow -\m i(\partial/\partial N)$.\label{sec:2}

    As such, they obey the uncertainty relation (\ref{NumberPhaseUn}). For a
    closed system with a fixed number of particles, the phase is completely
    undetermined. For the phase to be determined within $\inc \varphi \ll 1$,
    $N$ must be allowed to vary, that is, the condensate must be able to
    exchange particles with other parts of the complete physical system, which
    includes the non-condensate fraction of the bosons and the eventual
    measuring apparatus. 

    For the Josephson-effects experiments specifically considered by Anderson,
    the two weakly-coupled helium baths also exchange particles. For all these
    reasons, $N$ is allowed to fluctuate locally so that $\inc N$ takes a
    non-zero value. It can be shown \citep{Beliaev:58} that $\inc N$ is of
    order $N$ rather than unity, so that $\inc \varphi \sim {\cal O}(1/N)$ and
    $\varphi$ is well defined. }
\item{ A Hamiltonian $\cal H$ should therefore exist such that,
    $N$ being free to vary,
\begin{eqnarray}        \label{Hamilton1}
\hbar \frac{\partial N}{\partial t}&=& \mbox{ }\frac{\partial\cal
  H}{\partial\varphi}   \; ,
\\                      \label{Hamilton2}   
\hbar \frac{\partial \varphi}{\partial t}&=& -\frac{\partial\cal
  H}{\partial N}   \; .
\end{eqnarray}

Upon coarse-graining, the quantum operators become quasi-classical and their
coarse-grained average obey equations formerly identical to (\ref{Hamilton1})
and (\ref{Hamilton2}).  Eq.(\ref{Hamilton1}) defines the particle current $J=
\partial{ E}/\hbar\partial \varphi$ since $\partial{\cal H}/\partial \varphi
\Longrightarrow \partial E/\partial\varphi$ upon averaging.  Likewise with
$\partial{\cal H}/\partial N \Longrightarrow \partial E/\partial N = \mu +
\slantfrac{1}{2} m_4 v_\m s^2$ where $\mu$ is the chemical potential in the
fluid at rest, namely $\mu = m_4 P/\rho +m_4 gh+s_4 T$ with the usual
notations, Eq.(\ref{Hamilton2}) becomes
\begin{equation}        \label{acJosephson}
  \hbar \frac{\partial\varphi}{\partial t} = -(\mu + \slantfrac{1}{2} m_4 v_\m
  s^2)  \; .  
\end{equation}  

Eq.(\ref{acJosephson}) states that, whenever there exists a chemical
potential difference between two points 1 and 2 in a superfluid (or a
superconductor), the phase of the order parameter varies in time with a rate
proportional to $\mu_1 - \mu_2$: this {\it ac}-effect is quite detectable and
has nowadays many applications. It was first discussed by \citet{Josephson:62}
\citep{Josephson:64} for the tunnelling current between
superconductors coupled through a thin barrier. A full derivation for superfluid helium can be found in the monograph by Nozi\`eres and
Pines.\footnote{\citet{Nozieres:90} \S 5.7.}

Upon taking the gradient of both left and right-hand sides,
Eq.(\ref{Hamilton2}) becomes, using the definition (\ref{vs}) of $v_\m s$,
\begin{equation}        \label{SimpleEuler}
  \frac{\partial v_\m s}{\partial t} + \nabla\left(\frac{P}{\rho}+
  \frac{v_\m s^2}{2} \right) = 0\; ,
\end{equation}
which is nothing else but the Euler equation for an inviscid fluid with no
vorticity ($\bs{\omegaup} = \nabla\times \mb v_\m s = 0$). Equation
(\ref{SimpleEuler}) is precisely the same as that for the velocity of the
superfluid component in Landau's two-fluid hydrodynamics.  }
\item{ Anderson assumed that Eq.(\ref{acJosephson}) for the time variation of
    the order parameter phase holds with no solution of continuity between the
    classical inviscid fluid case and the quantum tunnelling one and, {\it i.e.},
    that it has universal applicability. This unifying approach is, in a broad
    sense, internally consistent but details are missing of how the normal
    component interacts with the superfluid component, which brings
    dissipative terms into Eq.(\ref{SimpleEuler}), and how the definition of
    $v_\m s$ as $(\hbar/m)\,\nabla\varphi$ breaks down at small distances
    where coarse-graining cannot be performed. These fine points have been
    raised in the discussion at the end of Anderson's communication at the
    Sussex Symposium on Quantum Fluids \citep{Anderson:66a}. His views are
    that the phase equation (\ref{acJosephson}), being more fundamental than
    (\ref{SimpleEuler}), always hold. This equation describes both simple
    superfluid acceleration, expressed by (\ref{SimpleEuler}), the ideal
    tunnelling situation envisioned by Josephson (see Sect.\ref{Josephson}),
    and when the variation of the phase is caused by the motion of
    vorticity.  }
\item{The last conjecture asserts that the dissipation of the kinetic energy
    of a superflow is, when averaged over time, proportional to the rate at
    which vorticity crosses the superflow streamlines. In fact, a stronger
    statement has been rigorously proved by \citet{Huggins:70}, which governs the
    detailed transfer of energy between the potential flow of the superfluid
    and moving vorticity. This process is pivotal to the understanding of
    superflow decay and, more generally, of vortex dynamics as discussed in
    Sec.\ref{VortexDynamics}.  }
\end{enumerate}

Anderson's ideas on phase slippage, linked to the motion of vortices, have
provided the conceptual framework for the experiments on the onset of
dissipation and the Josephson effects in superfluids, discussed below in
Sect.\ref{slippage} and \ref{CPR}. All facets of these experiments in
superfluid $^4$He and $^3$He can be very well accounted for with the help of
the macroscopic quantum phase $\varphi$. However,  these
ideas are still surrounded by an aura of mystery that lingers on in spite of the
facts that:~(1) the formal theoretical groundwork has been put on a firmer
basis;\footnote{Following for instance \citet{Lifshitz:80}, the density
  operator takes the form $\hat{\rho}(\mb r)=\sum_i{m_4\delta(\mb r_i-\mb r)}$
  and the current density operator the form %
  $\hat{\mb j}(\mb r) = \frac{1}{2}\sum_i{\hat{\mb p}_i \delta(\mb r_i-\mb r) +
    \delta(\mb r_i-\mb r)}\hat{\mb p}_i$ -- compare with
  Eqs.(\ref{NumberDensity}) and (\ref{Current}). The liquid velocity operator
  $\hat{\mb v}$ is in turn defined by $\hat{\mb j}(\mb r) =
  \frac{1}{2}(\hat{\rho}\hat{\mb v} + \hat{\mb v}\hat{\rho})$. These operators
  can then be shown to obey the commutation rule $\hat{\Phi}(\mb
  r)\hat{\rho}(\mb r') - \hat{\rho}(\mb r')\hat{\Phi}(\mb r) = -\m i\hbar \delta(\mb
  r -\mb r')$, $\hat{\Phi}$ being here the potential for the velocity operator
  $\hat{\mb v} = \nabla\hat \Phi$. The quantities $\hat\rho$ and $\hat\Phi$ are
  thus canonically conjugate. Their fluctuations obey an uncertainty relation
  of the form $\delta\hat\rho \; \delta\hat\Phi \geq \hbar/2$. Using the phase
  of the macroscopic wavefunction, Eq.(\ref{vs}), instead of the velocity
  potential, uncertainty relation (\ref{NumberPhaseUn}) is obtained.}~(2)~the
implications of the uncertainty relation to laboratory observations, as
well as of the other conjectures of Anderson, have been clarified by the
developments of the experiments in the past forty and so years since they were
formulated. 

This review will tackle some of these advances, in particular, by showing what
the phase slippage experiments really consist of, how phase slippage proceeds
from a dissipative regime governed by vortex dynamics to a true
dissipationless Josephson regime, and that this truly quantum behaviour
manifests itself in matter waves interferometric measurements. 


\section{Quantised vortex dynamics close to $T$=0}

\label{VortexDynamics}
                        
Vortex filaments are extended quasi-one-dimensional structures in the superfluid,
line vortices.  At the core of these defects, the superfluid order parameter
is either zero as in $^4$He or heavily distorted as in $^3$He. They
form the prevalent topological defects in superfluids.\footnote{There are a
  number of different topological defects in superfluid  $^3$He owing to the
  large number of degrees of freedom of the order parameter as briefly
  discussed in \S\ref{CPR}.}  

At distances larger than the core size, superfluid vortices behave according
to the laws of ideal fluid hydrodynamics, that is, as classical vortices with
a given, quantised, vorticity. Classical vortices have been studied for many
decades \citep{Lamb:45,Saffman:92}. Their properties have been the subject of
detailed studies in recent years in order to clarify in a number of standing
problems, their mass and impulse, the Magnus and Iordanski forces, the
eigenmodes of isolated vortices - the Kelvin waves, the collective behaviour
of vortex arrays - the Tkachenko waves, the reconnection of two vortices,
superfluid vortex tangles, and lastly, the formation of vortices and their
annihilation. This review is concerned mainly with the last topic but use will
be made of other properties of vortices, either single or few at a time,
mostly without consideration to the normal fluid background. These properties
bear a close resemblance to those of magnetic vortices in superconductors as
spelled out by \citet{Sonin:94}.

At temperatures below 1 K, vortices in superfluid $^4$He experience negligible
friction from the normal fluid, the fraction of which becomes very small. If
they deform only little and slowly, they constitute stable fluid eddies: their
velocity circulation is conserved (and furthermore, quantised) and they
cannot vanish to nothing. Their core radius, $a_0$, is of the order of the
superfluid coherence length, a few \AA\ in $^4$He
\citep{Glaberson:86,Donnelly:91}, one to two orders of magnitude larger in
$^3$He depending on pressure \citep{Vollhardt:90}.  As the temperature
increases, the scattering of phonons and rotons by the vortex cores causes
dissipation. Mutual friction between the superfluid vortices and normal fluid
sets in. Close to the superfluid transition temperature, the core size
increases and eventually diverges.

Some of the properties of vortices that have a relevance to phase slippage are
summarised below. Extended coverage of this topic can be found in the
monographs by \citet{Donnelly:91} and by \citet{Sonin:15}. Here, the dynamical
properties of superfluid vortices are derived directly from the existence of a
superfluid order parameter.  Some simplifying approximations are made in
order to get a simpler physical description of a vortex element, treated more
in the manner of a quasiparticle with mass, energy and impulse. The following
discussion then rests on physical concepts such as energy conservation or the
balance of forces. It follows largely the approach of \citet{Sonin:87}. It differs
from the more traditional and rigorous fluid-mechanical approach, as can be
found for instance in the monograph by \citet{Saffman:92}. It provides a more intuitive
feel for the behaviour of superfluid vortices that will prove useful in the
description of phase slips.

\subsection{Quantisation of circulation}

Superfluid vortices have quantised circulation. This property comes about
because their core is non-superfluid: it disrupts the order parameter field
and constitutes a topological defect in the superfluid. The circulation of the
superfluid velocity $\mb v_{\m s}$ on any path around such a defect, 
\begin{equation}         \label{VelocityCirculation}
  \oint {\bf v_{\m s}\cdot}{\m d}{\mb l} =
  \frac{\hbar}{m_4}\,\oint\nabla\varphi\cdot{\m d}{\mb l} 
\end{equation}
amounts to $\kappa_4 = 2\pi\hbar/m_4$\,\footnote{The quantum of circulation in
  $^4$He takes the value 9.97 $10^{-4}$ cm$^2$\,s$^{-1}$ and in $^3$He where the
  boson mass is $2m_3$, $\kappa_3 = \pi\hbar/m_3 =6.65\times10^{-4}$
  cm$^2$s$^{-1}$.} because the phase $\varphi$ of the order parameter can
change only by multiples of $2\pi$ along any closed contour {\it entirely}
located in the superfluid. This property holds for the true condensate
wavefunction as a basic requirement of quantum mechanics. It is not altered in
the coarse-graining average.

\begin{figure}[t]        
  \includegraphics[width=7cm]{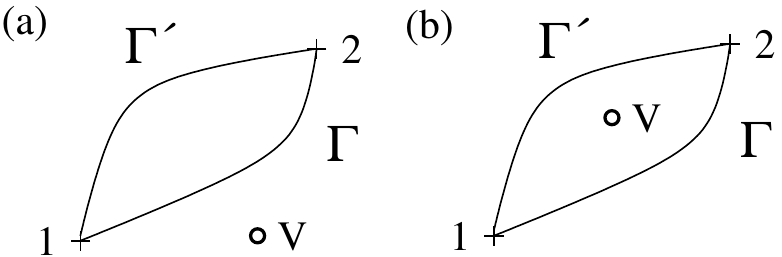}
  \caption{\label{paths}Two different situations for the path of integration
    in Eq.(\protect{\ref{Quantisation}}):~(a)~\greeksym{G} can be deformed
    continuously into \greeksym{G}'; both paths give the same phase difference
    between point 1 and point 2. (b)~ vortex V stands between the two paths;
    the phase differences along \greeksym{G} and \greeksym{G}' differ by
    $2\pi$.}
\end{figure}

Consider the velocity circulation from point 1 to point 2 in Fig.\ref{paths}
along a path $\Gamma$ entirely located in the superfluid:
\begin{equation}        \label{Quantisation}
  \kappa =  \int_1^2 {\bf v_{\m s}.}{\m d}{\bf l}
         =  \frac{\hbar}{m_4}\int_1^2 {\bf \nabla\varphi.}{\m d}{\bf l}
         =  \frac{\hbar}{m_4} (\varphi_2 - \varphi_1) \; .
\end{equation} 
\noindent
Along another path $\Gamma'$ also going from 1 to 2, as shown in
Fig.\ref{paths}, the circulation is $ (\hbar/m_4)(\varphi_2 - \varphi_1 +
2n\pi)$. If $\Gamma^\prime$ can be deformed into $\Gamma$ continuously while
remaining in the superfluid, then $n=0$ . If this cannot be done, $n$ may be a
non-zero integer, 1 in the case under consideration.

Thus, when path $\Gamma´$ crosses the core of a $^4$He vortex, in which
superfluidity is destroyed and the order parameter amplitude goes to zero, $n$
changes by 1 because $^4$He vortices carry a single quantum of circulation for
reasons discussed below. Conversely, when a vortex crosses a superfluid path
from 1 to 2, the circulation along that path changes by one quantum and the
phase difference by $2\pi$. This simple property forms the basis of the phase
slip phenomenon described in Sec.\ref{slippage}.

Experiments have confirmed to a high accuracy the quantisation of hydrodynamic
circulation both in $^4$He \citep{{Vinen:61,Whitmore:68,Karn:80}} and in $^3$He
\citep{Davis:91}. This feature constitutes a cornerstone of superfluid
physics, and evidence for the reality of the superfluid quantum phase.


\subsection{Vortex flow field and line energy}   
\label{VortexlineEnergy}


The flow velocity induced by a straight vortex filament, chosen along the unit
vector $\mb{\hat z}$, at a distance $\mb r$ measured in the plane perpendicular
to $\mb{\hat z}$ is easily expressed from the quantisation of the velocity
circulation and the symmetry around the vortex axis as
\begin{equation}        \label{VortexFlow}
\oint\mb v \cdot \m d\mb l = \kappa_4 \; \Longrightarrow \;
\mb v_{\m v} = \frac{\kappa_4 }{2\pi} \mb{\hat z}\times \frac{\mb{\hat r}}{r} \;,
\end{equation}
\noindent
provided that $r$ is larger than $a_0$. For $r \lesssim a_0$, the detailed
structure of the core becomes important.\footnote{See \citet{Fetter:76},
  \citet{Sonin:87}, \citet{Salomaa:88}, \citet{Dalfovo:92} for more extended
  discussions.} The quantity $\mb v_{\m v}$ is the vortical flow due to the
vortex element. The superfluid velocity $\mb v_{\m s}$ is the sum of an
eventual potential flow $\mb v_{\m p}$ existing independently of the vortex,
for instance applied externally, and of $\mb v_{\m v}$. The contribution of
$\mb v_{\m p}$ to the loop integral in Eq.(\ref{VortexFlow}) is nil and leaves
the circulation unchanged.  Straight vortex filaments are created by rotating
the helium container; they have been the object of very detailed
studies.\footnote{See \citet{Hall:60}, \citet{Andronikashvili:66},
  \citet{Sonin:87}, \citet{Krusius:93}, \citet{Finne:06}.}

\begin{figure}[t]        
  \begin{center}
    \includegraphics[width=3.8cm,height=4.5cm ]{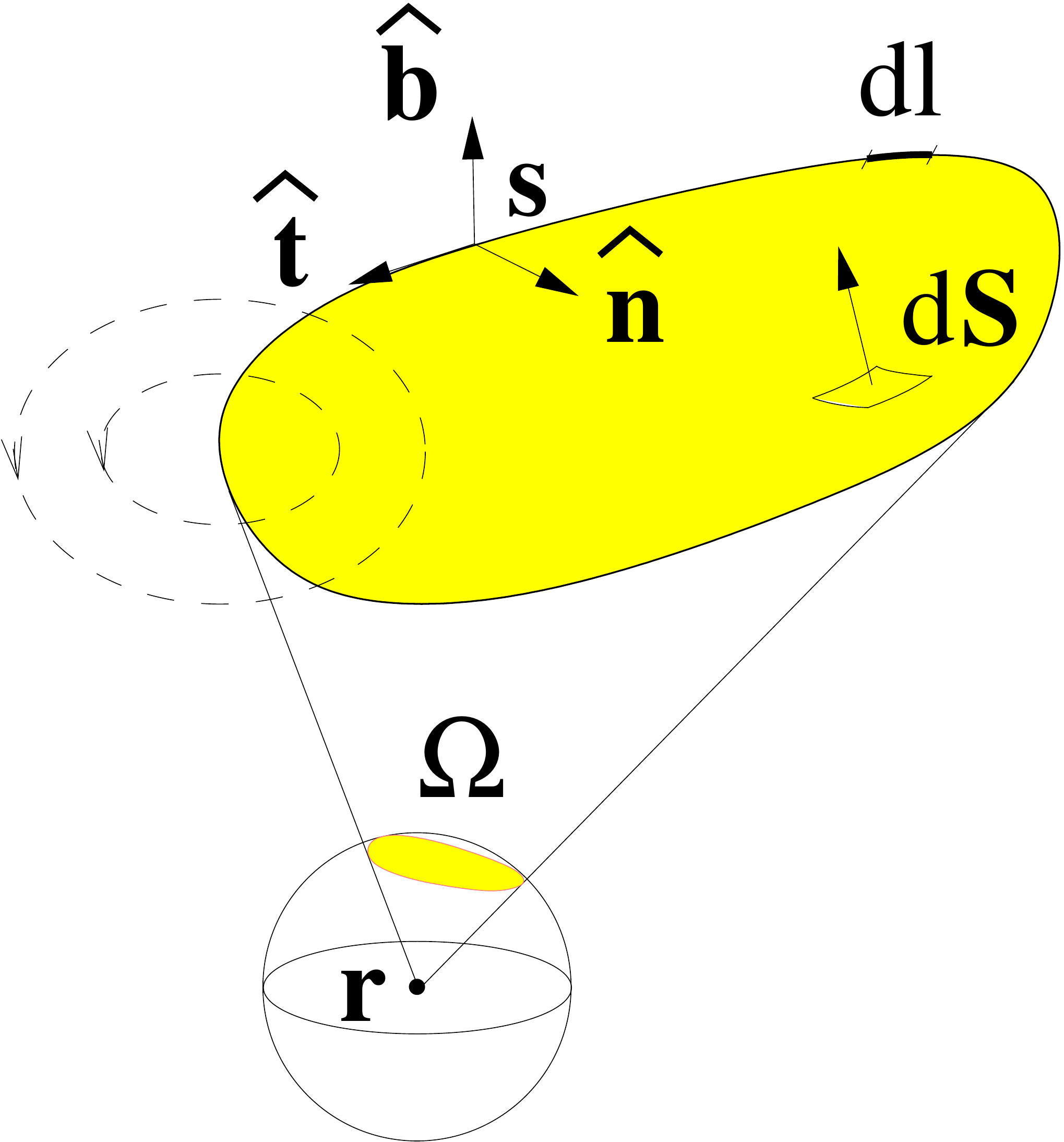}
    \caption{\label{SolidAngle}(Color online) The geometrical representation
      of a vortex loop, an arbitrary surface $S$ spanning the loop, with
      element $\m d \mb S$, and the solid angle $\Omega$ subtended by the loop
      from point $\mb r$, the tangent $\mb{\hat t}$, normal $\mb{\hat n}$, and
      binormal $\mb{\hat b}$ at point $\mb s$. The flow lines of the vortex
      velocity field (dashed line) thread surface $S$ where the phase changes
      determination by $2\pi$. Solid angle $\Omega$ offers a geometrical
      representation of the integrand in Eq.(\ref{ScalarPotential}).}
  \end{center}
\end{figure}

Equation (\ref{VortexFlow}) can be extended to {\it curved} vortices, provided
that their radii of curvature are much larger than the core radius $a_0$. It
bears a direct analogy with Amp\`ere's law, $\mb v$ standing for the magnetic
field and $\kappa$ for the electric current carried by the
conductor.\footnote{See, for example, \citet{Lamb:45}, \S 147.}
The velocity at point $\mb r$ induced by a {\it closed} vortex filament lying
along the curve $\mb s$ is then given by the analogue of the Biot-Savart law in
electrodynamics:\footnote{See  \citet{Saffman:92} \S 2.3.}
\begin{equation}        \label{BiotSavart}
\mb v_\m v(\mb r) = \frac{\kappa_4 }{4\pi} \oint\m d\mb l\times 
  \frac{\mb r -\mb s(l)}{|\mb r-\mb s(l)|^3} \;.  
\end{equation}
The geometrical representation of the vortex loop by $\mb s$ is such that $\m
d\mb l = \m d \mb s$ is a vector oriented along the tangent to the loop
$\mb{\hat t}$ of infinitesimal length $\m d l$, $l$ being the arc length of
the loop (see the sketch in Fig.\ref{SolidAngle}).  The tangent $\mb{\hat t}$
is the unit vector $\m d\mb s/\m d l =\m d\mb l/\m d l $. Its derivative with
respect to $l$ defines the normal to the loop $\mb{\hat n}$ and the radius of
curvature $R$: $\m d\mb{\hat t}/\m d l = \m d^2\mb s/\m d l^2 = \mb{\hat
  n}/R$. As noted above, the radius of curvature $R$ should be large -- and
the change of orientation of the tangent $\m d\mb{\hat t}/\m d l$ small --
with respect to the core radius for this representation of the vortex element
as a one-dimensional line to be valid.

The integrand in Eq.(\ref{BiotSavart}) gives the contribution of the
vortex element $\m d \mb l$ located at $\mb s$ on the loop to the full velocity
field. An integration by parts yields
\begin{equation}        \label{VectorPotential}
\mb v_\m v(\mb r) = \frac{\kappa_4}{4\pi}\nabla\times\oint \frac{\m d\mb l}
  {|\mb r-\mb s(l)|} = \nabla\times \mb A(\mb r)  \;,  
\end{equation}
which defines a vector potential for the vortex velocity field, $\mb v_\m
v = \nabla\times \mb A$.

Equation (\ref{VectorPotential}) fulfils the mantra of conventional mathematical
physics according to which a vector field can be split into an irrotational
part, which derives from a scalar potential, and a remainder, the solenoidal
part, which is not curl-free and which derives from a vector potential. 

While utterly correct in mathematical terms, this point of view may be
slightly misleading for the superfluid velocity fields.  The latter are a
subset only of the more general vector fields in the sense that vorticity is
localised in space to the vortex cores and that the vortex line can be treated
as a line singularity.  The Biot-Savart law (\ref{BiotSavart}) can then be put
under the following form \footnote{Stokes's theorem can be invoked to
  transform the line integral in Eq.(\ref{BiotSavart}) into an integral over
  the surface spanned by the vortex loop,
$$\oint \m d l\times\mb a = \iint
\big[(\nabla\mb a)\cdot\m d \mb S - \nabla\cdot\mb a\,\m d\mb S\big]
$$
with $\mb a = (\mb r -\mb R)/|\mb r -\mb R|^3$. Equation
(\ref{ScalarPotential}) then follows.}
\begin{equation}        \label{ScalarPotential}
\mb v_\m v(\mb r) = \frac{\kappa_4}{4\pi}\,\nabla_{\mb r}\,
\left\{\iint_S   \frac{\mb r-\mb R}{|\mb r-\mb R|^3} \cdot \m d\mb S \right\}
= \frac{\hbar}{m_4}\,\nabla \varphi_\m v \; ,  
\end{equation}
the infinitesimal surface element $\m d\mb S$ being located at position $\mb
R$. Thus the velocity fostered by the vortex derives from a scalar potential
as well as a vector potential.  Everywhere in the superfluid but at the
precise location of the vortex cores, the superfluid velocity $v_\m s$ is
indeed irrotational and derives from a scalar potential, the quantum
phase.\footnote{The situation in superfluid $^3$He-A is more complicated, as
  discussed in \S\ref{Peculiarities}.}

The velocity induced by a vortex loop decreases at large distance from the
loop as that of a dipole in the electromagnetic analogy, that is as $1/r^3$,
much faster than the $1/r$ dependence for straight vortex filaments (see
Eq.(\ref{VortexFlow})). The $1/r$ dependence can still be expected to hold at
a distance away from the core smaller than the local radius of curvature of
the vortex filament. At distances larger than the loop size, the velocity
field rapidly dies away. This property is well known for magnetic fields
generated by electric current loops. It means, for practical purposes, that
vortex loops far apart interfere very weakly and that distant boundaries have
negligible effect. These simplifying features will often be assumed in the
following.

\subsubsection{Vortex line energy}
            
The flow around the core of a vortex element carries kinetic energy, obtained
by integration of $\int (1/2) \rho_{\mathrm s} v^2_{\mathrm v} {\mathrm
  d}\tau\,$ over the volume $V$ in which this flow extends. The quantity
$\rho_{\mathrm s}$ is the superfluid density. This integral is evaluated by
introducing the vector potential $\mb A$, Eq.(\ref{VectorPotential}), from
which derives the vortical flow field, as follows:
\begin{displaymath}
  \begin{split}
    E_\m v &= \frac{\rho_\m s}{2}\,\int_\m V \!\!\m d\tau\,\mb v_\m v\cdot\mb v_\m v =
    \frac{\rho_\m s}{2}\,\int_\m V \!\! \m d\tau\, \mb v_\m
    v\cdot\nabla\times\mb A  \\
    &=  \frac{\rho_\m s}{2}\,\int_\m V \!\!\m d\tau\,\nabla\cdot(\mb A \times \mb v_\m v)  
      +  \frac{\rho_\m s}{2}\,\int_\m V \!\!\m d\tau\,\mb A \cdot(\nabla
      \times \mb v_\m v) \; .   
  \end{split}
\end{displaymath}
The last line is obtained with the help of vector identity $\nabla\cdot(\mb
a\times\mb b)=\mb b\cdot(\nabla\times \mb a) - \mb a\cdot(\nabla\times\mb b)$.
It consists of the sum of two volume integrals. The first can be changed
into a surface integral over $\mb A\times\mb v$ with the divergence
theorem. By taking the volume boundary sufficiently far from the vortex element,
supposed isolated in a large volume, the surface integral can be made
negligible.
In the second integral, the curl of $\mb v_\m v$ is zero everywhere but on the
vortex core, where it is singular: $\nabla\times\mb v_\m v = \kappa_4\,\mb{\hat
  t}\,\delta_2(\mb r-\mb s)$.  Integration over the two-dimensional delta
function $\delta_2(\mb x)$, defined in the plane normal to the tangent
$\mb{\hat t}$ to the loop, reduces this volume integral to a line
integral over the vortex element:
\begin{equation}        \label{ReducedEv}
  E_\m v = \frac{\rho_\m s \kappa_4}{2}\,\oint\m d\mb l\cdot\mb A \; .   
\end{equation}
The vortex kinetic energy is the circulation of the vector potential along the
the vortex filament.

By substitution of the expression (\ref{VectorPotential}) for the vector
potential $\mb A$ in Eq.(\ref{ReducedEv}), the vortex energy can be expressed by
a double contour integral over the vortex loop:\footnote{See
  \citet{Lamb:45}, \S 153 or \citet{Saffman:92}, \S 3.11.}
\begin{equation}        \label{ReducedEv2}
  E_\m v = \frac{\rho_\m s \kappa_4^2}{8\pi}\,\oint\!\!\oint \frac{\m d\mb s(l_1)}{\m d l_1}
    \cdot \frac{\m d\mb s(l_2)}{\m d l_2} 
    \frac{\m d l_1\,\m d l_2}{\left| \mb s(l_1)-\mb s(l_2)\right|}
    \; .   
\end{equation}

Because $E_\m v$ in Eq.(\ref{ReducedEv2}) varies as $\kappa_4^2$, loops
carrying two quanta of circulation would have four times the line energy of
single charge ones. Vortices with multiple quanta of circulation are thus
strongly disfavoured on energy grounds compared to separate singly-charged
vortices with the same total vorticity charge; they are energetically unstable
and decay spontaneously into several singly-charged entities. Only loops and
filaments carrying one quantum of circulation are considered here.

For a circular ring of radius $R$ the integral can be evaluated in terms of
elliptic functions\footnote{See \citet{Lamb:45} \S 163.} and expanded in terms
of the small parameter $a_0/R$.  The kinetic energy associated with the ring
velocity field is then given by
\begin{equation}        \label{RingEnergy}
 E_\m R = \frac{1}{2}\rho_\m s \kappa_4^2 R\,\ln \frac{8R}{a_0} 
     + {\cal O}\left(\frac{a_0}{R}\right) \; .
\end{equation}


For a straight vortex filament, the integral for the kinetic energy in the
volume comprised between two planes perpendicular to the filament stems out
directly from Eq.(\ref{VortexFlow}). For a unit length of vortex the result
reads:
\begin{equation}        \label{FlowEnergy}
{\mathsf{\epsilon}}_{\m f} = \frac{\rho_{\m s}\kappa_4^2}{4\pi}
                             \ln\left(\frac{r_{\m m}}{a_0}\right)\; .
\end{equation}
\noindent
The logarithmic divergence is cut at short distance to $a_0$, taken as the
definition of the core radius.  Its value, of the order of one \AA~ at low
pressure, is obtained from experiment \citep{Rayfield:64}. The far distance
cut-off $r_{\m m}$ is the minimum distance over which the vortex flow field is
undisturbed: it is the smallest of 1) the size of the container, 2) the
average radius of curvature of the vortex, 3) the distance to neighbouring
vortices.  For \aa ngstr\"om-size vortices, taking $r_{\m m}/a_0=10$,
${\mathsf{\epsilon}}_{\m v} \sim 2$ kelvin per \aa ngstr\"om: vortices are
high-energy excitations of the superfluid as compared to thermal excitations,
phonons or rotons. Changes in $r_{\m m}$ along the vortex line are disregarded
because they enter logarithmic terms and yield small corrections only for
$r_\m m \gg a_0$: when the vortex deforms, its energy changes mostly as its
length, and little with its shape.

The line energy of the core, usually taken as
$$
{\mathsf{\epsilon}}_{\m{sb}} = -\frac{7}{4}\,\frac{\rho_{\m s}\kappa_4^2}{4\pi} \; ,
$$
for a core rotating as a solid body,{\footnote{Using the Gross-Pitaevskii
    equation, \citet{Roberts:71} find that the prefactor 7/4 should be
    replaced by the not-so-different number 0.615 }} must be added to
Eq.(\ref{FlowEnergy}) to obtain the full vortex energy per unit length
\begin{equation}        \label{LineTension}
\mathsf{\epsilon}_{\m v} = \mathsf{\epsilon}_{\m f}\,+\,\mathsf{\epsilon}_{\m{sb}}
 = \frac{\rho_{\m s}\kappa_4^2}{4\pi} \left\{\ln\left(\frac{r_{\m m}}{a_0}
   \right) - \frac{7}{4}\right\}     \; .
\end{equation}
The full energy of a curved vortex line is thus approximated by
$\mathsf{\epsilon}_{\m v}$ times its total length.  For instance, the energy
of a vortex ring, Eq.(\ref{RingEnergy}), stems from Eq.(\ref{FlowEnergy}) if
$r_{\m m}$ is taken to be $8R$.

Expression (\ref{LineTension}) holds for straight vortex lines,   
rings, curved filaments or general loops provided than $r_{\m m} \gg a_0$. It can
be viewed as a force developing along the vortex axis, a line tension that
tends to shorten the vortex length. That is, the vortex line pulls on its
ends: if an end becomes loose it shrinks to zero. Stable vortices in finite
size containers either are closed on themselves in loops or connect to the
container walls.

\subsubsection{Stable vortices}

It follows from the existence of a positive line tension that a vortex loop
would tend to spontaneously reduce its length and minimise the line
energy. However, the energy so released by the vortex loop in its motion can be
disposed of into the surrounding fluid only in certain conditions of flow. The
line tension is opposed by other forces that arise from the vortex motion in
the fluid or from its interaction with the boundaries, namely, the Magnus
force and pinning forces.

As stand-alone loops or pinned filaments, their length is constant as long as
they cannot exchange energy with the rest of the fluid (or the external
world). In the presence of hard walls, their flow field must be such as to
satisfy the condition that no fluid can penetrate into the wall. A convenient
way of satisfying such a boundary condition is to continue the vortex filament
into the wall, forming an imaginary image vortex. Such a continuation
procedure can be shown to be possible and to yield a unique velocity
field.\footnote{See \citet{Saffman:92} \S 2.4.}  Vortices meeting with walls
usually satisfy the condition of no flow through a solid boundary by standing
perpendicular to it.\footnote{It is understood here that the boundary does not
  carry vorticity. A case of the contrary is discussed by \citet{Sonin:94b}.}
Thus, finite length vortices always close on themselves or end at walls. In
this latter case, they also form closed loops if their image is taken into
account. The opposite view, namely that vortices are most of the time
infinitely long as, for instance, vortices formed under rotation in a
cylindrical helium bucket, is also held.\footnote{Such a point of view is discussed by
  \citet{Saffman:92} \S 1.4.} The process of nucleation of vortices considered
below obviously requires that their size be finite (otherwise, the energy
involved would be infinite): the isolated vortex loops dealt with in the
following have a finite size, usually small.

\subsubsection{Vortex line impulse}
\label{ImpulseAndMass}

If an external potential flow with velocity $\mb v_\m p = (\hbar/m_4)\,
\nabla\varphi_\m p$ imposed by moving boundaries, a piston for instance, or by
nearby vortices, the kinetic energy of the combined flow $\mb v_\m p + \mb
v_\m v$ in a given volume $V$ is the sum of the kinetic energy of the remotely
applied superflow $\mb v_\m p$, that of the vortex loop, obtained from
Eq.(\ref{ReducedEv2}), and the volume integral of the cross term of the scalar
product of $\mb v_\m p$ and $\mb v_\m v$. This last term reads
\begin{equation}        \label{CrossIntegral}
  E_\m {\,I} = \rho_{\m s}\int_V \mb v_{\m p}\cdot\mb v_{\m v} \,\m d\tau 
      = \rho_\m s\,\frac{\hbar^2}{m_4^2} \int_V \nabla \varphi_{\m p}
                                   \cdot \nabla \varphi_{\m v} \,\m d\tau  \; ,
\end{equation}
and represents the energy of interaction between the
vortex and the applied flow. Making use of Green's first identity,\footnote{As expressed by
$$
  \int_V \nabla\Psi\cdot\nabla\Phi \,\m d\tau = 
    \int_S \Phi\,\nabla\Psi\cdot \m d\mb S - \int_V \Phi\nabla^2\Psi\, \m d\tau
    \; ,
$$
$S$ being the surface bounding volume $V$ and $\m d\mb S$ being the outward
pointing surface element, and taking into account mass conservation of the
fluid in incompressible flow ($\nabla^2\Psi=0$, $\Psi =
(\hbar/m_4)\,\varphi_{\m p}$ being the velocity potential of $\mb v_\m p$).} the
integral in Eq. (\ref{CrossIntegral}) can be rewritten as
\begin{equation}        \label{CrossIntegral2}
  E_\m {\,I} = \rho_{\m s} \frac{\hbar}{m_4}
    \int_S \varphi_{\m v}\, \mb v_\m p\cdot \m \,d\mb S \; ,
\end{equation}
where $\varphi_{\m v}$ is the phase change contributed by the vortex own flow field. 

The bounding surface $S$  yields not one but two contributions to the integral in
Eq.(\ref{CrossIntegral2}), the outer surface bounding $V$ and, quite
importantly, the cut spanning the vortex loop over which $\varphi_{\m v}$
changes discontinuously by $2\pi$ (see Fig.\ref{SolidAngle}).  If
$V$ can be chosen large enough, the velocity induced by the vortex on its
surface is negligible and $\varphi_{\m v}$ is a constant: the contribution to
Eq.(\ref{CrossIntegral2}) of the outer surface reduces to the net flux of $\mb
v_\m p$, which is zero. The contribution of the cut is $2\pi$ times the flux
of $\mb v_\m p$ through the vortex loop.  Introducing the mass flux of the
applied potential flow through the vortex loop, $J_\m p$, the contribution of
the cross term (\ref{CrossIntegral}) takes the very simple form
\begin{equation}        \label{CrossIntegral3}
  E_\m {\,I} = \rho_{\m s} \frac{2\pi\hbar}{m_4}
    \int_{\m{loop}} \mb v_\m p\cdot \,\m d\mb S = \kappa_4 J_{\m p} \; .
\end{equation}
Thus, an applied flow contributes to the vortex loop energy by the additional
mass flux $J_{\m p}$ that it causes through the loop times the quantum of
circulation. This result will be derived below in \S \ref{EnergyExchange}
from the more general phase-slippage theorem governing the exchange of energy
between potential and vortical flows.
 
In the event that $\mb v_\m p$ can be considered as constant over the surface
spanned by the vortex loop, Eq.(\ref{CrossIntegral3}) becomes even simpler:
\begin{equation}        \label{CrossIntegral4}
  E_\m {\,I} = \rho_{\m s} \kappa_4 \mb S\cdot \mb v_\m p \; ,
\end{equation}
in which $\mb S$ is the vectorial area of the loop, $\int \m d\mb S =
(1/2)\,\oint \mb r\times\m d \mb l$, $\m d\mb l$ being the line element at point
$\mb s$ of the oriented loop.

The total energy of the vortex immersed in an applied flow field is the sum of
its energy in the rest frame, $E_0$, given by Eq.(\ref{ReducedEv2}), and the
energy of interaction with the potential flow, $E_\m {\,I}$. For
Eq.(\ref{CrossIntegral4}), this reads
\begin{equation}        \label{Impulse}
  E_\m v = E_0 + \mb P\cdot \mb v_\m p\; , {\mbox{ with }} \mb P =
      \rho_\m s \kappa_4 \mb S \; ,
\end{equation}
where $\mb P$ ican be defined as the impulse of the vortex loop. 

For a circular loop of radius $R$, a vortex ring, Eq.(\ref{Impulse}) gives the
well-known result \citep{Lamb:45}:
\begin{equation}        \label{RingImpulse}
  P_\m R =  \pi \rho_\m s \kappa_4 R^2 \; .
\end{equation}

It emerges from this derivation (and the various approximations made along the
way) that, under a Galilean boost, vortex loops do behave as Landau
quasiparticles, with an energy proportional to their length and an impulse
proportional to their area. This approach puts some flesh on the bare bones of
the conventional (and exact) fluid-mechanical vortex dynamics; it gives
substance to the intuitive view than they can be treated as independent
elementary entities. This physically meaningful manner of separating the
vortical flow from the local value of the remotely potential superflow $v_{\m
  p}$ will prove quite useful in the following.

\subsubsection{Vortex self-velocity}

The impulse is not simply a plain geometrical quantity as
Eqs.(\ref{CrossIntegral4}) or (\ref{Impulse}) would let think. It is the
resultant of the impulsive pressures that must be applied to the fluid at rest
to create the vortex loop from rest.\footnote{See \citet{Lamb:45}, \S 152.} It
possesses some of the properties of a true momentum. For instance, the
propagation velocity of the vortex ring, Eq.(\ref{RingVelocity}), can be
expressed as the group velocity associated with the energy (\ref{RingEnergy})
and impulse (\ref{RingImpulse}) \citep{Langer:70,Roberts:70}:
\begin{equation}        \label{GroupVelocity}
  v_\m R = \frac{\m d E_\m R}{\m d P_\m R} = 
      \frac{\kappa_4}{4\pi R}\,\left(\ln
    \frac{8R}{a_0}-\frac{3}{4}\right) \; , 
\end{equation}
Expression (\ref{GroupVelocity}) tends asymptotically to the actual velocity
of a ring with a hollow core as computed directly from the Biot and Savart
law,\footnote{See \citep{Lamb:45}, \S 163.} which moves along its symmetry
axis $\hat{\mb n}$ with velocity
\begin{equation}        \label{RingVelocity}
  \frac{\m d\mb s}{\m d t} = \mb v_\m R = 
    \frac{\kappa_4}{4\pi R}\,\left(\ln
    \frac{8R}{a_0}-\frac{1}{4}\right)\hat{\mb n} \; .
\end{equation}
\noindent

However, these simple properties do not imply that a vortex has actual linear
momentum. The vortical impulse is more elusive. For instance, it can be shown
that a vortex ring moving freely under its own force at velocity $v_\m R$ and
impinging on a wall exerts no force on it \citep{Fetter:72}.  This somewhat
counter-intuitive result arises from the distribution of the flow around the
vortex loop \citep{Cross:74}. The contribution of the flow that goes in the
forward direction, and which causes the ring free motion, does impart a
momentum impulse into the wall equal to $P_\m R$, but the returning fluid away
from the ring, the backflow, yields an opposite contribution that leads to
full cancellation of the momentum transfer recorded over an infinitely
extended wall for the complete collision event. This push and pull action
constitutes a reminder that actual momentum is carried by the individual fluid
elements and that a vortex is a hydrodynamical object made up of many of those
elements.

Isolated circular rings propagate undistorted under their own velocity field
in the superfluid at rest for symmetry reasons.. Only a few vortex shapes propagate
undistorted in their own velocity field. Straight vortex pairs and helical
vortices are other examples \citep{Langer:70}.

For an arbitrarily curved vortex, the self-velocity of each curve element can
be approximated by Eq.\,(\ref{RingVelocity}), $R$ being replaced by the local
radius of curvature, $r_\m m= |\m d^2 \mb s/ \m d\xi^2|^{-1}$, parameter $\xi$
being the line length of the curve represented by $\mb s(\xi)$.  The validity
of this ``local induction'' approximation, which requires that $r_\m m$ be
large with respect to the vortex core radius, has been discussed in particular
by \citet{Schwarz:78,Schwarz:85} who has used it in extensive numerical
simulations of 3D vortex motion.

\subsubsection{The vortex mass}
\label{VortexMass}

The impulse of a vortex discussed above is in no way related to the vortex
self-velocity as the product of this velocity by an inertial mass.  The
problem of the mass of a vortex has been a long lasting riddle, which has now
been resolved in a satisfactory way in superfluid $^4$He.\footnote{Notably
  from the work of \citet{Baym:83}, \citet{Duan:94}, and \citet{Sonin:98}.}

This mass arises from several contributions.  If it is assumed that the vortex
has a hollow core of radius $a_0$ and that the compressibility of the
surrounding superfluid in rapid rotation can be neglected, the vortex mass is
simply the mass of the displaced fluid. For a cylindrical body, this amounts
to $\pi \rho_\m s a_0^2$ per unit length, a standard result of classical fluid
dynamics~\cite{Lamb:45}. The minuteness of $a_0$ in $^4$He,
$1\sim2$ \AA, of the same order as the interparticle spacing, makes this
contribution very small.

However, compressibility cannot be neglected in the vicinity of the vortex
core because the peripheral velocity, Eq.(\ref{VortexFlow}), becomes
large. The corresponding pressure drop is given by the Bernoulli equation:
\begin{equation}        \label{Bernoulli}
  \frac{\inc P}{\rho_\m s} = -\frac{1}{2} \inc(v^2_\m s) = 0 \; .
\end{equation}
The change in density at distance $r$ from the core where the velocity is
$\kappa_4/2\pi r$ is then:
\begin{equation}         \label{DensityChange}
  \inc \rho_\m s = \frac{\inc P}{c_1^2} =
  \frac{\kappa_4^2}{8\pi^2}\,\frac{\rho_s}{c_1^2}\, \frac{1}{r^2} \; ,
\end{equation}
using the relation between the (first) sound velocity and the compressibility
$c_1 = (\partial P/\partial \rho)^{-1/2}$,\footnote{See \citet{Landau:Hydro},
  \S 131.} which is justified when the normal fluid fraction is small ($\rho_\m
s \simeq \rho$). The overall change of mass about a unit length of vortex
filament arising from the fluid compressibility is obtained by integrating
Eq.(\ref{DensityChange}) over space:
\begin{equation}         \label{CompressibilityMass}
  \mu_\m v = \int_{a_0}^{r_\m m}\!\!\int_0^{2\pi}\!\!\int_0^1  
      \inc \rho_\m s r \m d r\m d \theta\m d z =
  \frac{\kappa_4^2}{4\pi}\,\frac{\rho_s}{c_1^2}\, \ln\frac{r_\m m}{a_0} \; .
\end{equation}
The vortex mass per unit length $\mu _\m v$diverges logarithmically with $r_\m
m$ and ranges from negligible for $r_\m m \sim$ a few core radii to important
for large vortices, $r_\m m/a_0 \gtrsim 10^3$. However, in most cases, the
mass of the vortex remains small and can be neglected except for high
frequency phenomena \citep{Baym:83,Sonin:87} and, possibly, for quantum
tunnelling \citep{Volovik:97}.

The Bernoulli effect, Eq.(\ref{DensityChange}), also causes $^3$He impurities
and ions to be trapped on the vortex cores because their chemical potential
decreases with the $^4$He density. They prefer to sit in low density regions
of the fluid. Trapped impurities add their own inertial mass $m_\m I$ to that
of the core. In superfluid $^3$He, the core is large and yields the dominant
contribution to the vortex mass
\citep{Kopnin:78,Kopnin:95,Duan:92,Volovik:97}.

\subsection{Energy exchange between potential and vortical flows}
    \label{EnergyExchange}
%

\begin{figure}[t]       
  \begin{center}
    \includegraphics[width=4cm]{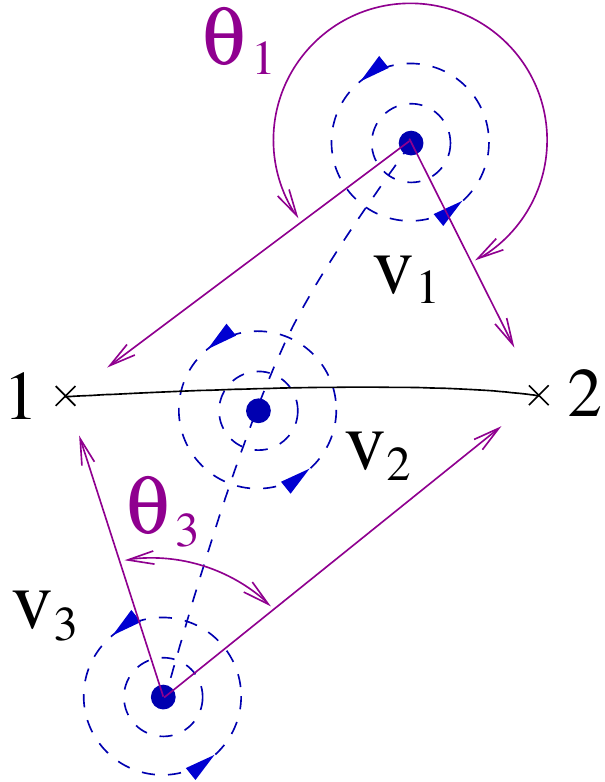}
    \caption{\label{LiftForce}(Color online) Stream of vortices $\m v_1$, $\m
      v_2$, $\m v_3$, \ldots\, crossing a path between point 1 and point 2,
      moving from bottom to top. The vortices are represented by dots and are
      assumed perpendicular to the figure and nearly straight in the vicinity
      of 1-2. Each contributes angle $\theta_i$ to the phase difference
      recorded between 1 and 2. In its travel from far below to far above line
      1-2, each vortex contributes $2\pi$ to the phase difference. According
      to the {\it ac} Josephson effect, a pressure difference develops in the
      superfluid due to the stream of vortices (see
      Eq.(\ref{AndersonPhaseSlippage}).  }
  \end{center}
\end{figure}

Following the insight of \citet{Anderson:66a}, the idea that phase slippage by
moving vorticity causes dissipation in superfluids and superconductors has
become conventional wisdom. If, referring for instance to the situation of
Fig.\ref{LiftForce}, there is not just one vortex as in Fig.\ref{paths} but a
constant stream of vortices crossing the path 1-2 at a rate of $n$ per second,
driven by some external force, a pressure difference develops in the
superfluid according to the Josephson {\it ac}-relation (\ref{acJosephson}).
When the superfluid is free to move, it is accelerated by the cross stream of
vortices: work is done onto the superfluid by the applied external force, for
instance an electric field acting on charges trapped in the vortex cores. This
Section dwells on the mechanism for this exchange of energy between the purely
potential superflow and vorticity.

\citet{Anderson:66a}  noted in an appendix entitled ``A `new' corollary in classical
hydrodynamics'' that, whenever there exists a steady stream of
vortices, for instance at the mouth of an orifice, the quantum phase in the
superfluid would change there at a constant rate and, according to
Eq.(\ref{acJosephson}), the following chemical potential difference would build up
\begin{equation}         \label{AndersonPhaseSlippage}
  \hbar \frac{\m d n_{\m{vortices}}}{\m d t}\Big |_1^2 = \bigg < \hbar \,\frac{\m
    d(\varphi_2 - \varphi_1)}{\m d t} \bigg > = \big < \mu_2 - \mu_1 \big > \; .
\end{equation}
In Eq.(\ref{AndersonPhaseSlippage}), the brackets stand for time-averaging and
the quantity $\m d n_{\m{vortices}} / \m d t$ for the rate of passage of
vortices across a line joining points 1 and 2, as depicted in Fig.\ref{LiftForce}.

This result is of no special importance in classical hydrodynamics because the
velocity circulation carried by each vortex, albeit constant, can take any
value, while in the superfluid it is directly related to the phase of the
macroscopic wavefunction and quantised . A formal proof of this conjecture,
based on the standard decomposition of any vector field into an irrotational
contribution and a solenoidal one was given by
\citet{Huggins:70}.\footnote{See also \citep{Zimmermann:96} and
  \citet{Greiter:05}) for alternate derivations.}  The following derivation is
based on the more physical approach to vortex dynamics, which makes use of the
concepts of force and energy.

\subsubsection{The Magnus force}

\begin{figure}[t]          
  \begin{center}
    \includegraphics[width=2.5cm]{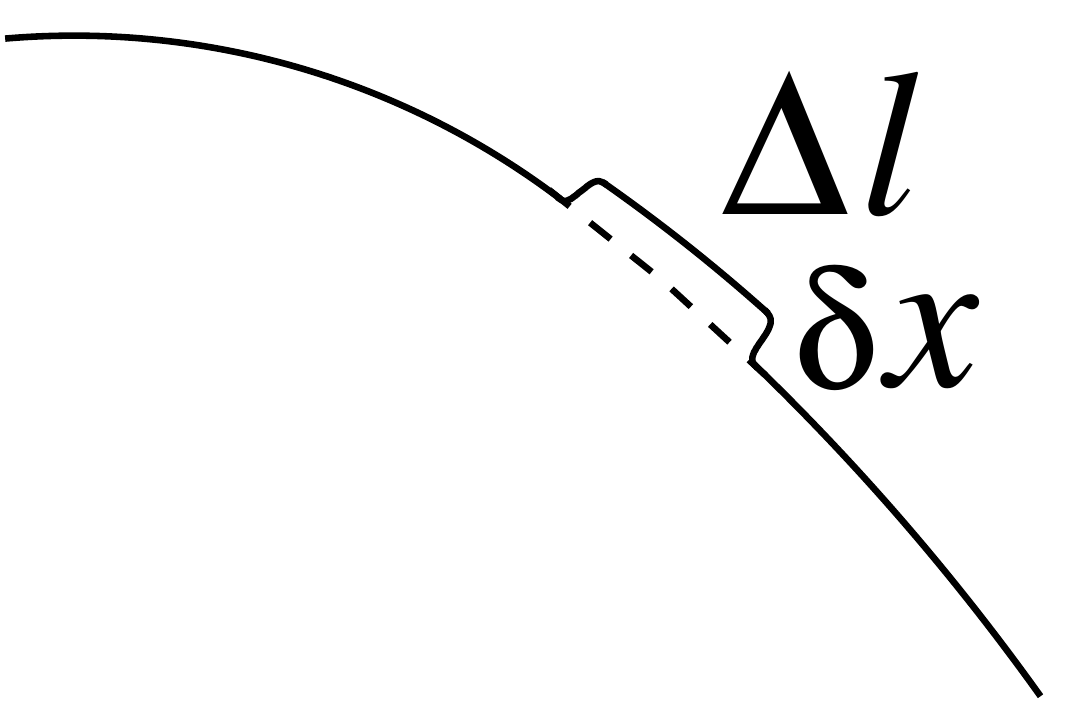}
    \caption{ \label{VirtualDisplacement} Virtual displacement by $\inc x$ of
      a small line element $\Delta l$. }
  \end{center}
\end{figure}

Consider the interaction energy between a vortex loop and a potential flow
$\mb v_\m p$, Eq.(\ref{CrossIntegral4}). Under an infinitesimal displacement
$\inc \mb x$ of a small line element $\Delta \mb l$, as shown in
Fig.\ref{VirtualDisplacement}, the energy of the vortex loop changes according
to
\begin{equation}        \label{DerivativeEI}
  \inc(\Delta E_\m I) = \kappa_4 \rho_\m s\,  \inc\mb x\times \Delta \mb l\cdot 
    \mb v_\m l 
    = \kappa_4 \rho_\m s\, \Delta \mb l\times \mb v_\m l \cdot\inc\mb x \; ,        
\end{equation}  
where $\mb v_\m l$ is the local superflow velocity as seen by the vortex
element standing still.  The local flow velocity $\mb v_\m l$ is the sum of
the applied superflow $\mb v_{\m p}$ and the flow induced by the other parts
of the vortex loop, $\mb v_\m v$. Equation (\ref{DerivativeEI}) expresses the
functional derivative $\inc(\Delta E_\m I)/ \inc\mb x$ of the energy with
respect to an infinitesimal deformation of the vortex line.

If the vortex loop moves along at velocity $\mb v_{\m{loop}}$ together with the
element under consideration in the rest frame of the observer, $\mb v_\m l$ in
Eq.(\ref{DerivativeEI}) becomes $\mb v_\m l - \mb v_{\m{loop}}$ and this force takes
the same form as the Magnus force for a line vortex in
classical hydrodynamics with a fluid density $\rho_\m s$:\,\footnote{See
  \citet{Sonin:97} for a complete discussion of the Magnus force in classical
  fluids, neutral superfluids and charged superfluids.} 
\begin{equation}        \label{MagnusForce}
  \frac{\inc(\Delta E_\m I)}{\inc\mb x} =
  \Delta \mb F_\m M =  \kappa_4 \rho_\m s (\mb v_{\m{loop}} - \mb v_\m l)\times
  \Delta\mb l \; ,   
\end{equation}

\begin{figure}[t]         
  \begin{center}
    \includegraphics[width=7cm]{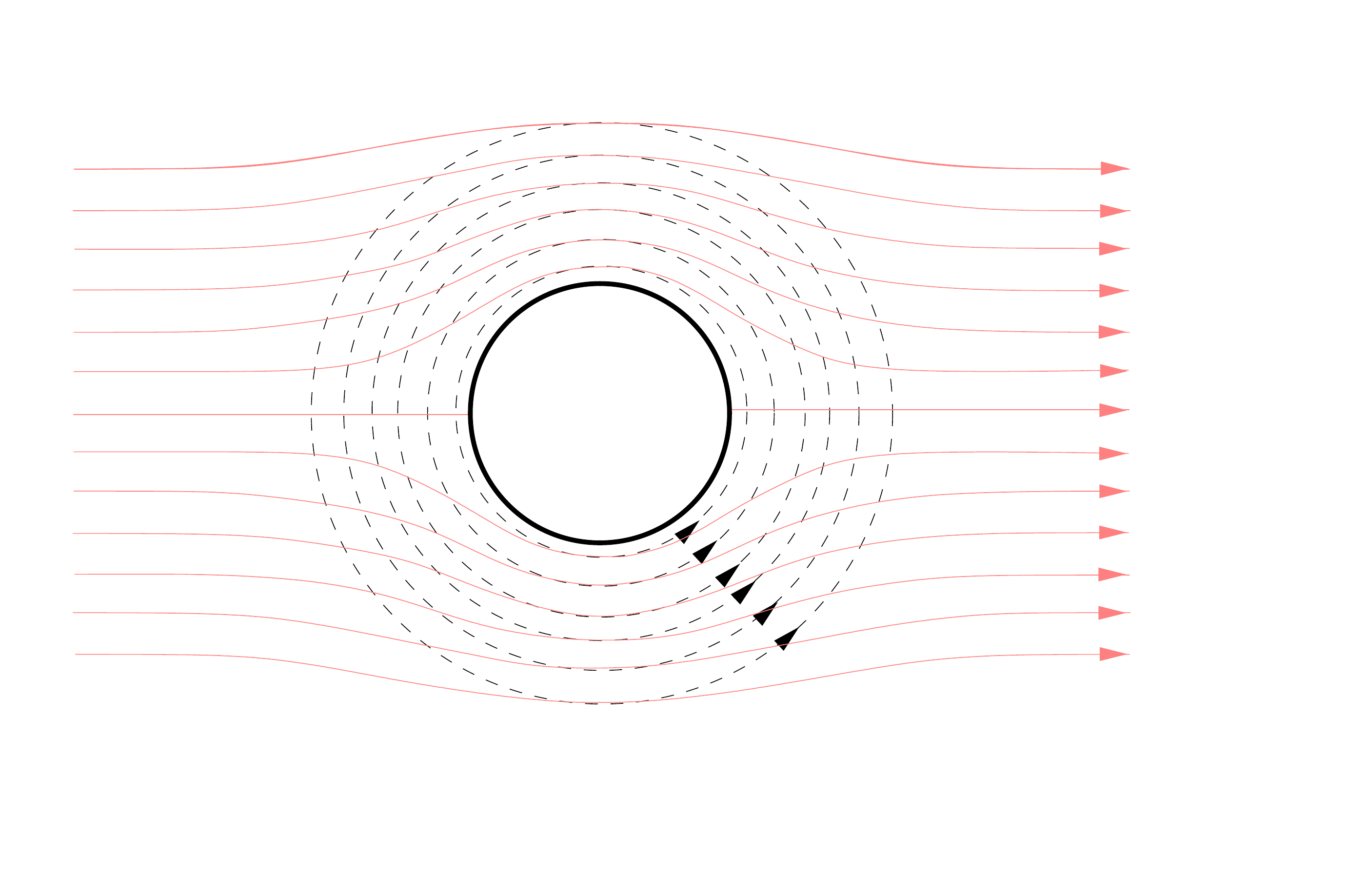}
    \caption{ \label{MagnusBernoulli} (Color online) Potential streamlines past the
      cylindrical vortex core flowing from left to right in thin plain lines.
      The flow from the vortex, in dashed lines, subtracts from the potential
      flow in the upper half and adds to it in the lower half, inducing a
      Bernoulli pressure difference between top and bottom. }
  \end{center}
\end{figure}

The Magnus force, Eq.(\ref{MagnusForce}), has a simple physical origin. It is
due to the Bernoulli effect that arises from the rotational flow around the
vortex core. As shown in Fig.\ref{MagnusBernoulli}, this flow adds to the
potential flow in the lower half-plane and subtracts from it in the upper
half-plane.  Integrating the resulting pressure difference obtained from the
Bernoulli Eq.(\ref{Bernoulli}) over the cylinder yields a downward force
expressed by Eq.(\ref{MagnusForce}).
The Magnus force on each element of the vortex line arises ultimately from
momentum conservation in the fluid and comes into play whenever the vortex
trajectory differs from that of the local fluid particles. When no other force
acts on the vortex core (such as, e.g., an electric field on charges trapped
in the core, or friction from the normal fluid component, \ldots) $\mb F_\m M$
must be zero, hence $\mb v_{\m{loop}} = \mb v_\m l$: the vortex core moves
with the local superfluid velocity.  The velocity of the core at point ${\mb
  s}$ is the sum of the velocity of the local potential flow ${\mb v}_\m p$ at
${\mb s}$ when there is no vortex, and of the velocity ${\mb v}_{\m v}$
induced at ${\mb s}$ by the other parts of the vortex.  If no flow is applied,
$\mb v_\m p =0$, then $\mb v_{\m{loop}} = \mb v_\m v$: the vortex loop moves
under its own flow field in the superfluid at rest at large distance. The
vortex thus appears to behave as a quasiparticle in its own right although it
stands only for the vortical part of the total flow. The physical picture that
emerges from this approach rings a familiar bell to condensed matter
physicists.

\subsubsection{Quantised vorticity and the Kelvin-Helmholtz theorem}

That free vortex loops moves with the local fluid particles conforms to the
Kelvin-Helmholtz theorem. This result has been obtained here as a consequence
of the quantisation of circulation, Eq.(\ref{VortexFlow}). The
Kelvin-Helmholtz theorem is usually derived from the Euler equation and the
implicit assumption that the motion of the fluid is isentropic
\citep{Landau:Hydro}.\footnote{See \S 8.} A further implicit assumption is
that the velocity field and the loop deformation are well-behaved
analytically, that is, continuous in space and time.\footnote{For a
  discussion, see \citet{Saffman:92} \S 1.6.} The relevance of these remarks
will become apparent in Sect.\ref{VortexNucleation} on vortex nucleation,
which deals with the spontaneous appearance of vorticity, in other words, the
violation of the Kelvin-Helmholtz theorem. The derivation given above does not
hide these fine points under the rug; it explicitly rests on the quantisation
of circulation, hence its conservation, and also implies isentropic and
continuous superfluid motion. When this fails new phenomena occur: vortices
may be nucleated.

As the effect of external forces and mutual friction has been set aside for
simplicity, no work is done on the vortex itself except by the interaction
with the local superflow. Thus any gain or loss of energy by the
vortex balances that lost or gained by the potential flow. The way by which
this conservation of energy proceeds is instructive; the detailed analysis is
given in the following.

\subsubsection{The phase slippage theorem}

If $\inc \mb x$, used in Eq.(\ref{DerivativeEI}) as a virtual displacement
to compute the forces acting on $\Delta \mb l$, becomes a real displacement $\mb
v_{\m{loop}}  \Delta\m t$, actual work during the time $ \Delta\m  t$ is done by the
applied potential flow on the vortex loop. The energy balance is expressed by
rewriting Eq.(\ref{DerivativeEI}) as
\begin{equation}        \label{EnergyBalance}
  \begin{split}
    \inc(\Delta E_\m I) &= \kappa_4 \rho_\m s\, \Delta \mb l\times 
    (\mb v_\m p+\mb v_\m v)\cdot \inc \mb x \\
    &= \kappa_4\rho_\m s\,\Delta\mb l\times \mb v_\m p\cdot \mb v_{\m{loop}} \Delta t +
    \kappa_4 \rho_\m s\,\Delta \mb l\times \mb v_\m v\cdot \mb v_{\m{loop}} \Delta t 
    \; .     
  \end{split}
\end{equation}  

In free motion -- disregarding friction of the core on the normal component
and with no force applied externally -- the vortex loop follows the fluid
stream: $\mb v_{\m{loop}} = \mb v_\m l=\mb v_\m p+\mb v_\m v$. The triple
products are equal in magnitude and opposite in sign. The energy increment
expressed by Eq.(\ref{EnergyBalance}) is equal to zero. 
Total energy is conserved in the course of the vortex motion by the balance of
the two terms in the last equality (\ref{EnergyBalance}). The first, rewritten as
\begin{equation}         \label{PotenialFlowCutEnergyChange}
  \inc(\Delta E_\m I)_1    =  \kappa_4 \rho_\m s\, ( \mb
    v_{\m{loop}} \Delta t \times \Delta \mb l) \cdot \mb v_\m p \; ,
\end{equation}
is readily seen proportional to the rate at which the potential flow
streamlines are crossed by the vortex element $\Delta \mb l$. It expresses the
change of the potential flow kinetic energy when its streamlines are crossed by the
vortex line, causing a change of the phase difference of $2\pi$ along them.

The second term requires a little more formal work to be recognised as a
contribution to the vortex self-energy $E_\m v$. What needs to be shown is
that it corresponds to the energy variation for a small, local deformation of
the vortex loop. This is established in Appendix A with the following result,
\begin{equation}         \label{VortexSefEnergyChange}
  \inc(\Delta E_\m I)_2 = \kappa_4\rho_\m s\,\Delta\mb l\times \mb v_\m v\cdot
  \mb v_{\m{loop}} \Delta t 
   =  \Delta E_\m v(\inc \mb x,\mb v_{\m{loop}}) \; ,
\end{equation}
for the displacement $\inc \mb x = \mb v_{\m{loop}} \Delta t$ of the loop element $\Delta \mb l$. 

The energy balance expressed by Eq.(\ref{EnergyBalance}) between the potential
flow kinetic energy and the vortex self-energy constitutes the fundamental
relation governing phase slippage.  In integral form, it yields
  Eq.(\ref{CrossIntegral3}). It shows the way by which a vortex loop of
arbitrary shape can form by expanding from an infinitesimal loop. 

The gist of Eq.(\ref{VortexSefEnergyChange}) is that whenever a vortex cuts
potential flow streamlines, it reversibly exchanges energy with the potential
flow and it concurrently changes the velocity circulation along these
streamlines by one quantum unit, causing slippage of the quantum phase. This
process takes place in real time and locally, not only in a time-averaged
fashion as in Anderson's conjecture, Eq.(\ref{AndersonPhaseSlippage}).  If the
potential flow is divergent -- for instance outward the mouth of a duct where
the streamlines flare out, the vortex expands in length, collects energy from
the flow and slows it down. If the vortex runs away from that point to a far
off distance and never comes back, this energy is irreversibly lost for the
potential flow: dissipation of superflow energy has occurred.  Reversing the
flow direction, which then becomes convergent, results in the vortex shrinking
and the potential flow picking up energy: a collapsing vortex dumps its energy
into the potential flow and speeds it up.

These processes alter the quantum phase and will be discussed in
Sec.\ref{HalfRingModel} on the phase slip mechanism. But before turning to the
inner details of the phase slips, their experimental observations will be
briefly sketched in the following Section.


\section{Phase slippage experiments}\label{slippage}


As the {\it dc} and {\it ac} effects predicted in the early sixties by Brian
Josephson \citep{Josephson:62,Josephson:64,Josephson:65} to take place between
two suitably coupled superconductors were quickly observed
\citep{Anderson:63,Shapiro:63}, the search for analogous effects in
superfluids also begun, with the tantalising goal of observing unique
quantum-mechanical effects in hydrodynamics. This search for a long time gave
inconclusive results,\footnote{See the work of \citet{Richards:65},
  \citet{Khorana:67}, \citet{Khorana:69}, \citet{Richards:70},
  \citet{Guernsey:71}, \cite{Gregory:72}, \citet{Hulin:72}.}  or led to blind
alleys.\footnote{As mentioned by \citet{Schofield:71}, \citet{Musinski:72},
  \citet{Musinski:73}, \citet{Gamota:74}.}  It was only in the mid-eighties
that decisive steps forward were taken.\footnote{The work of
  \citet{Avenel:85}, \citet{Avenel:86b}, \citet{Varoquaux:87},
  \citet{Amar:90}, \citet{Amar:92}, \citet{Zimmermann:93b},
  \citet{Zimmermann:96} is described below.}

\subsection{The Richards-Anderson experiment}

\begin{figure}[t]       
  \begin{center}
    \includegraphics[height=9cm]{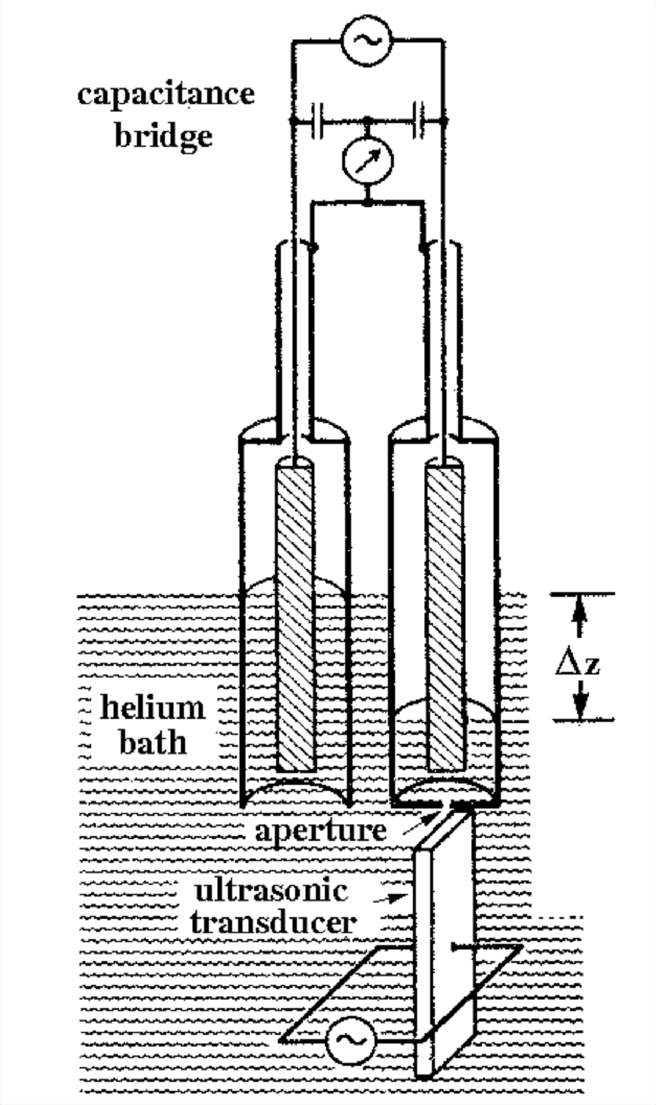}
    \caption{ \label{RichardsAndersonCell} The Richards-Anderson cell
      (1965).} 
  \end{center}
\end{figure}

In order to observe the Josephson {\it ac}-effect in superfluid helium,
\citet{Richards:65} designed an experiment based on the beat note expected to
form between the sound wave emitted by a quartz transducer immersed in the
superfluid and the internal pressure fluctuation due to the {\it ac}-effect.  In this
historical setup, shown in
Fig.\ref{RichardsAndersonCell},\footnote{Later refined by
  \citet{Richards:70}} two identical coaxial capacitors are suspended over a
liquid helium bath cooled at a temperature below the lambda point (of the
order of 1.15 K). One of the capacitors is fully open-ended, the other is
partially closed at the bottom by nickel foil with a very small aperture. The
foil is 25 micrometres thick, in which a 15 micrometre aperture had been
punched with a sharp needle: the pinhole thus manufactured constitutes the
``weak link'' between the two superfluid pools.

If a helium level difference $\Delta z$ between the two coaxial capacitors is
created by lowering and raising the whole assembly over the liquid helium
bath, the return to hydrostatic equilibrium is impeded by the pressure head of
the steady stream of vortices corresponding to
Eq.(\ref{AndersonPhaseSlippage}). The level difference can be precisely
monitored by a capacitance bridge. When an ultrasound wave is shone by a
quartz transducer facing the micro-aperture as shown in
Fig.\ref{RichardsAndersonCell}, it can couple to the stream of vortices
and modulate the flow.

Steps in the return to equilibrium were indeed observed at level differences
which were multiples and submultiples of the fundamental head difference
frequency expected from the Josephson {\it ac} relation:
$
\Delta z  = n\,\hbar\omega\, /\,n'\,m_4 g \; 
$
where $n$ and $n'$ are integers, and $g$ the acceleration of gravity.
Richards and Anderson's results were reproduced by other researchers using
similar setups, notably \citet{Khorana:67}, \citet{Khorana:69}, \citet{Hulin:71}, and
\citet{Hulin:72}.  Different setups, involving rotating or oscillating
toroidal cells \citep{Guernsey:71,Gregory:72}, vortices accelerated by ions
\citep{Carey:73}, a two-orifice flow arrangement \citep{Gamota:74} were also
tried but with mixed success at best, suffering from lack of reproducibility
and poised with numerous unexpected features.

It eventually became clear that the early claims of observation of the
Josephson {\it ac} effect by synchronisation of the pressure head on the sound
frequency did not meet universal acceptance. On the contrary, an alternate
explanation in terms of acoustic standing waves in the cell was put forward on
experimental grounds by \citet{Leiderer:73}, as well as \citet{Musinski:72}
\citep{Musinski:73}, and on theoretical grounds by \citet{Rudnick:73}. It was
nonetheless argued by \citet{Anderson:75a} that, although acoustic resonances
in the cell could be a concern, they could not account for all of the features
observed in their experiments.

These efforts directed toward the demonstration of the hydrodynamic Josephson
effects, together with direct studies of the critical velocity itself
\citep{Trela:67,Gamota:73}, did bring experimental confirmation of the views
of Feynman and Anderson that vortices were associated with the appearance of
dissipation in superfluid flow. However, quantitative studies leading to a
clear picture of how these vortices were created and how they interacted with
the superflow were lacking. A consensus grew that somehow their formation and
evolution had a chaotic character, presumably due to random pre-existing
vorticity in the superfluid and to a probable evolution toward some form of
turbulent motion of the quantised vortices, a belief confirmed in part by the
more recent studies described in Sec.\ref{AllThat}. The flurry of activity
stirred by the initial reports of observations related to the Josephson
effects in helium receded almost completely.

With the hindsight gained from the experiments described further on, it can
now be concluded that the synchronisation envisioned by \citet{Richards:65}
would be nigh impossible to achieve. A particularly clear exposition of this
synchronisation experiment is given by \citet{Anderson:84} in terms of
parametric effects due to the system non-linearities, of the same kind as
frequency pulling in radiofrequency oscillators. These effects require that
energy be stored reversibly in a non-linear element, here the Josephson
junction or, for {\it rf}-devices, a non-linear inductance. In $^4$He far from
the \greeksym{l}-point the relation between the current and the phase
difference across the weak link shows no non-linearities. The energy that the
vortices gather from the potential flow is carried swiftly away from the
orifice and is irreversibly lost. It can be used to pull or push the flow in
synchronism with the sound excitation for a very brief lapse of time only,
much shorter than the period of the audiofrequencies used in these
experiments.\footnote{As will become clear in the discussion of phase slip
  mechanism in \S\ref{HalfRingModel}} Furthermore, for the comparatively large
orifices used then, vortices appear in a rather irregular fashion, not
individually but in lumps with varying numbers, as discussed in
Sec.\ref{LargeSlips}. These peculiarities hamper the eventual synchronisation
to a regular pattern of steps.

\subsection{The hydromechanical resonator}

In the early eighties, several groups went on striving to improve the
detection techniques used in the search for the hydrodynamical Josephson
effects. The use of a diaphragm-driven hydromechanical resonator fully
immersed in the superfluid was pioneered by Zimmermann, Jr., and his
students.\footnote{Namely \citet{Wirth:81,Anderson:84,Beecken:87a}.} A similar
device with two chambers was built by \citet{Manninen:83} for critical velocity
measurements in superfluid $^3$He, and used by \citet{Lounasmaa:83} for the
search of an {\it ac} Josephson effect in superfluid $^3$He, a topic that will
be covered in Sect. \ref{CPR}. The expertise developed at Cornell on torsional
oscillators was put to use in superfluid $^3$He by Reppy and his students
\citep{Crooker:84}. Again, the hydrodynamic Josephson effects could not be
observed in these various experiments, for one or several of the following
reasons:
\begin{trivlist}
  \item{$\bullet$~the apertures used as weak links were too large;}
  \item{$\bullet$~the mass flow rate sensitivity was marginally adequate only;}
  \item{$\bullet$~the superfluid motion was driven from current sources that were too
      stiff to let the response of the weak link be seen;}
  \item{$\bullet$~and, last but not least, the cells were too bulky and too
      sensitive to external mechanical vibrations to allow for non-invasive
      measurements.} 
\end{trivlist}  

The first reason was clearly perceived as essential. Efforts  shifted
from superfluid $^4$He to the newly discovered superfluid $^3$He because the
coherence length is two orders of magnitude larger, putting the
fabrication of a genuinely-weak superfluid link within reach of
experimental low temperature physicists. Work was carried out in that
direction by \citet{Wirth:81}, who were the first to use sub-micronic orifices
in free-standing ultra-thin foils, and others
\citep{Sudraud:87},\citep{Amar:90}.

The detection of the minute mass currents that would flow in micro-apertures
improved markedly in the early eighties as reliable {\it rf}-SQUIDs became
available.\footnote{SQUID is an acronym for Superconducting QUantum
  Interferometric Device. The present sensitivity of {\it dc}-SQUID based
  displacement sensors used in the phase slippage described further on is
  $\sim 10^{-15}$ m/$\sqrt{\m Hz}$, or one fermi per root hertz.}
Ultra-sensitive pressure and displacement gauges could then be developed
\citep{Avenel:86}.

\subsection{Early  phase slippage experiments }

\begin{figure}[!t]      
  \begin{center}
    \includegraphics[width=45 mm,angle=0]{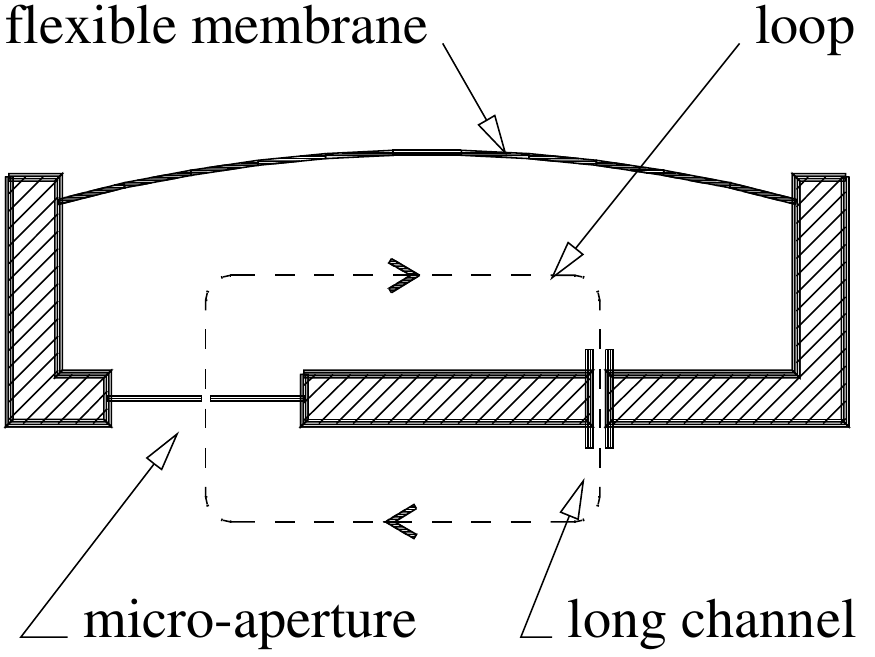}
    \caption{\label{Cell} Schematic drawing of the flexible-diaphragm
      double-hole hydromechanical resonator. The dashed line shows a closed
      loop threading the two holes -- micro-aperture and long channel -- 
      located entirely in the superfluid.  }
  \end{center}
\end{figure}

The phase slippage experiments that were carried out starting from the
mid-eighties using these refined techniques \cite{Varoquaux:87,Avenel:85}
confirmed Feynman and Anderson's views on dissipation in superflows and
brought a large measure of clarification to the critical velocity problem
\cite{Varoquaux:91} and to the formation of vortices in superfluid $^4$He
\cite{Avenel:93}. These experimental results and their interpretation have
since been largely confirmed by other workers.\footnote{See the accounts of
 \citet{Zimmermann:96}, and \citet{Packard:98}. }

The design of the first weak link in which hydrodynamical Josephson effects
were seen \citep{Sudraud:87} struck a compromise between to two conflicting
requirements, that it be weak enough to effectively depress the wavefunction
amplitude while preserving the macroscopic coherence of the superfluid, and
that it be big enough to let a measurable flow of liquid go through. A slit
geometry was chosen for the micro-aperture, the smaller dimension of which was
comparable with the coherence length in superfluid $^3$He, $\xi_0$, which is
in the sub-micron range (see \S\ref{CPR}). This orifice was micro-machined by
ion-milling in a free standing foil the thickness of which was also comparable
to $\xi_0$. The third dimension of the slit was made large to provide a
substantial cross sectional area through which the superfluid would flow.

Phase slippage was studied with the help of a miniature hydromechanical
device, represented schematically in Fig.\ref{Cell}, which is basically a
flexible-wall Helmholtz resonator with two vents, immersed in the superfluid
bath.\footnote{Calling the device a ``Helmholtz'' resonator has been
  criticised as the compressibility of the fluid inside the chamber has a
  negligible effect at the low frequencies of the experiments, hence the
  little more convoluted appellation used here. The term ``hydromechanical
  resonator'' is also used in this paper} The flexible wall is constituted by
a Kapton membrane coated with aluminium. In the version shown in
Fig.\ref{Cell}, there are two openings connecting the resonator chamber to the
main superfluid bath. One is a micro-aperture in which the critical velocity
is reached.  The critical event consists in a sudden change of the resonance
amplitude corresponding to a departure from the expected classical
hydrodynamics response of the flow velocity through the micro-aperture as
discussed in the following.  The other opening is a relatively open duct and
provides a parallel path to the superfluid, along which the quantum phase
remains well determined in all circumstances. The velocity circulation along
the superfluid closed loop threading the two openings shown in Fig.\ref{Cell}
changes by an integral number of quanta for each critical event.
\begin{figure}[!t]      
  \begin{center}
    \hskip -3mm
    \includegraphics[width=80mm,angle=180]{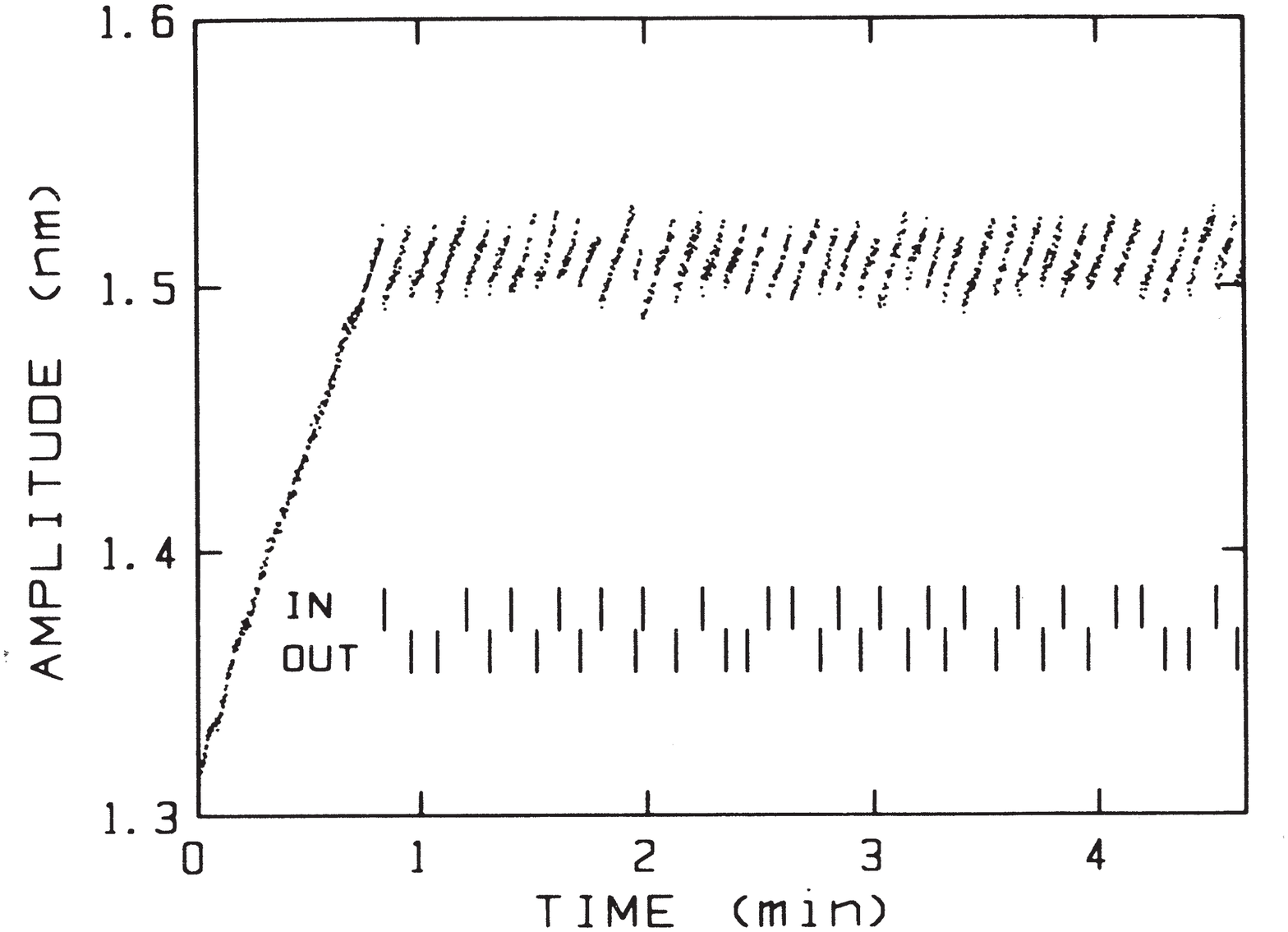}
    \caption{\label{PhaseSlipsCanJPhys} Time chart of the peak resonance
      amplitude of the resonator membrane, in nm, for positive and for
      negative excursions. Each dot represents a measurement. The time
      interval between individual measurements is half a period, 177 ms here.
      The ticks labelled ``in'' and ``out'' indicate whether the jumps occur
      when the liquid flows in or out of the cell. The drive power applied to
      drive the resonance is very small ($2.4 \; 10^{-18}$ W). When the peak
      amplitude is subcritical, its value builds up linearly with time as seen
      at the left of the chart. The critical events are sharply defined and
      quite reproducible but occur at a threshold that varies slightly from
      event to event.  The superfluid $^4$He contained 100 to 300 ppb of
      $^3$He and was cooled to about 10 mK under a very small hydrostatic
      pressure head. From \protect\cite{Avenel:85}.  }
  \end{center}
\end{figure}

The operation of these resonators is described in detail in the
literature.\footnote{See, for instance,
  \citet{Varoquaux:87}, \citet{Avenel:87}, \citet{Beecken:87},
  \citet{Varoquaux:94}, \citet{Avenel:95}.}  Flow is
driven in and out of the resonator by an electrostatic {\it ac}-drive applied
to the aluminium-coated Kapton membrane. The membrane is mounted in such way
as to be as flexible as possible; it provides the restoring elastic force in
the resonator.  The ``common mode'' flow of liquid in and out of the cell body
through the two vents of the resonator provides a force of inertia to the
hydromechanical device. These inertial and elastic terms determine the
resonance frequency.  The device is usually driven at or close to resonance in
continuous mode. The raw data consist of peak amplitude charts as represented
in Fig.\ref{PhaseSlipsCanJPhys}. A strong impulsive force may also be applied
to the membrane; the large transient response of the resonator reveals
additional features when it becomes non-linear.

In the absence of dissipation, the resonance motion under a small steady {\it
  ac} drive increases linearly in amplitude under the action of the drive as
energy gradually gets stored in the resonator. This linear rise on the left of
the trace in Fig.\ref{PhaseSlipsCanJPhys} proceeds until the flow velocity in
the micro-aperture becomes critical. Sudden drops of the peak amplitude from
one half-cycle to the next then appear. These drops signal that a lump of
resonator energy has been lost between two successive recordings of the
absolute peak amplitude. Quite importantly, these lumps are identical -- to
experimental uncertainty -- from event to event.

These events are interpreted as the footprint left by vortices crossing the
potential flow pattern in the vicinity of the micro-aperture. In
their course, they cut all potential flow lines, pick up energy at
the expense of the potential flow of the resonator and change the phase
difference along these flow lines by $2\pi$. Hence the name ``phase slips''.
These slips of the phase are sudden and take place at a fairly well defined
threshold, which defines the phase slip critical velocity. It will become
apparent in the following that this critical velocity, while signalling a
breakdown of superfluidity, differs from other quantities also called
``critical velocities''. These features are discussed in detail further on, in
Secs.\ref{PhaseSlipCriticalVelocity} and \ref{HalfRingModel}.

\subsection{Phase-slippage experimental results}

The observation of phase slips in $^4$He has led to a number of quite
significant results. They brought a confirmation of Anderson's ideas, much welcome in view
of the controversies about previous experiments. And quite importantly, they
have shed light on the previously indecipherable problems of the critical
velocity and of vortex nucleation. Below are summarised their most important
qualitative features and some of their implications.
\begin{enumerate}
\item{ The critical velocity threshold, which can be seen on time charts such
    as that shown in Fig.\ref{PhaseSlipsCanJPhys}, is markedly
    temperature-dependent down to below 200 mK and reaches a well-defined
    plateau below 150 mK.  These features are shown in Fig.\ref{vc} and
    are analysed in Sec.\ref{PhaseSlipCriticalVelocity} below. As the
    thermodynamic properties of superfluid $^4$He are very nearly independent
    of temperature below 1 K, this observation indicates that the critical
    process in action is not governed solely by hydrodynamics. Statistical
    mechanics may well play the leading role.  }
\item{ Aperture size is not found to be a relevant factor, as long as it is
    ``small enough'', roughly below a few \umu m. This feature and the temperature
    dependence mentioned above are in sharp contrast with the Feynman critical
    velocity, which, as discussed below in \S \ref{FeynmanCriticalVelocity},
    exhibits a well-characterised dependence on size and none on temperature
    (except very close to the $\ulambda$ transition).}
\item{ The actual velocity threshold for phase slips shows significant scatter
    from one slip to the next in a given sequence, as can be seen in
    Fig.\ref{TimeChart}.  This scatter lies much above the instrumental noise
    level of peak amplitude detection. It represents a genuine stochastic
    property of the physical process at work, which turns out to display a
    temperature dependence similar to that of the critical velocity itself, as
    shown in Fig.\ref{Delta-vc}.  }
\item{ The phase slip pattern shows quite reproducible properties in the
    course of a given cool-down as long as the experimental cell is kept at a
    temperature below 10$\sim$15 K. If the temperature is cycled up to liquid
    nitrogen temperature and down again, small changes to the critical threshold and
    to the pattern itself can occur. These changes reveal the importance of minute
    alterations in the surface state of the cell, e.g., contamination of the
    micro-aperture walls by solidified gases during thermal cycling. }
\item{ Quite importantly, phase slips are the signature that {\it quantised
      vortices} are created in aperture flow above a well-defined threshold of
    flow velocity.  This statement is based on the measured value of the phase
    change, found to be $2\pi$ to the accuracy of the experiment
    \citep{Avenel:85}. This amounts to a change of precisely one quantum of
    circulation in the superfluid loop threading the micro-aperture and the
    long parallel channel.\footnote{See Fig.\ref{Cell}, or Fig.\ref{Gyrometer}
      for a more realistic cell.} A detailed scenario for the occurrence and
    development of these phase slips has been described by \citet{Burkhart:94}
    and is discussed below in Sec.\ref{HalfRingModel}.  }
\end{enumerate}

Critical velocities and phase slips in the superfluid phases of $^3$He show
different features that will be briefly touched upon in Sect.\ref{AllThat}.

%
\section{Critical velocities in superfluids} 
%
\label{CriticalVelocities}

The critical velocity in a superfluid is defined as the threshold above which
the flow of the superfluid component becomes dissipative, that is, the
property of superfluidity is lost. This rather broad definition encompasses a
number of different physical situations. This Section begins with an overview
of the different brands of ``critical'' velocities that comply with this
definition. It will end up by focusing on that which involves the phase slip
phenomenon, namely, the nucleation of superfluid vortices.

Neither the problem of critical velocities in superfluids nor that of the
nucleation of vortices are new. The former is as old as the discovery of
superfluidity (see the monograph by \citet{Wilks:67}). The latter, first
discussed by Vinen in the early sixties \cite{Vinen:63}, has met an even more
tortuous fate. It was first thought, still was in some quarters not so long ago,
to be nigh impossible: such an extended hydrodynamical
object as a vortex line with a finite circulation involving the collective
motion of a large number of helium atoms would have a vanishingly small probability
of occurring spontaneously. More recent experiments probing superflow on a
finer scale of length have shown otherwise.\footnote{As reported by
  \citet{Muirhead:84}, \citet{Varoquaux:87}, \citet{Varoquaux:03}.} 

\subsection{The Landau criterion}

As discussed previously in \S\ref{TwoFluidModel},
\citet{Landau:41}\footnote{For complete accounts of Landau's work,
  see \citet{Khalatnikov:65} and \citet{Wilks:67}..} explained the absence of
dissipation in the flow of helium-4 by the existence of a sharply defined
dispersion curve for elementary excitations, the phonons and the rotons. This
property is now associated with the phenomenon of Bose-Einstein condensation
\citep{Griffin:87,Griffin:93}, as has long been suspected
\citep{London:54}. Elementary excitation energy levels $\epsilon(\mb p)$ are
well defined. They have a negligible spread in energy. States with very low
energy are thus extremely rare. This scarcity of low-lying states is held
responsible for the inviscid property of $^4$He.

An impurity, or a solid obstacle, can only exchange an energy $\epsilon(\mb
p)$ at momentum $\mb p$ that exactly matches the energy of an elementary
excitation of the fluid.  If the superfluid moves at velocity $\mb v_\m s$,
the energy of elementary excitations in the frame of reference at rest becomes
$\epsilon-\mb v_\m s\cdot \mb p$ \cite{Wilks:67,Baym:69}. The same holds for a
moving obstacle, by Galilean invariance. If this energy turns negative,
elementary excitations proliferate and superfluidity is lost. The condition on
the superfluid velocity for this to happen reads:
\begin{equation}        \label{LandauVelocity}
  v_\m s \geqslant  v_\m L = \frac{\epsilon(p)}{p}\Big|_\m{min} \simeq 
    \frac{\epsilon(p)}{p}\Big|_\m{roton} \; .
\end{equation}
Unless this condition is met, there is no dissipative interaction between
the fluid and its surroundings: the flow is viscousless.

The minimum value of $\epsilon/p$ for helium lies very close to the roton
minimum, as shown in Fig.\ref{DispersionCurve}. This means that rotons are
created when the Landau critical velocity is reached in $^4$He.  At low
pressure, the roton minimum parameters are such that $v_\m L \simeq 60$ m/s.
The Landau critical velocity has been observed under certain conditions in the
propagation of ions in which rotons are created in $^4$He as reviewed by
\citet{McClintock:95}.  The much less dense Bose-Einstein Condensed gases
sustain a phonon-like energy spectrum at low momentum \citep{Bogolyubov:47}
and no roton-like features; the Landau velocity is the sound velocity,
$c=\epsilon(p)/p|_{p=0}$ and phonons are emitted.  The Landau critical
velocity $v_\m L$ in superfluid $^4$He is smaller than the sound velocity
($c=220$ m/s at low pressure) but is still quite larger than the critical
velocities observed in most experiments.


\subsection{Feynman's approach}
\label{FeynmanCriticalVelocity}

\citet{Feynman:55} realised, following Onsager, that not only
would vorticity be quantised in $^4$He in units of the quantum of circulation
$\kappa_4=2\pi \hbar/m_4$\, but, preceding Anderson, that these
vortices would be responsible for the onset of dissipation and for a critical
velocity in the superfluid. In Feynman's views, vortices would be puffed out
of the mouth of orifices much in the way of smoke rings -- or von Karmann
alleys past obstacles in classical (Navier-Stokes) fluids.

Such vortex rings can be treated as elementary excitations of the superfluid,
which they rightfully are from the vantage point taken in
Sec.\ref{VortexDynamics}. Hence Landau's criterion applies.  The limiting
velocity associated with these vortex rings, assumed to be circular, can be
evaluated from the expressions for the energy $E_\m R$ and impulse $P_\m R$,
Eqs.(\ref{RingEnergy}) and (\ref{RingImpulse}). The critical value set by
Eq.(\ref{LandauVelocity}) is reached for a radius $R$ such that $E_\m R/P_\m
R$ is at a minimum, which occurs when $R$ is as large as feasible, that is, of
the order of the orifice size $d$. This minimum value sets the lowest velocity
at which vortices can start to appear and defines the Feynman critical
velocity:
\begin{equation}        \label{FeynmanVelocity}
  v_\m F \simeq \frac{\kappa_4}{2\pi d} \ln\left(\frac{d}{a_0}\right) \; .
\end{equation}
As discussed below, $v_\m F$ is much closer to experimental values than the
Landau critical velocity for rotons. Although this agreement is heartening, it
also raises questions: how do these vortices come about and how do they
evolve?

\subsection{Several kinds of critical velocities}

\begin{figure}[t]       
  \begin{center}
    \includegraphics[height=65 mm]{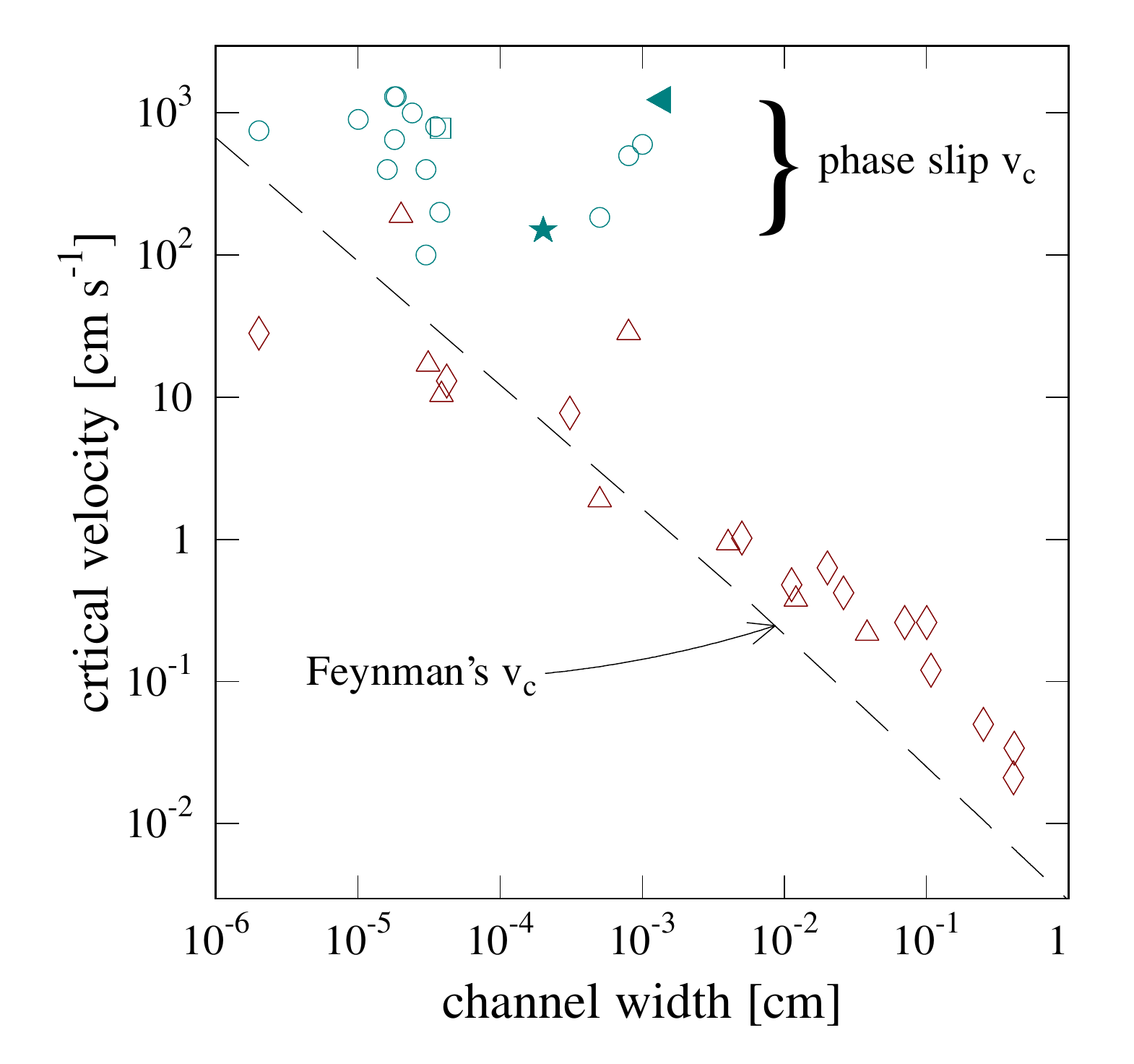}
    \caption{ \label{TwoCriticalVelocities} (Color online) Critical velocity
      data {\it vs} channel width ($\diamond$) -- \protect\citet{Wilks:67};
      ($\circ$) and ($\scriptstyle\triangle$) -- temperature-dependent and
      temperature-independent data as compiled by\protect\citet{Varoquaux:91};
      ($\blacktriangleleft$) -- \protect\cite{Shifflett:95}; ($\star$) --
      \protect\cite{Flaten:06a}; ($\square$) -- \protect\citet{Steinhauer:95} as
      reanalysed by \protect\citet{Varoquaux:96}. For the
      temperature-dependent data, the highest value, {\it i.e.}, that at the
      lowest temperature, has been retained. The dashed line is obtained from
      Eq.\protect{(\ref{FeynmanVelocity})} for the Feynman critical velocity.}
  \end{center}
\end{figure}

The compilation of the critical velocity data in various apertures and
channels from various sources available at the time of the Exeter Meeting in 1990
\cite{Varoquaux:91} is shown in Fig.\ref{TwoCriticalVelocities} together with
more recent data. Two different critical velocity regimes appear clearly on
this graph, a fast regime for small apertures, of the phase-slip type, which is
temperature-dependent, and a slower regime for larger channels, of the Feynman
type, which is temperature-independent.  The data points from various sources
for these two different types of critical velocity do not fall on well-defined
lines but merely bunch into clusters of points. As already stated, critical
velocities in apertures and capillaries are not very reproducible from
experiment to experiment indicating that, besides size, temperature, and
pressure, some less-well-controlled parameters also exert an influence.  In
some occasions, switching between these two types of critical velocity has
been observed in the course of the same cool-down
\cite{Hulin:74,Zimmermann:93}.

The critical velocity that depends on channel size does follow on average
relation (\ref{FeynmanVelocity}) for the Feynman mechanism. The higher
critical velocities, bunched around 5 to 10 m/s, faster than the Feynman
$v_{\m F}$ even for the smallest apertures but still considerably slower than
Landau's $v_{\m L}$ relate to the phase slip phenomenon and are discussed
below.

As a basis for comparison, it is worthwhile to summarise the findings of the
ion propagation studies in superfluid $^4$He at various pressures, which have
been reviewed by \citet{McClintock:91,McClintock:95}.  Ions can be created in
liquid helium and accelerated by electric field until they reach a limiting
velocity. The resulting drift velocities are measured by time-of-flight
techniques.  For negative ions, hollow bubbles 30 \AA\ in diameter with an
electron inside, two different behaviours are observed:
\begin{list}{$\bullet$}{}
\item Below about 10 bars, vortex rings are created, on the core of which
  single electrons get trapped: the drift velocity suddenly drops from that of
  the negatively charged bubbles to that of the much slower vortex rings
  \citep{Rayfield:64}.
\item Above 10 bars, the accelerated ion runs into the roton emission mechanism
  before vortex rings can be created.  The Landau critical velocity is
  observed to be $\approx$ 46 m/s at 24 bars down from a calculated value of
  60 m/s at SVT as the roton parameters change with pressure while the vortex
  creation velocity increases with pressure \citep{Varoquaux:96a}.
\item
Around 10 bars, both critical velocities, the Landau critical
velocity for the formation of rotons and that for the formation of vortex
rings can be observed to occur simultaneously because ions can be accelerated
above the threshold for roton emission. 
\end{list}

These ion propagation measurements provide a vivid illustration not only of
the existence of a critical velocity obeying the Landau criterion but also
that roton creation and vortex formation constitute different phenomena and
can exist concurrently.\footnote{A noteworthy attempt to by-pass this
  experimental finding is that of \protect\citet{Andreev:04}} The vortex
emission threshold displays other noteworthy features. It depends on
temperature in a non-trivial way, comparable to that of the phase-slip and
also shows the marked dependence on $^3$He impurity concentration observed for
phase slips in micro-aperture flows but not in larger channels. In both ion
propagation and aperture flow measurements, vortex formation displays similar
features.

Altogether, a study of the experimental data in superfluid $^4$He reveals
three different, well-defined, types of critical velocities; one,  the
celebrated Landau critical velocity, $v_\m L$, observed in $^4$He only for ion
propagation; another, $v_\m F$, that appears to follow the Feynman criterion
as shown in Fig.\ref{TwoCriticalVelocities} with all the uncertainties on the
hydrodynamical process of vortex creation in larger channels; and a third,
$v_\m c$, for phase slips, in want of an explanation: how are the
vortices of phase slips in aperture flow created, and how does the situation
differ from that in larger channels?

The short answer, based on qualitative evidence, is that the temperature
dependence of $v_\m c$ and its stochastic properties clearly point toward a
process of nucleation by thermal activation above $\sim$150 mK or so and by
quantum tunnelling below. This conclusion contradicts the common-place daily
observations of the formation of whirlpools and eddies, and also the widely
held belief that large scale topological defects with a quantum charge of
circulation cannot appear out of nowhere in the superfluid. It will be seen to
hold in $^4$He only because the nucleated vortices have nanometric size, a fact
that came to be appreciated because of the detailed analysis of phase slippage
observations related below.


\section{Phase slip critical velocity: a stochastic process}    
\label{PhaseSlipCriticalVelocity}


A more firmly established answer to the question formulated above comes from a
quantitative analysis of the experimental data for phase slips. The clues
given below conclusively show that, in small apertures, vortices are nucleated
by thermal activation above about 150 mK, and by quantum tunnelling
below.\footnote{This Section is based on the work of \citet{Varoquaux:01a},
  \citet{Varoquaux:03}, and \citet{Varoquaux:06}.}

\begin{figure}[!t]      
  \begin{center}
    \hskip -5 mm
    \includegraphics[width=85 mm,angle=0]{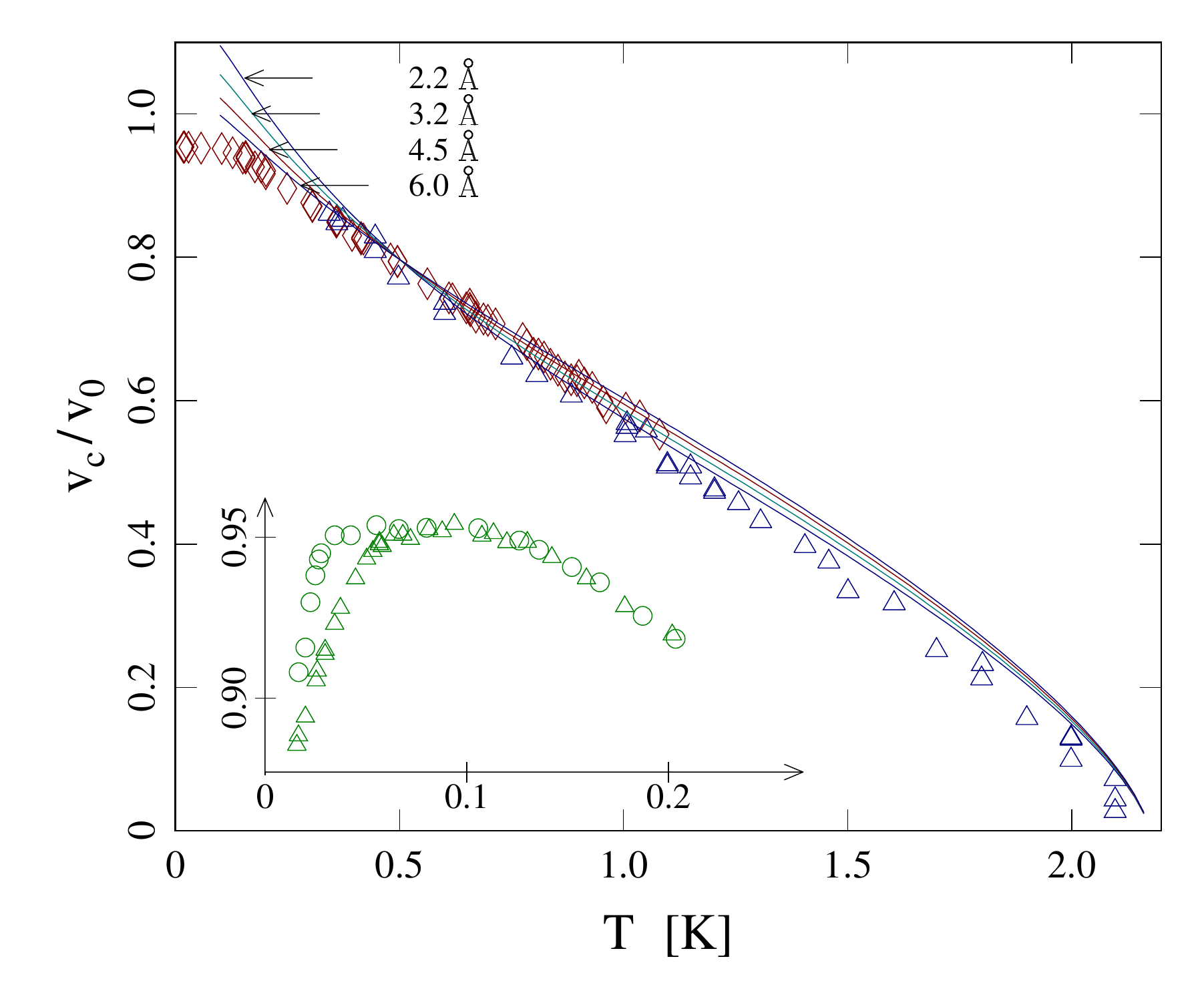}
    \caption{\label{vc}(Color online) Critical velocity, normalised to the
      zero temperature linear extrapolation value $v_0$, {\it vs} $T$, in
      kelvin: ($\diamond$), \citet{Avenel:93}, for ultra-pure $^4$He;
      ($\scriptstyle\triangle$), \citet{Zimmermann:98}.  The plain curves are
      computed from the half-ring model (see \S\ref{HalfRingModel}) for $a_0$
      = 2.2, 3.2, 4.5, 6.0 \AA\ and are normalised to match the experimental
      value at 0.5 K. The inset shows the influence of $^3$He impurities on
      $v_\m c$: ($\circ$), 3 ppb $^3$He in $^4$He; ($\scriptstyle\triangle$),
      45 ppb, from \protect\citet{Varoquaux:93}. Adapted from
      \protect\citep{Varoquaux:01a}. }
  \end{center}
\end{figure}

\begin{figure}[!t]      
  \begin{center}
    \includegraphics[width=75mm,angle=0]{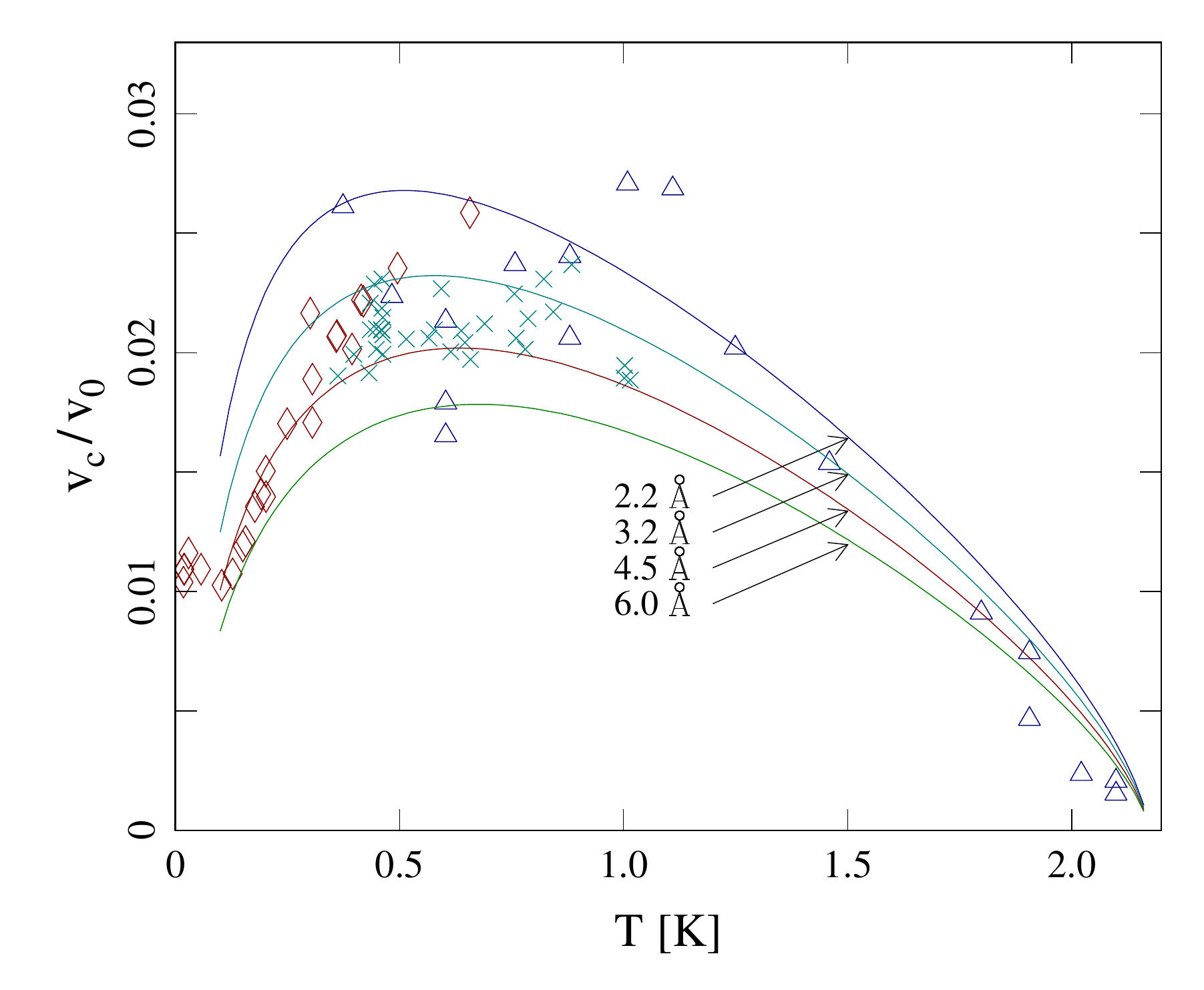}
    \caption{ \label{Delta-vc}(Color online) Statistical width of the critical
      velocity transition, normalised to the linear extrapolation limit at
      $T=0$, $v_0$, in terms of temperature: ($\diamond$),
      \protect\citet{Avenel:93}; ($\scriptstyle\triangle$),
      \protect\citet{Zimmermann:98}; ($\times$),
      \protect\citet{Steinhauer:95}. The plain curves are computed as in
      Fig.\ref{vc} for four values for $a_0$. Adapted from
      \protect\citet{Varoquaux:01a}.}
  \end{center}
\end{figure}

The first piece of evidence for the nucleation of vortices, that is their
creation {\it ex nihilo}, rests on the temperature dependence of the phase-slip
critical velocity shown in Fig.\ref{vc}. This figure, as
Fig.\ref{TwoCriticalVelocities} (and Fig.\ref{Delta-vc} to be discussed
further on), represents an attempt to compare data from
different groups. The data points are scattered but a general trend
emerges. The phase slip critical velocity increases in a near-linear manner
when the temperature decreases from 2 K to $\sim\,$0.2 K. That is, the
functional dependence of $v_{\m c}$ upon $T$ goes as $v_{\m c} =
v_0(1-T/T_0)$. The data depart from this linear dependence below 200 mK,
where they reach a plateau, and above 2 K because the critical velocity goes
to zero at $T_{\ulambda}$.

This temperature dependence, first observed in 1985 at Orsay
\citep{Varoquaux:87,Varoquaux:01a} and now a well-established experimental
fact \citep{Zimmermann:98}, \citep{Steinhauer:95} is very telling. It came as
a surprise at first because the critical velocities observed before were
temperature-independent below $\sim$1 K. As the quantum fluid is nearly fully
in its ground state below 1~K -- the normal fluid fraction becomes less than
1~\% -- one is led to suspect that a Arrhenius-type process must come into
play.  If such is the case, that is, if thermal fluctuations in the fluid
with an energy of at most a few $k_\m B T$ can trigger the appearance of
fully-formed vortex out of nowhere, the energy of this vortex must also be of
the order of a few $k_\m B T$: it must be a very small vortex. But very small
vortices require rather large superfluid velocities to sustain themselves --
as seen on Eq.(\ref{RingVelocity}). That these requirements can be met emerges
from a detailed quantitative analysis of the experimental data in the
framework of nucleation process.

The nucleation rate for a thermally activated process is expressed by
Arrhenius's \,law:
\begin{equation}        \label{Kramers}
{\mathit\Gamma}_{\mathrm K} =
\frac{\omega_0}{2\pi}\,\left[(1+\alpha^2)^{1/2}-\alpha\right] 
          \exp\left\{-\frac{E_\m a}{k_{\m B}T}\right\}\;.
\end{equation}
where $\omega_0/2\pi$ is the attempt frequency and $E_\m a$ the activation
energy of the process, which depends on the velocity $v_\m p$ and, more
weakly, on the pressure $P$ and the temperature $T$. The correction for
dissipation in the square brackets has been introduced by \citet{Kramers:40}
to describe the escape of a particle trapped in a potential well and
interacting with a thermal bath in its environment. The particle undergoes
Brownian motion fluctuations and experiences dissipation. This dissipation is
characterised by a dimensionless coefficient $\alpha = 1/(2\omega_0\tau)$,
$\tau$ being the time of relaxation of the system toward equilibrium. In
superfluid helium, dissipation is small. Although some dissipation is
necessary for the system to reach equilibrium with its environment, its
influence on the thermal activation rate is very small and will be neglected
in the following. However, this will not be anymore so in the quantum
regime, considered below, because dissipation causes decoherence.

The expression for the critical velocity that stems from the Arrhenius rate,
Eq.(\ref{Kramers}), is derived as follows.  In experiments such as those shown
in Fig.\ref{PhaseSlipsCanJPhys}, the velocity varies periodically at the
resonance frequency as $v_\m p\,\cos(\omega t)$, $v_\m p$ being the peak
velocity of the potential flow. The probability that a phase slip takes place
during the half-cycle $\omega t_\m i = -\pi/2\, ,\; \omega t_\m f = \pi/2$ is
\begin{eqnarray}        \label{probability}
  p &=& 1 - \exp\left\{-\int_{t_\m i}^{t_\m f}
    \mathit\Gamma_{\m K}(P,T,v_\m p\cos(\omega t')\right\} dt'
    \\                         
    &=& 1 - \exp\left\{
       -\frac{\omega_0}{2\pi\omega}\sqrt{\frac{-2\pi k_\m B T}
       {v_\m p \left. \left. \partial{E_\m a} \right/ \partial{v} \right|_{\m t=0}}}
      \exp \left\{ -\frac{E_\m a}{k_\m B T} \right\}
      \right\} .     \nonumber 
\end{eqnarray}   
The last equation (\ref{probability}) results from an asymptotic evaluation of the
integral at the saddle point $t=0$.
%

The critical velocity $v_\m c$ is defined as the velocity for which $p=1/2$. This
definition is independent of the experimental setup, except for the occurrence in
Eq.(\ref{probability}) of the natural frequency of the hydromechanical resonator
$\omega$. The implicit relation between $v_\m c$ and $E_\m a$ then reads:
\begin{equation}        \label{CriticalVelocity}
\frac{\omega_0}{2\pi\omega}\sqrt{\frac{-2\pi k_\m BT}
       {v_\m c \left. \left. \partial{E_\m a} \right/ \partial{v}
         \right|_{v_\m c}}}
      \exp \left\{ -\frac{E_\m a(P,T,v_\m c)}{k_\m B T} \right\} = \ln 2 \;  .
\end{equation}      

In Eq.(\ref{CriticalVelocity}), the attempt frequency is normalised by the
resonator drive frequency: the Brownian particle attempts to escape from the
potential well at rate $\omega_0/2\pi$ but an escape event is likely only in
the time window in a given half-cycle of the resonance during which the energy
barrier stays close to its minimum value $E_\m a(v_\m c)$.  This time interval
is inversely proportional to $\omega$, which explains why an instrumental
parameter gets its way into Eqs.(\ref{probability}) and
(\ref{CriticalVelocity}).

The velocity at which each individual critical event takes place is a stochastic
quantity. Its statistical spread can be characterised by the \`\!\`\!width'' of the
probability distribution defined \cite{Avenel:93,Zimmermann:90} as the inverse
of the slope of the distribution at $v_\m c$, $\left({\partial p/\partial
    v|_{v_\m c}}\right)^{-1}$. This critical width is found to be expressed
by:
\begin{equation}        \label{Width}
\Delta v_\m c = - \frac{2}{\ln 2}
       \left[
       \frac{1}{2}\left\{
       \frac{1}{v_\m c}+\left.\left. \frac{\partial^2E_\m a}{\partial
         v^2}\right|_{v_\m c}
       \right/\left.\frac{\partial E_\m a}{\partial v}\right|_{v_\m c}
       \right\}
       + \frac{1}{k_\m B T}\left.\frac{\partial E_\m a}{\partial
           v}\right|_{v_\m c}
       \right]^{-1} \; .
\end{equation}

\begin{figure}[!t]      
  \begin{center}
    \includegraphics[width=65mm]{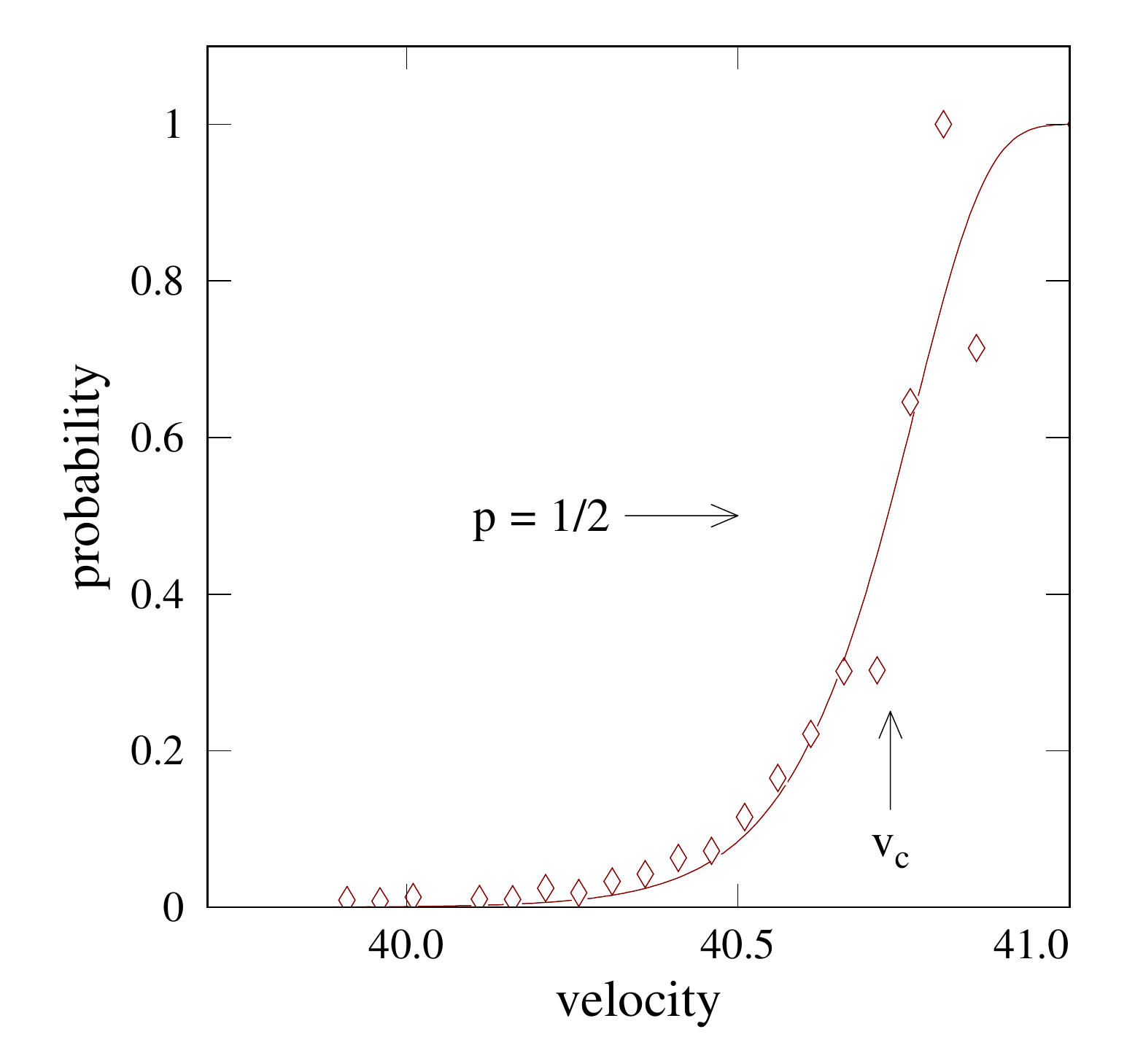}
    \caption{ \label{PR02-pure-proba} (Color online) 
      Probability $p$ {\it vs} phase slip velocity
      in winding number.  The plain curve is a non-linear least square fit to
      the analytic form Eq.(\ref{probability}), which contains two adjustable
      parameters, $v_\m c$ and $\Delta v_\m c$. The critical velocity
      resulting from this distribution of the measured values is
      defined as the fit value for $p=1/2$. The critical velocity distribution
      width is given by the slope at $p=1/2$. From \protect\citet{Varoquaux:03}.}
  \end{center} 
\end{figure}

The quantities $v_\m c$ and $\Delta v_\m c$ are derived from $p$, itself
obtained by integrating the histograms of the number of nucleation events
ordered in velocity bins. The outcome of this procedure is illustrated in
Fig.\,\ref{PR02-pure-proba}: $p$ shows an asymmetric-{\sl S} shape
characteristic of the double exponential dependence of $p$ on $v$,
Eq.(\ref{probability}), a consequence of Arrhenius's \,law,
Eq.(\ref{Kramers}), being plugged into a Poisson probability distribution. The
observation of this asymmetric-{\sl S} probability distribution constitutes
an additional experimental clue for the existence of a nucleation process. The
quantities $v_\m c$ and $\Delta v_\m c$ are easily extracted from the
probability curves $p(v)$, but the inverse path from $v_\m c$ and $\Delta v_\m
c$ back to $E_\m a(v)$ and $\omega_0$ by
numerical integration of the differential equation (\ref{CriticalVelocity})
leads to inaccurate results.

\begin{figure}[!t]      
  \begin{center}
    \includegraphics[width=70mm]{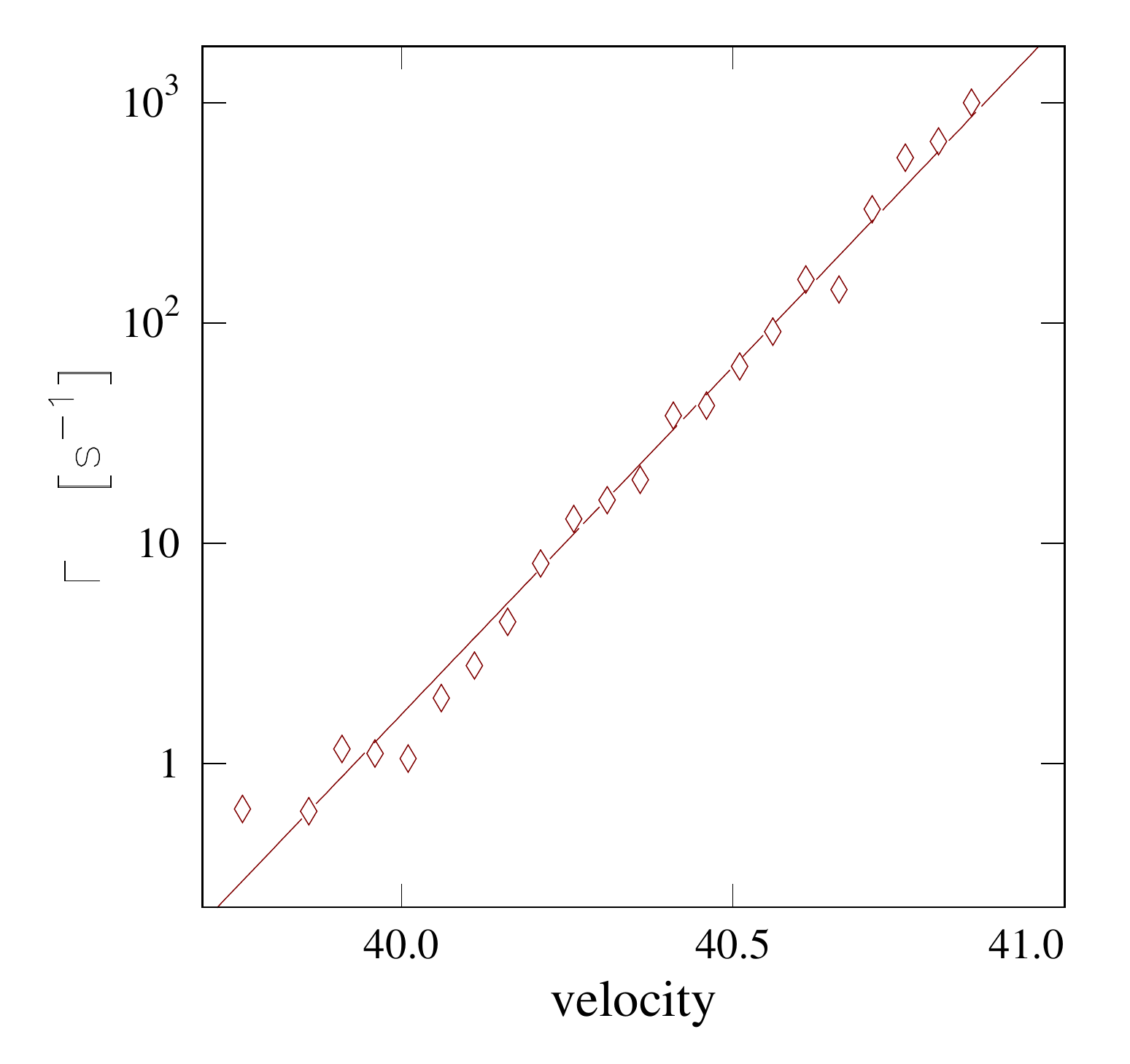}
    \caption{ \label{PR02-pure-rate} Nucleation rate $\mathit\Gamma$ expressed
      in s$^{-1}$ on a semi-logarithmic scale {\it vs} slip velocity in winding
      number in ultra-pure $^4$He at 17.70 mK and saturated vapour
      pressure. The line is a linear fit to the data. From \protect\citet{Varoquaux:03}. }
  \end{center} 
\end{figure}

In view of these difficulties \citet{Varoquaux:03} improved the data
analysis by obtaining the escape rate $\mathit\Gamma(v)$ directly from the
phase slip data.  This rate is the ratio in any velocity bin of the number of
slips that have occurred at that velocity to the total time spent by the
system at that given velocity.  The outcome of this procedure is illustrated
in Fig.\,\ref{PR02-pure-rate}.  The slope of $\ln \mathit\Gamma(v)$ directly
yields $\left.{\partial E_\m a/\partial v} \right|_{v_\m c}$; the value of
$\ln \mathit\Gamma$ at $v_\m c$ gives a combination of $\ln\omega_0$ and $E_\m
a(v_\m c)$. 

The need to solve Eq.(\ref{CriticalVelocity}) has been alleviated
but to cleanly disentangle these two quantities and solve this inverse problem
is still error prone. At this point experiment itself offers help as will shortly be shown.

%
\subsection{Vortex nucleation: thermal {\bf{\textit{vs}}} quantum } 
%
\label{VortexNucleation}

Below $\sim$ 0.15 K, the phase slip critical velocity $v_\m c$ suddenly ceases
to vary with $T$, as seen in Fig.\,\ref{vc} for ultra-pure $^4$He, $v_\m c(T)$
remains flat down to the lowest temperatures ($\sim\,$12 mK). The effect of
$^3$He impurities, shown in the inset, could mimic the appearance of such a
plateau but is ruled out because of the extreme purity of the $^4$He sample,
which contains less than 1 part in 10$^9$ of $^3$He impurities. The crossover
between these two regimes is very sharp. At the same crossover temperature
$T_\m q$, $\Delta v_\m c$ also levels off sharply. It is believed on
experimental grounds that this saturation is intrinsic and is not due to stray
heating or parasitic mechanical vibrations \cite{Avenel:93}.

If the nucleation barrier were undergoing an abrupt change at $T_\m q$, for
instance because of a bifurcation toward a vortex instability of a different
nature \cite{Josserand:95}, in all likelihood $\Delta v_\m c$ would jump to a
different value characteristic of the new process (presumably small since
$v_\m c$ reaches a plateau). Such a jump is not observed in
Fig.\,\ref{Delta-vc}. Furthermore, $v_\m c$ levels off below $T_\m q$, which
would imply through Eq.(\ref{CriticalVelocity}) that $E_a$ becomes a very
steep function of $v$, but $\Delta v_\m c$ also levels off, which, through
Eq.(\ref{Width}), would imply the contrary. This discussion leads one to
investigate the possibility that, below $T_\m q$, thermally-assisted escape
over the barrier gives way to quantum tunnelling under the barrier
\cite{Ihas:92}. This switch from thermal to quantum does induce plateaus below
$T_\m q$ for both $v_\m c$ and $\Delta v_\m c$.

Independently of these phase slippage studies, the group of Peter McClintock
at Lancaster had also reached the conclusion from their ion propagation
experiments of the existence of a crossover around 300 mK from a thermal to a
quantum regime for the nucleation of vortices \cite{Hendry:88}, as predicted
by Muirhead, Vinen, and Donnelly \cite{Muirhead:84}. There certainly are
significant differences between the ion limiting drift velocity and aperture
critical flow -- in particular, the latter is nearly one order of magnitude
smaller -- but the qualitative similarities are strikingly telling. The two
completely different types of experiments indicate that vortices would appear
as a result of a nucleation process on a nanometric scale, both in a thermal
regime above $T_\m q$ and in a quantum one below.

\subsection{The macroscopic quantum tunnelling rate}

Taking this hint at face value, zero point fluctuations are now assumed to
overtake thermal fluctuations below $T_\m q$: the potential barrier is not
surmounted with the assistance of a large thermal fluctuation, it is tunnelled
under quantum-mechanically; the quantum-tunnelling event is ``assisted'' by
the zero point fluctuations, so to speak \cite{Martinis:88}, in the same
manner as the Arrhenius process is assisted by thermal fluctuations.  What is
remarkable here, and not so easy to admit at first, is that such a tunnelling
process affects a macroscopic number of atoms, those necessary to form a
vortex of about 50 \AA\ in length, as turns out to be the case.

These ``macroscopic quantum tunnelling'' (MQT) processes have been the object
of numerous experimental and theoretical studies, mainly in superconducting
Josephson devices \cite{Caldeira:83a} . The case for vortices in helium can be
worked out in a similar manner.

The quantum tunnelling rate of escape of a particle out of a potential well
$V(q)$ is a textbook problem.\footnote{See for instance \citet{Landau:Quantum}
  \S 50.} The rate is proportional to $\exp{-S/\hbar}$, $S$ being, in the WKB
approximation, the action of the escaping particle along the saddle-point
trajectory at the top of the potential barrier, the so-called ``bounce''
\cite{Coleman:77}. For a particle of mass $m$ and energy $E$ escaping from a
one-dimensional barrier $V(q)$, this action reads
\begin{equation}        \label{Action}
  S = 2 \int_{q_1}^{q_2} \m d q \sqrt{2m[V(q)-E]} \; .
\end{equation}
The determination of the bounce yields the generalised coordinates $q_1$ and
$q_2$ of the points at which the particle enters and leaves the barrier.  A
discussion of the quantum tunnelling of vortices in terms of Eq.(\ref{Action})
thus requires a Lagrangian formulation of vortex dynamics. Such a formulation
has been carried out in particular by \citet{Sonin:95} \footnote{See also
  Sonin's book (2015), \S 12.2.} and by \citet{Fischer:00}. However, analytic
  results can be obtained only at the cost of approximations that yield a less
  than fair comparison with experiments as discussed by \citet{Varoquaux:01a}.

A simplified and more productive approach can be borrowed from the literature
for Josephson devices. Extending the work of \citet{Caldeira:83a}, and
\citet{Larkin:84} to vortices in helium, \citet{Varoquaux:06} has used for
$V(q)$ a simple analytic form reduced to a sum of two terms, respectively
parabolic and cubic in $q$:
\begin{equation}        \label{CubicParabolic}
  V(q) = V_0 + \frac{1}{2}m\omega_0^2 q^2 
    \big(1-\frac{2q}{3q_\m b}\big)\; ,
\end{equation}
where $\omega_0$ is the angular frequency of the lowest mode of the trapped
particle (that will be found comparable to the attempt frequency) and $q_\m b$ the
generalised coordinate of the barrier top location. The barrier height $E_\m
b$ is expressed in terms of these two parameters by $m\omega_0^2 q_\m b^2/6$.

Equation (\ref{CubicParabolic}) expresses the vanishing potential barrier
height when the applied velocity reaches the limiting velocity, $v_{\m c0}$ at
which the system ``runs away'', the so-called ``lability'' point.\footnote{For
  an illustration, see Fig.2 in \citet{Anderson:66a} for the ``tilted
  washboard'' model.}  At this point where the system becomes labile, the
critical velocity is reached even in the absence of thermal or quantum
fluctuations.  Such a hydrodynamic instability threshold at which vortices
appear spontaneously has been shown to occur in numerical simulations of flows
past an obstacle using the Gross-Pitaevskii equation by \citet{Frisch:92} and
others.\footnote{For instance, \citet{Nore:00}, \citet{Berloff:01},
  \citet{Rica:01}.}

The zero-temperature WKB tunnelling rate for the phenomenological cubic-plus-parabolic
potential $E_\m b$, Eq.(\ref{CubicParabolic}), is found to be \cite{Caldeira:83a} 
\begin{equation}        \label{ZeroTemperature}
  \mathit\Gamma_0 =\frac{\omega_0}{2\pi}\, \left(120\pi\, \frac{S_0}{\hbar}\right)^{1/2}
    \exp -\frac{S_0}{\hbar} \; ,   
\end{equation}
the action $S_0$ being equal to $36 E_\m b/5\omega_0$. 

From this result, it can be anticipated that the crossover between the quantum
and the thermal regime lies around a temperature close to that for which the
exponents in Eqs.(\ref{Kramers}) and (\ref{ZeroTemperature}) are equal, namely
$T = 5\omega_0/36 k_\m B$ -- assuming that the activation energy in
Eq.(\ref{Kramers}), $E_\m a$, reduces to the simple cubic-plus-parabolic form,
$E_\m b$. A more precise study of the mathematical properties of the quantum
channel for escape leads to the following relation \cite{Melnikov:91}
\begin{equation}        \label{Crossover}
  \hbar \omega_0 = 2 \pi k_B T_\m q \; .
\end{equation}
Thus, from the experimental knowledge of the temperature of the crossover
from thermal to quantal, $\omega_0$ is fixed to pinpoint accuracy by
Eq.(\ref{Crossover}). Its value agrees with that (less precisely determined)
obtained from the analysis of the Arrhenius regime outlined in the previous
paragraph: some degree of self-consistency has been achieved.  The values of
the barrier height $E_\m b$ at each given velocity then follow easily, using
the full expressions for the rate in terms in terms of $E_\m b$, $\omega_0$
and, also, for the damping parameter $\alpha$ as discussed below.

\subsection{Friction in MQT}
\label{Friction}

Damping turns out to matter significantly for quantum tunnelling of
semi-macroscopic objects, contrarily to the thermal regime. The relevance and
applicability of the concept of quantum tunnelling to macroscopic quantities
such as the electric current through a Josephson junction or the flow of
superfluid through a micro-aperture, although still sometimes questioned, have
been checked in detail for the electrodynamic Josephson effect
\citep{Martinis:87}. One of the conceptual problems is that, when a
macroscopic quantum system is coupled to an environment that acts as a thermal
bath, the coupling gives rise to a source of classical fluctuations and
friction. The quantum process suffers decoherence and is
profoundly affected.

This issue was tackled by Caldeira and Leggett \cite{Caldeira:83a}, and a
number of other authors.\footnote{See, for instance, \citet{Melnikov:91} and
  also \citet{Varoquaux:03} for more references and details on this Section.}
For weak frequency-independent damping ($\alpha \ll 1$) and the
cubic-plus-parabolic potential, the tunnelling rate takes the form:
\footnote{As explained by \citet{Caldeira:83a}, \citet{Waxman:85}, and
  \citet{Grabert:87}}
\begin{equation}        \label{QuantumRate}
  \begin{split}
    {\mathit\Gamma}_{\m{qt}}= 
    & \frac{\omega_0}{2\pi}
    \,\left(864\pi\,\frac{E_\m b}{\hbar\omega_0}\right)^{1/2} 
    \times\exp\left\{-\,\frac{36}{5}\,\frac{E_\m b}{\hbar\omega_0}
    \left[1+\frac{45\zeta(3)}{\pi^3}\alpha\right]^{} \right.
    \\
    &\left. +\frac{18}{\pi}\alpha\,\frac{T^2}{T_\m q^2}+
    {\mathcal O}\left(\alpha^2,\,\alpha\,\frac{T^4}{T_\m q^4}\right)\right\}
    \; .
   \end{split} 
\end{equation}
According to Eq.(\ref{QuantumRate}), damping depresses the MQT escape
rate at $T=0$ -- $\alpha$ being a positive quantity -- and introduces a
temperature dependence that increases the rate as $T$ increases. These effects
are large, even for weak damping, because they enter the exponent of the
exponential factor in Eq.(\ref{QuantumRate}).  Relation (\ref{Crossover})
between $T_\m q$ and $\omega_0$ is nearly unaffected by damping: $\omega_0$ is
simply changed into $\omega_0[(1+\alpha^2)^{1/2} -\alpha]$ according to
Eq.(\ref{Kramers}), a minor modification for $\alpha\ll 1$.

\begin{figure}[t]       
  \begin{center}
    \includegraphics[width=40 mm,angle=90]{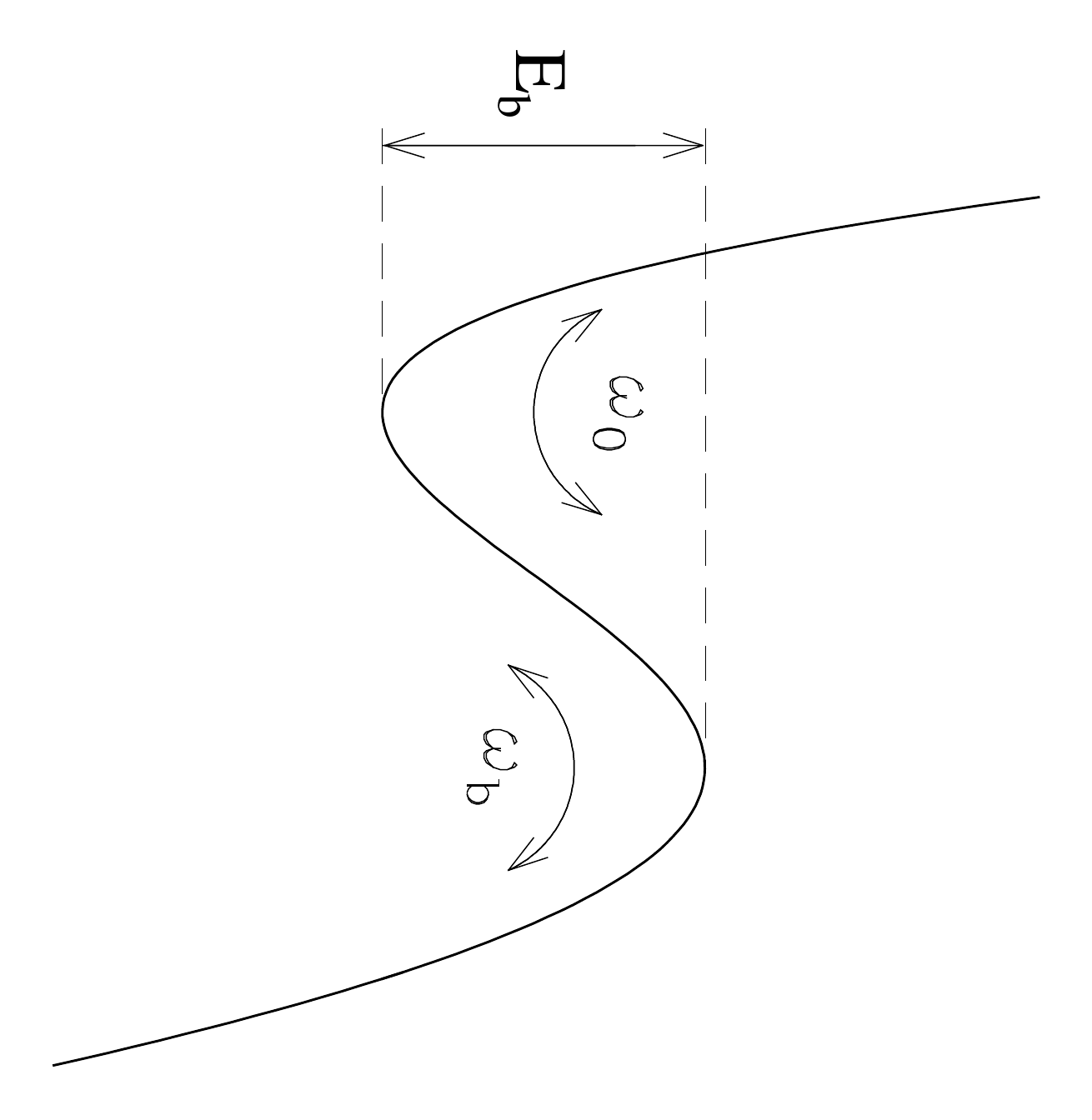}
    \caption{ \label{PotentialWell} Potential well trapping a particle in one
      dimension. The particle can escape to the continuum of states to the
      right. The lowest mode at the bottom of the well has angular frequency
      $\omega_0$; $\omega_\m b$ would be the corresponding quantity if the
      potential was inverted bottom over top. For the simple case of
      Eq.(\ref{CubicParabolic}), $\omega_0=\omega_\m b$. There can exist
      intermediate energy levels within the well, which are populated
      according to the Boltzmann factor. Particle escape can take place by
      quantum tunnelling ``under'' the barrier or by thermal activation
      ``over'' the barrier. The intermediate energy levels can be used as
      ladder rungs by the particle attempting to escape. These processes are
      embedded in Eq.(\protect\ref{GeneralKramers}).}
  \end{center}
\end{figure}

Equation (\ref{QuantumRate}) is valid up to about $T_\m q/2$. From $T_\m q/2$
to $\sim T_\m q$, one has to resort to numerical calculations
\cite{Grabert:87}. In the thermal activation regime, $T\gtrsim T_\m q$,
quantum corrections affect the Kramers escape rate up to about $3\,T_\m q$ and
can be evaluated analytically. These high-temperature quantum corrections
depend only weakly on friction.  A complete solution of the problem of the
influence of friction, weak, moderate or strong, has first been worked out in
the classical regime ($T\gg T_\m q$) by \citet{Grabert:88} and extended
to the temperature range $T\gtrsim T_\m q$ by \citet{Rips:89} who showed that
the rate for arbitrary damping can be factorised in three terms,
\begin{equation}         \label{GeneralKramers}
{\mathit\Gamma} = \,f_\m q\,{\mathit\Upsilon}\,\mathit\Gamma_\m K \;,
\end{equation}
each term having a well-defined physical meaning: ${\mathit\Gamma}_\m K$ is
 the classical Kramers rate, $f_\m q$ the quantum correction
factor, and $\mathit\Upsilon$ the depopulation factor. The high
temperature limit of $f_\m q$ is
\begin{equation}             \label{LeadingQuantumCorrection}
f_\m q  =
\exp \left\{\frac{\hbar^2}{24} \frac{(\omega_0^2+\omega_\m b^2)}{(k_{\m B}T)^2}
     +{\mathcal O}(\alpha/T^3,1/T^4)\right\}  \; ,
\end{equation}
in which $\omega_0$ and $\omega_\m b$ are the confining potential parameters
depicted in Fig.\ref{PotentialWell}. Analytic results for $f_\m q$ are known
to slightly below $T_\m q$ \cite{Grabert:87,Hanggi:90}.

The depopulation factor $\mathit\Upsilon$ arises from the depletion of the
occupancy of the energy levels inside the potential well in the course of the
escape process. This depletion occurs when the intermediate levels, if they
exist, are not replenished fast enough by the thermal fluctuations.  For the
nucleation of vortices, friction turns out to always be both sufficient and
not too large so that depopulation corrections remain small and
$\mathit\Upsilon \sim 1$.

\begin{figure}[t]       
 \begin{center}
    \includegraphics[width=55mm]{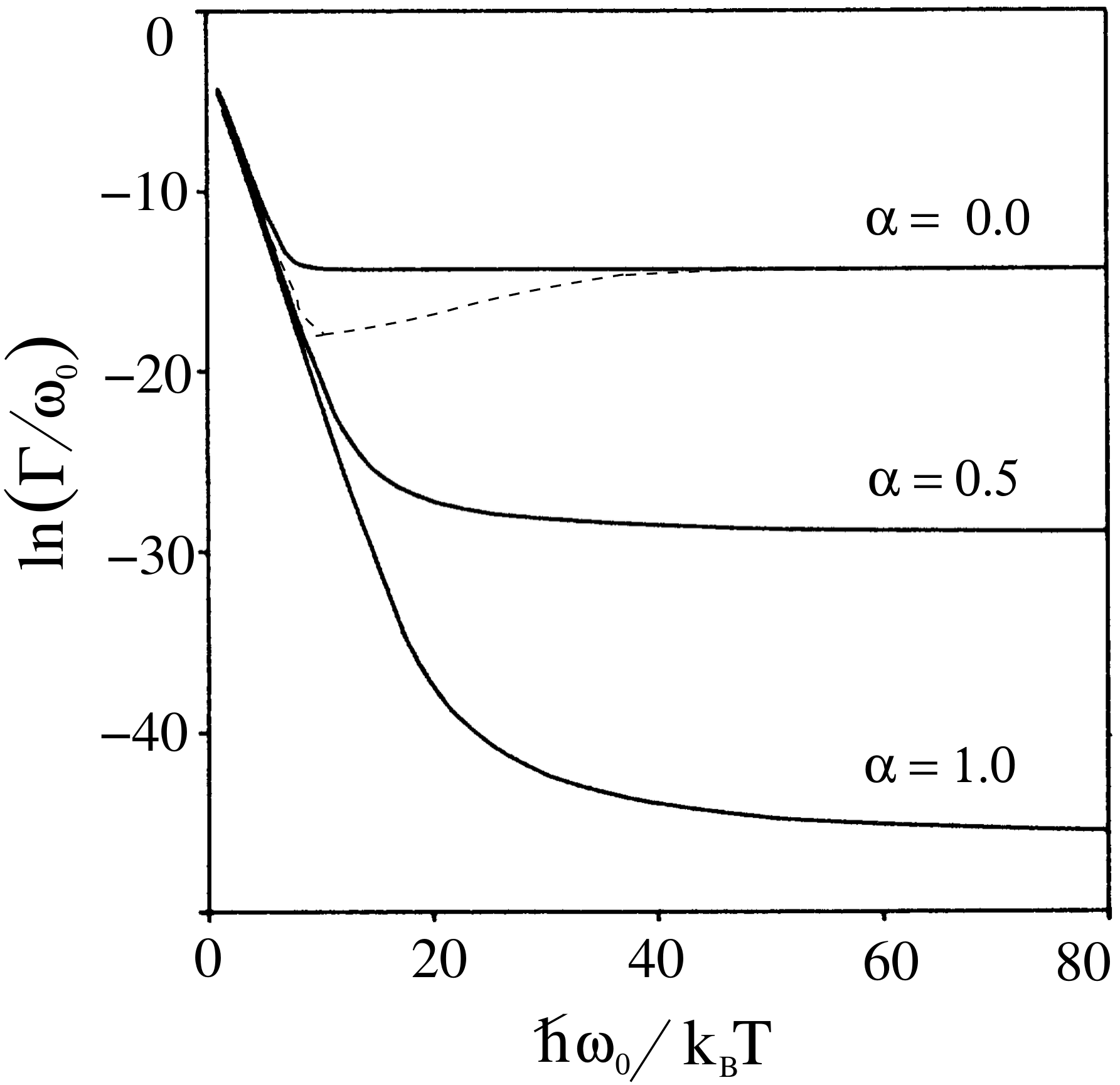}
    \caption{ \label{EscapeRate} Logarithm of the escape rate normalised to
      the attempt frequency in terms of inverse temperature, also normalised
      to $\omega_0$ for various values of the damping parameter $\alpha$
      \citep{Grabert:87}. The dotted line is a hand-sketch of the situation
      where $\alpha$ increases with temperature, starting from zero at $T=0$
      (see \S \ref{Friction}). From \protect\citet{Varoquaux:03}. }
  \end{center} 
\end{figure}
 
The escape rate can be calculated over the full temperature range by piecing
together Eqs.(\ref{Kramers}), (\ref{QuantumRate}), (\ref{GeneralKramers}) and
(\ref{LeadingQuantumCorrection}). The results for three values of the damping
parameter $\alpha$ are shown in Fig.\ref{EscapeRate}. A hand-sketched line
pictures the escape rate for $\alpha$ varying from zero at $T=0$ to 0.5 above
$T_\m q$: it is seen to actually decrease when the temperature increases from
absolute zero. This unique situation is found in the nucleation of vortices in
$^4$He as will now be described.

\subsection{Experimental energy barrier and damping coefficient}
\label{ExperimentalEnergyBarrier}

The knowledge of the rate $\mathit\Gamma$ makes it possible to extract from
the measured nucleation rate and crossover temperature the values of the
energy barrier in terms of $v_\m c$.  The value of $\omega_0$ given by
Eq.(\ref{Crossover}) ($\omega_0/2\pi = 2\times10^{10}$ Hz for $T_q$=0.147 K)
is consistent with the attempt frequency appropriate to the
thermally-activated regime \cite{Varoquaux:86} and that found directly from
the fits to the probability $p$ as shown in Fig.\,\ref{PR02-pure-proba}. This
agreement has been mentioned above.

This value of $\omega_0$ is comparable to the highest Kelvin mode
eigenfrequency that a vortex filament in $^4$He can sustain.  The Kelvin modes
are helical waves with a dispersion relation expressed for a straight isolated
vortex by
\begin{equation}        \label{KelvinMode}
\omega^{\ts\pm} = \frac{\kappa_4}{\pi\,a_0^2}\left[1\pm
           \left\{1 +ka_0\left[\frac{\mbox{K}_0(ka_0)}{\mbox{K}_1(ka_0)}
           \right]\right\}^{\frac{1}{2}}\right] \;,                        
\end{equation}
\noindent
where $\mbox{K}_0$ and $\mbox{K}_1$ are the modified Bessel functions of
zeroth and first orders.\footnote{See, for instance \citet{Fetter:65},
  \citet{Glaberson:86}, \citet{Sonin:87}, or \citet{Donnelly:91}.} In the short wavelength
limit, $k^{-1} \rightarrow 0$, the high frequency mode reduces to:
\begin{equation}        \label{HighFrequency}
\omega^{\ts +} = \frac{\kappa_4}{\pi a_0^2} \;.
\end{equation}
\noindent
Equation (\ref{HighFrequency}) sets the shortest time scale on which vortices
can be expected to respond.

By analogy with the 2D-motion of point electric charges subjected to a
{\it rf} magnetic field~\cite{Muirhead:85}, this frequency is sometimes called the
``cyclotron'' frequency.  This frequency is that of the cycloidal motion taken
by a long hollow cylinder impulsively pulled sideways in an inviscid fluid
\citep{Donnelly:91}. The cylinder stands for the vortex core, assumed to be
hollow and with radius $a_0$. The displaced mass per unit length of such a
cylinder is $\rho \pi a_0^2$. For high frequency motions, the vortex mass is
modified as discussed in \S\ref{ImpulseAndMass}, and Eq.(\ref{HighFrequency})
is renormalised to $\omega^{+}= \kappa_4/[\pi a_0^2 {\m{ln}}(r_\m m/a_0)]$,
$r_\m m$ being defined below Eq.(\ref{FlowEnergy}). With
$a_0=2.5$\,\AA\ and $r_\m m/a_0 \sim 10$, $\omega^{\ts +}/2\pi = 3.5\, 10^{10}$
Hz, a value comparable to the attempt frequency given by
Eq.(\ref{Crossover}). That the attempt frequency be linked to the highest
frequency that the nucleating vortex can sustained makes good physical sense.

\begin{figure}[t]        
  \begin{center}
    \includegraphics[width=68mm]{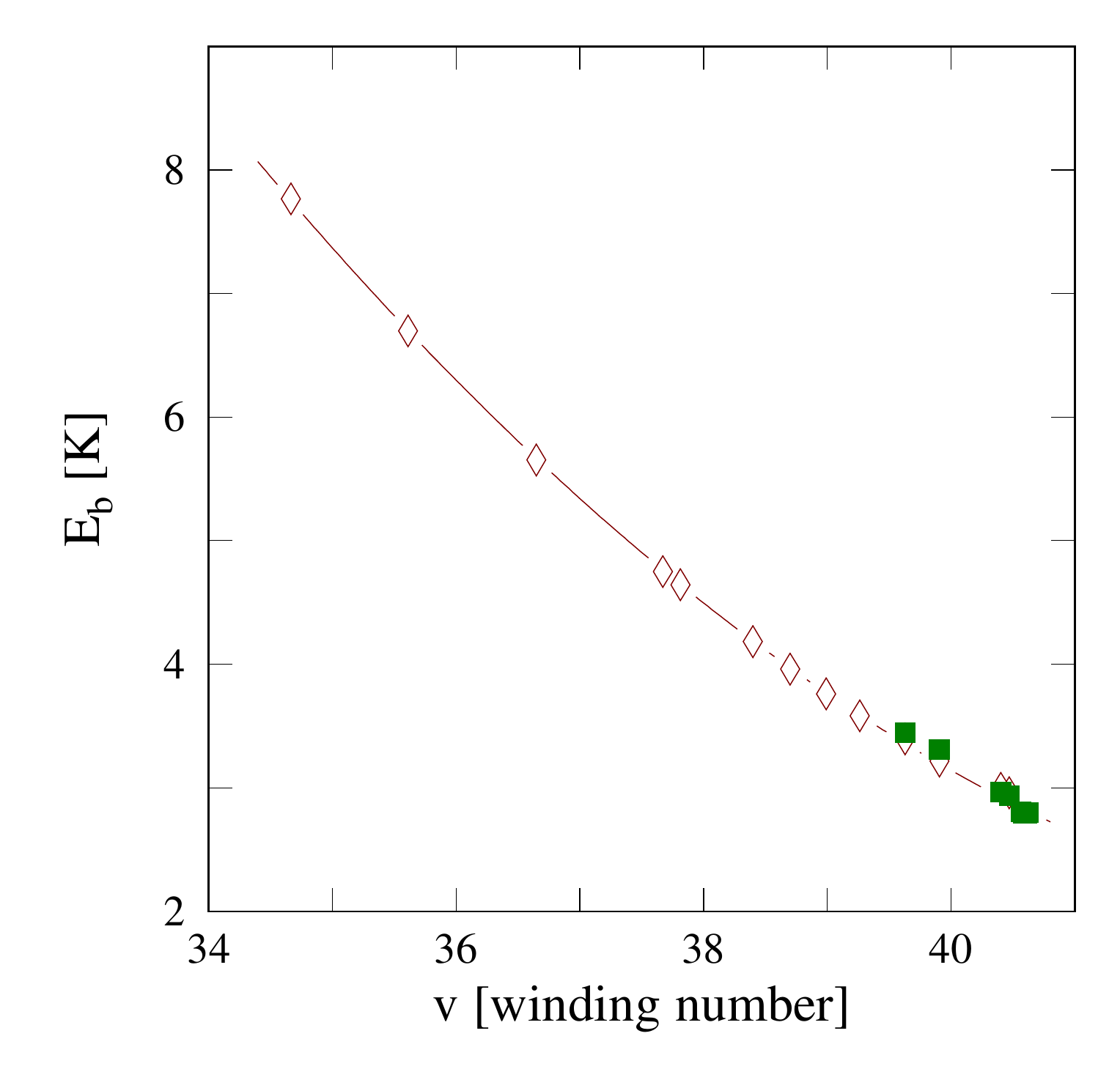}
    \caption{ \label{ExperimentalEb}(Color online)  The barrier energy
      $E_\m b$ in kelvin  {\it vs} $v$, the mean velocity in the aperture expressed
      in phase winding numbers and obtained from the nucleation rate data of
      \protect\citet{Varoquaux:03}: ($\scriptstyle{\blacksquare}$), low
      temperature data transformed using the numerical tables of
      \citet{Grabert:87}; ($\lozenge$), high temperature
      data. From  \protect\citet{Varoquaux:03}. }
  \end{center} 
\end{figure}

With the known value of $\omega_0$, the energy barrier $E_\m b$ can be
extracted from the measured rate with the help of Eqs.(\ref{QuantumRate}) and
(\ref{GeneralKramers}). These values of $E_\m b$ for the experiments on
ultra-pure $^4$He analysed by \citet{Varoquaux:03} are shown in
Fig.\ref{ExperimentalEb}.  The high $T$ and low $T$ analyses are seen to yield
consistent results in the region where they overlap.

\begin{figure}[t]       
  \begin{center}
    \includegraphics[width=85mm,height=65 mm]{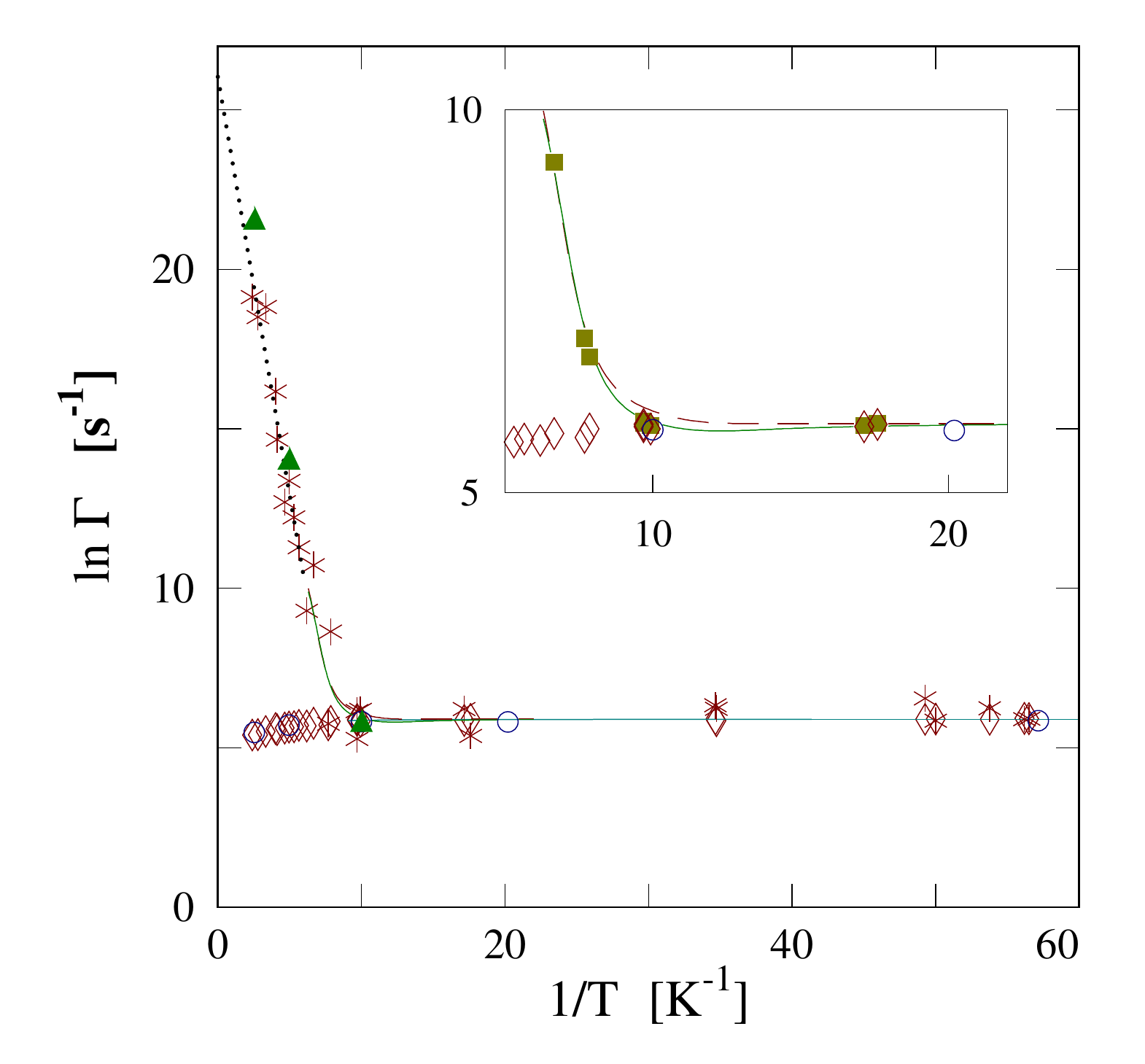}
    \caption{ \label{ArrheniusPlot}(Color online) Arrhenius plot $\ln
      \mathit\Gamma(v)$ {\it vs} $1/T$, $\mathit\Gamma$ being expressed in
      s$^{-1}$ and $T$ in kelvin, as measured at varying $T$ and $v_\m c$ for two
      ultra-pure $^4$He samples ($\lozenge$) and ($\triangle$);
      ($\ast$) and ($\blacktriangle$) after correction for the change of the
      velocity with $T$. The plain and the dot-dot curves are calculated from
      the experimental barrier energy $E_\m b$ represented in
      Fig.\ref{ExperimentalEb}. From \protect\citet{Varoquaux:03}. 
      In the inset, $\ln \mathit\Gamma(v_\m q)$,
      ($\scriptstyle\blacksquare$), has been obtained with smoothed values of
      $v_\m c$.  The curves represent the calculated values of $\ln {\mathit
        \Gamma(v_\m q)}$ with $\alpha= 0$ (dash-dash) or varying with $T$
      (plain). The latter gives a better representation of the data and
      illustrates the influence of damping depicted by the dot-dot line in
      Fig.\ref{EscapeRate}. }
  \end{center} 
\end{figure}

The quantitative analysis can be carried out one step further using the
variation of the barrier energy $E_\m b$ in terms of $v$ to construct a
Arrhenius plot -- the logarithm of the escape rate $\mathit\Gamma$ in terms of the inverse
temperature {\it for a fixed potential well} -- from the experimental data and
comparing directly the outcome to the results from theory. Arrhenius plots are
drawn at constant $E_\m b$ and varying temperature but the experimental
results are obtained at velocities that vary with temperature, hence at
varying $E_\m b$. The correction can be computed from the experimentally
determined $E_\m b$ given in Fig.\ref{ExperimentalEb}. The final outcome for
the nucleation rate data of \citet{Varoquaux:03} in ultra pure $^4$He is
plotted in Fig.\ref{ArrheniusPlot}.

As can be noted in Fig.\ref{ArrheniusPlot}, the raw experimental,
velocity-dependent rates exhibit little variation over the range of
parameters: escape rates are only observed in a certain window determined by
the measuring technique.  At low temperatures, $T< T_\m q$, the critical
velocity is close to its zero temperature limit $v_\m q$ and the corrections
to $\mathit\Gamma$ are small.  As $T$ increases above $T_\m q$, $v_\m c$
decreases and $\mathit\Gamma$ has to be determined by piece-wise integration
of $\m d \ln\mathit\Gamma / \m d v$.  The high temperature extrapolation for
$\mathit\Gamma$ obtained in such a manner does display the usual $1/T$
dependence.

In the intermediate temperature range, the corrected $\mathit\Gamma$ shows, as
can be seen in the inset of Fig.\ref{ArrheniusPlot}, a small but real drop
below its zero temperature limit as the temperature is raised.  This drop
reveals the influence of damping on the escape rate illustrated in
Fig.\ref{EscapeRate}. A damping coefficient $\alpha$ that increases from 0 at
$T=0$ to $\sim\,0.1$ around $T_\m q$ and more slowly above accounts for the
observed drop \cite{Varoquaux:03}. This $T$-dependent dissipation also makes
the crossover between the thermal and the quantum regimes even sharper than
for $\alpha = 0$, and closer to observations. The nucleation of vortices in
$^4$He thus offers a rare observation of the effect of damping on MQT.

\subsection{The vortex half-ring model}\ 
\label{HalfRingModel}

The case has been put so far  for the nucleation of vortices,
thermal or quantal. The nucleation barrier $E_\m b$ is of the
order of a few kelvins (see Fig.\ref{ExperimentalEb}) and the attempt
frequency $\sim 2\times10^{10}$ Hz, close to that of the highest Kelvin wave
mode.

A simple model accounts for these features.  This model -- the nucleation of
vortex half-rings at a prominent asperity on the walls -- finds its roots in
the work of  \citet{Langer:67b}, \citet{Langer:70}, \citet{Volovik:72}, and
\citet{Muirhead:84}. It was further developed and put on the firm experimental
findings described above by \citet{Avenel:93}.

The model premises are the following. Consider, as done by \citet{Langer:70},
the homogeneous nucleation of a vortex ring in a homogeneous flow $v_\m s$
extending over large distances. When the ring has grown to reach radius $R$ in a
plane perpendicular to the flow, its energy in the laboratory frame, where the
observer is at rest and sees the superfluid moving at velocity $v_\m s$, is
expressed by
\begin{equation}        \label{VortexFreeEnergy}
  E_\m v = E_\m R - P_\m R v_\m s \; .
\end{equation} 
The rest energy $E_\m R$ and impulse $P_\m R$ of the vortex ring are given by
Eqs.(\ref{RingEnergy}) and (\ref{RingImpulse}). The minus sign in the right
hand side of Eq.(\ref{VortexFreeEnergy}) arises because the vortex opposes the
flow, that is, its impulse $P_\m R$ points straight against $v_\m s$:
this configuration minimises $E_\m v$.
%
\begin{figure}[t]        
  \begin{center}
    \includegraphics[width=70mm]{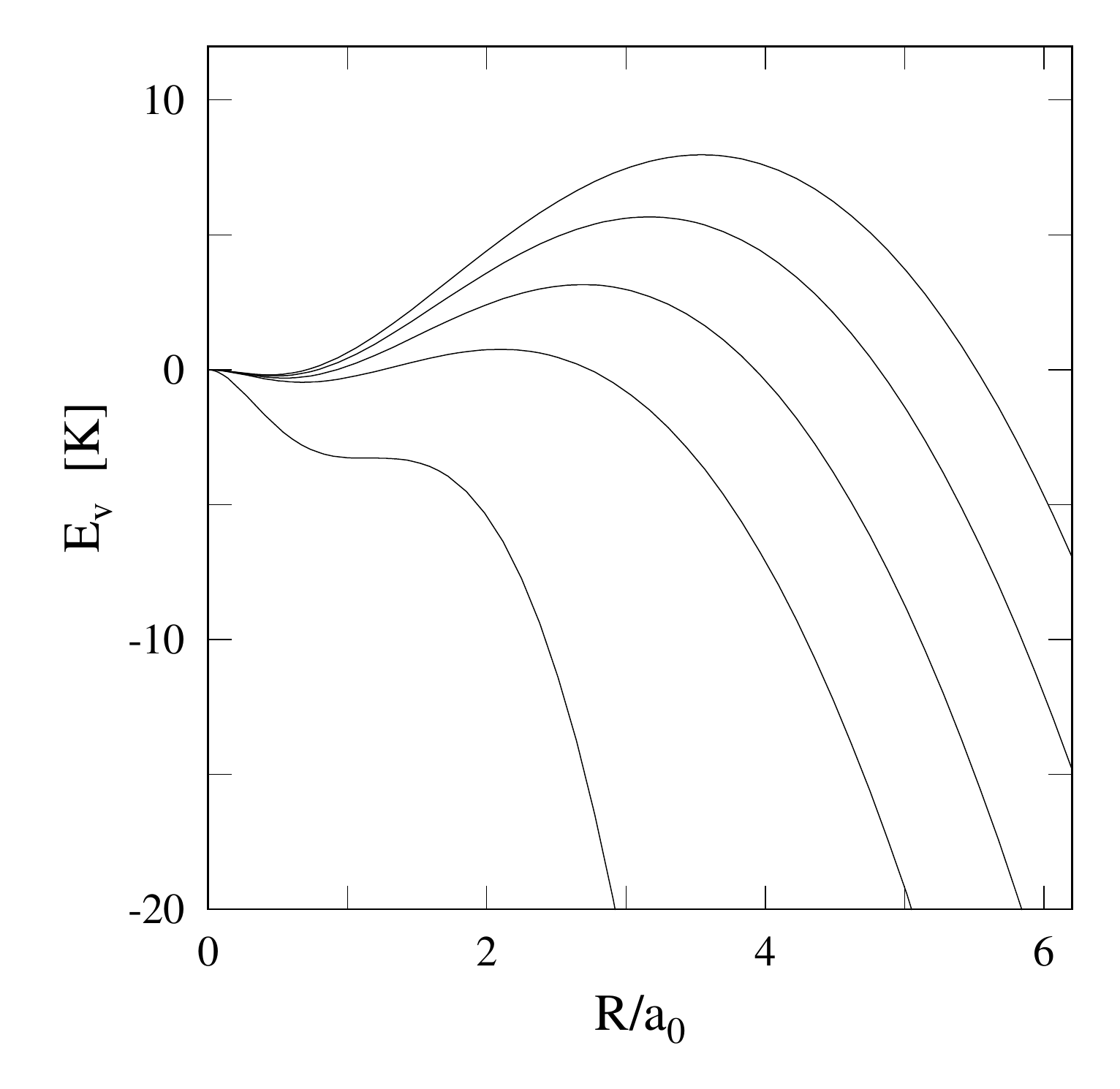}
    \caption{ \label{Ev} Energy barrier $E_\m v$ normalised by $\rho_\m s
      \kappa_4^2 a_0$ and expressed in kelvin {\it vs} the vortex radius
      $R/a_0$ for the vortex half-ring model.  This figure illustrates various
      potential forms given by Eq.(\ref{VortexFreeEnergy}) at various
      superfluid flow velocities taken from the data in Fig.\ref{vc} at, from
      top down, $T\rightarrow T_{\ulambda}$, 1 K, 0.5 K and at the quantum
      crossover $T_\m q$. The lowest curve illustrates the situation in which the
      potential barrier vanishes and the trapped particle runs away, the
      so-called lability point. Adapted from \protect\citet{Avenel:93}.  }
  \end{center} 
\end{figure}
%

The rest energy $E_\m R$ increases with vortex size as $R\ln R$ and the
impulse $P_\m R$ as $R^2$: the impulse term becomes dominant at large radii
and causes $E_\m v$ to become negative. The variation of $E_\m v$ in terms of
$R$ has the shape of a confining well potential, which becomes shallower and
shallower with increasing $v_\m s$, as depicted in Fig.\ref{Ev}. The barrier
height can easily be computed numerically and plugged into the expression
for $v_\m c$, Eq.(\ref{CriticalVelocity}).  An analytic approximation for
$v_\m c$ involving the neglect of logarithmic terms and valid for large
vortices ($R\gg a_0$) has been given by \citet{Langer:70}.

This critical process would yield a mist of vortices in the bulk of the
superfluid.  This sort of vorticity condensation does not take place for two
reasons. First, the velocity of potential flows, which follows from the
Laplace equation, reaches its maximum value at the boundaries, not in the
bulk. Secondly, the nucleation of a vortex half-ring at the boundary itself
involves, for the same radius hence the same self-induced velocity, a half of
the energy given by Eq.(\ref{VortexFreeEnergy}); for that reason alone,
half-ring nucleation at walls is always much more probable at the same
velocity $v_\m s$ than full ring nucleation in the bulk. 

Halving the full-ring energy for the half-ring holds for classical
hydrodynamics, the other half being taken care of by the image in the plane
boundary. For a superfluid vortex, the actual energy of a half-ring is smaller
than in the classical ideal fluid because the superfluid density is depleted
at the solid wall and the core radius increases.

The half-ring model for the nucleation of vortices has been proposed for ion
critical velocity by \citet{Muirhead:84} and for aperture flows by
\citet{Burkhart:94}.\footnote{See also
  \citet{Zimmermann:94,Burkhart:95,Varoquaux:96}.}  A variant, based on a
different accounting of the vortex core energy, has been studied by
\citet{Zimmermann:98}.  Other mechanisms have been discussed
\cite{Josserand:95,Josserand:95b,Andreev:04} for which it is unclear that the
end product of the nucleation process is actually a vortex.

The barrier height for the vortex half-ring nucleation can easily be computed
and plugged into the expressions for $v_\m c$ and $\Delta v_\m c$,
Eqs.(\ref{CriticalVelocity}) and (\ref{Width}).  Critical velocities $v_\m c$
and statistical widths $\Delta v_\m c$ computed in such a manner are shown as
a function of temperature by the solid lines in Figs.\ref{vc} and
\ref{Delta-vc} for several values of the vortex core parameter $a_0$.  A value
of 4.5 \AA\ gives near-quantitative agreement with the experimental
observations over the entire temperature range. This value exceeds that in the
bulk ($a_0\simeq 2.5$ \AA). This is thought to reflect the proximity of the
wall as discussed by \citet{Varoquaux:01a}. With this value, the nucleating
half-ring has a radius of approximately 15 \AA\ at the top of the barrier and
a self-velocity of $v_\m R = 13.5$ m/sec from Eq.(\ref{GroupVelocity}); this
value compares well with the values shown by open circles in
Fig.\ref{TwoCriticalVelocities}.
%

    \begin{figure}[h]   
      \includegraphics[width=43mm,angle=90]{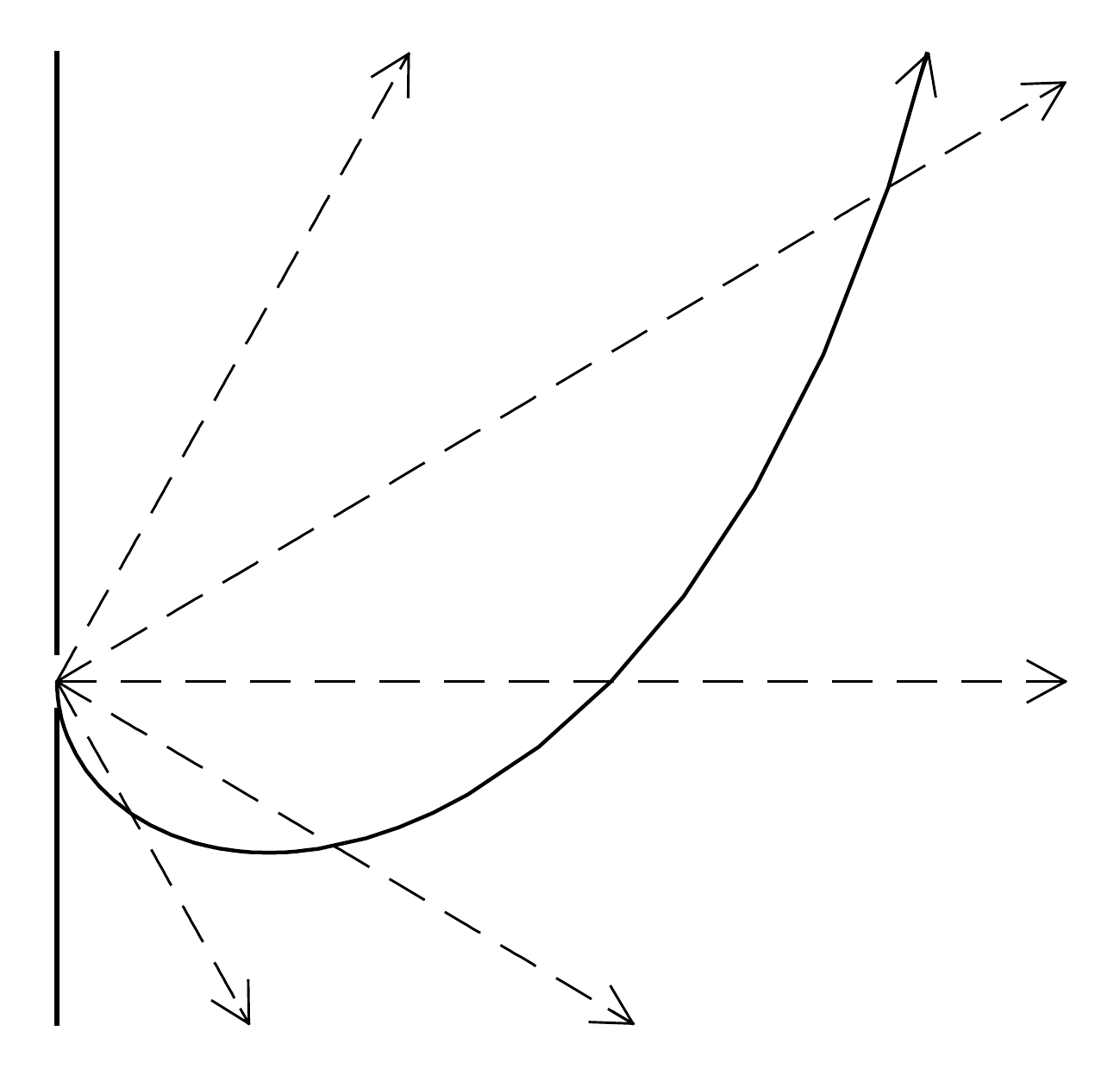}
      \includegraphics[width=57mm,angle=-0]{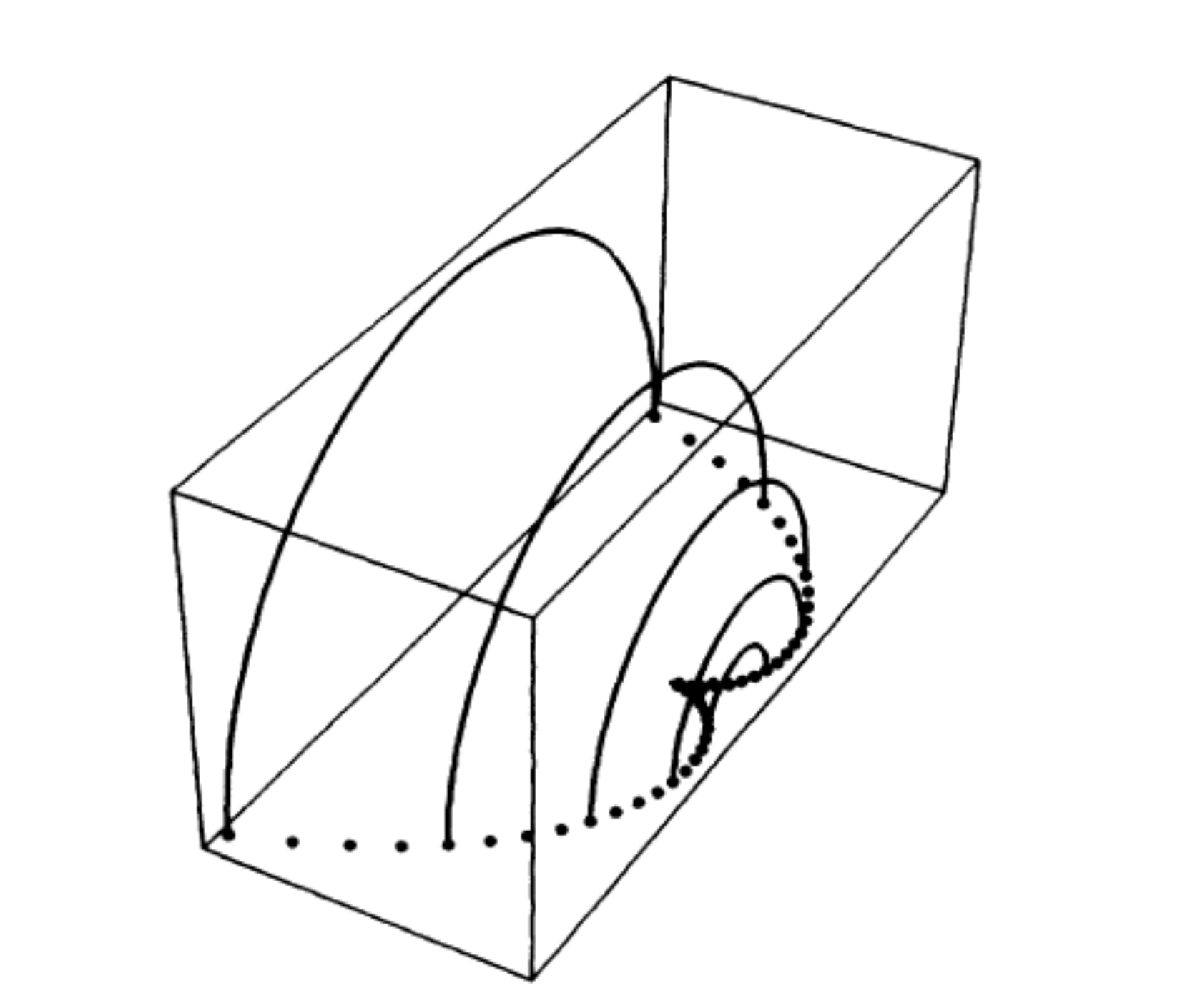}
      \caption{ \label{trajectory} Schematic views in 2D (top) and 3D (bottom)
        of the vortex half-ring trajectory over a point-like pinhole punched
        in an infinite horizontal plane. The dashed lines on the 2D plot are
        the potential flow streamlines that emerge from the pinhole. The 3D
        view shows the vortex half-ring being first pushed by the potential
        flow to the right and then flying back over the pinhole to finally
        drift away to the upper left. Adapted
        from \protect\citep{Varoquaux:06}.}
    \end{figure}

Once nucleated, the vortex half-ring floats away, carried out by the
superfluid stream at the local superfluid velocity and by its own velocity,
$v_\m R = \partial E_\m R/\partial P_\m R$. It can be noted that, at the top
of the barrier, $\partial E_\m v/\partial R =0$: the vortex self-velocity
$v_\m R$ exactly balances the applied $v_\m s$; the nucleating vortex is at a
near standstill.

If the flow were uniform, with parallel streamlines, nothing much would
happen. Downstream from the aperture however, the flow is divergent, as
pictured in Fig.\ref{trajectory}. The vortex half-ring tends to follow the
local streamlines and grow under the combined action of the potential flow and
its own self-velocity: it then gains energy at the expense of the potential
flow as explained in \S\ref{EnergyExchange}. In such a way, it  expands from
nanometric to micrometric sizes and above, and wanders away. Interaction with
the normal fluid, encounters with other vortices and friction on the solid
boundaries cause a loss of vortex energy that eventually leads to its
disappearance. The vortex in its motion away from the micro-aperture takes a
given finite lump of energy to remote places of the cell and never returns.

This scenario for a phase slip produces a change of the phase difference
between the two sides of the micro-aperture of exactly $2\pi$ because the
vortex ends up crossing all the streamlines, as pictured in
Fig.\ref{trajectory}. This crossing causes the velocity circulation to change
by exactly one quantum $\kappa_4$ on all the superfluid paths extending from
one side of the aperture to the other. Such a dissipative event gives the
signature of single phase slips that are seen in Fig.\ref{PhaseSlipsCanJPhys}.
Extensive numerical simulations by \citet{Schwarz:93,Schwarz:93c} and
\citet{Flaten:06b} fully confirm the above scenario for phase slips.

The effect of $^3$He impurities on the phase slip critical velocity at low
temperature is striking, as seen in the inset of Fig.\ref{vc}. It is due to
the condensation of these impurities on the vortex cores, which changes their
line energy and the potential barrier for nucleation. This impurity
dependence, studied in detail by \citet{Varoquaux:93}, was used as a local
probe of the superfluid velocity at which vortices nucleate. This velocity was
found to be 22 m/s, an important piece of information in reasonable agreement
with the value derived from the vortex half-ring model.

In all, the model parameters hang fairly well together, showing that specific
finer details are probably not very relevant and that the model
simplifications are reasonably well founded. It nonetheless remains that the
nature and geometry of a typical nucleation site are wholly unspecified and
that the enhancement factor between the mean aperture velocity and the
velocity at the nucleation site is not under control, as shown in particular
by \citet{Shifflett:95}.

%
\section{Vortex pinning, mills and flow collapse}
                     
%
\label{AllThat} 
                       
Single phase slips are observed in experimental situations that may be loosely
characterised as ``clean'', broadly speaking for uncontaminated apertures of
relatively small sizes (a few micrometres at the most), with low background of
mechanical and acoustical interferences ..., and with probing techniques
that do not manhandle the superfluid, namely, with low frequency
hydromechanical resonators.

When these conditions are not met, flow dissipation occurs in a more erratic
manner in large bursts -- multiple phase slips or ``collapses'' of the
superflow.  Such collapses of the superflow through an orifice were first observed
by Sabo and Zimmermann, Jr. \footnote{As quoted by \citet{Hess:77}.} and by
\citet{Hess:77}.
 
Multiple phase slips and collapses constitute an apparent disruption of the
vortex nucleation mechanism described in the previous Section. Their
properties have been studied in detail by \citet{Avenel:95} and are briefly
mentioned below, together with possible mechanisms for their formation. It is
likely that these events provide a bridge between the ``clean'' single phase
slip case and the usual situation of the Feynman-type critical velocities that
are temperature-independent below 1 K and dependent on the channel
size. Preexisting vorticity is widely suspected to come into play in
these ``extrinsic'' critical velocities.

\subsection{Pinned vorticity}

\citet{Awschalom:84} have directly shown the existence of remanent vorticity
in $^4$He, which had long been assumed, and have given an estimate of its
background level. These authors studied the propagation of ions in the presence of
vortex lines. Ions get trapped in the vortex cores and completely change
course, revealing the presence of these vortex lines. These vortices, presumably
nucleated at the \greeksym{l} transition when the critical velocity is low
and critical fluctuations large, remain stuck in various places of the
superfluid sample container. This trapped vorticity, according to
\citet{Adams:85}, either is quite loosely bound to the substrate and
disappears rapidly, or is strongly pinned and dislodged only by strong
perturbations.

To achieve a stable configuration, a pinned vortex has to take on a shape such
that its local radius of curvature results in a self-velocity that exactly
opposes at each of its points the local value of the superflow.  This dynamic
equilibrium is what is meant here by pinning.  Vortex pinning exists in bulk
$^4$He as discussed here,\footnote{See \citet{Varoquaux:98}, \citet{Donev:01},
  and \citet{Neumann:14} for more references.} in films,\footnote{See
  \citet{Ellis:93}.} in $^3$He,\footnote{See, among others,
  \citet{Hakonen:87},\citet{Zieve:92}, or \citet{Krusius:93}.} in neutron stars \footnote{See \citet{Packard:72},
  \citet{Alpar:80}, and \citet{Langlois:00}.} \ldots

To account for laboratory observations and with the outcome of extensive
numerical simulations of vortex dynamics, Schwarz has proposed the following
formula for the velocity at which such strongly pinned vortices unpin
\cite{Schwarz:81,Schwarz:85},\footnote{See also \citet{Tsubota:94},
  \citet{Neumann:14}.}
\begin{equation}        \label{Unpin}
  v_\m u \lesssim  \frac{\kappa_4}{2\pi D}\,\ln\left(\frac{b}{a_0}\right) \;,
\end{equation}
$D$ being the size of the pinned vortex and $b$ being a characteristic size of
the pinning asperity. Equation (\ref{Unpin}) bears a strong resemblance with
that for the Feynman critical velocity, Eq.(\ref{FeynmanVelocity}).  

As a rule of thumb, the pinning energy of the vortex line on such an asperity
with radius $b$ is approximately equal to $b$ times the line tension of the
vortex given by Eq.(\ref{LineTension}).  Long vortices unpin at very low
velocities unless they are perched on a tall pedestal, but very small vortices
pinned on microscopic defects can survive a wide range of superflow
velocities; according to Eq.(\ref{Unpin}), a straight vortex filament pinned at
both ends on 20 \AA\ asperities 200 nm apart resists transverse flows of
velocities up to 20 cm/s.

Such pinned vorticity has long been thought to play a role in critical
velocities. The long standing suggestion by \citet{Glaberson:66} of vortex
mills had its time of fame \citep{Amar:92}.  In these authors's \,views,
imposing a flow on a vortex pinned between the opposite lips of an aperture
would induce deformations such that the vortex would twist on itself,
self-reconnect, and mill out fresh vortex loops. Upon scrutiny however, vortex
mills are not so easy to set up.

The first thing to realise is that such a mill must involve a pinned vortex of
sub-micrometric size so that it is not washed away by any flow velocity above
a few cm/s. Pinned vorticity in large channels cannot withstand the
Feynman-type critical velocities shown in Fig.\ref{TwoCriticalVelocities}.

Less obviously, vortices are not prone to twist on themselves and foster
loops.  As shown by numerical simulations of 3D flows involving few vortices
only,\footnote{K.W. Schwarz: private communication to the author (1989).}
vortex loops and filaments are found to be stable against large deformations:
it takes the complex flow fields associated with fully developed vortex
tangles to produce small rings \citep{Svistunov:95,Tsubota:00}.\footnote{See
  the review by \citet{Tsubota:09}} And it takes some quite special vortex
pinning geometry to set up a mill that actually works.

\begin{figure}[!t]       
  \begin{center}
    \includegraphics[angle=0,width=45 mm,height=67 mm]{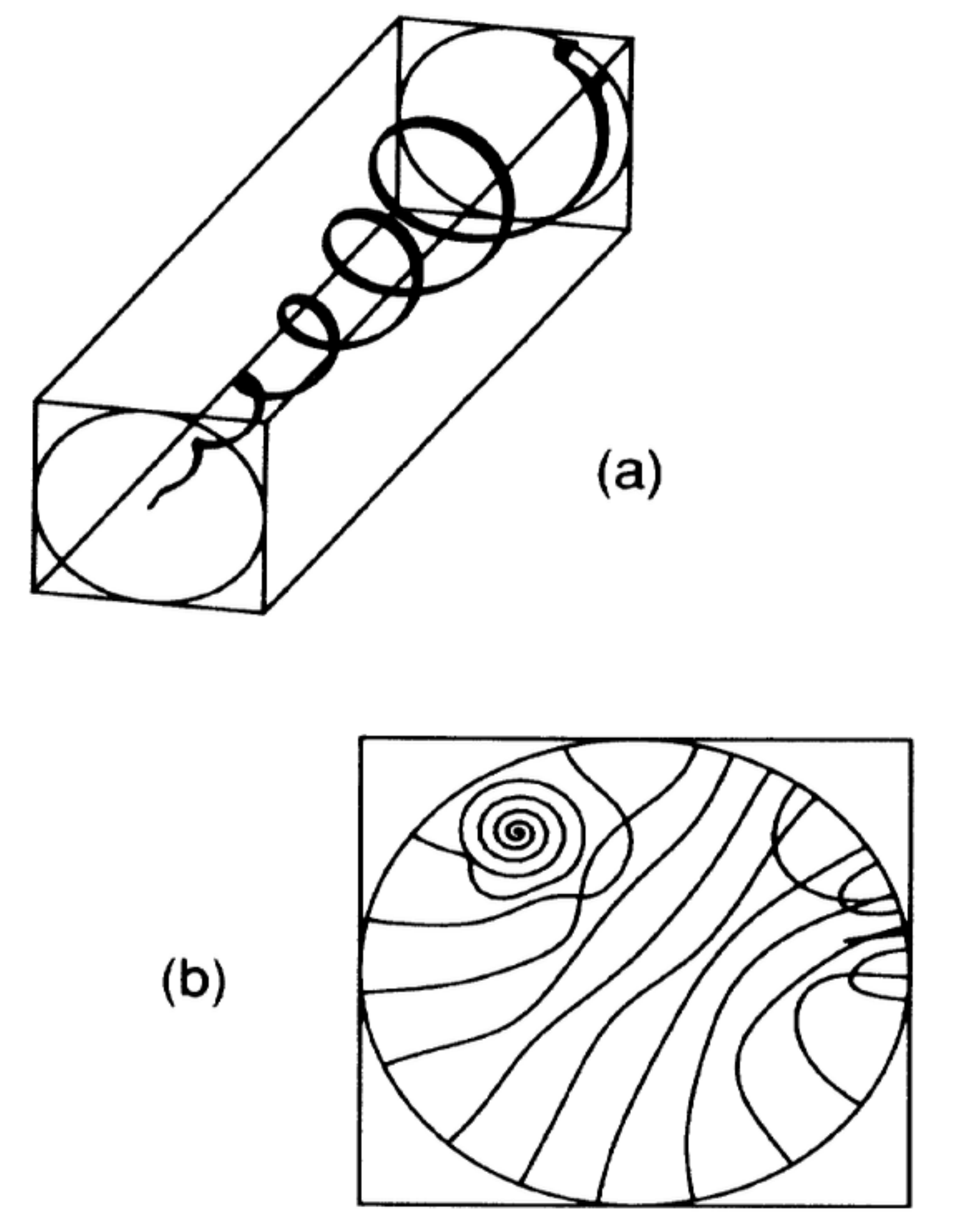}  %
    \caption{\label{Schwarz:90-Fig3} Operation of a helical vortex mill:~(a)
      Spiral-helix configuration of a streamwise vortex filament pinned at the
      centre of the bottom left section of the channel.  The helical vortex
      self-velocity opposes the superflow while the vortex filament grows in
      size and spirals on itself. (b) End-on view of subsequent
      reconnections. The vortex is pinned near the upper left of the channel
      cut view. The outwardly growing spiral sporadically reconnects to the
      wall and releases a new line segment, which then moves to the lower
      right. These numerical simulations are carried out in the presence of a
      sizeable mutual friction with the normal component in order to stabilise
      the numerical algorithms, which is why the generated half-rings rapidly
      decrease in size. From \protect\citet{Schwarz:90}.}
  \end{center}              
\end{figure}

\citet{Schwarz:90} has demonstrated the existence of such a mill by numerical
simulations. Imagine a vortex filament pinned at one end in a region close to
the aperture mouth or the channel entrance where it bends sharply to withstand
the local superflow. This end of the vortex is at a near-stagnation point. Its
other end is being carried away by the flow along the streamlines; it moves
freely with its end sliding perpendicular to the wall. The filament develops a
helical instability as depicted in Fig.\ref{Schwarz:90-Fig3}, a sort of driven
Kelvin wave, and reconnects sporadically to the wall when the amplitude of the
helix grows large enough. The freed bit immediately stands against the flow
and forms a vortex half-ring: such a helical vortex mill, which has to be of
sub-micrometric size to withstand the near-by flow, does churn out fresh
vortices.

The occurrence of multiple slips such as those shown in Fig.\ref{TimeChart},
is probably caused by such a form of vortex mill on a microscopic size.
Before coming to this topic, a description of multiple slips in greater
details must be provided.

%

\subsection{The two types of large slips}
\label{LargeSlips}

Examples of multiple slips are shown on the peak amplitude charts in
Fig.\ref{TimeChart} and in Fig.\ref{collapse} for two different runs in the
same experimental cell. They display rather different patterns. In
Fig.\ref{TimeChart}, multiples slips are fairly frequent and their winding
number multiplicity remains moderate. As the probability for a one-slip event
per half-cycle is not large, that for a double slip is small, and it becomes
negligible for higher multiples. A separate mechanism for their formation must
be found. The event shown in Fig.\ref{collapse} is quite spectacular as it
leads to a near-extinction of the resonance.

In Figs. \ref{TimeChart} and \ref{collapse} aperture velocities are expressed
by the number of turns $\delta \varphi /2\pi$ by which the quantum-mechanical
phase winds across the aperture. Phase winding numbers are related to
  mean flow velocities in cm/s by multiplication by $\ell_{\m h}/\kappa_4$, the
  ``hydraulic'' length $\ell_{\m h}$ characterising the extension of the aperture
  along the flow. For a phase slip by 2$\pi$, the phase winding number changes
  by one unit and the trapped circulation in the resonator loop by one
  quantum.The hydraulic length is defined by lumping the (classical) kinetic
  energy of the fluid moving with velocity $\mb v$ inside the aperture of
  cross-section $s_{\m h}$ and in its vicinity according to: 
\begin{equation}        \label{HydraulicLength}
 \ell_\m h = s_{\m h}\,\int_V \mb
  v(\mb r)^2 \m d^3 \mb r \,\big/\, \big(\int_S\mb v(\mb r)\cdot \m d^2 \mb
  r\big)^2 \; . 
\end{equation}
The actual flow velocity averaged over the cross-section of the
micro-aperture is proportional to $\delta \varphi$, the multiplying factor
being $\hbar /m_4 \ell_\m h$.  The hydraulic length $\ell_\m h$ of the
micro-aperture is of the order of 1 \umu m in the experiments shown in
Fig.\ref{PhaseSlipsCanJPhys}.

Some degree of understanding of the formation of multiple slips can be gained
by plotting the mean value of the phase slip multiplicity, expressed in number
of quanta, against the flow velocity at which the slips take place
\cite{Varoquaux:95}. This flow velocity is close to the critical velocity for
single phase slips, {\it i.e.} the vortex nucleation velocity; it varies with
temperature, pressure, and resonator drive level. A plot summarising these
variations is shown in Fig.\,\ref{MultiSlip} for $<\!n_{_{+}}\!\!>$, {\it i.e.},  in
the flow direction conventionally chosen as the $(+)$ direction.  Slips in the
opposite $(-)$ direction behave qualitatively in the same manner but the
phenomenon displays a clear quantitative asymmetry. 

As can be seen in Fig.\,\ref{MultiSlip}, the mean slip multiplicity decreases,
as does the nucleation velocity, on either side of the quantum plateau -- a
$^3$He impurity effect on the low--$T$ side -- a thermal effect on the high--$T$
side.  At 16 bars, $<\!n_{_{+}}\!\!>$ increases from $\approx 1$ at 12 mK
where the velocity is 43.2 in winding number to $15 \sim 20$ on the plateau of
$v$ with a winding number of 48; $<\!n_{_{+}}\!\!>$ drops back to 1 at 225 mK
and the velocity to $\sim 46.2$. The same trend is observed at 24 bars, with a
lower velocity on the plateau of 46.4 and a higher mean multiplicity of $\sim$
23.  Rather oddly, the slip multiplicity $<\!n_{_{+}}\!\!>$ increases with
pressure while the mean velocity at which these multiples occurs
decreases. One would have expected -- naively it seems -- that more slips
would be ushered into faster streams. The opposite may show that the
phenomenon under study is not purely ruled by hydrodynamics in the fluid bulk.

Another important feature of the data shown in Fig.\,\ref{MultiSlip} concerns
the dependence of the velocity threshold for the appearance of multiple slips
on hydrostatic pressure. The $P$-dependence of the upturn of
$<\!n_{_{+}}\!\!>$ in terms of $v$ exactly tracks that of $v_\m c$, the
critical velocity for single phase slip nucleation: multiple slips occur when
single slips occur. Would multiple slips appear because of an alteration, or
as a consequence, of single slip nucleation?

The very large drops in the resonance amplitude of the resonator such as the
event shown in Fig.\ref{collapse} and in the inset sometimes result in a
complete collapse of the resonance. Under the conditions of this particular
exper-
%
\begin{widetext}
\begin{center}
\begin{figure}[h]       
\includegraphics[height=160mm,angle=90]{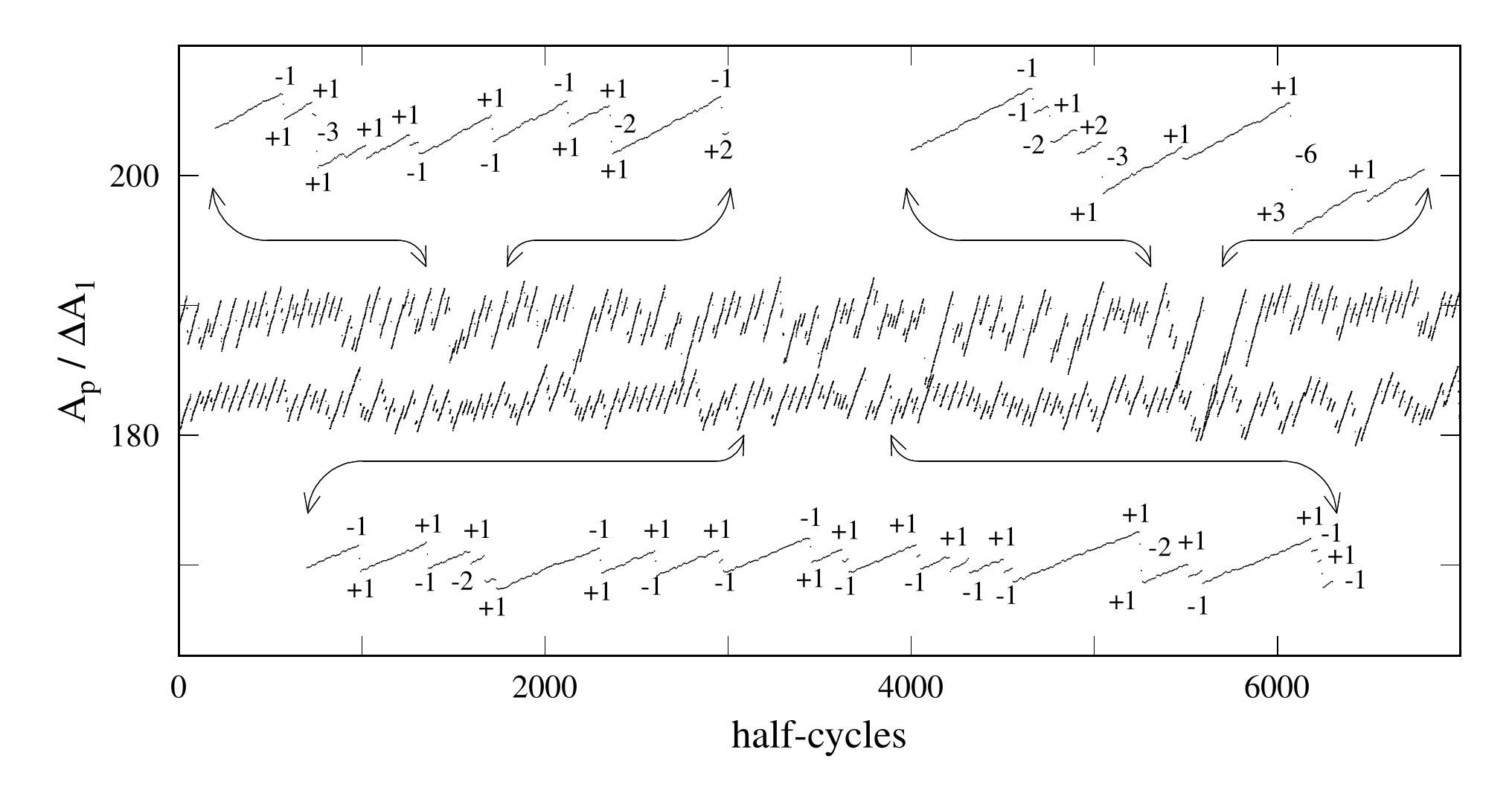}  
\caption{ \label{TimeChart} Absolute peak amplitude $A_\m p$ normalised to the
  amplitude jump of a single slip $\Delta A_1$ during successive half-cycles
  of the resonance plotted as a function of half-cycle index (time runs from
  left to right) at 100.7 mK (top) and 201.5 mK (bottom) in nominal purity
  helium at SVP.  The expanded traces at the very
  top and bottom of the graph show the slip sizes in signed winding numbers
  (according to flow direction, in and out of the resonator chamber).  The
  numerous multiple slips seen on these charts occur at or close to the
  critical velocity for single slips. From \protect\citep{Varoquaux:03}. }
\end{figure}
\end{center}
\end{widetext}
%
iment \citep{Avenel:95}, these events were rare (one in $10^4 \mbox{ to }
10^5$ slips). A striking feature is that they may occur at velocities much
below the vortex nucleation threshold, down to less than a third of $v_c$, the
critical velocity for phase slips.  These intriguing ``singular'' collapses,
first studied by \citet{Hess:77}, differ from the multiple slips of
Fig.\ref{MultiSlip}. The underlying mechanisms responsible for each are bound
to be different, as discussed below \cite{Varoquaux:01a}.

The pattern of formation of multiple slips and collapses changes on cycling
the cell from room temperature and back but remains stable during each given
cool-down. It depends on the degree of contamination of the cell, degree which
cannot easily be controlled experimentally. The detailed microscopic
configuration of the aperture wall where nucleation takes place has a strong
influence on multiple slip formation.

\begin{figure}[h]        
  \begin{center}
    \hskip -5 mm
    \includegraphics[height=70 mm]{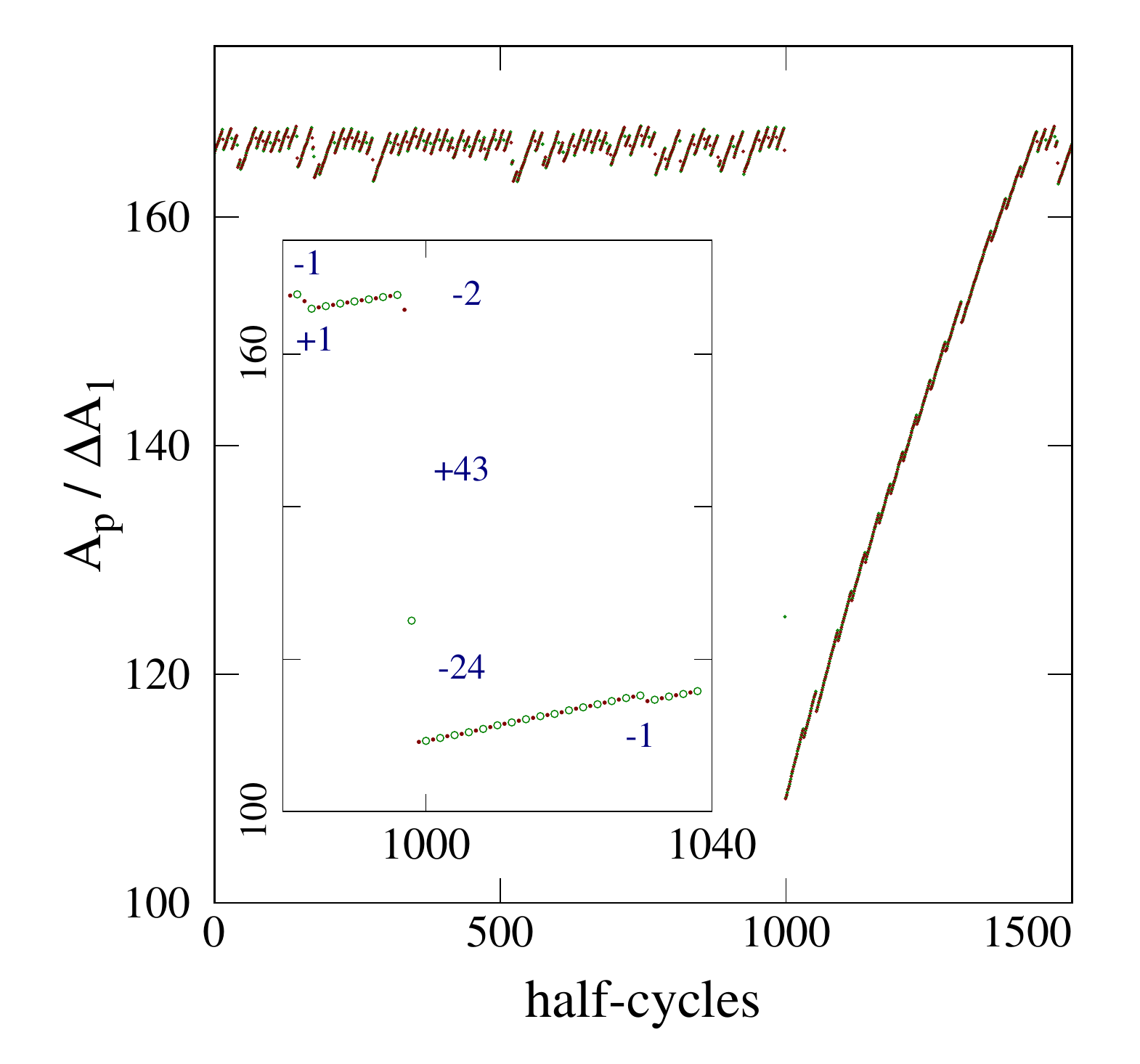}  
    \caption{\label{collapse} Absolute peak amplitudes at successive
      half-cycles of the resonator motion, normalised as in
      Fig.\ref{TimeChart}, {\it vs} half-cycle index in a $^4$He sample
      containing 100 ppb of $^3$He impurity, at 24.0 bars and 12.5 mK. The
      resonance half-period is 31.8 msec. The top trace shows a succession of
      amplitude drops which correspond, for its main part, to phase slips by
      $2\pi$ of opposite sign, with occasional larger slips -- the multiple
      slips of Fig.\ref{TimeChart}. The large feature around the 1000$^{th}$
      half-cycle is a ``singular'' collapse, as defined in this Section. The
      insert shows the details of this collapse, ($\bullet$) being for
      positive peaks, ($\circ$) for negative peaks. It is preceded by a slip
      by -2 ($\times 2\pi$) and followed by a slow recovery of the peak
      amplitude caused by the applied drive, punctuated by single and double
      slips. The time-resolved evolution of this collapse has actually been
      tracked. See \protect\citet{Avenel:95}.  }
  \end{center}
\end{figure}
\begin{figure}[t]        
  \begin{center}
    \includegraphics[width=80 mm,height=70 mm,angle=0]{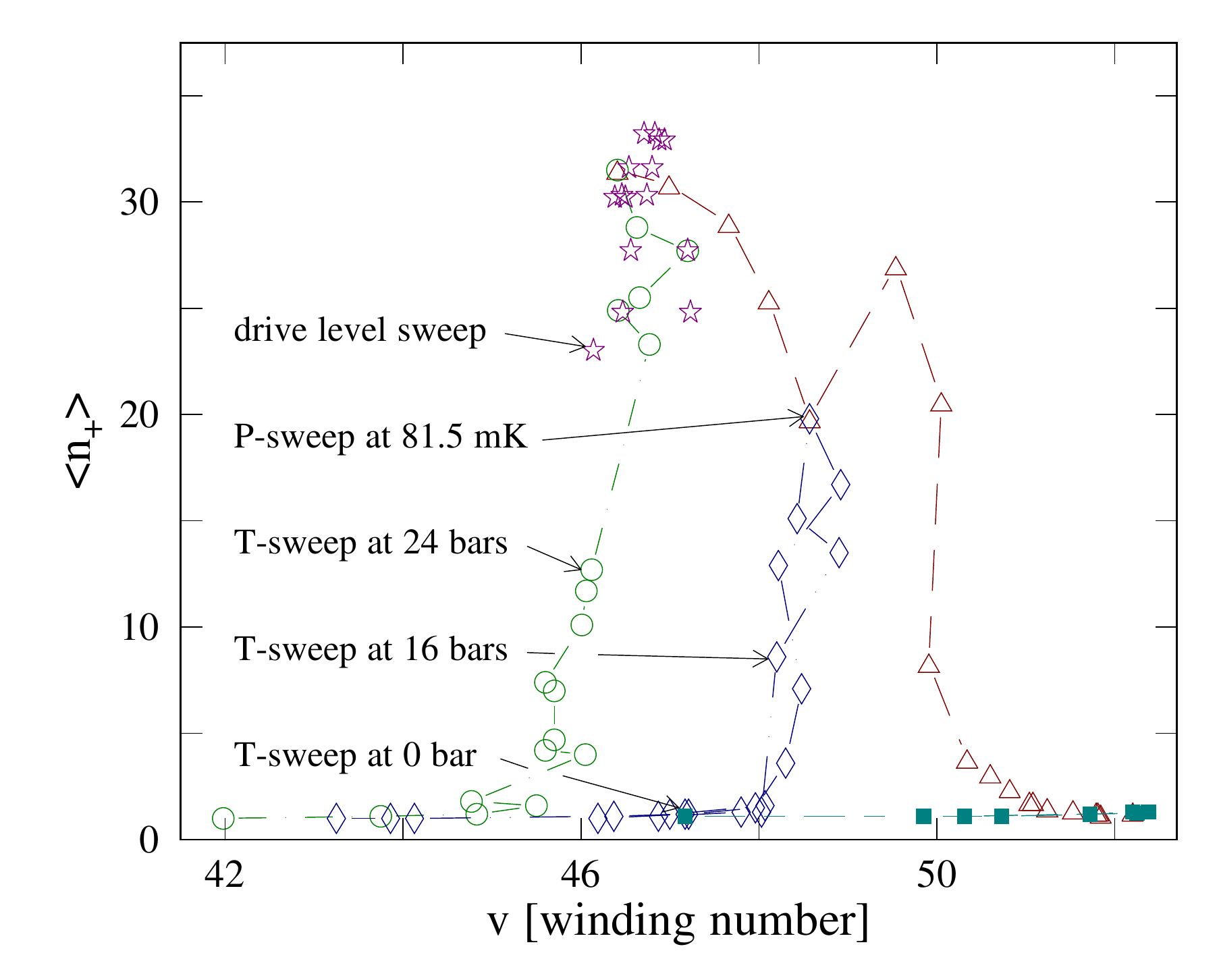}  
    \caption{\label{MultiSlip} Mean size of (positive) multiple slips {\it vs}
      velocity in phase winding number in nominal purity $^4$He (100 ppb
      $^3$He): ($\triangle$) pressure sweep from 0.4 to 24 bars at 81.5 mK
      (for all even values of the pressure $P$, and 0.4, 1, 3, 5, and 7 bars)
      -- ($\diamond$) temperature sweep at 16 bars -- ($\circ$) temperature
      sweep at 24 bars - ($\ast$) drive level sweep at 24 bars, 81.5 mK --
      ($\scriptscriptstyle\blacksquare$) temperature sweep at 0 bar.  Lines
      connect successive data points in the temperature and pressure
      sweeps. For the temperature sweeps, from 14 to 200 mK approximately, $v$
      first increases when the $^3$He impurities evaporate from the vortex
      core, reaches the quantum plateau and then decreases, following the same
      pattern as shown in the insert of Fig.\ref{vc} for $v_\m c$. Adapted from
      \protect\citet{Varoquaux:01a}. }
  \end{center}              
\end{figure}

  
\subsection{Extrinsic critical velocities} 

To try and clarify these matters, a series of experiments was conducted by
\citet{Hakonen:98b}\cite{Varoquaux:98}, in which the experimental cell was
deliberately heavily contaminated by atomic clusters of air and of H$_2$ in
order to favour the pinning of vortices. Numerous multiple slips and collapses
of the ``singular'' type occurred. The peak amplitude charts of the resonator
became very difficult to interpret except in few instances. In one of these,
two apparent critical velocities for single phase slips were observed. The
higher critical velocity corresponded to the one observed in the absence of
contamination. The lower critical velocity was thought to reveal the influence
of a vortex pinned in the immediate vicinity of the prevailing nucleation
site.

Following this interpretation, the pinned vortex induces a local velocity
which adds to that of the applied flow and causes an apparent decrease in the
critical velocity for phase slips. Because of this change, the presence of the
pinned vortex can be monitored. The lifetime in the pinned state and the
unpinning velocity can be measured, yielding precious information on the
pinning process, reported in detail by \citet{Hakonen:98b}.\footnote{This
  topic is also covered by \citet{Varoquaux:98}, \citet{Varoquaux:00a},
  \citet{Varoquaux:00b}, \citet{Varoquaux:01a}}.  This observation also brings
evidence that pinned vorticity can alter the vortex nucleation process
responsible for phase slips.

With such pinned vortices hanging around, multiple slips could formed
according to the following scheme \citep{Varoquaux:01a}. First, a vortex
half-ring is nucleated at a prominent nucleation site. It pins shortly after
nucleation when its velocity relative to the boundary is still small and the
capture by a pinning site easy. A micro-mill is thus formed, which remains
active as long as the flow is sufficient to maintain the helical
instability. As it is set up to withstand one flow direction, it is destroyed
when the flow velocity reverses itself in the resonance motion. It eventually
re-establishes itself during a subsequent resonance cycle, causing a new
multiple slip.  This process depends on the precise details of the pinning
site configuration and of the primordial vortex trajectory, factors which
allow for the variableness of multiple slips on contamination and pressure.

In the same experiments by \citet{Hakonen:98b}, a large number of unpinning
events were also observed to take place at an ``anomalously low'' unpinning
velocity. A parallel can be made \citep{Varoquaux:98} with the singular
collapses that also occur at ``subcritical'' velocities and that were also
quite frequent in the same experiments, suggesting that the two effects might
have a common cause.  Noting furthermore that pinning and unpinning processes
were also quite frequent, releasing a fair amount of vagrant vorticity, it
appears quite plausible that both singular collapses and low velocity
unpinning events are caused by vagrant vortices hopping from pinning sites to
pinning sites, eventually hovering over a pinned vortex or a vortex nucleation
site.  The transient boost to the local velocity may push the pinned vortex
off its perch, or may cause a burst of vortices to be shed.

These observations, albeit incidental, have important consequences for the
critical velocity problem: existing vortices, either pinned or free-moving,
can contribute to the nucleation of new vortices at the walls of the
experimental cell at apparent velocities much lower than the critical velocity
for phase slips.  A mechanism is thus provided by which superflow dissipation
sets in at large scale for mean velocities much smaller than the velocity for
vortex nucleation on the microscopic scale, possibly bridging the gap between
phase slip and Feynman-type critical velocities.

To conclude this Section, the critical velocities in superfluids that are true
and proven include the Landau critical velocity for roton creation in ion
propagation \cite{McClintock:95}, the formation of vortices by a
hydrodynamical instability in BEC gases \cite{Madison:01} and in $^3$He
\cite{Eltsov:05}, the nucleation of vortices by thermal activation and quantum
tunnelling in $^4$He, both for ion propagation and in aperture flow. 

There is also rather compelling experimental evidence for the interplay on a
microscopic scale between vortex nucleation and pinned vorticity; this
evidence points toward the existence of helical vortex micro-mills that can
generate bursts of vortices, even, in some cases, at fairly low flow velocities.
Finally, vagrant vortices interacting with these mills, or with the vortex
nucleation sites are found to generate enough vorticity to completely kill the
superflow and explain singular collapses.  

How these different events occur is illustrated in detail by the numerical
simulations of the onset and decay of vortex tangles in large channels
\cite{Schwarz:83,Schwarz:91}, of the influence of surface roughness on the
critical velocity for a self-sustaining vortex tangle \cite{Schwarz:92}, and
of the evolution of phase-slip cascades from a single remnant vortex as a
function of channel size \cite{Schwarz:93}. These processes depend on the cell
geometry but not on temperature.

A fair degree of understanding of the possible mechanisms behind the Feynman critical
velocity has thus been achieved by the study of phase slippage signatures of
these various large slips.


\section{Josephson-type effects in superfluids}\label{Josephson}


Anderson's conjectures, seen in the previous Sections to be fully confirmed in
the hydrodynamic (macroscopic) limit of quantised vortex dynamics, have also
been carried over to the microscopic limit of quantum tunnelling, as described
below.

The reasoning goes that Eqs.(\ref{Hamilton1}) and (\ref{Hamilton2}) are {\it
  fundamental}~\footnote{See the discussion following Anderson's talk at the
  Sussex University Symposium in 1965 \citep{Anderson:66a}.} enough to carry
the day both at large and short distances, namely when the coherence length is
either small or large with respect to characteristic dimensions of the
hydrodynamic weak link. The former case has been covered in the previous
Sections. In the latter case --  weak quantum coupling between two
superfluid baths -- the contention is that effects analogous to the famed
Josephson effects between two weakly-coupled bits of superconducting
material must also exist between two loosely-connected pools of superfluid
provided that superfluid coherence is not entirely lost through the
connection. These Josephson-type effects in superfluids are dealt with below.
 
\subsection{A simple model}
\label{SimpleModel}

The Hamilton equations (\ref{Hamilton1}) and (\ref{Hamilton2}) express in a
quite general way the time evolution of $\varphi$ and $N$, as discussed in
Sec.\ref{OrderParameter}.  These equations hold in fact for the operators
$\hat{N}$ and $\hat{\varphi}$ but their coarse-grained averages can be treated
as {\it c}-numbers to a very good approximation because their relative quantum
uncertainties are very small. Averaging over a volume of superfluid small
compared to the size of the sample but still containing a large number of
atoms leads to Eq.(\ref{acJosephson}):
\begin{displaymath} \hbar \frac{\partial\varphi}{\partial t} = -(\mu +
\slantfrac{1}{2} m_4 v_\m s^2) \; .
\end{displaymath}
Equation (\ref{acJosephson}) describes the Josephson {\it ac} effect.\footnote{The
contribution of the entropy to the chemical potential, ${\mathcal S}T$, should
also be taken into account in Eq.(\ref{acJosephson}) if the temperature is not very low.}
\begin{figure}           
  \begin{center} \leavevmode
    \includegraphics[angle=0,width=27 mm]{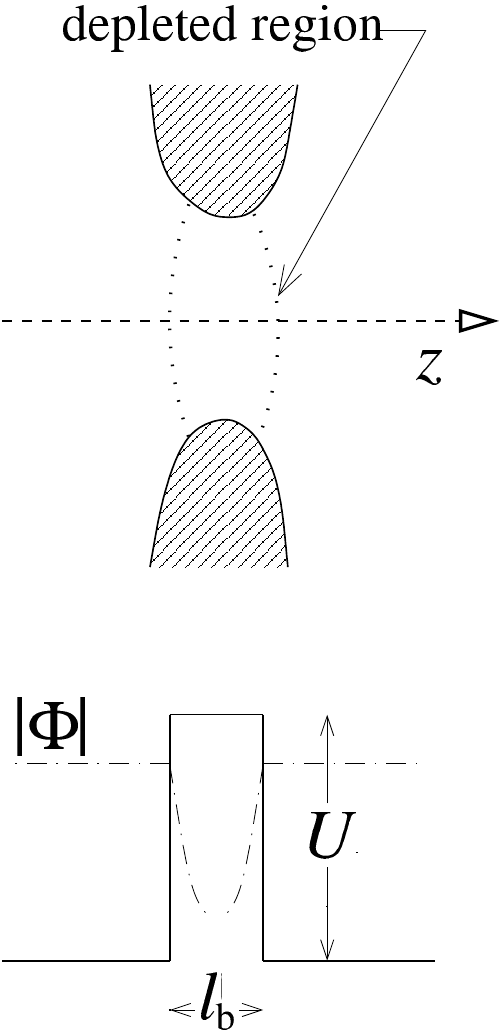}
    \caption{\label{Weak link} Schematic representation of a two--dimensional
      weak link: (top) cut view of the elongated slit in the partitioning
      wall; (bottom) amplitude of the macroscopic wavefunction $\Phi$ in the
      region in which it is depleted by the energy barrier of height $U$ and
      extent $l_\m b$ due to the constricting walls. }
  \end{center}
\end{figure}
When applied to the gradient of the phase,
it can be cast, using Eq.(\ref{vs}), into the Euler equation
(\ref{SimpleEuler}):
\[ \frac{\partial \mb v}{\partial t} + \nabla (v_{\mathrm{ a}} P +
\frac{1}{2} m_{\mathrm{ a}}v^2) = 0 \;,
\] $m_{\mathrm a}$ being the atomic mass of the effective boson.

Equations (\ref{acJosephson}) and (\ref{SimpleEuler}) look plainly classical
enough. Quantum mechanics hides in the possible multiple determinations of the
overall phase $\varphi(t,\mb r)$ of the order parameter, yielding a quantised
circulation of the fluid velocity, and, in a subtler manner, when $\hat{N}$
cannot be coarse-grained averaged because the amplitude of the order parameter
vanishes or varies too rapidly over short distances.  A (quantum) mechanism is
then provided for $\varphi$ to vary discontinuously from one
determination to another, violating the Kelvin-Helmholtz theorem.

The second Heisenberg equation of motion, that for $\dot N$, expresses
particle number conservation:
\begin{equation}       \label{numberconservation} 
  \hbar\,\frac{\partial N}{\partial t} = \frac{\partial
  E}{\partial \varphi} \;.
\end{equation}
As stressed by \citet{Anderson:66b}, the range of validity of
Eqs.(\ref{acJosephson}) and (\ref{numberconservation}) is quite wide. They will
still hold when hydrodynamics breaks down as for tunnelling supercurrents. In
this kind of situations, the internal energy $E$ depends in a non-trivial way
on $\varphi$, as may be expected from Eq.(\ref{numberconservation}).

When applied between two regions of the superfluid, Eqs.(\ref{acJosephson})
and (\ref{numberconservation}) describe the supercurrent flowing from one
region to the other.  This situation becomes especially interesting when the
two regions, the two superfluid baths, are sufficiently well separated so that
they only {\it  weakly coupled}: a well-defined phase difference between
them, $\udelta \varphi$, can then be sustained.

Such a situation can be modelled by a potential barrier, as in Fig.\ref{Weak
  link}. The thin partition separating the two baths presents a thin elongated
slit through which a trickle flow only of superfluid can leak. If the two
smaller dimensions of the slit are comparable to the superfluid coherence
length --- the distance over which its wavefunction can heal --- the amplitude
of the wavefunction is reduced in the narrow passage, as pictured in the
bottom panel of Fig.\ref{Weak link}. In superconductivity, such weak links, or
micro-bridges, are known to lead to the same kind of effects as tunnel
junctions \cite{Likharev:79,Golubov:04}. 

For superflows through such a micro-aperture, the problem can be restricted to
one dimension along $z$ and, to simplify further, the barrier (the weak link)
can be taken as a square potential wall of height $U$ over length $l_\m b$.
\footnote{See the discussion given by \citet{Vinen:68}, still very relevant 40
  and more years later, also \citet{Varoquaux:92}.}

In the bulk of the fluid, the wave function corresponding to a state with
energy $E$ is taken as a plane wave with identical amplitude
$|\mi\Phi|=(\rho_\m s/m_\m a)^{1/2}$ on both sides of the barrier ($ m_\m
a = 2 m_3$ for superfluid $^3$He), but with
phases that differ by $\udelta \varphi$: these are the boundary conditions at
the weak link walls at $z=0$ and $z=l_\m b$.

Inside the barrier $|\mi\Phi(z)|$ is assumed to be severely depressed: the
interactions within the fluid can be neglected.\footnote{In the framework of
  the Gross-Pitaevskii equation, the interaction term
  $V_0(\mi\Phi^*\mi\Phi)\mi\Phi$ becomes small in the depleted region and can
  be neglected - see \S \ref{p-wave} for superfluid $^3$He} With this
approximation of weak coupling, the tenuous fluid inside the weak link behaves
as a simple non-interaction gas and the equation of motion reduces to a
one-particle Schr\"odinger equation:
\[ \m{i}\hbar\frac{\partial \mi\Phi}{\partial t} =
-\frac{\hbar^2}{2m_{\mathrm{ a}}} \nabla^2\mi\Phi + U\mi\Phi \; , \hskip 2mm U
> E \;,
\] 
and also has a plane wave solution $\exp\{-\mathrm{i}(Et/\hbar -
kz)\}$. The momentum takes two values corresponding to the two possible
directions of (damped) propagation:
\[ k_{\ts\pm} = \pm (\m{i}/\hbar)\sqrt{2m_{\mathrm{ a}}(U-E)} \;.
\] 
Let $b_\m b=\hbar/\sqrt{2m_{\mathrm{ a}}(U-E)}$\,: the barrier height is
characterised by a penetration length.  The wave function inside the barrier
is found by standard methods:
\[ \mi\Phi(z) = \frac{|\mi\Phi|}{{\sinh}(l_\m b/b_\m b)}
  \left\{{\sinh} \left(\frac{z}{b_\m b}\right)\,{\mbox e}^{\ts
  {\m i\udelta\varphi}} - {\sinh}\left(\frac{z-l_\m b}{b_\m b}\right)\right\}
\; .
\] 
The modulus of $\mi\Phi$ midway in the barrier is expressed by
\begin{equation} \label{Modulus} 
  \mi\Phi^*(l_\m b/2)\,\mi\Phi(l_\m b/2) =
  \frac{\rho_\m s/m_\m a}{2 \cosh^2(l_\m b/2b_\m b)}[1+\cos \udelta\varphi] \;,
\end{equation}
and is a $2\pi$-periodic function that vanishes for $\udelta\varphi =
\pi\,\pm 2n\pi$. The weak coupling condition is satisfied in superfluid
helium for $l_\m b \gtrsim b_\m b$.

 Knowing the wavefunction, the current density, Eq.(\ref{Current}), can be
straightforwardly computed. The total current through a
micro-aperture of effective cross-section $s_\m b$ is found to be:
\begin{eqnarray} 
  J &=& \frac{\hbar s_\m b}{2\m i b_\m b}
    \frac{\rho_\m s/m_\m a}{{\sinh}^2\left(l_\m b/b_\m b\right)}
   \left\{\left[{\sinh}\left(\frac{z}{b_\m b}\right)\,
   {\mbox{e}}^{\ts{ -\m{i}\udelta\varphi}} - {\sinh}
   \left(\frac{z-l_\m b}{b_\m b}\right)\right]\times\right.  \nonumber 
  \\ &&\hskip -5mm
     \left. \left[{\cosh}\left(\frac{z}{b_\m b}\right) {\mbox{e}}^{\ts{
     \m{i}\udelta\varphi}} - {\cosh}\left(\frac{z-l_\m b}{b_\m
    b}\right)\right] - {\mbox{complex conjugate}}\right\} \nonumber 
  \\ &=& 
     J_\m c\, \sin\left(\udelta\varphi\right)\,, \;\;{\mbox{with}}\;
     J_\m c = \frac{\hbar}{m_\m a}\,\frac{s_\m b}{b_\m b}\,
       \frac{\rho_\m s}{{\sinh(l_\m b/b_\m b)}} \;.
\label{dcJosephson}
\end{eqnarray}

Equation (\ref{dcJosephson}) describes the Josephson {\it dc} effect.
Although this equation has been obtained here in a drastically simplified
manner, it is nearly identical to the result of more involved theories, each
with its own set of approximations -- the Ginzburg-Landau model
\cite{Monien:86}, an ideal tunnel junction \cite{Rainer:87}, or a strictly
point-like orifice \cite{Kurkijarvi:88}.

The supercurrent $J$ is periodic by $2\pi$ in $\udelta\varphi$ as it must be
since changing the phase by $2\pi$ on one side of the barrier must leave the
overall physical situation unchanged.  It vanishes for $\udelta \varphi = \pm
\pi$ not because the velocity, proportional to $\udelta\varphi$, goes to zero
but because the superfluid density, which is proportional to $|\m\Phi|^2
\sin(\udelta\varphi)/\udelta \varphi$ inside the barrier, does; the modulus of
the wave function at midpoint in the barrier, Eq.(\ref{Modulus}), vanishes:
superfluidity is actually destroyed at that point, which is why the
supercurrent goes to zero and the phase can slip by $2\pi$ (or lumps of
$2\pi$).

If the coupling is not weak, a more elaborate calculation is necessary: the
sine function is replaced by a general $2\pi$--periodic function
$f_{2\pi}(\udelta\varphi)$, the current-phase relation, or CPR, for a
``non-ideal'' weak link.  Often, this relation is not even single-valued and,
when the phase is varied, the current may jump discontinuously from one
determination to another: the weak link is then said to be hysteretic. This
behaviour is due to the nucleation of topological defects such as vortices as
seen in the previous Section. It is accompanied by dissipation while the ideal
Josephson case (when $f_{2\pi}(\udelta\varphi)$ is a sine function) is
dissipation-less \cite{Likharev:79,Thuneberg:05,Viljas:05}.

In the transition between the ``ideal'', non-hysteretic, purely sinusoidal
CPR's and the mostly linear CPR seen, for example, in
Fig.\ref{PhaseSlipsCanJPhys}, a slanted sine function is often observed. Part
of this distortion arises from purely classical fluid flow in the vicinity of
the micro-aperture.  The full phase difference across the weak link,
$\varphi_\m w$, includes, besides the phase difference across the barrier
$\varphi_\m b$, the rather trivial velocity potential drop in the vicinity of
the weak link where the superfluid velocity $v_\m s$, and the corresponding
phase gradient, behave in accord with classical ideal fluid dynamics.

In order to account in a simple manner for this classical contribution, it is
convenient to introduce the equivalent "hydraulic" length and cross-sectional
area of these regions, $\ell_\m h$ and $s_\m h$, in such a way that the flow is
described in a `rod-like'' manner.\footnote{The hydraulic length $\ell_\m h$
  is defined by Eq.(\ref{HydraulicLength}. The thickness of the tunnel barrier
  is neglected.}. The superfluid velocity is then expressed
simply by
$v_\m s = (\hbar/m_\m a)\inc\varphi_\m h/\ell_\m h $ 
and the current by $J = \rho_\m s s_\m h v_\m s$.

The total phase difference $\inc \varphi$ is the sum of
the phase drop through this "hydraulic" region and through
the barrier acting as the weak link, assumed ideal, {\it i.e.}, such that:
\begin{equation}        \label{TotalPhaseDrop}
  \inc\varphi =  \inc\varphi_\m h + \inc\varphi_\m b
              =  \frac{m_\m a \ell_\m h}{\rho_\m s \hbar s_\m h} \, J \; .
\end{equation}
The same mass current also flows through the depletion region and varies,
following  Eq.(\ref{dcJosephson}), as a sine function of the phase difference
$\inc \varphi_\m b$ as long as the coupling is weak.

Combining Eqs.(\ref{dcJosephson}) and (\ref{TotalPhaseDrop}), and renaming the
hydraulic part $\inc \varphi_\m h$ of the phase difference $\zeta$ to stress
its ancillary role yields the relation between the current and the phase of a
(slightly more) realistic micro-aperture:
\begin{equation}        \label{DeaverPierce} 
\varphi = \zeta + \alpha \sin\zeta \;,\;\;\;\;\;\ J=J_\m c \sin\zeta \; ,
\end{equation}
with $\alpha = (m_\m a \ell_\m h / \rho_\m s s_\m h \hbar)\, J_\m c$ and $J_\m
c$ expressed from Eq.(\ref{dcJosephson}).\footnote{Relations
  (\ref{DeaverPierce}) were proposed by \citet{Deaver:72} for
  superconducting junctions. See also \citet{Likharev:79}.} The non-ideality
  parameter $\alpha$ and the critical current through the junction $J_\m c$
  are given a meaning in terms the geometrical details of the micro-aperture. They
  can be derived from experiments and compared with the expected values.
 \footnote{See \citet{Avenel:88} for an example of this procedure and
   \citet{Varoquaux:92} for a more complete analysis.}

 Since the healing length is of atomic dimensions for $^4$He, a near-ideal
 Josephson effect cannot be expected to be found in the micro-apertures that
 can be manufactured at present, except very close to the ${\ulambda}$
 transition when this length diverges.  Experiments close to $T_{\ulambda}$
 have been conducted successfully by \citet{Sukhatme:01} and
 \citet{Hoskinson:06-Nature} and are described in \S\ref{LambdaPoint}. The
 experiments that have first shown the existence of the Josephson {\it
   dc}-effect in superfluids have been carried out in $^3$He
 \citep{Avenel:88}.

\subsection{Current and phase in  superfluid $^{\mb 3}$He}
\label{CPR}

The helium-3 nucleus is made up of two protons and one neutron: $^3$He is 
a fermion. As for the abundant and heavier isotope, $^4$He, its
zero-point energy in the condensed phase is large and it remains in the liquid
phase down to absolute zero at pressures below about 35 bars. It thus forms a
Fermi liquid with a Fermi sphere over which Landau quasiparticles
float. Because the interatomic potential is attractive at large distance, these
quasi-particles can form Cooper pairs and $^3$He was long suspected to become
a BCS superfluid below some hard-to-predict temperature. The discovery by
Osheroff, Richardson and Lee of the transition to not one but two superfluid
phases \citep{Osheroff:72} fixed the transition temperature to 2.49 mK on the
melting curve, at a pressure of 34.34 bars and opened an exciting new chapter of
low temperature physics.

As the experimental properties of these new superfluid phases were quickly
unravelled \citep{Wheatley:75a,Wheatley:75b,Lee:78}, they were identified from
their nuclear susceptibility properties observed by NMR as resulting from the
formation of Cooper pairs in a {\it spin-triplet} state \citep{Leggett:75}. A
new breed of superfluid was just born. The overall antisymmetry of the
wavefunction under the exchange of two fermions then requires an odd angular
momentum state $l=1,\,3\,\ldots$.  The available experimental data, mainly the
phase diagram, the specific heat, and the nuclear susceptibility, led to the
identification of the A and B phases as p-wave Cooper-pair superfluids with
total spin $S=1$ and total angular momentum $L=1$.

The formalism describing the properties of these anisotropic superfluid phases
was quickly developed.\footnote{As related by \citet{Leggett:75} and
  \citet{Anderson:75} -- see the monograph by \citet{Vollhardt:90}.} It
extended the Bardeen-Cooper-Schrieffer theory of s-wave superconductivity to
the neutral triplet-spin-state superfluid.  The most general pair wavefunction
with three possible substates for the spin and the orbital parts is an
arbitrary superposition of these 3x3 substates, involving nine complex
parameters. Assuming weak coupling between the pairs -- a surprisingly good
assumption at low pressure -- this extension of the BCS theory
\citep{Leggett:75} leads to a 3x3 order parameter for B-phase of the form
\begin{equation}         \label{BPhaseOrderParameter}
  A_{\mu i} = \Delta(\hat{\bs{\m k}}) \m{e}^{\ts{\m i\varphi}} \, 
    R_{\mu i}(\hat{\bs{\m n}},\theta)  \; .
\end{equation}
The gap parameter $\Delta(\hat{\bs{\m k}})$ and the phase factor
$\m{e}^{\m i\varphi}$ have the same interpretation as for s-wave
superconductors. The B-phase contains the $S_{\m z}$ = 0, +1, and -1 pairs
($\ket{\uparrow\downarrow + \downarrow\uparrow}$, $\ket{\uparrow\uparrow}$, 
and $\ket{\downarrow\downarrow}$) in equal amounts in
zero applied magnetic field; $\Delta(\hat{\bs{\m k}})$ is isotropic and
independent of $\hat{\bs{\m k}}$, the direction on the Fermi sphere.
 
The matrix $R_{\mu i}(\hat{\bs{\m n}},\theta)$ describes the rotation bringing
the spin quantisation axis along the orbital quantisation axis. This rotation
is characterised by a unit vector $\hat{\bs{\m n}}$ and an angle
$\theta$. Both the gap parameter $\Delta(\hat{\bs{\m k}})$ and the rotation
are real quantities independent of the overall phase $\varphi$. Therefore, the
B-phase order parameter (\ref{BPhaseOrderParameter}) takes the same form
as that for $^4$He, namely, the product of a phase factor ${\m{exp}}(\m i\varphi)$
with a well-defined phase $\varphi$ and a phase independent modulus. The same
reasoning as for the $^4$He case applies when performing a Galilean
transformation: mass transport in the pseudo-isotropic B-phase is related to
the gradient of $\varphi$.

The modulus of order parameter (\ref{BPhaseOrderParameter}), or gap parameter
$\Delta$, can be thought of as the binding energy of a Cooper pair at $T=0$;
it is of the order of $k_\m B T_\m c$,\footnote{ For $^3$He-B in weak coupling
  theory, $\Delta(0)=ak_\m B T_\m c$ with $a\leq 1.75$ \citet{Leggett:75}.}
$T_\m c$ being the superfluid transition temperature. The smallest time lapse
over which this energy can be defined is limited by the uncertainty relation
for time--energy: $\inc t\simeq \hbar / \Delta$. During that time, the pair
spreads over a length $\xi_0 = \hbar v_\m F/\Delta \;, $ where $v_\m F$ is the
velocity of the $^3$He quasiparticles over the Fermi surface.  \footnote{The
  zero temperature coherence length of the B-phase is given by $\xi_0=
  [7\zeta(3)/48\pi^2]^{1/2}\,[\hbar v_\m F/k_\m B T_\m c]$
  (\citet{Vollhardt:90}, \S 3.4). The temperature dependent coherent length
  diverges as $\xi(T)=\xi_0(1-T/T\m c)^{-1/2}$ in the Ginzburg-Landau regime.}
It can be seen from this heuristic argument (\citet{Lounasmaa:83},
\citet{Davis:02}) that properties of the superfluid are well-defined only over
distances larger than the coherence length $\xi_0$, of the order of 600 \AA~at
$T=0$ and low pressure -- 120 \AA~ at melting pressure. The prospect to
observe quantum departures from classical hydrodynamics in $^3$He appears much
more favourable than in the case of $^4$He:\footnote{The state of the art in
  aperture manufacturing evolved with time from submicronic slits
  \citep{Sudraud:87} to nanometric holes \citep{Pereverzev:01}.} the coherence
length is no longer very small compared to the size of apertures that can be
micro-machined; genuine hydrodynamic Josephson effects can be expected to take
place in the B-phase in submicron size apertures, or pinholes.

\subsection{Weak links in p-wave superfluids}
\label{p-wave}

Weak links for superfluids come in two breeds, single micro-apertures in thin
wall partitions and larger scale arrays of such apertures geometrically
arranged a few microns apart, actual pinholes for the
former, a mock-up for tunnel junctions for the latter. The flow patterns in
the vicinity of each kind lead to different weak link behaviours.

The first successful Josephson-type experiments were carried out by
\citet{Avenel:88}, using the same micro-resonator and single aperture as for
their experiments on phase-slippage in $^4$He.\footnote{See \citet{Sudraud:87}
  for an account of earlier attempts and \citet{Sato:12} for later
  developments.}  Their observations spurred intense theoretical interest in
the description of phase slippage in $^3$He. 

Analytic calculations of the current through a pinhole orifice with all
dimensions smaller than $\xi_0$ and with specular reflection of the
quasiparticles on the walls by \citet{Kurkijarvi:88} in the framework of
quasiclassical theory, following earlier work by \citet{Kopnin:86} and
\citet{Monien:86}, lead to the following current-phase relation:
\begin{equation}          \label{CPRKurki}
  J = \mbox{\tt a} s_{\m h} v_{\m F} N(E_{\m F})  \Delta(T)
   \,\sin\big(\frac{\varphi}{2}\big) \,\tanh\bigg[\mbox{\tt b}
    \frac{\Delta(T)}{k_\m B T} \,\cos\big(\frac{\varphi}{2}\big)\bigg] \, ,
\end{equation}
where $N(E_{\m F})$ is the density of states at the Fermi surface, $v_{\m F}$
the Fermi velocity, $\mbox{\tt a}= \pi /2 $, and $\mbox{\tt b} = 1/2$ for the
B-phase. Equation(\ref{CPRKurki}) takes the same form as for a s-wave
supercurrent through a superconducting micro-bridge.\footnote{As obtained by
  Kulik and Omel'yanchuk in 1977, see for instance \citet{Likharev:79} or
  \citet{Golubov:04}.} A similar form also holds approximately for the A-phase
with $\mbox{\tt a}= \pi /\sqrt{6}$ and $\mbox{\tt b} = (3/8)\sqrt{3/2}$, and
for the planar phase, a phase which may possibly be stabilised within the
micro-aperture by the walls.
\begin{figure}[t]        
  \begin{center}
    \includegraphics[width=70 mm]{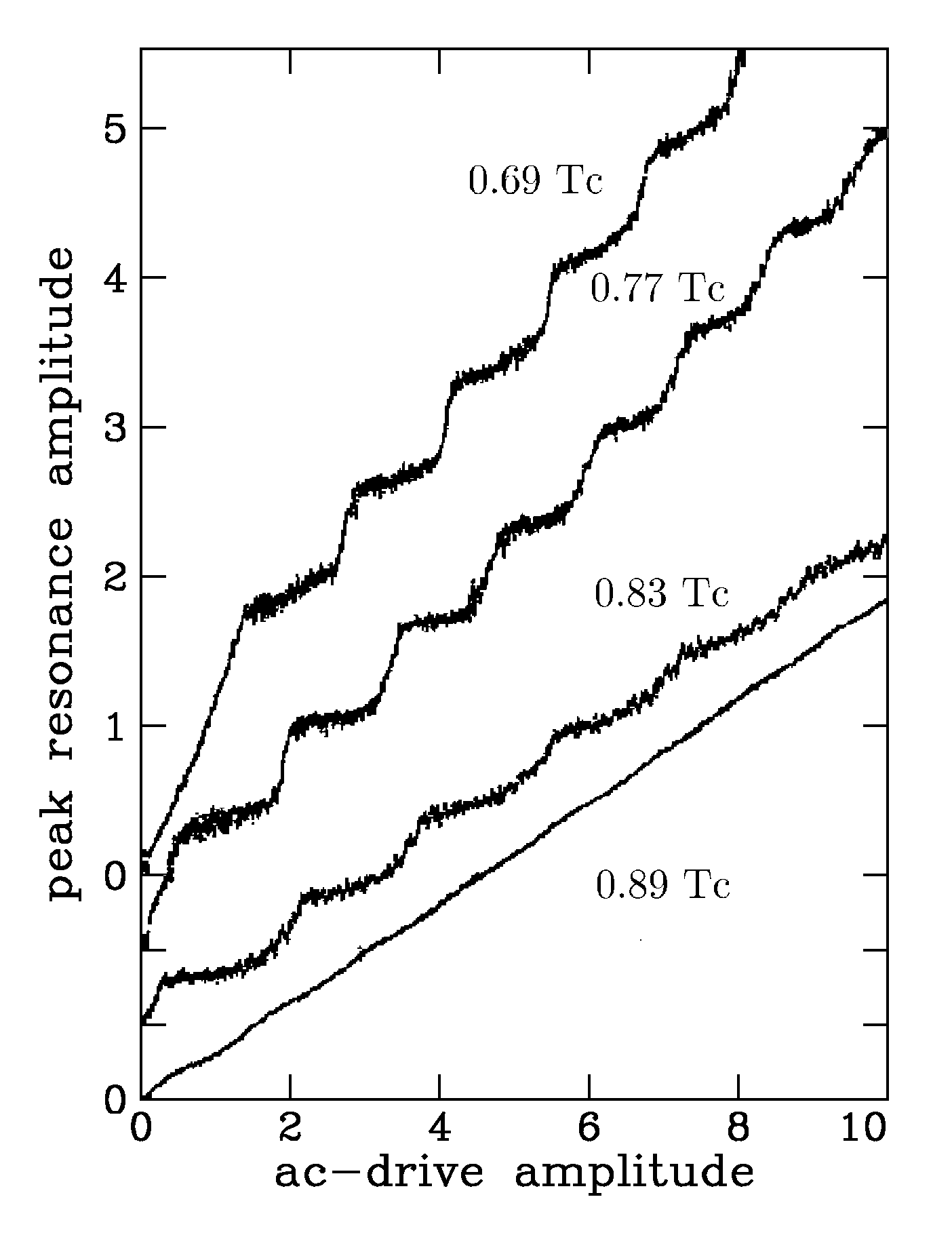}
    \caption{\label{PRL88-fig2-staircase}
      Staircase patterns (peak resonance amplitude versus applied drive level,
      both in arbitrary units) in $^3$He-B at a pressure of about 0.2 bar. The
      curves at various temperatures are shifted vertically for
      readability. Drive frequencies at the various temperatures are shifted
      from resonance so as to set comparable resonance conditions. From
      \protect\citet{Avenel:88}. }
  \end{center}
\end{figure}

Expression (\ref{CPRKurki}) reduces in the limit $\Delta(T)/k_\m B T \ll1$ to
the sinusoidal dependence of Eq.(\ref{dcJosephson}) for the current in terms
of the phase difference across the barrier $\varphi_\m b$. This result has
been obtained by a number of authors using a variety of
techniques.\footnote{See, in
  particular,\citet{Monien:86}, \citet{Hook:87}, \citet{Ullah:89},
  \citet{Kopnin:90}, \citet{Thuneberg:90}, \citet{Soininen:91}, \citet{Kopnin:91}}
It is no real surprise that the details of the structure of the order
parameter disappear when the dimensions of the orifice are small with respect
to the coherence length and that s-wave-like results are found for both the A
and B phases. Superfluid coherence is effectively weakened by the
micro-orifice because the length over which it heals becomes larger than the
physical size of the connecting duct; however, if the length of that duct is
short enough, a sizeable supercurrent can still exist, sustained by the
quantum tunnelling of quasi-particle pairs through the weak link.\footnote{A
  different situation has been examined by \citet{Rainer:87}, that of
  tunnelling through a very thin $^3$He-$^4$He film spanning a micro-hole,
  much like a soap bubble. The barrier parameters $l_\m b$ and $\xi_\m b (T)$
  can be also evaluated explicitly in this idealised case as well as the
  critical current through the weak link.}

At temperatures such that $\Delta(T)/k_\m B T$ is no longer a small quantity,
Eq.(\ref{CPRKurki}) becomes increasingly slanted with an abrupt slope close to
$\varphi=\pi$ when ${\m{cos}}\,\varphi/2$ changes sign while retaining the
periodicity by $2\pi$ in the phase difference. It displays a discontinuity for
$\varphi = \pi$ at $T=0$. This behaviour of the weak link comes on top of the
effect of the hydraulic inductance of the Deaver-Pierce model: the CPR is
bound to become hysteretic and multivalued even for an extremely small pinhole
in the limit $T\longrightarrow 0$.

This simple theoretical description accounts well for the experiments of
\citet{Avenel:88}, whose results are reproduced in
Fig.\ref{PRL88-fig2-staircase}.  Similar findings have been reached by
\citet{Backhaus:97} in an array of pinholes.  Close to the superfluid
transition, weakly-coupled reservoirs of superfluid $^3$He-B exhibit a
behaviour that involves the direct analogues of both Josephson
{\it ac} and {\it dc} effects in superconductors. The relation between the
superfluid current and phase -- or CPR -- is well represented by a slanted
sine function. In the range of applicability of Eq.(\ref{CPRKurki}), single
micro-apertures and arrays behave alike. The existence in superfluids of the
analogues of the Josephson effects in superconductors was thus established in
1988 on firm grounds, both experimentally and theoretically, but more features
were soon revealed by further studies.

\subsection{Multivalued CPR's, {\mbox{\bf\greeksym{p}}} states and
  {\mbox{\bf\greeksym{p}}} defects}

While near-ideal Josephson behaviour prevails in $^3$He-B at low pressure
close to $T_\m c$, departures from a sinusoidal current-phase
relation were observed by \citet{Avenel:88,Avenel:89} in a single orifice and
later by \citet{Backhaus:97,Marchenkov:99} in an array of 0.1$\mu$m--diameter
apertures, evolving to a near-straight line relation below 0.6\,$T_\m c$. The
latter case is reminiscent of the situation in $^4$He, in which vortices are
nucleated.

As the temperature is lowered further below $T_\m c$, the superfluid coherence
length becomes smaller than the aperture size used in present-day
experiments. This trend is even more pronounced at higher pressure, where
$T_\m c$ is higher (and $\xi_0$ smaller). Room is thus left for a wall
dominated order parameter texture within the weak link or its immediate
proximity: the p-wave nature of superfluid $^3$He can then reveal itself.

Very detailed numerical simulations based on the time-dependent
Ginzburg-Landau equations have been carried out by \citet{Soininen:92}
\footnote{Also by \citet{Kopnin:90,Kopnin:92}.} for a finite size aperture
quite similar to the one used by \citet{Avenel:88}. These simulations show in
colourful figures the time evolution, when the $^3$He superfluid is set to
flow through the micro-slit, of the components of the superfluid density
tensor parallel to the two short dimensions of the slit.  Both the mass and
the spin degrees of freedom of the spin-triplet p-wave order parameter take
part in the phase slippage process. The various components of the order
parameter evolve separately in space and time and do not go to zero
simultaneously at the same location in the micro-aperture. The regions of
space, in which the order parameter is depressed and about which the phase
slips, peel off from the walls and traverse the slit at right angle with the
flow direction. Thus, phase slips in the p-wave superfluid exhibit a fairly
complex spatial and temporal evolution both in the pseudo-isotropic B phase
and in the anisotropic A phase.  In addition, the A phase may sustain
core-less phase slippage as suggested by \citet{Anderson:77} and as discussed
below.  These simulations also illustrate the details of operation of a vortex
mill \citep{Soininen:92} in which phase slip avalanches and multiple vortex
creation take place. The state of sophistication of 

\begin{widetext}
  \begin{center}
  \begin{figure}[h]      
    \includegraphics[width=150 mm,height=70mm]{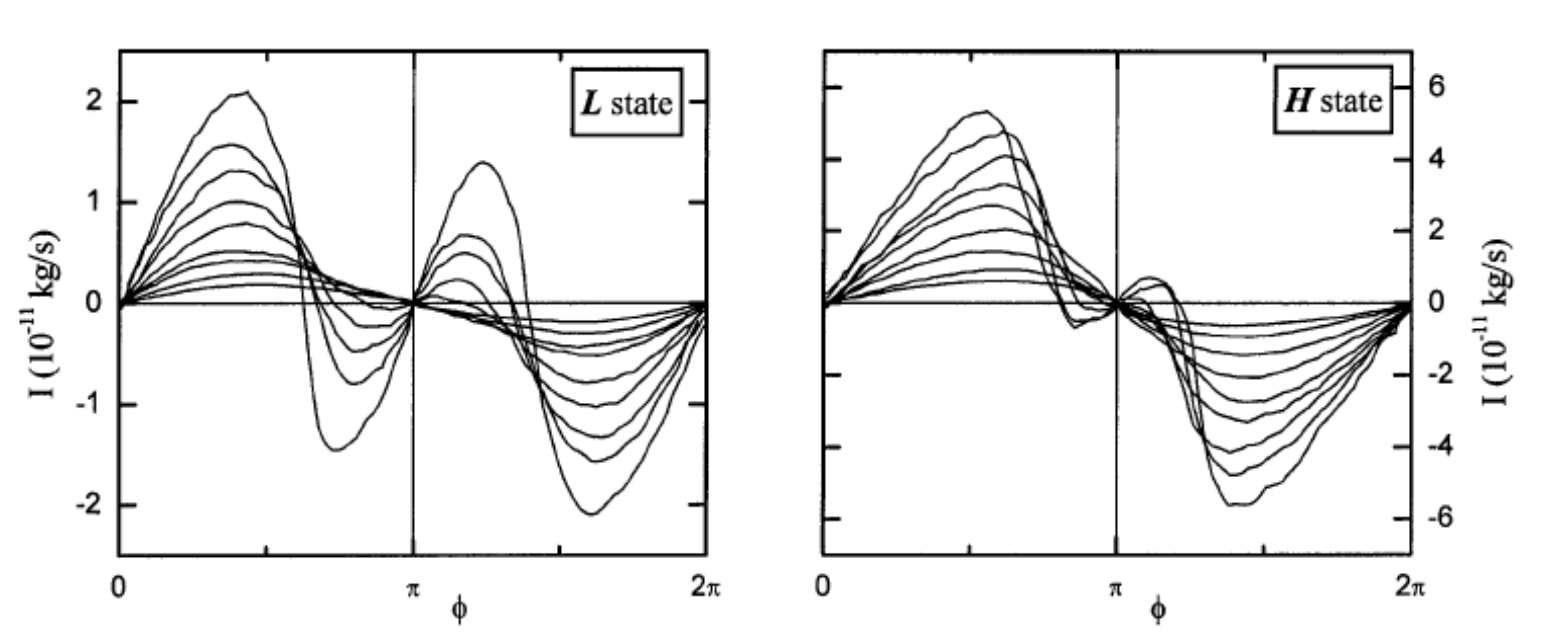}
    \caption{\label{Marchenkov:99-PiState}
      Current-phase relations in $^3$He-B observed in an array weak link. The
      two panels show CPR's for the low current state (left panel) and the
      high current state (right panel) for temperatures ranging from $T/T_\m c
      = 0.850$ down to 0.450 $T/T_\m c$ in steps of approximately
      0.05$\,T/T_\m c$. The mass current through the weak link increases as
      the temperature is lowered. At temperatures close to $T_\m c$, the
      current-phase relations can be fit well with the Deaver-Pierce model. At
      the temperature decreases (and the critical current increases), this
      model becomes inadequate as a $\pi$-periodic component gradually sets
      in. From \protect\citet{Marchenkov:99}.}
  \end{figure}
  \end{center}
\end{widetext}
the microscopic
description of superfluid $^3$He makes it possible to obtain such detailed
information.

The phase slippage observations in superfluid $^3$He reflects this wealth of
riches.  While near-ideal Josephson behaviour prevails in $^3$He-B at low
pressure close to $T_\m c$, more complicated staircase patterns than those
shown in Fig.(\ref{PRL88-fig2-staircase}) develop below $0.7 T_\m c$
\citep{Avenel:89}, which cannot be described by the Deaver-Pierce model. These
patterns are not even reproducible from one cool-down through $T_\m c$, or
through the A to B transition, to the next. It is likely that different order
parameter textures and topological defects are coming into play.

Among those features, two notable ones were reported by
\citet{Backhaus:98,Marchenkov:99} and are shown in
Fig.\ref{Marchenkov:99-PiState} in $^3$He-B at saturated vapour
pressure. These authors used a two-hole micro-resonator with a weak link made
of a 65x65 array of 100 nm round holes micro-machined in a 50 nm thick silicon
nitride free-standing membrane. The 4225 holes in parallel offer a large
enough flow path for the mass flow rate under an applied pressure head to be
closely monitored. They operated the resonator in free ringing mode and
recorded the transient response following a large impulsive drive
excitation. The phase is derived from the measurement of the pressure head
between the two sides of the weak link by integration of
Eq.(\ref{acJosephson}). The current-phase relations displayed in
Fig.\ref{Marchenkov:99-PiState} are obtained with this direct
technique.\footnote{More experimental details and further references can be
  found in the reviews by \citet{Davis:02} and \citet{Sato:12}.}

The first feature shown in Fig.(\ref{Marchenkov:99-PiState}) is the existence
of two possible CPR's at the same temperature, one with a larger critical
current than the other, the second, the appearance in the $2\pi$-periodic CPR
of an increasingly strong $\pi$-periodic admixture as the temperature is
lowered.  \citet{Avenel:99} pointed out that this admixture could simply arise
from the unavoidable dispersion between the sizes of the micro-holes in the
array.\footnote{An alternate explanation for the existence of $\pi$-states is
  offered by \citet{Eska:10} and is based on the built-in non-linearities of
  the single-hole resonator used in the experiments of
  \citet{Backhaus:98,Marchenkov:99},} This rather trivial explanation
holds in part under all circumstances but is not the end of the story, as was
soon shown by \citet{Avenel:00}'s own observations using a single
micro-aperture for which there is obviously no scatter in critical currents or
transit times.

\citet{Avenel:00} and \citet{Mukharsky:04} took advantage of the Sagnac effect
(see \S \ref{interferometry}) to ramp up and down in a precise manner the
macroscopic phase difference $\delta \varphi$ applied across the weak link.
They reported the observation of several different CPR branches -- usually
more than two -- most with $\pi$-components and with different critical
velocities at the same temperature but in different cooldowns through the
superfluid transition temperature. Each of these several $J(\varphi)$'s was
usually robustly fixed in each run and the general trend was to go from $2\pi$
periodicity to $\pi$ periodicity as the temperature is lowered, although,
occasionally, no $\pi$ periodicity was observed even at the lowest
temperature.  At higher pressure (10 bars), hysteretic behaviour in the single
micro-aperture was prevalent and up to three simultaneous branches for the CPR
were observed. Switching between these different branches could be triggered
by applying strong transient drive voltage to the resonator, indicating that
textural effects were most likely at play.

Some of these features were actually predicted long before their
observation by \citet{Thuneberg:88} who worked out a numerical solution for the
Ginzburg-Landau equations of the state of $^3$He-B confined inside a
micro-aperture. This author found two different CPR's according to whether the
$\hat{\bs{\m n}}$-vector of the B-phase order parameter, assumed to lie
perpendicular to solid walls, is in a parallel or antiparallel configuration
on both sides of the membrane carrying the micro-aperture. In the antiparallel
configuration, the spin and mass currents are out of phase, resulting in a
lower critical current. Eventually, the decoupling between mass and spin
currents leads to the admixture of a $\pi$-periodic component to the
2$\pi$-periodic CPR.

The experimental discovery of these effects by \citet{Backhaus:98} and
\citet{Marchenkov:99} spurred theoretical interest. Thuneberg's
numerical findings were soon confirmed and sharpened by the analytic
investigations of \citet{Yip:99} and \citet{Viljas:99,Viljas:02-PRB} and
extended numerical simulations for two-dimensional geometries by
\citet{Viljas:02-JLTP}. The upshots of these studies are the
following:\footnote{See also Janne Viljas's Thesis (Espoo 2004) available at
  http://lib.hut.fi/Diss/\,, \citet{Smerzi:01}, \citet{Zhang:01},
  \citet{Nishida:02} and the very clear review by \citet{Viljas:04-JLTP}. A
  related situation, that of ``$\pi$--junctions'', has been much studied in
  electrodynamic junctions \citep{Golubov:04}. }
\begin{trivlist}
\item{$\bullet$~Following \citet{Yip:99} and \citet{Viljas:99},\footnote{Also
      \citet{Zhang:01}.} the $\pi$ states in $^3$He-B are due to the
    interference of currents carried by quasiparticles with different spins
    that acquire different excess phases from the internal spin structure of
    the order parameter while travelling through the weak link. More
    specifically, the $\ket{\uparrow\uparrow}$ and $\ket{\downarrow\downarrow}$
    Cooper pair populations may be viewed as independent superfluids, the phases
    of which may be slightly shifted with respect to one another because of a
    differing spin-orbit coupling. Summing the corresponding mass currents,
    given by Eq.(\ref{CPRKurki}), represented by slanted sine CPR's shifted in
    phase by $\pm \delta \varphi$ leads, if the shift is large enough, to a
    positive-slope branch in the CPR at $\pi$: this $\pi$-state mechanism
    relies on different spin-orbit orientations on both sides of the weak link
    and operates at the single pinhole level \citep{Viljas:02-JLTP}.}

\item{$\bullet$~ The Josephson coupling between two baths of $^3$He-B mixes
    the phase difference to the spin-orbit texture of the order parameter: the
    equilibrium configuration of the texture then depends on the phase bias
    applied to (hence on the current carried by) the weak link. The texture is
    assumed fixed in the simpler calculations: this is the {\it isotextural}
    case, which offers only a coarse agreement with observations. If the
    texture is allowed to adjust to the local mass and spin currents by
    expressing the balance between its stiffness and its interactions with the
    walls and with the mass current, a $\pi$-state can also arise: this {\it
      anisotextural} effect requires a self-adjusting string of calculations
    and provides quantitative agreement with pinhole array experiments
    \citep{Viljas:02-PRB}.}
\item{$\bullet$~These refined calculations led to the realisation that
    multiple Andreev reflections and sub-gap structures also played a role in
    the transmission of the supercurrent through the weak link
    \cite{Asano:01},\footnote{See also the contributions of
      \citet{Smerzi:01}, \citet{Viljas:05} and \citet{Thuneberg:05}.}
    and that a A-like phase inside the superfluid junction could also result
    in the existence of a $\pi$-state \citep{Nishida:02}. }
\item{$\bullet$~Dissipation in pressure-driven {\it dc}-supercurrents
    \citep{Simmonds:00} could also be explained by multiple Andreev
    reflections \citep{Mukharsky:04-JLTP,Viljas:05} or by time-dependent
    anisotextural effects and spin-wave emission,
    \citep{Viljas:04-PRL,Viljas:05}: if a pressure difference is applied
    across the weak link, an {\it ac}-oscillation (at the {\it ac}-Josephson
    frequency) of the texture ensues, causing dissipation by spin-wave
    radiation. The two dissipation mechanisms, sub-gap processes and textural
    losses, can come on top of one another. \footnote{A related mechanism
      governing the vortex dynamics in Fermi superfluids at temperatures well
      below $T_\m c$, reported by \citet{Silaev:12}, arises from the kinetics
      of localised excitations bound to the vortex cores and driven out of
      equilibrium by vortex motion. The local heating of the vortex cores
      results in an energy flux carried by nonequilibrium quasiparticles and
      in a dissipation mechanism that can operate even at zero temperature.}}
\end{trivlist}

Observations related to these topics are those of \citet{Mukharsky:04} who, in
the course of high-precision CPR measurements using the Sagnac effect
described in \S\ref{interferometry}, found the signature of a stable textural
defect that sustains a change of the phase by $\pi$ away from the weak link.
This differs from the $\pi$-state discussed above. ``Cosmic-like'' solitons,
proposed by \citet{Salomaa:88}, could constitute such a defect but they are
thought to be unstable in the bulk of the superfluid.

A comprehensive study of the possible planar interfaces
between two domains of superfluid $^3$He-B has been conducted by
\citet{Silveri:14}. Of all the possible planar structures allowed by the
symmetries of the B-phase order parameter, only one is found to be
energetically stable in the presence of walls. This particular interface is
characterised by the vanishing of one of the components of the interfacial order
parameter along a {\it gap-node direction} contained in the plane of the
domain wall. It sustains a phase change by $\pi$ and can appear as a remnant
of the A to B interface during cool-down through the transition.  

In the perspective of the present article (see also the review by
\citet{Davis:02}), these complex features of the Josephson supercurrents
illustrate quite vividly the nature of the superfluid order parameter and of
the phase coherence it entails. But their detailed studies are complicated because
they are entangled with order parameter textures, as mentioned above, but also
because the state of the superfluid inside the micro-junction may not
be precisely accounted for, as discussed in the next Section.

\subsection{The peculiarities of the A-phase}
\label{Peculiarities}

The A-phase takes over from the B-phase at the superfluid transition
temperature above a pressure of 21.2 bars. Strong coupling effects
resulting from atomic localisation increase with density. Part of
these enhanced interactions is mediated by spin-spin exchange, the so-called
paramagnons. Because of these effects, the A-phase condensate only consists of
$S_{\m z}$ = +1 and -1 pairs, ($\ket{\uparrow\uparrow}$ and
$\ket{\downarrow\downarrow}$), and the energy gap above the Fermi surface
$|{\m{\bs \Delta}}(\hat{\m{\bs k}})|$ is strongly anisotropic while retaining
the $L=1$ symmetry: it vanishes at a node in the direction of $\hat{\bs{\m
    l}}$, the orbital quantisation axis.  As for its spin part, the A-phase
behaves in some respect as an antiferromagnet with a spin quantisation axis
$\hat{\bs{\m d}}$. Its stability with respect to the B-phase is enhanced by an
external magnetic field.

The A-phase order parameter in zero magnetic field is expressed in terms of
three unit vectors, the spin quantisation axis $\hat{\bs{\m d}}$, and the
orthonormal vectors $\hat{\bs{\m m}}$ and $\hat{\bs{\m n}}$ forming a triad
with $\hat{\bs{\m l}}$, the direction of the orbital angular momentum of the
pairs. It is written in tensorial notation as:
\begin{equation}        \label{APhaseOrderParameter}
  A_{\mu i} = \Delta_{\m A} \hat{d_\mu} (\hat{m_i} + \m i\, \hat{n_i}) \; .
\end{equation}

In $^4$He, the Bose order parameter is a simple complex number and the phase
comes in quite naturally as it does for the BCS order parameter in s-wave
superconductors, for ultracold atoms, and for the B-phase order parameter,
Eq.(\ref{BPhaseOrderParameter}), as discussed above.

No single phase factor appears spontaneously in expression
(\ref{APhaseOrderParameter}) for the A-phase order parameter. However, the
single-particle wave function in the condensate still possesses an overall
phase among other components. This phase goes over to the macroscopic Bose
order parameter, which inherits of a global $U(1)$ phase rotation broken
symmetry.

But this is not the whole story: a rotation of the triad $\hat{\bs{\m
    l}},\,\hat{\bs{\m m}},\,\hat{\bs{\m n}}$ about the angular momentum
directrix $\hat{\bs{\m l}}$ by angle $\gamma$ also contributes an overall
phase factor to the A-phase order parameter. This property can be seen readily
by considering the complex plane perpendicular to $\hat{\bs{\m l}}$ containing
the complex vector $\hat{\bs{\m m}}+\m i\,\hat{\bs{\m n}}$ that appears in
Eq.(\ref{APhaseOrderParameter}): a rotation by angle $\gamma$ in the complex
plane transforms $\hat{\bs{\m m}}+\m i\,\hat{\bs{\m n}}$ into ${\m{exp}}(-\m
i\gamma)\,(\hat{\bs{\m m}}+\m i\,\hat{\bs{\m n}})$. The Galilean invariance
argument as used to derive the two-fluid model for $^4$He leads, when
everything is told, to the following expression for the velocity of superfluid
mass transport:\protect\footnote{See \citet{Vollhardt:90}, \S 7.1.}
\begin{equation}        \label{AphaseSuperfluidVelocity}
{\mb v_\m s} = -\, \frac{\hbar}{2m_3}\,(\nabla\gamma + 
  \cos \beta\,\nabla\alpha) \;.
\end{equation} 
The Euler angles $\alpha,\beta,\gamma$ fix the orientation of the orbital
triad $\hat{\bs{\m l}},\,\hat{\bs{\m n}},\,\hat{\bs{\m m}}$ in the chosen
reference frame, $\gamma$ expressing 
a rotation about $\hat{\bs{\m l}}$ as
already mentioned.  Two independent phase gradients appear in
Eq.(\ref{AphaseSuperfluidVelocity}).  In one, $-\nabla\gamma$, the angle plays
the role of the usual phase $\varphi$. This feature arises because the $U(1)$
phase rotation broken symmetry as already mentioned. The other stems from the
bending of the $\hat{\bs{\m l}}$-texture.  Superflow is not simply governed by
the gradient of the global phase alone. The velocity field $\bs{v_\m s}$ is no
longer irrotational in general, hence the circulation of $\bs v_\m s$ over a
closed loop is no longer necessarily quantised.

In the presence of non-uniform $\hat{\bs{\m l}}$ textures, the change of
orientation of  $\hat{\bs{\m l}}$ in space may also contribute to the 
supercurrent.  The contour integral of Eq.(\ref{AphaseSuperfluidVelocity})
along a closed loop $\Gamma$ can be put under the form
\begin{equation}        \label{HoCirculationTheorem}
{\oint_{\Gamma}\,\bs v_\m s\cdot\m d\bs r = \frac{\hbar}{2m_3}\,\big[ 2\pi n + 
\sigma(\cal D}\big ] \;.
\end{equation} 
The first contribution to the rhs of Eq.(\ref{HoCirculationTheorem} is
recognised as the quantised velocity circulation around line singularities, as
found in superfluid $^4$He, the second is expressed \citep{Ho:78} as the area
circumscribed on the unit sphere by unit vector $\hat{\bs{\m l}}$ when
carrying the loop integral along contour $\Gamma$. This last contribution is
nil in the trivial case
%
\begin{widetext}
  \begin{center}
 \begin{figure}[h]       
    \includegraphics[width=120mm,height=60 mm]{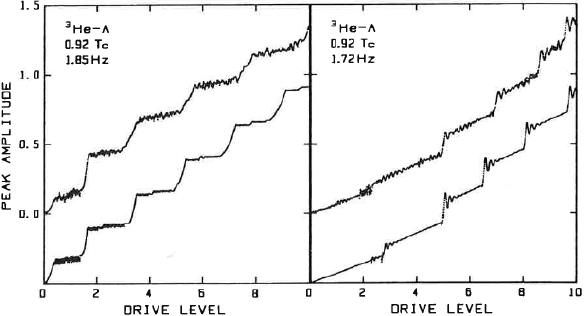}
    \caption{\label{StaircasesAphase}
      Staircase patterns in the A-phase at 28.4 bars and $T=0.92 T_\m c$
      ($T_\m c = 2.417$ mK). The left and right panels show
      the patterns slightly above and below the resonator resonance at
      $\omega_\m m=1.783$ Hz, as observed (upper curves) and as computed
      (lower curve) with the Deaver-Pierce model. From \protect\citep{Avenel:89}. }
 \end{figure}
  \end{center}
\end{widetext}
%
where $\hat{\bs{\m l}}$ keeps pointing in a fixed
direction. There exist other less-trivial cases with $\sigma({\cal D})=0$ as
discussed by \citet{Ho:78}, but, in general, this contribution is non-zero and
the velocity circulation is non-quantised.  

The interplay between superflow, vortices and textures of the order parameter
becomes quite complex.{\footnote{For a pointed but still gentle introduction
    to the intricacies of the A-phase hydrodynamics, see \citet{Hall:86} who
    also covered the effect of magnetic fields, not considered here. }}  In
particular, the A-phase persistent superflow can be relaxed by textural motion
alone without the creation of topological singularities of the order parameter
such as vortices. However, if large-scale motion of the texture is suppressed,
dissipation can be quenched and persistent currents stabilised, as shown by
the experiments of \citet{Gammel:85}.  The authors demonstrated the existence
of such currents in the annular space of a torsional oscillator packed with 25
\umu\; silicon carbide powder. The effect of the powder was to immobilise
$\hat{\bs{\m l}}$. The (small) supercurrent was detected indirectly through
its effect on the damping of the small-amplitude of the torsional
oscillator. This crafty experiment showed that the A-phase possesses, if to a
less convincing extent than the B-phase, the distinctive attribute of
dissipationless flow.

The phase slippage concept can also be extended to the A-phase, as proposed by
\citet{Anderson:77}. Josephson-type experiments can be contemplated
with some uncertainty as to their outcome because of the lack of quantisation
of the velocity circulation, and also because of the large dissipation
associated with the motion of the order parameter gap in the direction of
$\hat{\bs{\m l}}$, where its nodes lie.

These experiments were attempted by \citet{Avenel:89} with the same resonator
as for their B-phase experiments. They did observe staircase patterns in
the A-phase both close to the superfluid transition temperature with
a rather non-ideal current-phase relation, and further down in
temperature where new features occurred.
The patterns shown in Fig.\ref{StaircasesAphase} obtained in the A-phase at $T=0.92
T_\m c$ at frequencies slightly above and below the resonance frequency of the
flexible wall resonator are quite well defined but differ markedly: they
exhibit large dispersive effects, a sharply peaked resonance and low
dissipation. The outcome of numerical simulations of the resonator response
using the Deaver-Pierce model, Eq.(\ref{DeaverPierce}) are
shown below the experimental data curves. The resonance quality factor is
high, $Q= 80$. The non-ideality parameter $\alpha$ is equal to 5: the phase slips are
hysteretic and weakly dissipative. 
 
Given this observation that the A-phase phase-slip pattern seems to 
follow the ubiquitous Deaver-Pierce model at 0.92 $T_\m c$, it could be
expected to become more and more ideal when raising the
temperature closer to $T_\m c$. This trend could unfortunately not be ascertained
in these experiments because the operating frequencies, which decrease as
$\rho_\m s/\rho$, become too low and the useful signal gets lost in the
background $1/f$ mechanical noise of the detection device.

\begin{widetext}
  \begin{center}
 \begin{figure}[h]       
    \includegraphics[width=120mm,height=60 mm]{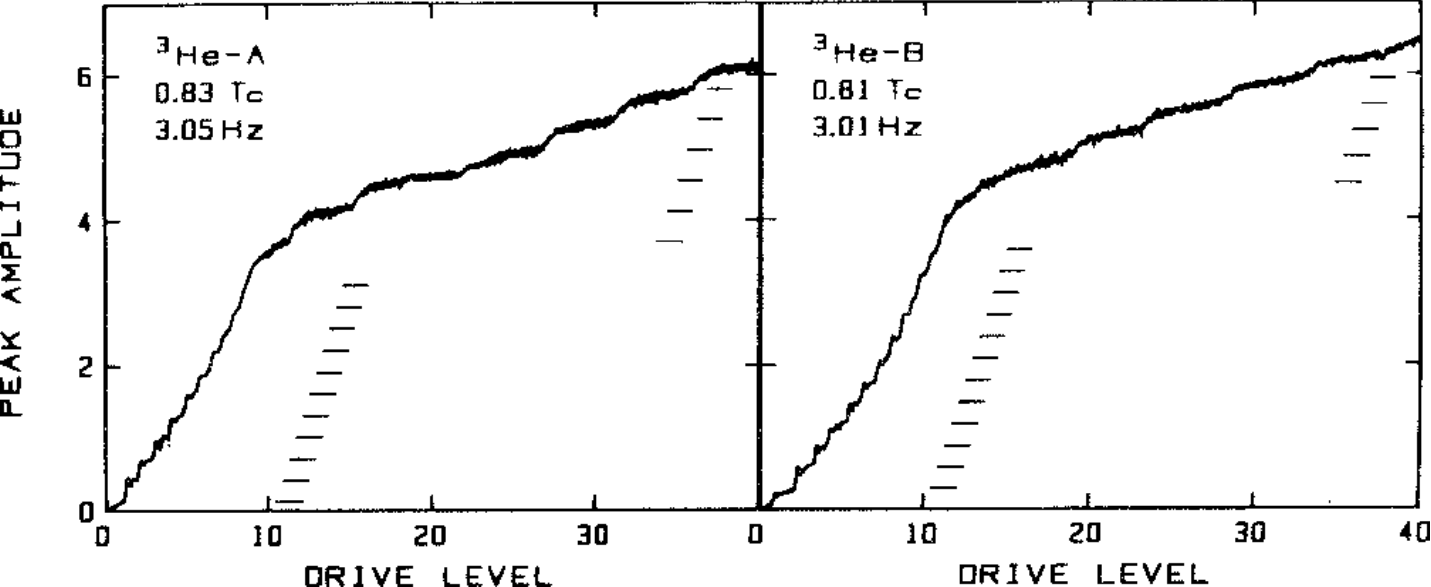}
    \caption{\label{ComparisonA-Bphase}
      Staircase patterns in the A and B phases at nearly identical frequencies
      and temperature. The horizontal ticks mark the periodicity of the
      staircase pattern (top) and of the low-level structure (bottom). This
      low-level structure has been attributed by the authors to solid friction
      of unknown origin somewhere in the resonator (but may also arise from
      sub-gap energy levels). From \protect\citep{Avenel:89}.}
 \end{figure}
  \end{center}
\end{widetext}
%
Further down in temperature, an instructive direct comparison between the
A-phase response and that of the B-phase at high pressure can be carried out
by taking advantage of the following circumstance: at 28.4 bars, the A to B
transition occurs at about 0.81 $T_\m c$; it is signalled by a sudden drop in
resonance frequency caused by the drop in superfluid density accompanying the
first order transition. As the A-phase can be supercooled into the domain of
stability of the B-phase, both phases can be studies at comparable
frequencies.

The outcome of this comparison is shown in Fig.\ref{ComparisonA-Bphase} and
reveals a remarkable similarity between the two phases. The response of the
superfluid in the weak link depends little on whether the bulk of the liquid
is in the A or the B phase. As discussed by \citet{Avenel:89}, the observed
behaviour inside the micro-slit corresponds well to the situation described by
\citet{Kurkijarvi:88} who finds that the current-phase relations for the A and
B phases differ only little (see Eq.(\ref{CPRKurki})).  It may also happen
that the state of the superfluid in the weak link remains the same
irrespective of the state in the bulk. It has been predicted by \citet{Li:88}
and \citet{Fetter:88} that the A-polar phase would be favoured by the
depletion of some of the components of the A-phase order parameter close to
the aperture walls.

The sub-gap structure, shown in Fig.\ref{ComparisonA-Bphase}, which develops
for resonance amplitudes below the critical threshold at which dissipative
phase slips start to occur, is quite intriguing. It is interpreted by
\citet{Avenel:89} as arising from possible (aniso)-textural effects inducing
solid friction. It could also possibly be revealing the existence of sub-gap
resonant levels.

These experiments establish very clearly that phase slippage takes place in
the A phase, that persistent currents can be trapped in the loop threading the
double-hole resonator, and that the velocity circulation along these trapped
currents changes by multiples of the quantum of circulation in the same manner
as in the B phase: it so turns out, as was the case in the persistent current
experiments by \citet{Gammel:85}, that the $\hat{\bs{\m l}}$-texture is
sufficiently well pinned in the regions where ${\bs v_\m s}$ picks up
significant speed.

Topological defects, seen above to play an important role in phase slippage
experiments, offer a vast and fascinating domain of study, both in the A and B
phases. The vortex core develops complex structures, as reviewed by
\citet{Salomaa:87}. Vortex sheets can form in rotating $^3$He-A, as observed
by \citet{Parts:94} using very sensitive NMR techniques, which have brought
about a wealth of information on vortices in superfluids under rotation. This
work has been reviewed by \citet{Finne:06}. Analogies can be drawn between the
formation of defects in superfluid $^3$He and that of cosmic strings in the
Early Universe because the order parameter symmetries that can be broken are
the same. These enticing prospects for experimental cosmology have been
reviewed by \citet{Eltsov:00}, \citet{Bauerle:00}, \citet{Bunkov:10} and
\citet{Volovik:03} in his wonderful book. Phase slippage is beyond doubt
relevant in these situations and will be used to study this vast new field.

\subsection{$^4$He close to the \greeksym{l}--point}
\label{LambdaPoint}

The existence of Josephson-like effects was thus clearly established in
superfluid $^3$He for the {\it dc}-effect and in both $^3$He and $^4$He
superfluids for the {\it ac}--effect by Year 2000. The remaining problem was
the possible observation of a quasi-sinusoidal current-phase relation in
$^4$He. The minuteness of the coherence length in $^4$He makes the fabrication
of a suitable weak link a tall order except in the immediate vicinity of the
${\ulambda}$-point, where it diverges as $\xi=\xi_0(1-T/T_{\ulambda})^{-2/3}$
with $\xi_0 \sim$~1 to 2 \AA\; \citep{Langer:70}. At the ${\ulambda}$-point
however, superfluidity is suppressed by thermal fluctuations. Would the
hydrodynamic Josephson effects not be even more readily washed out by the same
token?\footnote{As already mentioned by \citet{Anderson:64}, p. 120.}

This concern was formalised by Zimmermann (1987) whose
argument runs approximately as follows. The Josephson coupling energy is
obtained from the Josephson current, Eq.(\ref{dcJosephson}), by integration
with respect to $(\hbar/m_\m 4) \inc\varphi$. Its maximum value is therefore
$(\hbar/m_4)\,J_\m c$ and reads 
\begin{equation}        \label{JosephsonEnergy}
E_\m J =  \left(\frac{\hbar}{m_4}\right)^2\,\frac{s_\m b}{b_\m b}\,
       \frac{\rho_\m s}{\sinh(l_\m b/b_\m b)} \; .
\end{equation}

In the weak coupling limit for which Eq.(\ref{dcJosephson}) holds, the
wavefunction is strongly depleted within the barrier and the penetration
length is smaller than the length of the barrier, $b_\m b \lesssim l_\m b$.
Making use of the scaling relation between $\rho_\m s(T)$ and $\xi(T)$
\citep{Josephson:66},\footnote{See \citet{Halperin:76} and
  \citet{Hohenberg:77} for details.}
\begin{equation}        \label{ScalingRelation}
  \rho_\m s \xi = (m_4/\hbar)^2 k_\m B T  \; ,
\end{equation}
and since $\sinh(l_\m b/b_\m b) > l_\m b/b_\m b$ it stems from
Eq.(\ref{JosephsonEnergy}) that
\begin{equation}        \label{ZimmermannCondition}
E_\m J \lesssim \frac{s_\m b}{l_\m b} \,\frac{k_\m B T}{\xi(T)}
 \; .
\end{equation}
For the round pinhole with diameter $d$ considered by Zimmermann, $s_\m b =
\pi d^2/4$ and $l_\m b > \ell_\m h$, the hydraulic length $\ell_\m h$ being $\pi d /4$ for a
circular orifice,\footnote{As derived by \citet{Anderson:66a}, p.305} so that $E_\m J < k_\m
BT \,d/\xi(T)$.  Zimmermann concludes from this upper bound for the Josephson
energy that, as $\xi(T)$ diverges upon approaching $T_\m {\ulambda}$ from
below, the Josephson coupling energy will end up being less than the thermal
energy and that the Josephson {\it dc}-effect will be washed out by thermal
fluctuations.

Similar concerns were spelled out by \citet{Ullah:89} for their calculations
of weak link properties in $^3$He-B in the Ginzburg-Landau regime: ``{\it We
  do not address the important problem of thermal fluctuations destroying the
  superfluidity in the very small volume of the weak link. To our knowledge,
  there is no reliable, quantitative theory of the stability of the superfluid
  phase is severely confined geometries. We believe that this question can be
  convincingly answered only by experiment}''. This remark is even more
relevant to superfluid $^4$He close to $T_{\ulambda}$.

\begin{figure}[t!]        
  \begin{center}
   \vskip -5 mm \hskip 6 mm
    \includegraphics[width=90 mm]{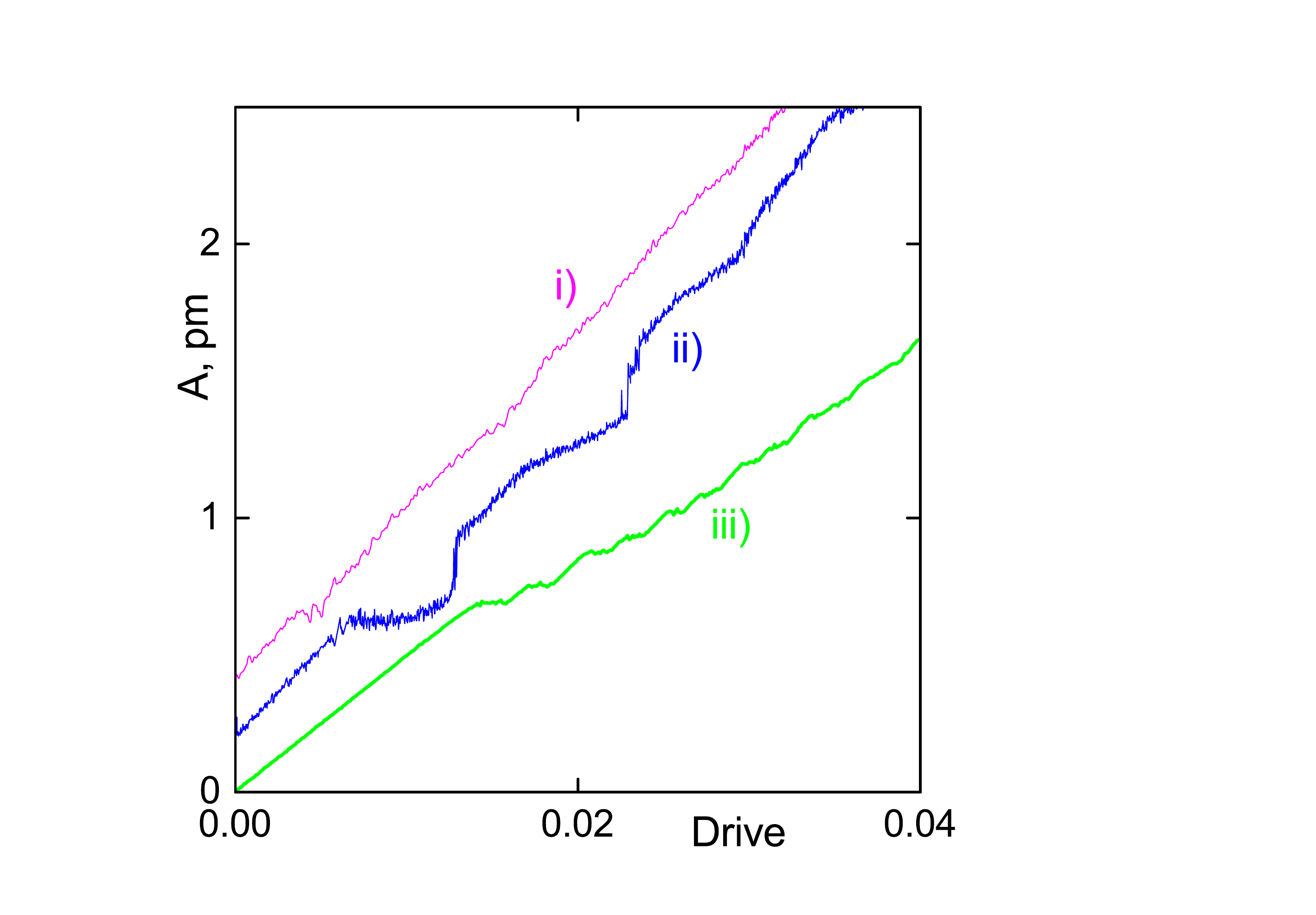}
    \caption{\label{Sukhatme-staircase} 
      (Color online) Staircase patterns in $^4$He close to $T_\m{\ulambda}$
      (amplitude, in picometres versus drive, in arbitrary units, from
      \protect\citet{Sukhatme:01}) for $T_{\ulambda}-T=$ 61 $\mu$K -- curve
      i), 154 $\mu$K -- curve ii), and 3.72 mK -- curve iii). The plots are
      shifted vertically by 2 pm with respect to one another for clarity. The
      data for curve iii) has been divided by 20 along the $x$-axis and by 10
      along the $y$-axis. As in the case of $^3$He-B -- shown in
      Fig.\ref{PRL88-fig2-staircase} -- the height of the first step
      corresponds to the critical current $J_\m c$. Each subsequent step
      corresponds to an additional phase difference of $2\pi$ (Courtesy of
      Yuri Mukharsky).  }
  \end{center}
\end{figure}

The first hint of a successful experimental observation was reported by
\citet{Sukhatme:01}. These authors used an array weak link of 24 micro-slits 3 \umu m x
0.17 \umu m about 10 \umu m apart in a 0.15 \umu m thick membrane. Their
findings are summarised in Fig.\ref{Sukhatme-staircase}. At 3.72 mK below
$T_{\ulambda}$ -- the bottom curve in the figure, the scale of which is
shrunk -- the critical velocity is well-marked, as well as the staircase steps,
indicating a dissipative phase slippage process. A phase-slip
regime has been reached. This was hoped for since the temperature-dependent
coherent length 3.72 mK below $T_{\ulambda}$, $\xi(T)=24$
nm,\footnote{Approaching $T_{\ulambda}$ from below, $\rho_\m s$ is known to
  vanish as \citep{Langer:70}
\[
\rho_\m s \simeq 2.4\, \rho_{\ulambda}(1-T/T_{\ulambda})^{2/3} \; ,
\]
where $\rho_{\ulambda}$ is the density at the ${\ulambda}$ transition, 0.1459
g/cm$^3$. From relation (\ref{ScalingRelation}), the temperature-dependent
coherence length becomes $\xi(T)= 0.338/(1-T/T_{\ulambda})^{2/3}$ in nm. } is
smaller than the micro-slit width, but by less than one order of magnitude.

Getting closer to $T_{\ulambda}$ from below, successively at $T_{\ulambda}-T = 154
$\umu K, and 61 \umu K, the expected trend toward a smoother, less dissipative
staircase pattern is observed, much like in Fig.(\ref{PRL88-fig2-staircase})
for $^3$He, supporting the conclusion that the same hydrodynamic ideal
Josephson effect can be observed in superfluid $^4$He close to the superfluid
transition temperature $T_{\ulambda} = 2.17$ K.

This conclusion raises questions, and possibly, some eyebrows as well:
\begin{trivlist}
\item{$\bullet$~ Would, for some reason, Zimmermann's argument be
    invalid?}
\item{$\bullet$~ How come that dissipative phase slippage, the mechanism for
    which seems to rely on the nucleation of a single vortex and its crossing
    of all streamlines of the superfluid flow through a single micro-slit also
    operates for an extended array of them?}
\end{trivlist}

Zimmermann's original argument, outlined above, was applied to a single round
hole. \citet{Sukhatme:01}'s 24 parallel slits yield an estimated enhancement
factor of 500 in the superflow passage area $s_\m b$ that appears in
Eq.(\ref{ZimmermannCondition}), provided that the supercurrents in the
apertures effectively sum up. The overall Josephson energy is increased by the
same factor and the disruptive effect of thermal fluctuations is pushed back
much closer to $T_{\ulambda}$. This line of reasoning was pursued by
\citet{Chui:03}, but its soundness depends on the answer to the second
question, which turns out to be trickier.

The Berkeley group carried out a number of studies\footnote{In particular, the
  work of \citet{Hoskinson:06,Sato:06}, and also of
  \citet{Sato:08,Narayana:10,Narayana:11}.} with aperture arrays similar to
those used for $^3$He-B by \citet{Marchenkov:99}. The size of the round
pinholes in these arrays, 90 nm in diameter, is comparable to the coherence
length one millikelvin away from the \ulambda-point but the distance
separating the pinholes, located on a 3 \umu m square lattice, is much larger.
Two phase-slippage regimes are identified when the temperature is lowered
below $T_\m{\ulambda}$ as already reported by \citet{Sukhatme:01}.  At
$\sim$~50 to 100 \umu K below $T_\m{\ulambda}$ (and slightly further down in
temperature in the experiments by \citet{Sato:06}), a reversible
(non-dissipative) Josephson regime is observed. In this regime, the
phase-slips occur in a fully {\it synchronous} manner.  Between approximately
0.3 to 15 mK below $T_\m{\ulambda}$, a transition toward a dissipative
phase-slip regime sets in as the synchronisation between the apertures gets
lost.  Further below $T_\m{\ulambda}$, the phase-slip regime becomes {\it
  asynchronous}. The amplitude of the resultant phase slippage signal from the
array does not sum up to what it should be. It also exhibits large slips and
collapses somewhat similar to those described in \S\ref{LargeSlips} for a
single orifice (but of a different sort, see below).

It is clear that inhomogeneities in aperture size and surface properties, the
edge effects at the periphery of the array, and local critical fluctuations
introduce a spread in the values of the critical current in the different
apertures. Phase slips occur at different times during resonator motion. The
summation of the currents through the various apertures, as attempted in the
numerical simulations of arrays of superfluid Josephson junctions by
\citet{Avenel:99}, \citet{Pekker:07}, and \citet{Sato:07} has to be exercised
with care.

It can be argued \citep{Chui:03} that the Josephson currents in the
micro-apertures are small and perturb little the quantum phases in the bulk on
both sides of the membrane supporting the weak link array. Phases are well
defined below $T_\m{\ulambda}$ -- for instance, the quantisation of
circulation is enforced -- and so should their difference.  This reasoning
would appear to leave only one degree of freedom to undergo fluctuations, with
a thermal energy of $k_\m B\,T/2$ to be shared among the $N$ apertures of the
array: the effect of fluctuations in each individual aperture would
effectively be quenched on taking the average over the whole array.
%
\begin{figure}[t]
     \includegraphics[width=55 mm]{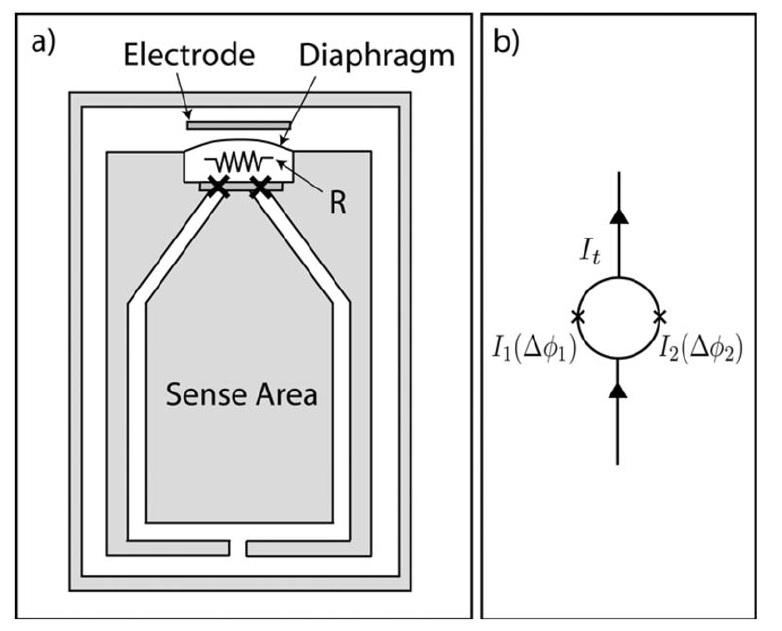}
     \includegraphics[width=70mm]{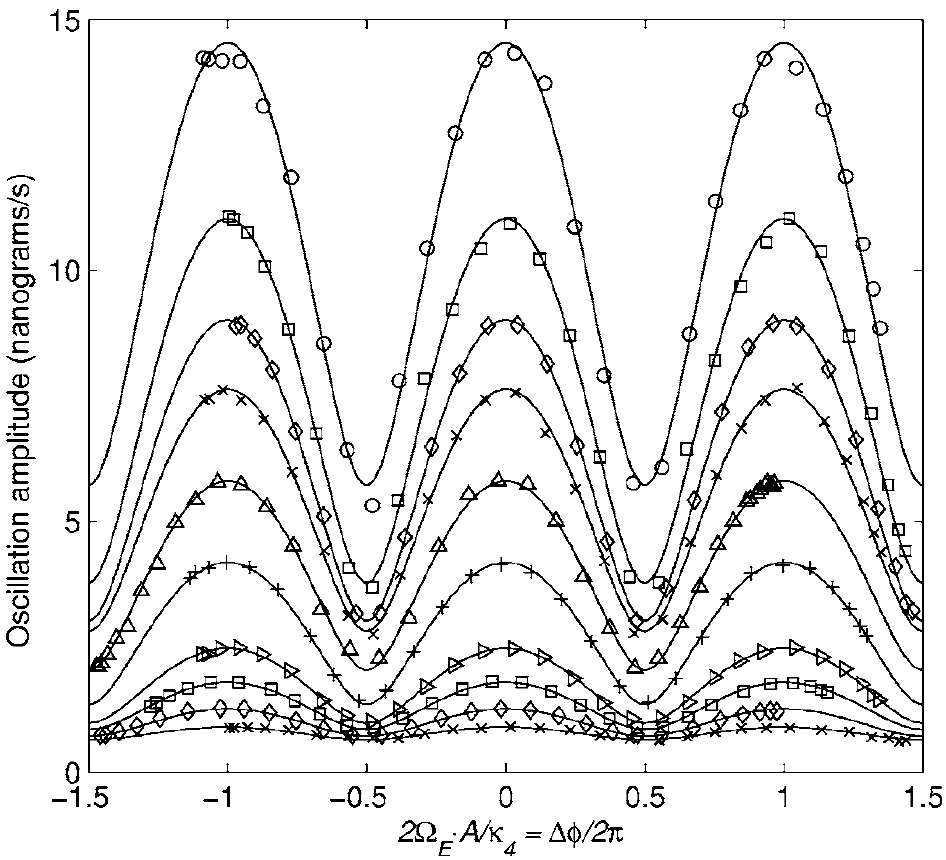}
    \caption{\label{Hoskinson:06-Interference}
      Superfluid Interferences in a two-aperture resonator.
      \\ 
      Top panel: (a) Sketch of the hydromechanical resonator.  The unshaded
      regions are filled with
      superfluid $^4$He. The flexible metallised diaphragm at the top of the
      upper chamber serves both as a microphone to detect the resonant
      oscillations and as a pressure pump to drive the flow across the two
      aperture arrays indicated by crosses. These arrays interrupt the
      superfluid channel enclosing the sense area.  (b) Diagram of the
      resonator equivalent circuit showing the analogy with the electrodynamic
      dc-SQUID. There are two superposed currents flowing through
      the weak links, one corresponding to the rotation flux
      picked-up by the sensing loop, Eq.(\protect\ref{Bias}), the other being
      the common-mode read-out current from the flexible diaphragm. 
      \\
      Bottom panel: Peak amplitude of the diaphragm displacement on resonance
      as a function of the rotation flux $2\, {\bf \Omega}\cdot{\bf A}$ picked
      up by the superfluid loop enclosing the sense area -- see
      \S\ref{interferometry} for a detailed description of these
      interferometer operation based on the Sagnac effect. The measured data
      is shown by the symbols; the solid lines are fits to the data of the
      equation modelling the resonator motion as described by
      \protect\citet{Hoskinson:06}. The modulation curves were taken at
      temperatures $T_{\ulambda}\,-\,T =$~12, 7.0, 4.0, 3.0, 2.0, 1.5, 0.9,
      0.6, 0.4, and 0.3 mK from top to bottom. This temperature span covers
      the coherent Josephson regime in the array discussed in
      \S\ref{LambdaPoint}. From \protect\citet{Hoskinson:06} }
\end{figure}

This argument has to be stretched a long way to explain the large span in
$T_{\ulambda} -T$ over which the synchronous phase-slippage regime subsides
both very close to the \greeksym{l}-point when quantum coherence should end up
being killed by thermal fluctuations, and quite a way below it where it should
be randomised by array imperfections.  In other words, the robustness of the
coherence effect mentioned above against dephasing by environmental effects
appears quite remarkable.
\citet{Perron:13} have pointed out that the superfluid onset in the
micro-slits used by \citet{Sukhatme:01} is expected to be depressed by size
effects to $T_{\ulambda} - T_\m c \simeq 430$~\umu K whereas the Josephson
effect could be tracked to as close as 28 \umu K below $T_{\ulambda}$.
Similarly, for the pinholes used by \citet{Sato:06}, $T_{\ulambda} - T_\m c
\simeq 2.3$~mK while the Josephson effect survived up to possibly 0.5
mK from $T_{\ulambda}$. As concluded by \citet{Perron:13}, ``{\it in both
  experiments one obtains superflow in a temperature region where the helium
  in the isolated weak links should be normal. Both of these experiments are
  thus relying on proximity effects, due to the surrounding bulk liquid, to
  maintain a non-zero order parameter in the weak links}''.

These authors draw their conclusion from studies of the interconnection
of an array of 2 \umu m$\times$2 \umu m micro-pools  linked through the film of
superfluid $^4$He. They have found from measurements of the specific heat and
the superfluid fraction in the vicinity of $T_{\ulambda}$ that correlation
effects are still effective at distances up to 100 times $\xi(T,L)$, the
finite-size correlation length suitably renormalised for confinement over the
distance $L$, the size of the micro-pools boxes. The unexpectedly
large extent of the correlation observed between micro-pools can be likened to the
robustness of the coherent behaviour of Josephson junction arrays close to
$T_{\ulambda}$.

\citet{Pekker:07}, besides their numerical studies referred to above, also
treated the problem of aperture current summation in an irregular array as an
order-disorder transition in a mean-field approximation approach. They
introduce a distribution of aperture critical currents and an effective
inter-aperture coupling parameter. They report qualitative agreement with the
experiments of \citet{Sato:06} including ``{\it system-wide avalanches}'',
both in the numerical simulations of the array behaviour and in the ordering
transition approach.

The observed long range of cross-aperture coupling may arise from a simple
classical hydrodynamics scheme, which is an extension to arrays of the
putative mechanism for single-aperture large slips discussed in
\S\ref{LargeSlips}. Suppose that, during the surge of the superflow through
the array, a (quantum) phase slip occurs early in one of the apertures,
releasing a vortex half-ring that starts drifting (classically) sideways along
the membrane supporting the array. Soon, this vortex half-ring runs into the
flow lines emerging from a nearby aperture, gaining energy from it to proceed
in its course and, possibly, triggering the nucleation of another vortex
half-ring, and so on. This multiplication process may die by itself at the ebb
of the flow. Or, if it overcomes the friction on the normal component, it may
trigger $2\pi$ slips over all the microholes of the array, or, possibly, swell
to  the system-wide avalanches \citep{Pekker:07} observed by
\citet{Sato:06}. This mechanism for avalanches are thus intrinsic to aperture array
dynamics and distinct from flow collapses in single apertures discussed in
\S\ref{LargeSlips}.

Even more so than in single apertures, ``macroscopic quantum coherence''
manifests itself in the aperture array in a dual manner. First, the condensate
acts as an ideal Euler fluid, maintaining orderly streamlines throughout the
superflow in accordance to the Kelvin-Helmholtz theorem. Then, when a
non-adiabatic process takes place, violating velocity circulation
conservation, it does so in a quantum manner, allowing the phase to change by
multiples of $2\pi$, for instance by the nucleation of a quantised vortex or by
the current source or sink provided by Josephson tunnelling through a thin
barrier of normal fluid.

\section{Concluding comments}
\label{conclusion}

\subsection{Matter waves and superfluid interferometry}
\label{interferometry}

The single-hole or two-hole hydromechanical resonators used in the phase
slippage experiments described above have been presented so far as the
analogues of {\it rf} or {\it dc} superconducting quantum devices (SQUIDs), a
useful analogy to help understand the way they operate. Another analogy is
used in this Section to illustrate the concept of coherent matter fields, or
{\it matter waves}, introduced for superfluid helium by Anderson in
1965. These devices are now considered as the likes of optical Sagnac
interferometers; as the latter, they can be used to measure absolute rotations
with very high sensitivity.
 
Consider a pool of superfluid in the shape of a conduit bending on itself as shown in
Figs.\ref{Hoskinson:06-Interference} or \ref{Gyrometer}.  The circulation of the velocity is
quantised {\it in the inertial frame} -- the reference frame fixed with
respect to the distant stars -- along any closed contour $\Gamma$ located
entirely in the superfluid:
$$                       
\oint_{\Gamma} {\bf v}_\m s \cdot{\m d}{\bf l}= 
       \frac{\hbar}{m_\m a}\,\oint_{\Gamma}\,\nabla\varphi \cdot{\m d}{\bf l}
      =  n\kappa_\m a   \; ,
$$ where $\kappa_\m a=2\pi\hbar/2 m_3$ for $^3$He, 3/2 times that quantity for
$^4$He, and $n$ is an integer.

If the cryostat housing the pool is set into rotation with rotation vector ${\bf
  \Omega}$, the velocity transforms in the new frame according to ${\bf v}'_\m
s ={\bf v}_\m s - {\bf \Omega} \times {\bf r}$ and the quantisation of
circulation condition now reads
\begin{equation}        \label{Circulation}
    \oint_{\Gamma}{\bf v}'_\m s\cdot{\m d}{\bf l} 
    =  n\kappa_\m a - \oint_{\Gamma} {\bf \Omega} \times {\bf r}\cdot{\m d}{\bf l}
    =  n\kappa_\m a - 2\,{\bf \Omega}\cdot{\bf S}_\Gamma\,, 
\end{equation}
${\bf S}_\Gamma$ being the geometrical (oriented) area of the closed
superfluid contour.  

For an actual conduit with finite cross section such as the one pictured in
Figs.\ref{Hoskinson:06-Interference} and \ref{Gyrometer}, there is a variety
of choices for the contour $\Gamma$. The mean circulation of the velocity
results from a suitable average over the various distinct superfluid contours
threading the conduit. Taking the average of Eq.(\ref{Circulation}) over all
the streamlines threading the conduit amidst stray thermal currents, pinned
vortices, and textures, weighed according to the (infinitesimal) mass current
that they carry, leads to \citep{Avenel:97}:
\begin{equation}        \label{Bias}
\left\langle\oint_{\Gamma} {\bf v}'_s\cdot{\m d}{\bf l}\right\rangle  
    = n\kappa_\m a + \kappa_\m b - 2\,{\bf \Omega}\cdot{\left\langle\bf S\right\rangle} \, ,
\end{equation}
where $\left\langle{\bf S}\right\rangle$ is the average of the contour areas
over the conduit. The average of the quanta of circulation carried by the
various streamlines, $\left\langle n\right\rangle\,\kappa_\m a$ has been
written as $n\,\kappa_\m a + \kappa_\m b$ to explicitly separate the
non-quantised phase bias $\delta\varphi_{\m b} = 2\pi \kappa_{\m b}/\kappa_{\m
  a}$ arising from pinned vorticity from the strictly quantised contribution
$2\pi n$.

The last term to the right of Eq.\,(\ref{Bias}) also amounts to a
non-quantised contribution to the phase bias, which varies with the flux of
the rotation vector ${\bf\Omega}$ through $\left\langle\bf S\right\rangle$:
the measurement of the corresponding phase difference with the interferometers
depicted in Figs.\ref{Hoskinson:06-Interference} or \ref{Gyrometer},
$\delta\varphi_\m S= (m_\m a/\hbar)\,2\,{\bf{\Omega \cdot \left\langle S
    \right\rangle}}$, gives access to the rotation velocity. Alternatively,
changing the orientation with respect to the North axis of the superfluid loop
picks up more or less of the rotation flux due to the Earth rotation
$\Omega_\oplus$. A known phase difference can be coupled to the
weak link. The experimenter is provided with a ``gauge wheel'' to steer the
phase.\footnote{The original ``gauge wheel'' proposal due to Liu has been
  discussed by \citet{Ho:80a}.}

Exploiting the properties of superfluids to detect very slow rotations has
been proposed even before the discovery of the Josephson effects in
superfluids, understandably with some lack of accuracy as to how the
experiment could be conducted. Cerdonio and Vitale clarified in 1984 the way
in which inertial and gravitational fields could be detected with superfluid
$^4$He analogues of the {\it rf}-SQUID\,\cite{Cerdonio:84,Bonaldi:90}. A
number of authors followed suit afterwards for superfluid $^3$He and
$^4$He\,\cite{Varoquaux:92,Packard:92b,Hess:92}, and for the Bose-Einstein
condensed gases\,\cite{Stringari:01}.

Detailed schemes for the actual implementation of superfluid $^4$He gyros have
been worked out with the help of numerical simulations\,\cite{Avenel:94} and
from the analysis of the operation of existing double-hole hydromechanical
resonators.\footnote{See \citet{Aarts:94,Schwab:96a,Schwab:96b} and the review
  by \citet{Sato:12}.} The first measurement of
$\Omega_\oplus$ with a superfluid device was performed using a
resonator operating in hysteretic mode in superfluid $^4$He with a
rotation-sensing loop of 4.0 cm$^2$ by \citet{Avenel:96}.  Soon after, the
Berkeley group reported the observation of the effect of the rotation of the
Earth with a similar device operated in the staircase mode, in much the same
way as conventional {\it rf}-SQUID magnetometers.

\begin{figure}[t]        
  \begin{center}
    \includegraphics[width=45 mm]{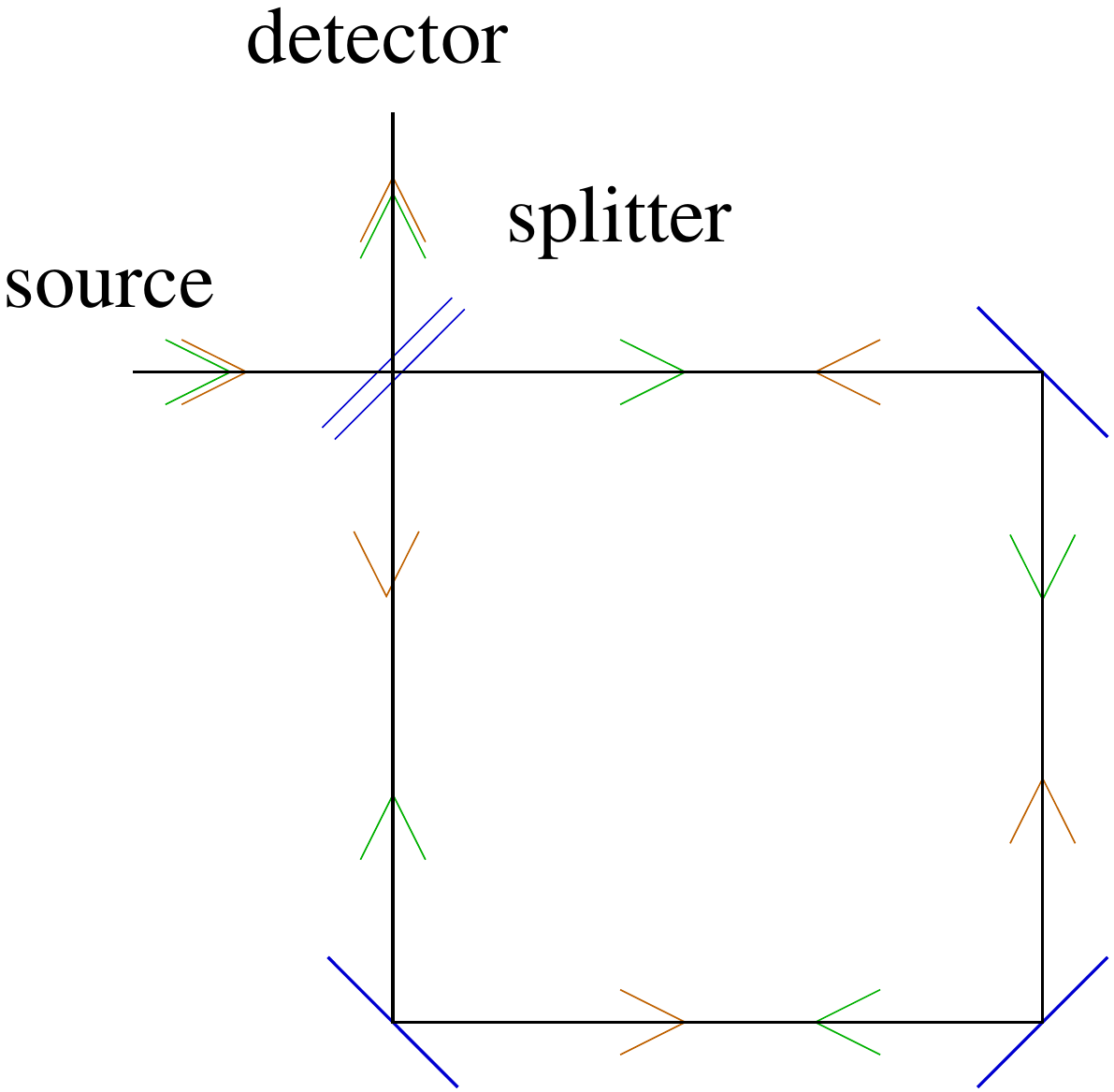}
    \caption{\label{SagnacInterferometer}
      (Color online) Schematics of the optical Sagnac interferometer. A
      collimated beam from the source enters the interferometer through a beam
      splitter and is separated into two beams travelling along the optical
      path defined by the three mirrors in opposite directions, as indicated
      by the arrows. Re-entering the beam splitter, they recombine and produce
      a fringe pattern on the detector plate.  }
  \end{center}
\end{figure}

One may mention for the record that early attempts to measure $\Omega_\oplus$
led to disappointing results to the dismay of
experimenters\,\citep{Schwab:96a,Schwab:96b,Avenel:98}. It was however quickly
realised that the currents in the bulk of the cell outside the
resonator\,\cite{Avenel:98,Schwab:98}, simply caused by the re-orientation of
the cryostat, were interfering with, and possibly overwhelming, the relatively
weak $\Omega_\oplus$--induced Sagnac current in the pick-up loop.  The
influence of these stray currents can be made negligible by a proper design of
the cell. A sheath on the port connecting the resonator to the main body of
the cell was used to that effect by \citet{Avenel:96}\citep{Avenel:97}.  The
absence of such a decoupling device between the Sagnac current in the pick-up
loop and the stray currents around the cell could cause uncontrolled
inaccuracies of several tens of \%\,\citep{Schwab:96a,Schwab:97,Schwab:98}.

The potentialities of superfluid gyros as extremely sensitive and stable
rotation sensors, able to track, {\it e.g.}, General Relativity effects, have
been considered by \citet{Avenel:98}, \citet{Chui:05}, \citet{Sato:12} and
\citet{Sato:14}. It appears that these gyros can compete with the most advanced
rotation sensors, in particular because they are inherently driftless at very
low temperatures.

These gyrometric devices are the direct superfluid analogues of the well-known
Sagnac optical interferometers, as can be seen by inspection of the sketches of
the latter in Fig.\ref{SagnacInterferometer} and of the superfluid device  in
Fig.\ref{Gyrometer}: the light source provides the incident light beam, the
flexible membrane the supercurrent; counter-rotating waves travel along the
square optical path, and along the coiled capillary, for the co-rotating part;
the waves interfere in the beam splitter in the optical case, in the Josephson
weak link in the superfluid case. The interferometer shown in
Fig.\ref{Hoskinson:06-Interference} is closer to a Mach-Zehnder interferometer
than to a Sagnac one but the analogy goes along in the same way.

But is this reasoning by analogy, or the display of clear fringe patterns such
as those shown in Fig.\ref{Hoskinson:06-Interference}, sufficient proof that
the Sagnac effect is involved in the operation of these
superfluid interferometers?  
\begin{figure}[t!]       
  \hskip -2mm
  \includegraphics[width=42mm,angle=0]{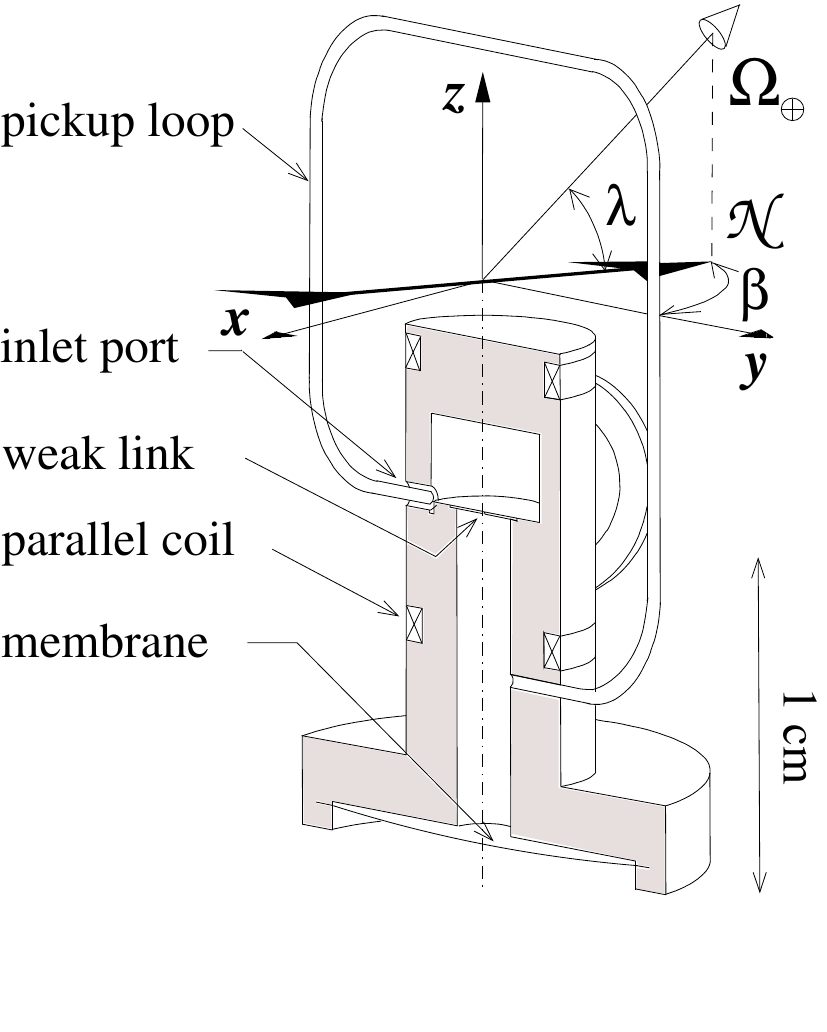}
  \includegraphics[width=33mm,angle=0]{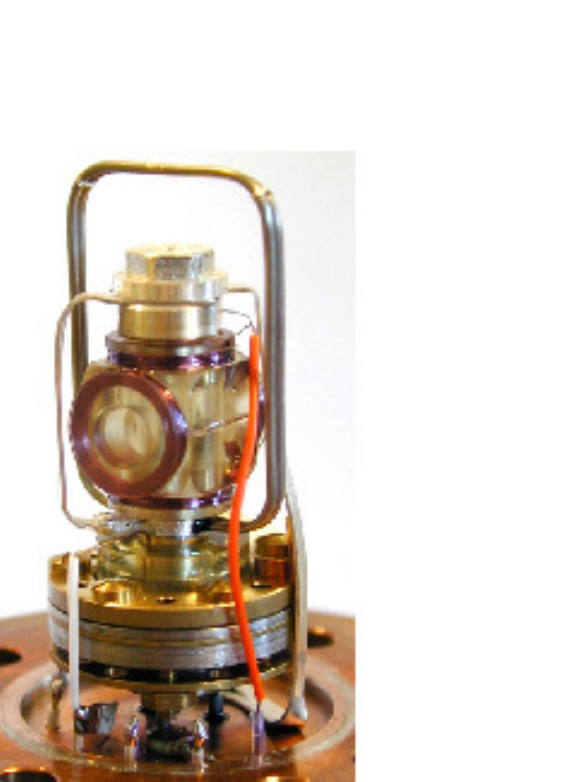}
\caption{ \label{Gyrometer}(Color online) 
  Photograph and schematic view of the cell of \citet{Avenel:04},
  approximately to scale for the inner parts, except for the loop, which is
  made of two turns of 0.4 mm internal diameter capillary (only one turn is
  shown) with total area $5.90\,\pm\,0.10$ cm$^2$.  The lower chamber of the
  resonator is a cylindrical duct, 1 mm in diameter, and connects the weak
  link to the flexible diaphragm (at the bottom) and to one end of the pickup
  loop (on the side); the upper chamber, a squat cylinder, is connected to the
  other end of the loop and to an inlet toward the main superfluid bath in
  which the resonator is immersed. Two pairs of coils produce fields parallel
  and perpendicular to the flow through the weak link to control locally the
  order parameter texture in $^3$He. The cryostat is rotated about the
  vertical axis $z$ by angle $\beta$ from the North, shown by a compass
  needle; $\lambda$ is the latitude, 48$^{\circ}$43$^\prime$ at Saclay. This
  cell has been used to detect the rotation of the Earth from a ``blind''
  laboratory with a sensitivity of $5\times
  10^{-3}\,\Omega_\oplus\mbox{ Hz}^{-1/2}$.}
\end{figure}

Apparently not; superfluid gyros are still sometimes mistaken for purely
inertial devices such as spinning tops, as discussed by \citet{Varoquaux:08}.
Clearly, the superfluid in a rotating bucket experiment is a dense medium. It
can be weighed on a scale. For large enough rotation velocities, when enough
vortex lines have been created, the fluid free surface eventually becomes
concave, as in Newton's rotating water bucket experiment. For small velocities
however, things are different: the superfluid does not even starts spinning
because of the absence of viscosity. Hence the common sense reluctance to
admit that the far-fetched analogy between the behaviour of this condensed
matter system and that of massless photons travelling at the velocity
of light, or elementary particles like electrons or neutrons, or even confined
ultra-cold atomic gases, holds any water.

A more formal approach is the following. As mentioned above, the Sagnac effect
has been observed in a number of different physical systems, ranging from
photon, electron, neutron and cold atomic gas interferometers to atomic clocks
and the Global Positioning System.\footnote{The literature on the Sagnac
  effect is extremely vast. See \citet{Hasselbach:93}, \citet{Stedman:97},
  \citet{Neutze:98} for recent reviews on the effect with matter waves.} The
unifying concept behind these widely different situations is provided by the
transportation of Einstein clocks from location A to location B on a rotating
platform \citep{Langevin:21,Rizzi:04}.

Consider how these clocks can be synchronised, first when they are infinitely
close to one another. The space-time metric is characterised in the
conventional notation by $ -\m d s^2= g_{00}\m d (x^0)^2 + 2g_{0i}\m dx^0 \m
dx^i+g_{ii}\m d(x^i)^2$. The infinitesimal time interval $\m d t$ between two
nearly simultaneous events taking place at this given location in space is
such that $\m ds^2 = g_{00}\m d (x^0)^2 = -\m c^2 \m d t^2$, $\m c$ being the
velocity of light. 

If the clocks are now separated in space by an infinitesimal amount $\m d x^i$
and the two events taken, say at location A, as the ticking of clock A
for one and as the signal transmitted by clock B of its ticking a small distance
away for the second, the two ticks occur with a time lag given by
$ 
  \m c \m d t = - g_{0i} \m d x^i\,/\,g_{00} \; ,
$
the repeated summation being on the space coordinates.  

If clock B is now transported over a finite path $\Gamma$ closing on itself in
a frame rotating with velocity $\mb\Omega$, the total time shift results from
an integration along path $\Gamma$:\footnote{As done in the book of
  \citet{Landau:ClassicalFields}, \S 90.}
\begin{equation}        \label{TimeDifference} 
  \Delta t = \frac{1}{\m c}\,\oint_{\Gamma} \frac{g_{0i}\m d x^i}
    {-g_{00}} 
    =\oint_{\Gamma} \frac{\mb\Omega \!\bs\times\!  \mb r \bs\cdot\m d\mb r}
    {\m c^2-(\mb\Omega\!\bs\times\!\mb r)^2}
    \simeq \frac{2}{\m c^2}\, \mb\Omega \bs\cdot \mb S  \; ,
\end{equation}
$\mb S$ being the vector area subtended by the loop $\Gamma$. 

Time delay (\ref{TimeDifference}) between the reading of the clock standing
still on the rotating platform and that of the transported clock lies at the
root of the Sagnac effect. As it depends on the rotation velocity and the
actual path $\mathit \Gamma$, absolute clock synchronisation cannot be
achieved.  Sagnac corrections, Eq.(\ref{TimeDifference}), must be performed as
done routinely for Global Positioning Systems \citep{Ashby:04}.  \footnote{It
  may be worth recalling that this clock transportation experiment was
  actually performed by \citet{Hafele:72} who boarded eastward and westward
  bound commercial jetliners taking as luggage a portable atomic clock.}

So much with clocks. For helium, a Lorentz invariant two-fluid model can be
built over the usual Landau superfluid hydrodynamics as done by
\citet{Carter:92}.\footnote{As also outlined by \citet{Ho:80b}.} The invariant
velocity circulation, the generalisation of Eq.(\ref{VelocityCirculation}),
reads
\begin{equation}        \label{CovariantCirculation}
  \int_\Xi\{ v'_0 \m dx^0 + v'_i \m dx^i\} = n\kappa \; ,
\end{equation}
where $(v'_o,v'_i)$ is the four-velocity in the rotating frame $(c^2+\mb v'_\m
n\bs\cdot \mb v'_{\m s}\,, -\mb v'_\m s)$.  The normal fluid velocity $v'_\m
n$ and the superfluid velocity $v'_\m s$ are very small compared to $\m c$ so
that the time-like component of the four-velocity reduces to $\m c^2$. The
integration over $\Xi$ is an actual loop integral only for the space-like
components. The corresponding world line is not closed because the time for
synchronised clocks varies according to Eq.(\ref{TimeDifference}). Upon
integration, Eq.(\ref{Circulation}) is recovered,
\begin{equation}        \label{Circulation2}
  \oint_\Gamma v'_i \m dx^i = n\kappa + \int\! c^2 g_{0i} \m  dx^i/g_{00} \simeq
  n\kappa - 2\,\mb\Omega
  \bs\cdot \mb S  \; , 
\end{equation}
which establishes the link between superfluid physics and the relativistic
clock approach: the true and honest Sagnac effect described by the transported
clocks, Eq.(\ref{TimeDifference}), and the circulation quantisation condition in
the rotating frame leading to Eq.(\ref{Bias}) are one and the
same.\footnote{See \citet{Volovik:03} for an alternate approach and
  \citet{Varoquaux:08} for additional references. Equation (11)
  in the latter reference is misprinted.}

Thus, Einstein-synchronised clocks provide the time standard by which phase
differences can be kept track of in all the studied physical systems. As
appropriately summarised by \citet{Greenberger:83} for neutron interferometry
experiments: ``{\it the phase
  shift} (in the rotating interferometer) {\it is seen to be caused by the
  different rates at which a clock ticks along each of the two beams}''.  The
rate at which that clock ticks for helium depends on the chemical potential
$\mu$, due to the molecular field of the condensate as shown by Beliaev, and
on the Sagnac phase shift.

The helium Sagnac experiments illustrate convincingly the reality of matter
wave interference in the superfluid heliums, a substantially massive coherent
field. Coherent, as shown at some length here, means coherence of the quantum
phase, giving a wave-like character to a bulky fluid. A few remarks on this
follow in  a manner of conclusion.
               
\subsection{Landau's two fluids, ODLRO, and macrorealism}
\label{Conclusion}

Anderson's introductory words to the reprint of his 1966 paper in his book of
1994 are the following \footnote{\citet{Anderson:94}, p. 165.}\,:``{\it I feel
  this is the clearest discussion of superfluidity available. Note that on
  many points this is contradictory or orthogonal to Landau orthodoxy as
  pronounced by Khalatnikov. Whether Landau would have agreed was never
  clarified because of his accident''.} This remark poses the problem of the
complementarity between the two-fluid model, the description of Bose
condensation by the non-vanishing off-diagonal terms of the density matrix
over a long range, and the macroscopic wavefunction approach.

Two crucial aspects of Landau's legacy have been invoked here. The two-fluid
model has ruled once and for good on the separation between normal fluid and
superfluid components, both for the thermodynamics and the hydrodynamics. As
used in this review in its reduced form for incompressible flows, which treats
both components as independent, it has allowed to basically disregard the
normal component. There are obviously limits to this high-handed
simplification, especially close to the \ulambda-point, but it has provided
the backbone of the simplified vortex dynamics of Sec.\ref{VortexDynamics} and
subsequent Sections on phase-slip processes.  The second pillar of Landau's
contributions to helium superflow is his criterion that many consider as the
genuine {\it intrinsic} critical velocity in superfluids. This criterion rests
on the existence of a sharply-defined phonon-roton excitation spectrum, which
allows for no low lying elementary excitations as Landau implicitly
postulated.

Landau's foreknowledge was soon put on firm grounds
 by the work of Bogolyubov, Beliaev, Penrose, Onsager and others. The formal
 description of the Bose condensate correlations ended up in the concept of
 off-diagonal long range order \citep{Yang:62} and a formal definition of the
 single wavefunction shared by the particles in the condensate.  For
 not-too-complicated superfluids -- $^4$He and the B-phase of $^3$He -- this
 wavefunction, or order parameter, has a definite overall phase.

 The role of this phase actually came to the fore when \citet{Anderson:66b}
 gave it dynamical variable status and universal applicability. In Appendix A
 of his paper, ``{\it ODLRO  vs macroscopic particle fields}'', he states
 explicitly that ``{\it recognising that in principle the relative phase of
   any two} (superfluid) {\it systems may always be measured by a
   Josephson-type experiment, one immediately has a usable local
   description}'' (of those systems). 

 This local description has been put to good use, as recounted in the
 foregoing at some length. It has opened the way to a full understanding of
 the interaction of quantised vortices and superflow, put on firm classical
 hydrodynamics footing by \citet{Huggins:70}. \citet{Sonin:95} and others have
 further expanded the vortex velocity field idea into a workable scheme for
 vortex dynamics.  More importantly, it has bridged the gap between a
 predominantly ``classical'' two-fluid hydrodynamics and the more intimately
 quantum Josephson effects. The experimental observation of these effects,
 and, in particular, the detailed way by which dissipative phase slippage,
 understood first as the nucleation and propagation of vortices, evolves into
 the purest brand of Josephson hydrodynamics effects, both for $^3$He and
 $^4$He superfluids, has brought fresh food for thought.

 Detailed numerical simulations of vortex dynamics have been conducted by
 Schwarz and others,\footnote{See, for example, the contributions of
   D. C. Samuels, of O. C. Idowu et al., and of M. Tsubota, T. Araki, and
   S. K. Nemirovskii in \citet{Barenghi:01}.} in particular for the problems
 of the vortex in-an-aperture, or trapped on a thin wire, or else,
 pinned. Collision between vortices, and the resulting reconnection, their
 multiplication by vortex mills churning out fresh quantised vorticity, the
 formation of vortex tangles, and the several ensuing critical velocities, are
 examples of the improved way of dealing with quantised vortices fostered one
 way or another by Anderson's Considerations, which thus appear
 complementary to Landau's views. Experimental observations have fully borne
 out over the years this central concept of a macroscopic quantum phase
 governing the dynamical behaviour of the superfluid.

 At least in {\it one} given pool of superfluid. The idea that a given pool,
 or bucket, or droplet, of superfluid has its own phase has become so common
 place that the question of knowing if the quantum phase of an isolated
 droplet of superfluid has a value of its own as compared to an other one
 seems out of place. As put by \citet{Anderson:86}: ``{\it{Do superfluids that
     have never seen each other have a well-defined relative phase}}''\,?
 
 Before answering by a qualified ``yes'', it may be useful to consider the new
 inputs to this problem of the meaning, or reality, of the phase in Bose
 condensates that have emerged after 1996 from investigations in ultra-cold
 atomic systems. This is not the topic of this review, but interference
 experiments with BEC gases do bear on some aspects of it. 

 A seemingly curious fact was noted by \citet{Javanainen:96} in their
 numerical simulations of the setting up of an interference pattern between
 two condensates formed in separate traps and left to overlap with
 one another.\footnote{Such an experiment was successfully performed soon
   after by \citet{Andrews:97} and others. See also below.}  The initial
 number of particles in each condensate is well-defined in this computer
 experiment. The phases of the condensate wavefunctions are in no instance
 invoked in the number crunching sequence describing the interference process.
 Yet it does appear: the simple statistical count of bosons in separate bins suffices.

 This finding is a manifestation for the special case of atoms in cold
 traps of the well-known tendency of bosons to ``bunch'' together. This
 phenomenon is very clearly illustrated by \citet{Castin:97} who
 tackled the same problem analytically. Namely, they studied the evolution of the relative
 phase of two separate BE condensates of like-species of atoms confined
 separate traps and left at some instant to interact by leaking two small beams
 of atoms to a beam splitter. The two outgoing split beams that have been mixed
 are read by two atom counters.

 Following these authors, consider first that only a single atom trap leaks atoms to
 the beam splitter. Assume the first counting event to occur in counter (+).
 The probability for this event to take place is 1/2. Because of the Bose
 statistics, the next event has a probability 3/4 to take place in counter (+),
 in which the first event was recorded, and 1/4 for counter (-) that has
 not seen an atom yet \citep{Feynman:65}. Iterating for $k$ such events in a row,
 \citet{Castin:97} find that the probability that the $k$ atoms end up
 registered by counter (+) -- and none by counter (-) -- takes the following
 form,
$$ 
   {\cal P}(k,0) = \frac{1}{2}\,  \frac{3}{4}\,\ldots  \frac{2k-1}{2k}\, ,
$$
which is quite different from the probability of the same final outcome for
atoms with no quantum-statistical correlations, namely $1/2^k$. The latter
becomes rapidly negligible for $k \gg 1$ while the former decreases only as
$k^{-1/2}$. Bosons have a strong tendency to crop or ``bunch''
together in states where they can find like-bosons.

\citet{Castin:97} proceed to study the case in which the two separate BE
condensates are now leaking beams to the beam splitter. If the two incident
beams are described by fields with well-defined phases and identical
amplitudes, $|\psi_0| \m e^{\m i\varphi_\m A}$ and $|\psi_0| \m e^{\m
  i\varphi_\m B}$, the mean intensities in the (+) and (-) output channels of the
beam splitter are given by
\begin{equation}         \label{BeatIntensities}
  I_{\ts +} =  2\, |\psi_0|^2  \cos^2(\inc\varphi/2) \;\mbox{ and }\;
      I_{\ts -}  = 2\, |\psi_0|^2  \sin^2(\inc\varphi/2)
\end{equation}
with $\inc\varphi = \varphi_\m B-\varphi_\m A$. This matter wave
interferometry measurement allows as expected the determination of the phase
difference between the two condensates.

What now if phases are not ascribed to the condensates but instead their
particle numbers $N_\m A$ and $N_\m B$ are? Assuming $N_\m B = N_\m A$ for
simplicity (as in the simulations of \citet{Javanainen:96}) the two-condensate
system is now described by a Fock state $\bra {N_\m A},\ket {N_\m A}$.
Performing the same interferometry measurement by counting atoms in the (+)
and (-) counters should end up yielding result (\ref{BeatIntensities}). It
does, but the interesting thing is how a phase difference arises from the
mere statistical count.

The detailed probability ${\cal P}(k_{\ts +},k_{\ts -})$ of counting $k_{\ts +}$ events in
counter (+) and $k_{\ts -}$ in counter (-) has been worked out analytically by
\citet{Castin:97}. The beat pattern given by Eq.(\ref{BeatIntensities}) is
found to emerge gradually from the successive counting events in the (+) and
(-) counters.  The phase difference can be obtained from the tallied
quantities $k_{\ts +}$ and $k_{\ts -}$: $\inc\varphi \simeq \arctan (\sqrt{k_{\ts -} /
  k_{\ts +}})$\,. The particle-like description turns into the wave-like
description as the counting proceeds. The sequence of measurements brings a
definite quantum phase to states for which none had been assumed to start
with: at the end of the day, one is left with a state containing $2N_\m A-k$
atoms and a phase difference known to an accuracy of order $1/k$.

The phase difference is an unpredictable random variable, which takes a
different value for any realisation of the counting experiment. It appears as
a mere by-product of the counting statistics. No direct interaction between
the atoms in the condensates has been assumed at any point. The BEC gases 
are taken as perfect gases. The effect purely originates from the quantum statistics of
bosons. As concluded by Castin and Dalibard, ``{\it the notion of phase-broken
  symmetry is therefore not indispensable in order to understand the beating
  of two condensates}''.

This conclusion, which has gained wide acceptance, \footnote{See,
    for example, \citet{Horak:99,Nienhuis:01}.} has been beautifully illustrated by the
experiments of \citet{Saba:05}. These authors dropped two condensates of
like-species out of their traps and, during the course of their free fall and
expansion, gently pushed with laser beams a few atoms from one to the
other. They dutifully observed the continuous emergence of a fringe pattern in
a quintessential form, without beam splitters and interferometers nor
destruction of the condensate clouds, thus realising a nearly non-invasive
measurement of their relative phase.

These cold atom gases constitute model systems. They can be studied to their
minutest details starting from the basic principles of quantum
mechanics. Photons in cavities provide another such instance. Quantum
statistical correlations between indistinguishable bosons play the leading
role. Particle interactions play a minor role of decoherence and are
neglected. Particle-wave duality is demonstrated to near perfection.

Superfluids differ in a number of respects. The macroscopic wavefunction
introduced by Anderson is defined on the premises that the particle number $N$
and its possible variation $\delta N$ are sufficiently large so that the
uncertainty in $\varphi$, expressed by Eq.(\ref{NumberPhaseUn}), $\delta N
\,\delta \varphi \sim 1$, is small in most instances. It is then neglected and
the operators $\hat N$ and $\hat\varphi$ are ``projected'' onto
$c$-numbers. They have acquired a value once and for good: the phase is forced
to ``exist'' even if its actual value is not determined by the same token and
its absolute meaning uncertain.

Actually, the above introduction of the superfluid phase resembles a leap of
faith: it is there because it is needed to reproduce the
hydrodynamics. \citet{Leggett:91} and \citet{Leggett:95} have paid careful
attention to this problem of the ``existence'' of the phase. One of the many
points raised by these authors is that, in order to attribute a well-defined
meaning to an {\it absolute} phase, one has first to consider the {\it
  relative} phase of one bucket of superfluid with respect to another --
presumably measured by performing a Josephson experiment. The various ways and
constraints of such an experiment have been expounded in the Sections above:
the ``relative'' phase between two weakly connected superfluid systems (1) an
(2), $\udelta\varphi_{12} = \varphi_2 - \varphi_1$, can be measured and indeed
possesses a well-defined meaning. Suppose now that systems (1) and (2) are
separated and a third system brought in. System (3) can be compared to (1),
with the result $\udelta\varphi_{13} = \varphi_3 - \varphi_1$: is it possible
to infer that the phase difference between (2) and (3) is $ \varphi_3 -
\varphi_2= \udelta\varphi_{13} - \udelta\varphi_{12}$\,? If yes, then phases
can be referred to a ``standard'' and acquire absolute meaning.

The not-so-trivial answer given by \citet{Leggett:95} is ``no'' if the systems
are let to settle to equilibrium with the environment and the two Josephson
phase measurements are independent, but ``yes'' if they are done
simultaneously. In other words, maintaining a superfluid phase standard across
the various Standards Laboratories of the planet would require connecting them
with a continuous superfluid duct. The phase information would have to be
tapped from this standard at the same time and place, or else a host of corrections,
such as that for local gravity or for the Sagnac effect -- the synchronisation
of the Einstein clocks -- \ldots, would have to be performed. If the phase readings
are not simultaneous, the correlation between phase measurements between
systems (1) and (2), and then (1) and (3) is upset by the sole act of
measuring, with a part played by the environment; decoherence takes its toll
and phases ultimately randomise \citep{Sols:94,Zapata:03}.

The situation in simple superfluids such as $^4$He and $^3$He-B, in which
macroscopic coherence holds over lengths of metres and more, and in which the
Planck constant hides at the nanometric scale provides a quite extreme example
of macroscopic matter field.  That the macroscopic field $\Phi(\mb r)$
standing as the single-particle wavefunction for condensate atoms possesses
quantum properties is indisputable -- the quantisation of velocity circulation
and the existence of persistent currents offer bullet-proof examples. However,
the sort of coherence shown by ultra-cold atom condensates does not stand out
readily for dense superfluids. The description of the superfluid dynamics in
terms the conjugate variables $N$ and $\varphi$ belongs more to thermodynamics
than to quantum mechanics. The correlation between atoms in the dense helium
fluid relies more on their hard-core repulsion than on their Bose statistics.

Yet, the quantum interferences by quantum tunnelling between macroscopically
distinct states in superfluid Josephson junctions, the quantum nucleation of
vortices clearly reveal the importance of the latter. In these situations, the
coarse-grained average fails and some other procedure, more in line with the
basic rules of quantum mechanics, and, in particular, the principle of
superposition, is in order. Would it be possible to envision experiments
showing actual macroscopic quantum {\it coherence} as discussed by
\citet{Leggett:80} and \citet{Leggett:02} \citep{Annett:03}? The superfluid
quantum phase would then gain a dual acception, actual coherence in the
superposition of different states at very small scale on the one hand and, on
the other, the ``rigidity'', in the language of F. London, of the velocity
potential of the ideal Euler fluid and the quantisation of the velocity
circulation.

The above remarks bring this ``essay on criticism'' of Anderson's
Considerations to a close. At this stage, but one hard conclusion can be
drawn: many offshoots have already sprung but more are to grow.

\section{Acknowledgement}

This work has been supported over many years by Centre National de la
Recherche Scientifique and by Commissariat \`a l'\'Energie Atomique et aux
\'Energies Alternatives. The author owes much to his many colleagues,
foremostly  Olivier Avenel and Yuri Mukharsky who have brought vital
contributions to many of the topics discussed above.

\appendix{\section{ }}
What need be shown is that the second term of the right-hand side of
Eq.(\ref{EnergyBalance}) corresponds to the variation of the vortex loop
self-energy for the infinitesimal deformation $\mb s(l)$ into $\mb s(l)+
{\mbox{\bf{\greeksym{e}}}}\, \delta(l-l_0)$ .  This energy variation can be
derived from the functional derivative of $E_\m v[\mb s]$, given by
Eq.(\ref{ReducedEv2}), with respect to the deformation $\inc \mb s$ :
%
\begin{eqnarray}        \label{FuncDeriv1}        
    \frac{\inc E[\mb s]}{\inc \mb s}\biggl|_{l_0} 
    &=& \frac{\rho_\m s \kappa_4^2}{8\pi}\,
    \Biggl\{-\oint \frac{\m d\mb s(l_0)}{\m d l_0} \cdot 
    \frac{\m d\mb s(l_2)}{\m  d l_2}  \,
    \frac{\mb s(l_0) -\mb s(l_2)}{\left| \mb s(l_0)-\mb s(l_2)\right|^3} \,\m d l_2 
    \nonumber
    \\ &+& 
    \;\lim_{\mbox{\bf{\greeksym{e}}} \rightarrow 0}\; 
    \oint\!\!\oint {\mbox{\bf{\greeksym{e}}}}\cdot\frac{\m d\mb
    s(l_2)}{\m  d l_2}\delta'(l_1-l_0)  \frac{\m d l_1\,\m d l_2}
    {\left| \mb s(l_1)-\mb s(l_2)\right|}
    \nonumber
    \\ &+&
    \; l_1 \Longleftrightarrow l_2 \Biggr\} \, .
\end{eqnarray}
The first term in the right hand side of Eq.(\ref{FuncDeriv1})
results from the differentiation of $1/|\mb s(l_1)- \mb s(l_2)|$ with respect to
$l_1$ and integration over the Dirac function representing the deformation at
$l_1=l_0$. The integral over the derivative of the Dirac $\delta$-function,
which comes from the differentiation of $ \m d\mb s(l_1) / \m d l_1$ yields the
derivative of the integrand evaluated at $l_0$.  The contribution of the
integration over $l_2$ over the same contour with the same deformation is
equal to that over $l_1$, expressed by the first two terms of the right hand
side of Eq.(\ref{FuncDeriv1}), and yields a factor of 2 in the
final result.

Using the notation $\mb{\hat t}(l)= \m d\mb s(l)/\m d l $ for the unit vector
tangent to the vortex loop at location $\mb s(l)$, Eq.(\ref{FuncDeriv1})
can be written as
\begin{equation}       \label{FunctionalDerivative2}
  \begin{split}        
    \frac{\inc E[\mb s]}{\inc \mb s}\biggl|_{l_0} 
    &= \frac{\rho_\m s \kappa_4^2}{4\pi}\,\oint 
    \Biggl\{\mb{\hat t}(l_0) \cdot\frac{\mb s(l_0)- \mb s(l_2)}
    {\left| \mb s(l_0)-\mb s(l_2)\right|^3} \,\mb{\hat t}(l_2) 
    \\& \hskip 1.5 cm 
    -\; \mb{\hat t}(l_0)\cdot\mb{\hat t}(l_2)\,\frac{\mb s(l_0)- \mb s(l_2)}
    {\left| \mb s(l_0)-\mb s(l_2)\right|^3} \Biggr\}
    \,\m d l_2
    \\& = \frac{\rho_\m s \kappa_4^2}{4\pi}\, \mb{\hat t}(l_0)\times\oint 
    \mb{\hat t}(l_2)\times\frac{\mb s(l_0)- \mb s(l_2)} 
    {\left| \mb s(l_0)-\mb s(l_2)\right|^3} \,\m d l_2 
    \\& = \rho_\m s\kappa_4\, \mb{\hat t}(l_0)\times \mb v_\m v(l_0)\,\; . 
  \end{split}
\end{equation}
The double cross product in the second equality of
(\ref{FunctionalDerivative2}) appears because of the vector relation $(\mb
a\cdot\mb c)\,\mb b -(\mb a\cdot\mb b)\,\mb c =\mb a\times(\mb b\times\mb
c)$. The last equality is obtained using the Biot-Savart law,
Eq.(\ref{BiotSavart}), for $\mb v_\m v$ the velocity induced by the vortex
loop on itself at $l_0$. The logarithmic divergences for $l_0 = l_2$ are
cut-off at the core radius $a_0$.\footnote{Compare with Eqs.(3.7) and (3.15)
  in Sonin's paper (1987).}  The change of the vortex self-energy is then
expressed by
\begin{displaymath}
  \begin{split}
  \Delta E_\m v &= \int_l^{l+\Delta l} \,\oint_{\m{loop}}\;  \frac{\inc E[\mb
    s]}{\inc \mb s}\biggl|_{l_0} \!\! \cdot \inc\mb x\, \delta(l-l_0) \m d l\,\m d l_0 \\
                       &=  \int_l^{l+\Delta l} \rho_\m s \kappa_4 \, 
   \hat{\mb t}(l_0) \times \mb v_\m v \cdot \mb v_\m p \Delta t\,\m d l_0 \;.
  \end{split}
\end{displaymath}
Assuming the integrand constant over the small element $\Delta \mb l$,
Eq.(\ref{VortexSefEnergyChange}) ensues.


\begin{thebibliography}{346}%
\makeatletter
\providecommand \@ifxundefined [1]{%
 \@ifx{#1\undefined}
}%
\providecommand \@ifnum [1]{%
 \ifnum #1\expandafter \@firstoftwo
 \else \expandafter \@secondoftwo
 \fi
}%
\providecommand \@ifx [1]{%
 \ifx #1\expandafter \@firstoftwo
 \else \expandafter \@secondoftwo
 \fi
}%
\providecommand \natexlab [1]{#1}%
\providecommand \enquote  [1]{``#1''}%
\providecommand \bibnamefont  [1]{#1}%
\providecommand \bibfnamefont [1]{#1}%
\providecommand \citenamefont [1]{#1}%
\providecommand \href@noop [0]{\@secondoftwo}%
\providecommand \href [0]{\begingroup \@sanitize@url \@href}%
\providecommand \@href[1]{\@@startlink{#1}\@@href}%
\providecommand \@@href[1]{\endgroup#1\@@endlink}%
\providecommand \@sanitize@url [0]{\catcode `\\12\catcode `\$12\catcode
  `\&12\catcode `\#12\catcode `\^12\catcode `\_12\catcode `\%12\relax}%
\providecommand \@@startlink[1]{}%
\providecommand \@@endlink[0]{}%
\providecommand \url  [0]{\begingroup\@sanitize@url \@url }%
\providecommand \@url [1]{\endgroup\@href {#1}{\urlprefix }}%
\providecommand \urlprefix  [0]{URL }%
\providecommand \Eprint [0]{\href }%
\providecommand \doibase [0]{http://dx.doi.org/}%
\providecommand \selectlanguage [0]{\@gobble}%
\providecommand \bibinfo  [0]{\@secondoftwo}%
\providecommand \bibfield  [0]{\@secondoftwo}%
\providecommand \translation [1]{[#1]}%
\providecommand \BibitemOpen [0]{}%
\providecommand \bibitemStop [0]{}%
\providecommand \bibitemNoStop [0]{.\EOS\space}%
\providecommand \EOS [0]{\spacefactor3000\relax}%
\providecommand \BibitemShut  [1]{\csname bibitem#1\endcsname}%
\let\auto@bib@innerbib\@empty
\bibitem [{\citenamefont {Aarts}\ \emph {et~al.}(1994)\citenamefont {Aarts},
  \citenamefont {Ihas}, \citenamefont {Avenel},\ and\ \citenamefont
  {Varoquaux}}]{Aarts:94}%
  \BibitemOpen
  \bibfield  {author} {\bibinfo {author} {\bibnamefont {Aarts}, \bibfnamefont
  {R.}}, \bibinfo {author} {\bibfnamefont {G.~G.}\ \bibnamefont {Ihas}},
  \bibinfo {author} {\bibfnamefont {O.}~\bibnamefont {Avenel}}, \ and\ \bibinfo
  {author} {\bibfnamefont {E.}~\bibnamefont {Varoquaux}}} (\bibinfo {year}
  {1994}),\ \href@noop {} {\bibfield  {journal} {\bibinfo  {journal} {Physica
  B}\ }\textbf {\bibinfo {volume} {194-196}},\ \bibinfo {pages}
  {493}}\BibitemShut {NoStop}%
\bibitem [{\citenamefont {Abrikosov}\ \emph {et~al.}(1961)\citenamefont
  {Abrikosov}, \citenamefont {Gorkov},\ and\ \citenamefont
  {Dzyaloshinski}}]{Abrikosov:61}%
  \BibitemOpen
  \bibfield  {author} {\bibinfo {author} {\bibnamefont {Abrikosov},
  \bibfnamefont {A.~A.}}, \bibinfo {author} {\bibfnamefont {L.~P.}\
  \bibnamefont {Gorkov}}, \ and\ \bibinfo {author} {\bibfnamefont {I.~E.}\
  \bibnamefont {Dzyaloshinski}}} (\bibinfo {year} {1961}),\ \href@noop {}
  {\emph {\bibinfo {title} {Methods of Quantum Field Theory in Statistical
  Physics}}}\ (\bibinfo  {publisher} {English edition, Prentice-Hall, Inc,
  Englewood Cliffs, N.J. 1963})\ \bibinfo {note} {ch. 5}\BibitemShut {NoStop}%
\bibitem [{\citenamefont {Adams}\ \emph {et~al.}(1985)\citenamefont {Adams},
  \citenamefont {Cieplak},\ and\ \citenamefont {Glaberson}}]{Adams:85}%
  \BibitemOpen
  \bibfield  {author} {\bibinfo {author} {\bibnamefont {Adams}, \bibfnamefont
  {P.}}, \bibinfo {author} {\bibfnamefont {M.}~\bibnamefont {Cieplak}}, \ and\
  \bibinfo {author} {\bibfnamefont {W.~I.}\ \bibnamefont {Glaberson}}}
  (\bibinfo {year} {1985}),\ \href@noop {} {\bibfield  {journal} {\bibinfo
  {journal} {Phys. Rev. B}\ }\textbf {\bibinfo {volume} {32}},\ \bibinfo
  {pages} {171}}\BibitemShut {NoStop}%
\bibitem [{\citenamefont {Alpar}\ \emph {et~al.}(1980)\citenamefont {Alpar},
  \citenamefont {Nandkumar},\ and\ \citenamefont {Pines}}]{Alpar:80}%
  \BibitemOpen
  \bibfield  {author} {\bibinfo {author} {\bibnamefont {Alpar}, \bibfnamefont
  {M.~A.}}, \bibinfo {author} {\bibfnamefont {R.}~\bibnamefont {Nandkumar}}, \
  and\ \bibinfo {author} {\bibfnamefont {D.}~\bibnamefont {Pines}}} (\bibinfo
  {year} {1980}),\ \href@noop {} {\bibfield  {journal} {\bibinfo  {journal}
  {Astrophys. J.}\ }\textbf {\bibinfo {volume} {288}},\ \bibinfo {pages}
  {191}}\BibitemShut {NoStop}%
\bibitem [{\citenamefont {Amar}\ \emph {et~al.}(1990)\citenamefont {Amar},
  \citenamefont {Davis}, \citenamefont {Packard},\ and\ \citenamefont
  {Lozes}}]{Amar:90}%
  \BibitemOpen
  \bibfield  {author} {\bibinfo {author} {\bibnamefont {Amar}, \bibfnamefont
  {A.}}, \bibinfo {author} {\bibfnamefont {J.~C.}\ \bibnamefont {Davis}},
  \bibinfo {author} {\bibfnamefont {R.~E.}\ \bibnamefont {Packard}}, \ and\
  \bibinfo {author} {\bibfnamefont {R.~L.}\ \bibnamefont {Lozes}}} (\bibinfo
  {year} {1990}),\ \href@noop {} {\bibfield  {journal} {\bibinfo  {journal}
  {Physica B}\ }\textbf {\bibinfo {volume} {165\&166}},\ \bibinfo {pages}
  {753}}\BibitemShut {NoStop}%
\bibitem [{\citenamefont {Amar}\ \emph {et~al.}(1992)\citenamefont {Amar},
  \citenamefont {Sasaki}, \citenamefont {Lozes}, \citenamefont {Davis},\ and\
  \citenamefont {Packard}}]{Amar:92}%
  \BibitemOpen
  \bibfield  {author} {\bibinfo {author} {\bibnamefont {Amar}, \bibfnamefont
  {A.}}, \bibinfo {author} {\bibfnamefont {Y.}~\bibnamefont {Sasaki}}, \bibinfo
  {author} {\bibfnamefont {R.}~\bibnamefont {Lozes}}, \bibinfo {author}
  {\bibfnamefont {J.~C.}\ \bibnamefont {Davis}}, \ and\ \bibinfo {author}
  {\bibfnamefont {R.~E.}\ \bibnamefont {Packard}}} (\bibinfo {year} {1992}),\
  \href@noop {} {\bibfield  {journal} {\bibinfo  {journal} {Phys.Rev. Lett.}\
  }\textbf {\bibinfo {volume} {68}},\ \bibinfo {pages} {2624}}\BibitemShut
  {NoStop}%
\bibitem [{\citenamefont {Anderson}\ \emph {et~al.}(1984)\citenamefont
  {Anderson}, \citenamefont {Beecken},\ and\ \citenamefont {{W. Zimmermann
  Jr}.}}]{Anderson:84}%
  \BibitemOpen
  \bibfield  {author} {\bibinfo {author} {\bibnamefont {Anderson},
  \bibfnamefont {J.}}, \bibinfo {author} {\bibfnamefont {B.}~\bibnamefont
  {Beecken}}, \ and\ \bibinfo {author} {\bibnamefont {{W. Zimmermann Jr}.}}}
  (\bibinfo {year} {1984}),\ \href@noop {} {\bibinfo  {journal} {Physica}\ ,\
  \bibinfo {pages} {313}}\BibitemShut {NoStop}%
\bibitem [{\citenamefont {Anderson}(1964)}]{Anderson:64}%
  \BibitemOpen
\bibfield  {journal} {  }\bibfield  {author} {\bibinfo {author} {\bibnamefont
  {Anderson}, \bibfnamefont {P.~W.}}} (\bibinfo {year} {1964}),\ in\ \href@noop
  {} {\emph {\bibinfo {booktitle} {Lectures in the Many-Body Problem}}},\
  \bibinfo {editor} {edited by\ \bibinfo {editor} {\bibfnamefont
  {E.}~\bibnamefont {Caianello}}}\ (\bibinfo  {publisher} {Academic Press,
  New-York})\ p.\ \bibinfo {pages} {113}\BibitemShut {NoStop}%
\bibitem [{\citenamefont {Anderson}(1965)}]{Anderson:65}%
  \BibitemOpen
  \bibfield  {author} {\bibinfo {author} {\bibnamefont {Anderson},
  \bibfnamefont {P.~W.}}} (\bibinfo {year} {1965}),\ in\ \href@noop {} {\emph
  {\bibinfo {booktitle} {Some Recent Definitions in the Basic Sciences}}},\
  Vol.~\bibinfo {volume} {2},\ \bibinfo {editor} {edited by\ \bibinfo {editor}
  {\bibfnamefont {A.}~\bibnamefont {Gelbert}}}\ (\bibinfo  {publisher} {Belfert
  Graduate School of Science, Yeshiva Univ., New-York})\ p.~\bibinfo {pages}
  {21},\ \bibinfo {note} {reprinted in Anderson (1984)}\BibitemShut {NoStop}%
\bibitem [{\citenamefont {Anderson}(1966{\natexlab{a}})}]{Anderson:66b}%
  \BibitemOpen
  \bibfield  {author} {\bibinfo {author} {\bibnamefont {Anderson},
  \bibfnamefont {P.~W.}}} (\bibinfo {year} {1966}{\natexlab{a}}),\ \href@noop
  {} {\bibfield  {journal} {\bibinfo  {journal} {Rev. Mod. Phys.}\ }\textbf
  {\bibinfo {volume} {38}},\ \bibinfo {pages} {298}}\BibitemShut {NoStop}%
\bibitem [{\citenamefont {Anderson}(1966{\natexlab{b}})}]{Anderson:66a}%
  \BibitemOpen
  \bibfield  {author} {\bibinfo {author} {\bibnamefont {Anderson},
  \bibfnamefont {P.~W.}}} (\bibinfo {year} {1966}{\natexlab{b}}),\ in\
  \href@noop {} {\emph {\bibinfo {booktitle} {Quantum Fluids}}},\ \bibinfo
  {editor} {edited by\ \bibinfo {editor} {\bibfnamefont {D.}~\bibnamefont
  {Brewer}}}\ (\bibinfo  {publisher} {North-Holland, Amsterdam 1966})\ p.\
  \bibinfo {pages} {146},\ \bibinfo {note} {this article, less the discussion
  at the Sussex University Symposium held in Brighton in 1965, is reproduced in
  a number of places
  \protect{\cite{Anderson:66b,Anderson:84b,Anderson:94}}}\BibitemShut {NoStop}%
\bibitem [{\citenamefont {Anderson}(1984)}]{Anderson:84b}%
  \BibitemOpen
  \bibfield  {author} {\bibinfo {author} {\bibnamefont {Anderson},
  \bibfnamefont {P.~W.}}} (\bibinfo {year} {1984}),\ \href@noop {} {\emph
  {\bibinfo {title} {Basic Notions of Condensed Matter}}}\ (\bibinfo
  {publisher} {Benjamin/Cummings, Menlo Park (CA)})\ \bibinfo {note} {also in
  P.W. Anderson \protect{\cite{Anderson:94}}}\BibitemShut {NoStop}%
\bibitem [{\citenamefont {Anderson}(1986)}]{Anderson:86}%
  \BibitemOpen
  \bibfield  {author} {\bibinfo {author} {\bibnamefont {Anderson},
  \bibfnamefont {P.~W.}}} (\bibinfo {year} {1986}),\ in\ \href@noop {} {\emph
  {\bibinfo {booktitle} {The Lesson of Quantum Theory}}},\ \bibinfo {editor}
  {edited by\ \bibinfo {editor} {\bibfnamefont {J.}~\bibnamefont {de~Boer}},
  \bibinfo {editor} {\bibfnamefont {E.}~\bibnamefont {Dahl}}, \ and\ \bibinfo
  {editor} {\bibfnamefont {O.}~\bibnamefont {Ulfbeck}}}\ (\bibinfo  {publisher}
  {Elsevier, New-York})\BibitemShut {NoStop}%
\bibitem [{\citenamefont {Anderson}(1994)}]{Anderson:94}%
  \BibitemOpen
  \bibfield  {author} {\bibinfo {author} {\bibnamefont {Anderson},
  \bibfnamefont {P.~W.}}} (\bibinfo {year} {1994}),\ \href@noop {} {\emph
  {\bibinfo {title} {A Career in Theoretical Physics}}}\ (\bibinfo  {publisher}
  {World Scientific, Singapore})\BibitemShut {NoStop}%
\bibitem [{\citenamefont {Anderson}\ and\ \citenamefont
  {Brinkman}(1975)}]{Anderson:75}%
  \BibitemOpen
  \bibfield  {author} {\bibinfo {author} {\bibnamefont {Anderson},
  \bibfnamefont {P.~W.}}, \ and\ \bibinfo {author} {\bibfnamefont
  {W.}~\bibnamefont {Brinkman}}} (\bibinfo {year} {1975}),\ in\ \href@noop {}
  {\emph {\bibinfo {booktitle} {The Helium Liquids}}},\ \bibinfo {editor}
  {edited by\ \bibinfo {editor} {\bibfnamefont {J.}~\bibnamefont {Armitage}}\
  and\ \bibinfo {editor} {\bibfnamefont {I.}~\bibnamefont {Farquhar}}}\
  (\bibinfo  {publisher} {Academic Press - London})\ p.\ \bibinfo {pages}
  {315}\BibitemShut {NoStop}%
\bibitem [{\citenamefont {Anderson}\ and\ \citenamefont
  {Richards}(1975)}]{Anderson:75a}%
  \BibitemOpen
  \bibfield  {author} {\bibinfo {author} {\bibnamefont {Anderson},
  \bibfnamefont {P.~W.}}, \ and\ \bibinfo {author} {\bibfnamefont {P.~L.}\
  \bibnamefont {Richards}}} (\bibinfo {year} {1975}),\ \href@noop {} {\bibfield
   {journal} {\bibinfo  {journal} {Phys. Rev. B}\ }\textbf {\bibinfo {volume}
  {11}},\ \bibinfo {pages} {2702}}\BibitemShut {NoStop}%
\bibitem [{\citenamefont {Anderson}\ and\ \citenamefont
  {Rowell}(1963)}]{Anderson:63}%
  \BibitemOpen
  \bibfield  {author} {\bibinfo {author} {\bibnamefont {Anderson},
  \bibfnamefont {P.~W.}}, \ and\ \bibinfo {author} {\bibfnamefont {J.~M.}\
  \bibnamefont {Rowell}}} (\bibinfo {year} {1963}),\ \href@noop {} {\bibfield
  {journal} {\bibinfo  {journal} {Phys. Rev. Lett.}\ }\textbf {\bibinfo
  {volume} {10}},\ \bibinfo {pages} {230}}\BibitemShut {NoStop}%
\bibitem [{\citenamefont {Anderson}\ and\ \citenamefont
  {Toulouse}(1977)}]{Anderson:77}%
  \BibitemOpen
  \bibfield  {author} {\bibinfo {author} {\bibnamefont {Anderson},
  \bibfnamefont {P.~W.}}, \ and\ \bibinfo {author} {\bibfnamefont
  {G.}~\bibnamefont {Toulouse}}} (\bibinfo {year} {1977}),\ \href@noop {}
  {\bibfield  {journal} {\bibinfo  {journal} {Phys. Rev. Lett.}\ }\textbf
  {\bibinfo {volume} {38}},\ \bibinfo {pages} {508}}\BibitemShut {NoStop}%
\bibitem [{\citenamefont {Andreev}\ and\ \citenamefont
  {Melnikovsky}(2004)}]{Andreev:04}%
  \BibitemOpen
  \bibfield  {author} {\bibinfo {author} {\bibnamefont {Andreev}, \bibfnamefont
  {A.}}, \ and\ \bibinfo {author} {\bibfnamefont {L.}~\bibnamefont
  {Melnikovsky}}} (\bibinfo {year} {2004}),\ \href@noop {} {\bibfield
  {journal} {\bibinfo  {journal} {J. Low Temp. Phys.}\ }\textbf {\bibinfo
  {volume} {135}},\ \bibinfo {pages} {411}}\BibitemShut {NoStop}%
\bibitem [{\citenamefont {Andrews}\ \emph {et~al.}(1997)\citenamefont
  {Andrews}, \citenamefont {Townsend}, \citenamefont {Miesner}, \citenamefont
  {Durfee}, \citenamefont {Kurn},\ and\ \citenamefont {Ketterle}}]{Andrews:97}%
  \BibitemOpen
  \bibfield  {author} {\bibinfo {author} {\bibnamefont {Andrews}, \bibfnamefont
  {M.~R.}}, \bibinfo {author} {\bibfnamefont {C.~G.}\ \bibnamefont {Townsend}},
  \bibinfo {author} {\bibfnamefont {H.-J.}\ \bibnamefont {Miesner}}, \bibinfo
  {author} {\bibfnamefont {D.~S.}\ \bibnamefont {Durfee}}, \bibinfo {author}
  {\bibfnamefont {D.~M.}\ \bibnamefont {Kurn}}, \ and\ \bibinfo {author}
  {\bibfnamefont {W.}~\bibnamefont {Ketterle}}} (\bibinfo {year} {1997}),\
  \href@noop {} {\bibfield  {journal} {\bibinfo  {journal} {Science}\ }\textbf
  {\bibinfo {volume} {275}},\ \bibinfo {pages} {637}}\BibitemShut {NoStop}%
\bibitem [{\citenamefont {Andronikashvili}\ and\ \citenamefont
  {Mamaladze}(1966)}]{Andronikashvili:66}%
  \BibitemOpen
  \bibfield  {author} {\bibinfo {author} {\bibnamefont {Andronikashvili},
  \bibfnamefont {E.}}, \ and\ \bibinfo {author} {\bibfnamefont
  {Y.}~\bibnamefont {Mamaladze}}} (\bibinfo {year} {1966}),\ \href@noop {}
  {\bibfield  {journal} {\bibinfo  {journal} {Rev. Mod. Phys.}\ }\textbf
  {\bibinfo {volume} {38}},\ \bibinfo {pages} {567}},\ \bibinfo {note}
  {erratum, Rev. Mod. Phys. {\bf 39}, 494 (1967)}\BibitemShut {NoStop}%
\bibitem [{\citenamefont {Annett}(2003)}]{Annett:03}%
  \BibitemOpen
  \bibfield  {author} {\bibinfo {author} {\bibnamefont {Annett}, \bibfnamefont
  {J.~F.}}} (\bibinfo {year} {2003}),\ \href@noop {} {\emph {\bibinfo {title}
  {Superconductivity, Superfluids and Condensates}}}\ (\bibinfo  {publisher}
  {Oxford University Press})\ \bibinfo {note} {ch. 3}\BibitemShut {NoStop}%
\bibitem [{\citenamefont {Asano}(2001)}]{Asano:01}%
  \BibitemOpen
  \bibfield  {author} {\bibinfo {author} {\bibnamefont {Asano}, \bibfnamefont
  {Y.}}} (\bibinfo {year} {2001}),\ \href@noop {} {\bibfield  {journal}
  {\bibinfo  {journal} {Phys. Rev. B}\ }\textbf {\bibinfo {volume} {64}},\
  \bibinfo {pages} {224515}}\BibitemShut {NoStop}%
\bibitem [{\citenamefont {Ashby}(2004)}]{Ashby:04}%
  \BibitemOpen
  \bibfield  {author} {\bibinfo {author} {\bibnamefont {Ashby}, \bibfnamefont
  {N.}}} (\bibinfo {year} {2004}),\ \enquote {\bibinfo {title} {Relativity in
  rotating frames},}\ Chap.~\bibinfo {chapter} {1}\ (\bibinfo  {publisher}
  {eds. Guido Rizzi and Matteo Ruggiero, Kluwer Academic Publishers, Dordrecht
  - The Netherlands})\BibitemShut {NoStop}%
\bibitem [{\citenamefont {Avenel}\ \emph {et~al.}(1994)\citenamefont {Avenel},
  \citenamefont {Aarts}, \citenamefont {Ihas},\ and\ \citenamefont
  {Varoquaux}}]{Avenel:94}%
  \BibitemOpen
  \bibfield  {author} {\bibinfo {author} {\bibnamefont {Avenel}, \bibfnamefont
  {O.}}, \bibinfo {author} {\bibfnamefont {R.}~\bibnamefont {Aarts}}, \bibinfo
  {author} {\bibfnamefont {G.~G.}\ \bibnamefont {Ihas}}, \ and\ \bibinfo
  {author} {\bibfnamefont {E.}~\bibnamefont {Varoquaux}}} (\bibinfo {year}
  {1994}),\ \href@noop {} {\bibfield  {journal} {\bibinfo  {journal} {Physica
  B}\ }\textbf {\bibinfo {volume} {194-196}},\ \bibinfo {pages}
  {491}}\BibitemShut {NoStop}%
\bibitem [{\citenamefont {Avenel}\ \emph {et~al.}(1995)\citenamefont {Avenel},
  \citenamefont {Bernard}, \citenamefont {Burkhart},\ and\ \citenamefont
  {Varoquaux}}]{Avenel:95}%
  \BibitemOpen
  \bibfield  {author} {\bibinfo {author} {\bibnamefont {Avenel}, \bibfnamefont
  {O.}}, \bibinfo {author} {\bibfnamefont {M.}~\bibnamefont {Bernard}},
  \bibinfo {author} {\bibfnamefont {S.}~\bibnamefont {Burkhart}}, \ and\
  \bibinfo {author} {\bibfnamefont {E.}~\bibnamefont {Varoquaux}}} (\bibinfo
  {year} {1995}),\ \href@noop {} {\bibfield  {journal} {\bibinfo  {journal}
  {Physica B}\ }\textbf {\bibinfo {volume} {210}},\ \bibinfo {pages}
  {215}}\BibitemShut {NoStop}%
\bibitem [{\citenamefont {Avenel}\ \emph {et~al.}(1997)\citenamefont {Avenel},
  \citenamefont {Hakonen},\ and\ \citenamefont {Varoquaux}}]{Avenel:97}%
  \BibitemOpen
  \bibfield  {author} {\bibinfo {author} {\bibnamefont {Avenel}, \bibfnamefont
  {O.}}, \bibinfo {author} {\bibfnamefont {P.}~\bibnamefont {Hakonen}}, \ and\
  \bibinfo {author} {\bibfnamefont {E.}~\bibnamefont {Varoquaux}}} (\bibinfo
  {year} {1997}),\ \href@noop {} {\bibfield  {journal} {\bibinfo  {journal}
  {Phys. Rev. Lett.}\ }\textbf {\bibinfo {volume} {78}},\ \bibinfo {pages}
  {3602}}\BibitemShut {NoStop}%
\bibitem [{\citenamefont {Avenel}\ \emph
  {et~al.}(1998{\natexlab{a}})\citenamefont {Avenel}, \citenamefont {Hakonen},\
  and\ \citenamefont {Varoquaux}}]{Avenel:98}%
  \BibitemOpen
  \bibfield  {author} {\bibinfo {author} {\bibnamefont {Avenel}, \bibfnamefont
  {O.}}, \bibinfo {author} {\bibfnamefont {P.}~\bibnamefont {Hakonen}}, \ and\
  \bibinfo {author} {\bibfnamefont {E.}~\bibnamefont {Varoquaux}}} (\bibinfo
  {year} {1998}{\natexlab{a}}),\ \href@noop {} {\bibfield  {journal} {\bibinfo
  {journal} {J. Low Temp. Phys.}\ }\textbf {\bibinfo {volume} {110}},\ \bibinfo
  {pages} {709}}\BibitemShut {NoStop}%
\bibitem [{\citenamefont {Avenel}\ \emph {et~al.}(1993)\citenamefont {Avenel},
  \citenamefont {Ihas},\ and\ \citenamefont {Varoquaux}}]{Avenel:93}%
  \BibitemOpen
  \bibfield  {author} {\bibinfo {author} {\bibnamefont {Avenel}, \bibfnamefont
  {O.}}, \bibinfo {author} {\bibfnamefont {G.}~\bibnamefont {Ihas}}, \ and\
  \bibinfo {author} {\bibfnamefont {E.}~\bibnamefont {Varoquaux}}} (\bibinfo
  {year} {1993}),\ \href@noop {} {\bibfield  {journal} {\bibinfo  {journal} {J.
  Low Temp. Phys.}\ }\textbf {\bibinfo {volume} {93}},\ \bibinfo {pages}
  {1031}}\BibitemShut {NoStop}%
\bibitem [{\citenamefont {Avenel}\ \emph
  {et~al.}(1998{\natexlab{b}})\citenamefont {Avenel}, \citenamefont
  {Mukharsky},\ and\ \citenamefont {Varoquaux}}]{Avenel:99}%
  \BibitemOpen
  \bibfield  {author} {\bibinfo {author} {\bibnamefont {Avenel}, \bibfnamefont
  {O.}}, \bibinfo {author} {\bibfnamefont {Y.}~\bibnamefont {Mukharsky}}, \
  and\ \bibinfo {author} {\bibfnamefont {E.}~\bibnamefont {Varoquaux}}}
  (\bibinfo {year} {1998}{\natexlab{b}}),\ \href@noop {} {\bibfield  {journal}
  {\bibinfo  {journal} {Nature}\ }\textbf {\bibinfo {volume} {397}},\ \bibinfo
  {pages} {484}}\BibitemShut {NoStop}%
\bibitem [{\citenamefont {Avenel}\ \emph {et~al.}(2000)\citenamefont {Avenel},
  \citenamefont {Mukharsky},\ and\ \citenamefont {Varoquaux}}]{Avenel:00}%
  \BibitemOpen
  \bibfield  {author} {\bibinfo {author} {\bibnamefont {Avenel}, \bibfnamefont
  {O.}}, \bibinfo {author} {\bibfnamefont {Y.}~\bibnamefont {Mukharsky}}, \
  and\ \bibinfo {author} {\bibfnamefont {E.}~\bibnamefont {Varoquaux}}}
  (\bibinfo {year} {2000}),\ \href@noop {} {\bibfield  {journal} {\bibinfo
  {journal} {Physica B}\ }\textbf {\bibinfo {volume} {280}},\ \bibinfo {pages}
  {130}}\BibitemShut {NoStop}%
\bibitem [{\citenamefont {Avenel}\ \emph {et~al.}(2004)\citenamefont {Avenel},
  \citenamefont {Mukharsky},\ and\ \citenamefont {Varoquaux}}]{Avenel:04}%
  \BibitemOpen
  \bibfield  {author} {\bibinfo {author} {\bibnamefont {Avenel}, \bibfnamefont
  {O.}}, \bibinfo {author} {\bibfnamefont {Y.}~\bibnamefont {Mukharsky}}, \
  and\ \bibinfo {author} {\bibfnamefont {E.}~\bibnamefont {Varoquaux}}}
  (\bibinfo {year} {2004}),\ \href@noop {} {\bibfield  {journal} {\bibinfo
  {journal} {J. Low Temp. Phys.}\ }\textbf {\bibinfo {volume} {135}},\ \bibinfo
  {pages} {745}}\BibitemShut {NoStop}%
\bibitem [{\citenamefont {Avenel}\ and\ \citenamefont
  {Varoquaux}(1985)}]{Avenel:85}%
  \BibitemOpen
  \bibfield  {author} {\bibinfo {author} {\bibnamefont {Avenel}, \bibfnamefont
  {O.}}, \ and\ \bibinfo {author} {\bibfnamefont {E.}~\bibnamefont
  {Varoquaux}}} (\bibinfo {year} {1985}),\ \href@noop {} {\bibfield  {journal}
  {\bibinfo  {journal} {Phys. Rev. Lett.}\ }\textbf {\bibinfo {volume} {55}},\
  \bibinfo {pages} {2704}}\BibitemShut {NoStop}%
\bibitem [{\citenamefont {Avenel}\ and\ \citenamefont
  {Varoquaux}(1986{\natexlab{a}})}]{Avenel:86}%
  \BibitemOpen
  \bibfield  {author} {\bibinfo {author} {\bibnamefont {Avenel}, \bibfnamefont
  {O.}}, \ and\ \bibinfo {author} {\bibfnamefont {E.}~\bibnamefont
  {Varoquaux}}} (\bibinfo {year} {1986}{\natexlab{a}}),\ \bibinfo
  {organization} {Proc. ICEC 11, Berlin -- 1986}\ (\bibinfo  {publisher}
  {Butterworths, Guilford})\ p.\ \bibinfo {pages} {587}\BibitemShut {NoStop}%
\bibitem [{\citenamefont {Avenel}\ and\ \citenamefont
  {Varoquaux}(1986{\natexlab{b}})}]{Avenel:86b}%
  \BibitemOpen
  \bibfield  {author} {\bibinfo {author} {\bibnamefont {Avenel}, \bibfnamefont
  {O.}}, \ and\ \bibinfo {author} {\bibfnamefont {E.}~\bibnamefont
  {Varoquaux}}} (\bibinfo {year} {1986}{\natexlab{b}}),\ \href@noop {}
  {\bibfield  {journal} {\bibinfo  {journal} {Phys. Rev. Lett.}\ }\textbf
  {\bibinfo {volume} {57}},\ \bibinfo {pages} {921}}\BibitemShut {NoStop}%
\bibitem [{\citenamefont {Avenel}\ and\ \citenamefont
  {Varoquaux}(1987)}]{Avenel:87}%
  \BibitemOpen
  \bibfield  {author} {\bibinfo {author} {\bibnamefont {Avenel}, \bibfnamefont
  {O.}}, \ and\ \bibinfo {author} {\bibfnamefont {E.}~\bibnamefont
  {Varoquaux}}} (\bibinfo {year} {1987}),\ \href@noop {} {\bibfield  {journal}
  {\bibinfo  {journal} {Jpn. J. Appl. Phys.}\ }\textbf {\bibinfo {volume}
  {26}},\ \bibinfo {pages} {1798}}\BibitemShut {NoStop}%
\bibitem [{\citenamefont {Avenel}\ and\ \citenamefont
  {Varoquaux}(1988)}]{Avenel:88}%
  \BibitemOpen
  \bibfield  {author} {\bibinfo {author} {\bibnamefont {Avenel}, \bibfnamefont
  {O.}}, \ and\ \bibinfo {author} {\bibfnamefont {E.}~\bibnamefont
  {Varoquaux}}} (\bibinfo {year} {1988}),\ \href@noop {} {\bibfield  {journal}
  {\bibinfo  {journal} {Phys. Rev. Lett.}\ }\textbf {\bibinfo {volume} {60}},\
  \bibinfo {pages} {416}}\BibitemShut {NoStop}%
\bibitem [{\citenamefont {Avenel}\ and\ \citenamefont
  {Varoquaux}(1989)}]{Avenel:89}%
  \BibitemOpen
  \bibfield  {author} {\bibinfo {author} {\bibnamefont {Avenel}, \bibfnamefont
  {O.}}, \ and\ \bibinfo {author} {\bibfnamefont {E.}~\bibnamefont
  {Varoquaux}}} (\bibinfo {year} {1989}),\ in\ \href@noop {} {\emph {\bibinfo
  {booktitle} {Quantum Fluids and Solids - 1989}}},\ \bibinfo {editor} {edited
  by\ \bibinfo {editor} {\bibfnamefont {G.}~\bibnamefont {Ihas}}\ and\ \bibinfo
  {editor} {\bibfnamefont {Y.}~\bibnamefont {Takano}}}\ (\bibinfo  {publisher}
  {American Institute of Physics})\ p.~\bibinfo {pages} {3}\BibitemShut
  {NoStop}%
\bibitem [{\citenamefont {Avenel}\ and\ \citenamefont
  {Varoquaux}(1996)}]{Avenel:96}%
  \BibitemOpen
  \bibfield  {author} {\bibinfo {author} {\bibnamefont {Avenel}, \bibfnamefont
  {O.}}, \ and\ \bibinfo {author} {\bibfnamefont {E.}~\bibnamefont
  {Varoquaux}}} (\bibinfo {year} {1996}),\ \href@noop {} {\bibfield  {journal}
  {\bibinfo  {journal} {Czech. J. Phys.}\ }\textbf {\bibinfo {volume}
  {46--S6}},\ \bibinfo {pages} {3319}}\BibitemShut {NoStop}%
\bibitem [{\citenamefont {Awschalom}\ and\ \citenamefont
  {Schwarz}(1984)}]{Awschalom:84}%
  \BibitemOpen
  \bibfield  {author} {\bibinfo {author} {\bibnamefont {Awschalom},
  \bibfnamefont {D.}}, \ and\ \bibinfo {author} {\bibfnamefont {K.~W.}\
  \bibnamefont {Schwarz}}} (\bibinfo {year} {1984}),\ \href@noop {} {\bibfield
  {journal} {\bibinfo  {journal} {Phys. Rev. Lett.}\ }\textbf {\bibinfo
  {volume} {52}},\ \bibinfo {pages} {49}}\BibitemShut {NoStop}%
\bibitem [{\citenamefont {Backhaus}\ \emph {et~al.}(1998)\citenamefont
  {Backhaus}, \citenamefont {Pereverzev}, \citenamefont {Simmonds},
  \citenamefont {Loshak}, \citenamefont {Davis},\ and\ \citenamefont
  {Packard}}]{Backhaus:98}%
  \BibitemOpen
  \bibfield  {author} {\bibinfo {author} {\bibnamefont {Backhaus},
  \bibfnamefont {S.}}, \bibinfo {author} {\bibfnamefont {S.}~\bibnamefont
  {Pereverzev}}, \bibinfo {author} {\bibfnamefont {R.~W.}\ \bibnamefont
  {Simmonds}}, \bibinfo {author} {\bibfnamefont {A.}~\bibnamefont {Loshak}},
  \bibinfo {author} {\bibfnamefont {J.~C.}\ \bibnamefont {Davis}}, \ and\
  \bibinfo {author} {\bibfnamefont {R.~E.}\ \bibnamefont {Packard}}} (\bibinfo
  {year} {1998}),\ \href@noop {} {\bibfield  {journal} {\bibinfo  {journal}
  {Nature}\ }\textbf {\bibinfo {volume} {392}},\ \bibinfo {pages}
  {687}}\BibitemShut {NoStop}%
\bibitem [{\citenamefont {Backhaus}\ \emph {et~al.}(1997)\citenamefont
  {Backhaus}, \citenamefont {Pereverzev}, \citenamefont {Loshak}, \citenamefont
  {Davis},\ and\ \citenamefont {Packard}}]{Backhaus:97}%
  \BibitemOpen
  \bibfield  {author} {\bibinfo {author} {\bibnamefont {Backhaus},
  \bibfnamefont {S.}}, \bibinfo {author} {\bibfnamefont {S.~V.}\ \bibnamefont
  {Pereverzev}}, \bibinfo {author} {\bibfnamefont {A.}~\bibnamefont {Loshak}},
  \bibinfo {author} {\bibfnamefont {J.~C.}\ \bibnamefont {Davis}}, \ and\
  \bibinfo {author} {\bibfnamefont {R.~E.}\ \bibnamefont {Packard}}} (\bibinfo
  {year} {1997}),\ \href@noop {} {\bibfield  {journal} {\bibinfo  {journal}
  {Science}\ }\textbf {\bibinfo {volume} {278}},\ \bibinfo {pages}
  {1435}}\BibitemShut {NoStop}%
\bibitem [{\citenamefont {Barenghi}\ \emph {et~al.}(2001)\citenamefont
  {Barenghi}, \citenamefont {Donnelly},\ and\ \citenamefont
  {Vinen}}]{Barenghi:01}%
  \BibitemOpen
  \bibinfo {editor} {\bibnamefont {Barenghi}, \bibfnamefont {C.}}, \bibinfo
  {editor} {\bibfnamefont {R.}~\bibnamefont {Donnelly}}, \ and\ \bibinfo
  {editor} {\bibfnamefont {W.}~\bibnamefont {Vinen}},\ Eds. (\bibinfo {year}
  {2001}),\ \href@noop {} {\emph {\bibinfo {title} {Quantized Vortex Dynamics
  and Superfluid Turbulence}}}\ (\bibinfo  {publisher} {Springer-Verlag,
  Berlin})\ \bibinfo {note} {{L}ecture Notes in Physics, Vol. 571}\BibitemShut
  {NoStop}%
\bibitem [{\citenamefont {B{\"a}uerle}\ \emph {et~al.}(2000)\citenamefont
  {B{\"a}uerle}, \citenamefont {Fisher},\ and\ \citenamefont
  {Godfrin}}]{Bauerle:00}%
  \BibitemOpen
  \bibfield  {author} {\bibinfo {author} {\bibnamefont {B{\"a}uerle},
  \bibfnamefont {C.}}, \bibinfo {author} {\bibfnamefont {Y.~B.~S.}\
  \bibnamefont {Fisher}}, \ and\ \bibinfo {author} {\bibfnamefont
  {H.}~\bibnamefont {Godfrin}}} (\bibinfo {year} {2000}),\ \enquote {\bibinfo
  {title} {Topological defects and the non-equilibrium dynamics of
  symmetry-breaking phase transitions},}\ \ (\bibinfo  {publisher} {Kluwer
  Academic Publishers})\ p.\ \bibinfo {pages} {105}\BibitemShut {NoStop}%
\bibitem [{\citenamefont {Baym}(1969)}]{Baym:69}%
  \BibitemOpen
  \bibfield  {author} {\bibinfo {author} {\bibnamefont {Baym}, \bibfnamefont
  {G.}}} (\bibinfo {year} {1969}),\ \enquote {\bibinfo {title} {Mathematical
  methods in solid state and superfluid theory},}\ Chap.~\bibinfo {chapter}
  {3}\ (\bibinfo  {publisher} {eds. R.C. Clark and G.H. Derrick, Oliver and
  Boyd Ltd, Edinburgh})\ p.\ \bibinfo {pages} {134}\BibitemShut {NoStop}%
\bibitem [{\citenamefont {Baym}\ and\ \citenamefont
  {Chandler}(1983)}]{Baym:83}%
  \BibitemOpen
  \bibfield  {author} {\bibinfo {author} {\bibnamefont {Baym}, \bibfnamefont
  {G.}}, \ and\ \bibinfo {author} {\bibfnamefont {E.}~\bibnamefont {Chandler}}}
  (\bibinfo {year} {1983}),\ \href@noop {} {\bibfield  {journal} {\bibinfo
  {journal} {J. Low Temp. Phys.}\ }\textbf {\bibinfo {volume} {50}},\ \bibinfo
  {pages} {57}}\BibitemShut {NoStop}%
\bibitem [{\citenamefont {Beecken}\ and\ \citenamefont {{Zimmermann,
  Jr}.}(1987{\natexlab{a}})}]{Beecken:87a}%
  \BibitemOpen
  \bibfield  {author} {\bibinfo {author} {\bibnamefont {Beecken}, \bibfnamefont
  {B.~P.}}, \ and\ \bibinfo {author} {\bibfnamefont {W.}~\bibnamefont
  {{Zimmermann, Jr}.}}} (\bibinfo {year} {1987}{\natexlab{a}}),\ \href@noop {}
  {\bibfield  {journal} {\bibinfo  {journal} {Phys. Rev.}\ }\textbf {\bibinfo
  {volume} {35}},\ \bibinfo {pages} {74}}\BibitemShut {NoStop}%
\bibitem [{\citenamefont {Beecken}\ and\ \citenamefont {{Zimmermann,
  Jr}.}(1987{\natexlab{b}})}]{Beecken:87}%
  \BibitemOpen
  \bibfield  {author} {\bibinfo {author} {\bibnamefont {Beecken}, \bibfnamefont
  {B.~P.}}, \ and\ \bibinfo {author} {\bibfnamefont {W.}~\bibnamefont
  {{Zimmermann, Jr}.}}} (\bibinfo {year} {1987}{\natexlab{b}}),\ \href@noop {}
  {\bibfield  {journal} {\bibinfo  {journal} {Phys. Rev.}\ }\textbf {\bibinfo
  {volume} {35}},\ \bibinfo {pages} {1630}}\BibitemShut {NoStop}%
\bibitem [{\citenamefont {Beliaev}(1958)}]{Beliaev:58}%
  \BibitemOpen
  \bibfield  {author} {\bibinfo {author} {\bibnamefont {Beliaev}, \bibfnamefont
  {S.}}} (\bibinfo {year} {1958}),\ \href@noop {} {\bibfield  {journal}
  {\bibinfo  {journal} {J. Exptl. Theoret. Phys.}\ }\textbf {\bibinfo {volume}
  {34}},\ \bibinfo {pages} {417}},\ \bibinfo {note} {{S}oviet Physics JETP {\bf
  7}, 289 (1958), reprinted in \citet{Pines:62}, p.313}\BibitemShut {NoStop}%
\bibitem [{\citenamefont {Berloff}\ and\ \citenamefont
  {Roberts}(2001)}]{Berloff:01}%
  \BibitemOpen
  \bibfield  {author} {\bibinfo {author} {\bibnamefont {Berloff}, \bibfnamefont
  {N.~G.}}, \ and\ \bibinfo {author} {\bibfnamefont {P.~H.}\ \bibnamefont
  {Roberts}}} (\bibinfo {year} {2001}),\ \bibinfo {note} {{\it loc.cit.}
  \citet{Barenghi:01}, p. 268,}\BibitemShut {NoStop}%
\bibitem [{\citenamefont {Bogolyubov}(1947)}]{Bogolyubov:47}%
  \BibitemOpen
  \bibfield  {author} {\bibinfo {author} {\bibnamefont {Bogolyubov},
  \bibfnamefont {N.~N.}}} (\bibinfo {year} {1947}),\ \href@noop {} {\bibfield
  {journal} {\bibinfo  {journal} {J. Phys. U.S.S.R.}\ }\textbf {\bibinfo
  {volume} {11}},\ \bibinfo {pages} {23}},\ \bibinfo {note} {reprinted in
  \citet{Pines:62} p. 292}\BibitemShut {NoStop}%
\bibitem [{\citenamefont {Bonaldi}\ \emph {et~al.}(1990)\citenamefont
  {Bonaldi}, \citenamefont {Vitale},\ and\ \citenamefont
  {Cerdonio}}]{Bonaldi:90}%
  \BibitemOpen
  \bibfield  {author} {\bibinfo {author} {\bibnamefont {Bonaldi}, \bibfnamefont
  {M.}}, \bibinfo {author} {\bibfnamefont {S.}~\bibnamefont {Vitale}}, \ and\
  \bibinfo {author} {\bibfnamefont {M.}~\bibnamefont {Cerdonio}}} (\bibinfo
  {year} {1990}),\ \href@noop {} {\bibfield  {journal} {\bibinfo  {journal}
  {Phys. Rev. B}\ }\textbf {\bibinfo {volume} {42}},\ \bibinfo {pages}
  {9865}}\BibitemShut {NoStop}%
\bibitem [{\citenamefont {Bowley}\ \emph {et~al.}(1992)\citenamefont {Bowley},
  \citenamefont {Kirk},\ and\ \citenamefont {King}}]{Bowley:92}%
  \BibitemOpen
  \bibfield  {author} {\bibinfo {author} {\bibnamefont {Bowley}, \bibfnamefont
  {R.}}, \bibinfo {author} {\bibfnamefont {A.}~\bibnamefont {Kirk}}, \ and\
  \bibinfo {author} {\bibfnamefont {P.}~\bibnamefont {King}}} (\bibinfo {year}
  {1992}),\ \href@noop {} {\bibfield  {journal} {\bibinfo  {journal} {J. Low
  Temp. Phys.}\ }\textbf {\bibinfo {volume} {88}},\ \bibinfo {pages}
  {73}}\BibitemShut {NoStop}%
\bibitem [{\citenamefont {Bunkov}(2010)}]{Bunkov:10}%
  \BibitemOpen
  \bibfield  {author} {\bibinfo {author} {\bibnamefont {Bunkov}, \bibfnamefont
  {Y.~M.}}} (\bibinfo {year} {2010}),\ \href@noop {} {\bibfield  {journal}
  {\bibinfo  {journal} {J. Low Temp. Phys}\ }\textbf {\bibinfo {volume}
  {158}},\ \bibinfo {pages} {118}}\BibitemShut {NoStop}%
\bibitem [{\citenamefont {Burkhart}(1995)}]{Burkhart:95}%
  \BibitemOpen
  \bibfield  {author} {\bibinfo {author} {\bibnamefont {Burkhart},
  \bibfnamefont {S.}}} (\bibinfo {year} {1995}),\ \href@noop {} {Ph.D. thesis}\
  (\bibinfo  {school} {Universit\'e Paris-Sud})\BibitemShut {NoStop}%
\bibitem [{\citenamefont {Burkhart}\ \emph {et~al.}(1994)\citenamefont
  {Burkhart}, \citenamefont {Bernard}, \citenamefont {Avenel},\ and\
  \citenamefont {Varoquaux}}]{Burkhart:94}%
  \BibitemOpen
  \bibfield  {author} {\bibinfo {author} {\bibnamefont {Burkhart},
  \bibfnamefont {S.}}, \bibinfo {author} {\bibfnamefont {M.}~\bibnamefont
  {Bernard}}, \bibinfo {author} {\bibfnamefont {O.}~\bibnamefont {Avenel}}, \
  and\ \bibinfo {author} {\bibfnamefont {E.}~\bibnamefont {Varoquaux}}}
  (\bibinfo {year} {1994}),\ \href@noop {} {\bibfield  {journal} {\bibinfo
  {journal} {Phys. Rev. Lett.}\ }\textbf {\bibinfo {volume} {72}},\ \bibinfo
  {pages} {380}}\BibitemShut {NoStop}%
\bibitem [{\citenamefont {Caldeira}\ and\ \citenamefont
  {Leggett}(1983)}]{Caldeira:83a}%
  \BibitemOpen
  \bibfield  {author} {\bibinfo {author} {\bibnamefont {Caldeira},
  \bibfnamefont {A.}}, \ and\ \bibinfo {author} {\bibfnamefont
  {A.}~\bibnamefont {Leggett}}} (\bibinfo {year} {1983}),\ \href@noop {}
  {\bibfield  {journal} {\bibinfo  {journal} {Ann. Phys. (N.Y.)}\ }\textbf
  {\bibinfo {volume} {149}},\ \bibinfo {pages} {374}},\ \bibinfo {note}
  {erratum, {153} (1984) 445(E)}\BibitemShut {NoStop}%
\bibitem [{\citenamefont {Carey}\ \emph {et~al.}(1973)\citenamefont {Carey},
  \citenamefont {Chandrasekhar},\ and\ \citenamefont {Dahm}}]{Carey:73}%
  \BibitemOpen
  \bibfield  {author} {\bibinfo {author} {\bibnamefont {Carey}, \bibfnamefont
  {R.}}, \bibinfo {author} {\bibfnamefont {B.~S.}\ \bibnamefont
  {Chandrasekhar}}, \ and\ \bibinfo {author} {\bibfnamefont {A.~J.}\
  \bibnamefont {Dahm}}} (\bibinfo {year} {1973}),\ \href@noop {} {\bibfield
  {journal} {\bibinfo  {journal} {Phys. Rev. Lett.}\ }\textbf {\bibinfo
  {volume} {31}},\ \bibinfo {pages} {873}}\BibitemShut {NoStop}%
\bibitem [{\citenamefont {Carter}\ and\ \citenamefont
  {Khalatnikov}(1992)}]{Carter:92}%
  \BibitemOpen
  \bibfield  {author} {\bibinfo {author} {\bibnamefont {Carter}, \bibfnamefont
  {B.}}, \ and\ \bibinfo {author} {\bibfnamefont {I.}~\bibnamefont
  {Khalatnikov}}} (\bibinfo {year} {1992}),\ \href@noop {} {\bibfield
  {journal} {\bibinfo  {journal} {Phys. Rev. D}\ }\textbf {\bibinfo {volume}
  {45}},\ \bibinfo {pages} {4535}},\ \bibinfo {note} {also V.V. Lebedev and
  I.M. Khalatnikov, Zh. Eksp. Teor. Fiz., {\bf 56}, 1601 (1982) [Sov. Phys.
  JETP {\bf 56}, 923 (1982)}\BibitemShut {NoStop}%
\bibitem [{\citenamefont {Castin}\ and\ \citenamefont
  {Dalibard}(1997)}]{Castin:97}%
  \BibitemOpen
  \bibfield  {author} {\bibinfo {author} {\bibnamefont {Castin}, \bibfnamefont
  {Y.}}, \ and\ \bibinfo {author} {\bibfnamefont {J.}~\bibnamefont {Dalibard}}}
  (\bibinfo {year} {1997}),\ \href@noop {} {\bibfield  {journal} {\bibinfo
  {journal} {Phys. Rev. A}\ }\textbf {\bibinfo {volume} {55}},\ \bibinfo
  {pages} {4330}}\BibitemShut {NoStop}%
\bibitem [{\citenamefont {Cerdonio}\ and\ \citenamefont
  {Vitale}(1984)}]{Cerdonio:84}%
  \BibitemOpen
  \bibfield  {author} {\bibinfo {author} {\bibnamefont {Cerdonio},
  \bibfnamefont {M.}}, \ and\ \bibinfo {author} {\bibfnamefont
  {S.}~\bibnamefont {Vitale}}} (\bibinfo {year} {1984}),\ \href@noop {}
  {\bibfield  {journal} {\bibinfo  {journal} {Phys. Rev. B}\ }\textbf {\bibinfo
  {volume} {29}},\ \bibinfo {pages} {481}}\BibitemShut {NoStop}%
\bibitem [{\citenamefont {Chui}\ \emph {et~al.}(2003)\citenamefont {Chui},
  \citenamefont {Holmes},\ and\ \citenamefont {Penanen}}]{Chui:03}%
  \BibitemOpen
  \bibfield  {author} {\bibinfo {author} {\bibnamefont {Chui}, \bibfnamefont
  {T.}}, \bibinfo {author} {\bibfnamefont {W.}~\bibnamefont {Holmes}}, \ and\
  \bibinfo {author} {\bibfnamefont {K.}~\bibnamefont {Penanen}}} (\bibinfo
  {year} {2003}),\ \href@noop {} {\bibfield  {journal} {\bibinfo  {journal}
  {Phys. Rev. Lett.}\ }\textbf {\bibinfo {volume} {90}},\ \bibinfo {pages}
  {085301}}\BibitemShut {NoStop}%
\bibitem [{\citenamefont {Chui}\ and\ \citenamefont {Penanen}(2005)}]{Chui:05}%
  \BibitemOpen
  \bibfield  {author} {\bibinfo {author} {\bibnamefont {Chui}, \bibfnamefont
  {T.}}, \ and\ \bibinfo {author} {\bibfnamefont {K.}~\bibnamefont {Penanen}}}
  (\bibinfo {year} {2005}),\ \href@noop {} {\bibfield  {journal} {\bibinfo
  {journal} {Phys. Rev. B}\ }\textbf {\bibinfo {volume} {71}},\ \bibinfo
  {pages} {132509}}\BibitemShut {NoStop}%
\bibitem [{\citenamefont {Cohen-Tannoudji}\ and\ \citenamefont
  {Robilliard}(2001)}]{Cohen-Tannoudji:01}%
  \BibitemOpen
  \bibfield  {author} {\bibinfo {author} {\bibnamefont {Cohen-Tannoudji},
  \bibfnamefont {C.}}, \ and\ \bibinfo {author} {\bibfnamefont
  {C.}~\bibnamefont {Robilliard}}} (\bibinfo {year} {2001}),\ \href@noop {}
  {\bibfield  {journal} {\bibinfo  {journal} {C. R. Acad. Sci. Paris}\ }\textbf
  {\bibinfo {volume} {2}}~(\bibinfo {number} {S\'erie IV}),\ \bibinfo {pages}
  {445}}\BibitemShut {NoStop}%
\bibitem [{\citenamefont {Coleman}(1977)}]{Coleman:77}%
  \BibitemOpen
  \bibfield  {author} {\bibinfo {author} {\bibnamefont {Coleman}, \bibfnamefont
  {S.}}} (\bibinfo {year} {1977}),\ \href@noop {} {\bibfield  {journal}
  {\bibinfo  {journal} {Phys. Rev. D}\ }\textbf {\bibinfo {volume} {15}},\
  \bibinfo {pages} {2929}}\BibitemShut {NoStop}%
\bibitem [{\citenamefont {Crooker}(1984)}]{Crooker:84}%
  \BibitemOpen
  \bibfield  {author} {\bibinfo {author} {\bibnamefont {Crooker}, \bibfnamefont
  {B.}}} (\bibinfo {year} {1984}),\ \href@noop {} {Ph.D. thesis}\ (\bibinfo
  {school} {Cornell Univ.})\BibitemShut {NoStop}%
\bibitem [{\citenamefont {Cross}(1974)}]{Cross:74}%
  \BibitemOpen
  \bibfield  {author} {\bibinfo {author} {\bibnamefont {Cross}, \bibfnamefont
  {M.}}} (\bibinfo {year} {1974}),\ \href@noop {} {\bibfield  {journal}
  {\bibinfo  {journal} {Phys. Rev.A}\ }\textbf {\bibinfo {volume} {10}},\
  \bibinfo {pages} {1442}}\BibitemShut {NoStop}%
\bibitem [{\citenamefont {Dalfovo}(1992)}]{Dalfovo:92}%
  \BibitemOpen
  \bibfield  {author} {\bibinfo {author} {\bibnamefont {Dalfovo}, \bibfnamefont
  {F.}}} (\bibinfo {year} {1992}),\ \href@noop {} {\bibfield  {journal}
  {\bibinfo  {journal} {Phys. Rev. B}\ }\textbf {\bibinfo {volume} {46}},\
  \bibinfo {pages} {5482}}\BibitemShut {NoStop}%
\bibitem [{\citenamefont {Dalfovo}\ \emph {et~al.}(1999)\citenamefont
  {Dalfovo}, \citenamefont {Giorgini}, \citenamefont {Pitaevskii},\ and\
  \citenamefont {Stringari}}]{Dalfovo:99}%
  \BibitemOpen
  \bibfield  {author} {\bibinfo {author} {\bibnamefont {Dalfovo}, \bibfnamefont
  {F.}}, \bibinfo {author} {\bibfnamefont {S.}~\bibnamefont {Giorgini}},
  \bibinfo {author} {\bibfnamefont {L.~P.}\ \bibnamefont {Pitaevskii}}, \ and\
  \bibinfo {author} {\bibfnamefont {S.}~\bibnamefont {Stringari}}} (\bibinfo
  {year} {1999}),\ \href@noop {} {\bibfield  {journal} {\bibinfo  {journal}
  {Rev. Mod. Phys.}\ }\textbf {\bibinfo {volume} {71}},\ \bibinfo {pages}
  {463}}\BibitemShut {NoStop}%
\bibitem [{\citenamefont {Davis}\ \emph {et~al.}(1991)\citenamefont {Davis},
  \citenamefont {Close}, \citenamefont {Zieve},\ and\ \citenamefont
  {Packard}}]{Davis:91}%
  \BibitemOpen
  \bibfield  {author} {\bibinfo {author} {\bibnamefont {Davis}, \bibfnamefont
  {J.~C.}}, \bibinfo {author} {\bibfnamefont {J.~D.}\ \bibnamefont {Close}},
  \bibinfo {author} {\bibfnamefont {R.}~\bibnamefont {Zieve}}, \ and\ \bibinfo
  {author} {\bibfnamefont {R.~E.}\ \bibnamefont {Packard}}} (\bibinfo {year}
  {1991}),\ \href@noop {} {\bibfield  {journal} {\bibinfo  {journal} {Phys.
  Rev. Lett.}\ }\textbf {\bibinfo {volume} {66}},\ \bibinfo {pages}
  {329}}\BibitemShut {NoStop}%
\bibitem [{\citenamefont {Davis}\ and\ \citenamefont
  {Packard}(2002)}]{Davis:02}%
  \BibitemOpen
  \bibfield  {author} {\bibinfo {author} {\bibnamefont {Davis}, \bibfnamefont
  {J.~C.}}, \ and\ \bibinfo {author} {\bibfnamefont {R.~E.}\ \bibnamefont
  {Packard}}} (\bibinfo {year} {2002}),\ \href@noop {} {\bibfield  {journal}
  {\bibinfo  {journal} {Rev. Mod. Phys.}\ }\textbf {\bibinfo {volume} {74}},\
  \bibinfo {pages} {741}}\BibitemShut {NoStop}%
\bibitem [{\citenamefont {Deaver}\ and\ \citenamefont
  {Pierce}(1972)}]{Deaver:72}%
  \BibitemOpen
  \bibfield  {author} {\bibinfo {author} {\bibnamefont {Deaver}, \bibfnamefont
  {B.~S.}}, \ and\ \bibinfo {author} {\bibfnamefont {J.~M.}\ \bibnamefont
  {Pierce}}} (\bibinfo {year} {1972}),\ \href@noop {} {\bibfield  {journal}
  {\bibinfo  {journal} {Phys. Lett.}\ }\textbf {\bibinfo {volume} {38\,A}},\
  \bibinfo {pages} {81}}\BibitemShut {NoStop}%
\bibitem [{\citenamefont {Dobbs}(2000)}]{Dobbs:00}%
  \BibitemOpen
  \bibfield  {author} {\bibinfo {author} {\bibnamefont {Dobbs}, \bibfnamefont
  {E.}}} (\bibinfo {year} {2000}),\ \href@noop {} {\emph {\bibinfo {title}
  {Helium Three}}}\ (\bibinfo  {publisher} {Oxford University Press})\ \bibinfo
  {note} {sec. 27.3, and Refs. therein}\BibitemShut {NoStop}%
\bibitem [{\citenamefont {Donev}\ \emph {et~al.}(2001)\citenamefont {Donev},
  \citenamefont {Hough}, ,\ and\ \citenamefont {Zieve}}]{Donev:01}%
  \BibitemOpen
  \bibfield  {author} {\bibinfo {author} {\bibnamefont {Donev}, \bibfnamefont
  {L.~A.~K.}}, \bibinfo {author} {\bibfnamefont {L.}~\bibnamefont {Hough}}, , \
  and\ \bibinfo {author} {\bibfnamefont {R.~J.}\ \bibnamefont {Zieve}}}
  (\bibinfo {year} {2001}),\ \href@noop {} {\bibfield  {journal} {\bibinfo
  {journal} {Phys. Rev. B}\ }\textbf {\bibinfo {volume} {64}},\ \bibinfo
  {pages} {180512}}\BibitemShut {NoStop}%
\bibitem [{\citenamefont {Donnelly}(1991)}]{Donnelly:91}%
  \BibitemOpen
  \bibfield  {author} {\bibinfo {author} {\bibnamefont {Donnelly},
  \bibfnamefont {J.}}} (\bibinfo {year} {1991}),\ \href@noop {} {\emph
  {\bibinfo {title} {Quantized Vortices in Helium}}}\ (\bibinfo  {publisher}
  {C.U.P., Cambridge})\BibitemShut {NoStop}%
\bibitem [{\citenamefont {Duan}(1994)}]{Duan:94}%
  \BibitemOpen
  \bibfield  {author} {\bibinfo {author} {\bibnamefont {Duan}, \bibfnamefont
  {J.-M.}}} (\bibinfo {year} {1994}),\ \href@noop {} {\bibfield  {journal}
  {\bibinfo  {journal} {Phys. Rev.}\ }\textbf {\bibinfo {volume} {B49}},\
  \bibinfo {pages} {12381}}\BibitemShut {NoStop}%
\bibitem [{\citenamefont {Duan}\ and\ \citenamefont {Leggett}(1992)}]{Duan:92}%
  \BibitemOpen
  \bibfield  {author} {\bibinfo {author} {\bibnamefont {Duan}, \bibfnamefont
  {J.-M.}}, \ and\ \bibinfo {author} {\bibfnamefont {A.}~\bibnamefont
  {Leggett}}} (\bibinfo {year} {1992}),\ \href@noop {} {\bibfield  {journal}
  {\bibinfo  {journal} {Phys. Rev. Lett.}\ }\textbf {\bibinfo {volume} {68}},\
  \bibinfo {pages} {1216}}\BibitemShut {NoStop}%
\bibitem [{\citenamefont {Eckart}(1938)}]{Eckart:38}%
  \BibitemOpen
  \bibfield  {author} {\bibinfo {author} {\bibnamefont {Eckart}, \bibfnamefont
  {C.}}} (\bibinfo {year} {1938}),\ \href@noop {} {\bibfield  {journal}
  {\bibinfo  {journal} {Phys. Rev.}\ }\textbf {\bibinfo {volume} {54}},\
  \bibinfo {pages} {920}}\BibitemShut {NoStop}%
\bibitem [{\citenamefont {Ellis}\ and\ \citenamefont {Li}(1993)}]{Ellis:93}%
  \BibitemOpen
  \bibfield  {author} {\bibinfo {author} {\bibnamefont {Ellis}, \bibfnamefont
  {F.~M.}}, \ and\ \bibinfo {author} {\bibfnamefont {L.}~\bibnamefont {Li}}}
  (\bibinfo {year} {1993}),\ \href@noop {} {\bibfield  {journal} {\bibinfo
  {journal} {Phys. Rev. Lett.}\ }\textbf {\bibinfo {volume} {71}},\ \bibinfo
  {pages} {1577}}\BibitemShut {NoStop}%
\bibitem [{\citenamefont {Eltsov}\ \emph {et~al.}(2005)\citenamefont {Eltsov},
  \citenamefont {Krusius},\ and\ \citenamefont {Volovik}}]{Eltsov:05}%
  \BibitemOpen
  \bibfield  {author} {\bibinfo {author} {\bibnamefont {Eltsov}, \bibfnamefont
  {V.}}, \bibinfo {author} {\bibfnamefont {M.}~\bibnamefont {Krusius}}, \ and\
  \bibinfo {author} {\bibfnamefont {G.}~\bibnamefont {Volovik}}} (\bibinfo
  {year} {2005}),\ in\ \href@noop {} {\emph {\bibinfo {booktitle} {Prog. Low
  Temp. Physics}}},\ Vol.~\bibinfo {volume} {XV},\ \bibinfo {editor} {edited
  by\ \bibinfo {editor} {\bibfnamefont {W.}~\bibnamefont {Halperin}}},\
  Chap.~\bibinfo {chapter} {1}\ (\bibinfo  {publisher} {Elsevier})\ p.~\bibinfo
  {pages} {1}\BibitemShut {NoStop}%
\bibitem [{\citenamefont {Eltsov}\ \emph {et~al.}(2000)\citenamefont {Eltsov},
  \citenamefont {Kibble}, \citenamefont {Krusius}, \citenamefont {Ruutu},\ and\
  \citenamefont {Volovik}}]{Eltsov:00}%
  \BibitemOpen
  \bibfield  {author} {\bibinfo {author} {\bibnamefont {Eltsov}, \bibfnamefont
  {V.~B.}}, \bibinfo {author} {\bibfnamefont {T.~W.~B.}\ \bibnamefont
  {Kibble}}, \bibinfo {author} {\bibfnamefont {M.}~\bibnamefont {Krusius}},
  \bibinfo {author} {\bibfnamefont {V.~M.~H.}\ \bibnamefont {Ruutu}}, \ and\
  \bibinfo {author} {\bibfnamefont {G.~E.}\ \bibnamefont {Volovik}}} (\bibinfo
  {year} {2000}),\ \href@noop {} {\bibfield  {journal} {\bibinfo  {journal}
  {Phys. Rev. Lett.}\ }\textbf {\bibinfo {volume} {85}},\ \bibinfo {pages}
  {4739}}\BibitemShut {NoStop}%
\bibitem [{\citenamefont {Enz}(1974)}]{Enz:74}%
  \BibitemOpen
  \bibfield  {author} {\bibinfo {author} {\bibnamefont {Enz}, \bibfnamefont
  {C.~P.}}} (\bibinfo {year} {1974}),\ \href@noop {} {\bibfield  {journal}
  {\bibinfo  {journal} {Rev. Mod. Phys.}\ }\textbf {\bibinfo {volume} {46}},\
  \bibinfo {pages} {705}}\BibitemShut {NoStop}%
\bibitem [{\citenamefont {Eska}\ \emph {et~al.}(2010)\citenamefont {Eska},
  \citenamefont {Gladchenko},\ and\ \citenamefont {Pereverzev}}]{Eska:10}%
  \BibitemOpen
  \bibfield  {author} {\bibinfo {author} {\bibnamefont {Eska}, \bibfnamefont
  {G.}}, \bibinfo {author} {\bibfnamefont {S.}~\bibnamefont {Gladchenko}}, \
  and\ \bibinfo {author} {\bibfnamefont {S.~V.}\ \bibnamefont {Pereverzev}}}
  (\bibinfo {year} {2010}),\ \href@noop {} {\bibfield  {journal} {\bibinfo
  {journal} {Europhys. Let.t.}\ }\textbf {\bibinfo {volume} {89}},\ \bibinfo
  {pages} {26009}}\BibitemShut {NoStop}%
\bibitem [{\citenamefont {Ess{\`e}n}\ and\ \citenamefont
  {Fiolhais}(2012)}]{Essen:12}%
  \BibitemOpen
  \bibfield  {author} {\bibinfo {author} {\bibnamefont {Ess{\`e}n},
  \bibfnamefont {H.}}, \ and\ \bibinfo {author} {\bibfnamefont {M.~C.~N.}\
  \bibnamefont {Fiolhais}}} (\bibinfo {year} {2012}),\ \href@noop {} {\bibfield
   {journal} {\bibinfo  {journal} {Am. J. Phys.}\ }\textbf {\bibinfo {volume}
  {80}},\ \bibinfo {pages} {164}}\BibitemShut {NoStop}%
\bibitem [{\citenamefont {Fetter}(1965)}]{Fetter:65}%
  \BibitemOpen
  \bibfield  {author} {\bibinfo {author} {\bibnamefont {Fetter}, \bibfnamefont
  {A.}}} (\bibinfo {year} {1965}),\ \href@noop {} {\bibfield  {journal}
  {\bibinfo  {journal} {Phys. Rev.}\ }\textbf {\bibinfo {volume} {138A}},\
  \bibinfo {pages} {429}}\BibitemShut {NoStop}%
\bibitem [{\citenamefont {Fetter}(1972)}]{Fetter:72}%
  \BibitemOpen
  \bibfield  {author} {\bibinfo {author} {\bibnamefont {Fetter}, \bibfnamefont
  {A.}}} (\bibinfo {year} {1972}),\ \href@noop {} {\bibfield  {journal}
  {\bibinfo  {journal} {Phys. Rev. A}\ }\textbf {\bibinfo {volume} {6}},\
  \bibinfo {pages} {402}}\BibitemShut {NoStop}%
\bibitem [{\citenamefont {Fetter}\ and\ \citenamefont
  {Ullah}(1988)}]{Fetter:88}%
  \BibitemOpen
  \bibfield  {author} {\bibinfo {author} {\bibnamefont {Fetter}, \bibfnamefont
  {A.}}, \ and\ \bibinfo {author} {\bibfnamefont {S.}~\bibnamefont {Ullah}}}
  (\bibinfo {year} {1988}),\ \href@noop {} {\bibfield  {journal} {\bibinfo
  {journal} {J. Low Temp. Phys.}\ }\textbf {\bibinfo {volume} {70}},\ \bibinfo
  {pages} {515}}\BibitemShut {NoStop}%
\bibitem [{\citenamefont {Fetter}(1976)}]{Fetter:76}%
  \BibitemOpen
  \bibfield  {author} {\bibinfo {author} {\bibnamefont {Fetter}, \bibfnamefont
  {A.~L.}}} (\bibinfo {year} {1976}),\ in\ \href@noop {} {\emph {\bibinfo
  {booktitle} {The Physics of Liquid and Solid Helium - Part 1}}},\
  Chap.~\bibinfo {chapter} {3}\ (\bibinfo  {publisher} {eds. K.H. Benneman, J.
  B. Ketterson, Wiley-Interscience, N.Y.})\ p.\ \bibinfo {pages}
  {238}\BibitemShut {NoStop}%
\bibitem [{\citenamefont {Feynman}(1955)}]{Feynman:55}%
  \BibitemOpen
  \bibfield  {author} {\bibinfo {author} {\bibnamefont {Feynman}, \bibfnamefont
  {R.}}} (\bibinfo {year} {1955}),\ in\ \href@noop {} {\emph {\bibinfo
  {booktitle} {Prog. Low Temp. Phys.}}},\ Vol.~\bibinfo {volume} {I},\
  Chap.~\bibinfo {chapter} {II}\ (\bibinfo  {publisher} {ed. C. J. Gorter,
  North-Holland - Amsterdam})\ p.~\bibinfo {pages} {17},\ \bibinfo {note} {see
  also R. P. Feynman, Phys. Rev. {\bf 91}, 1291, 1301 (1953) and {\bf 94}, 262
  (1964)}\BibitemShut {NoStop}%
\bibitem [{\citenamefont {Feynman}(1972)}]{Feynman:72}%
  \BibitemOpen
  \bibfield  {author} {\bibinfo {author} {\bibnamefont {Feynman}, \bibfnamefont
  {R.}}} (\bibinfo {year} {1972}),\ \href@noop {} {\emph {\bibinfo {title}
  {Statistical Mechanics}}}\ (\bibinfo  {publisher} {Benjamin/Cummings, Reading
  MS})\ \bibinfo {note} {ch.\,11}\BibitemShut {NoStop}%
\bibitem [{\citenamefont {Feynman}\ \emph {et~al.}(1965)\citenamefont
  {Feynman}, \citenamefont {Leighton},\ and\ \citenamefont
  {Sands}}]{Feynman:65}%
  \BibitemOpen
  \bibfield  {author} {\bibinfo {author} {\bibnamefont {Feynman}, \bibfnamefont
  {R.}}, \bibinfo {author} {\bibfnamefont {R.}~\bibnamefont {Leighton}}, \ and\
  \bibinfo {author} {\bibfnamefont {M.}~\bibnamefont {Sands}}} (\bibinfo {year}
  {1965}),\ \href@noop {} {\emph {\bibinfo {title} {The Feynman Lectures on
  Physics: Quantum Mechanics}}}\ (\bibinfo  {publisher} {Addison-Wesley})\
  \bibinfo {note} {vol. III, \S 4.3}\BibitemShut {NoStop}%
\bibitem [{\citenamefont {Finne}\ \emph {et~al.}(2006)\citenamefont {Finne},
  \citenamefont {Eltsov}, \citenamefont {H{\"a}nninen}, \citenamefont {Kopu},
  \citenamefont {Krusius}, \citenamefont {Tsubota},\ and\ \citenamefont
  {Volovik}}]{Finne:06}%
  \BibitemOpen
  \bibfield  {author} {\bibinfo {author} {\bibnamefont {Finne}, \bibfnamefont
  {A.~P.}}, \bibinfo {author} {\bibfnamefont {V.~B.}\ \bibnamefont {Eltsov}},
  \bibinfo {author} {\bibfnamefont {R.}~\bibnamefont {H{\"a}nninen}}, \bibinfo
  {author} {\bibfnamefont {N.~B. K.~J.}\ \bibnamefont {Kopu}}, \bibinfo
  {author} {\bibfnamefont {M.}~\bibnamefont {Krusius}}, \bibinfo {author}
  {\bibfnamefont {M.}~\bibnamefont {Tsubota}}, \ and\ \bibinfo {author}
  {\bibfnamefont {G.~E.}\ \bibnamefont {Volovik}}} (\bibinfo {year} {2006}),\
  \href@noop {} {\bibfield  {journal} {\bibinfo  {journal} {Rep. Prog. Phys.}\
  }\textbf {\bibinfo {volume} {69}},\ \bibinfo {pages} {3157}}\BibitemShut
  {NoStop}%
\bibitem [{\citenamefont {Fischer}(2000)}]{Fischer:00}%
  \BibitemOpen
  \bibfield  {author} {\bibinfo {author} {\bibnamefont {Fischer}, \bibfnamefont
  {U.~R.}}} (\bibinfo {year} {2000}),\ \href@noop {} {\bibfield  {journal}
  {\bibinfo  {journal} {Ann. Phys. (Leipzig)}\ }\textbf {\bibinfo {volume}
  {9}},\ \bibinfo {pages} {523}}\BibitemShut {NoStop}%
\bibitem [{\citenamefont {Flaten}\ \emph
  {et~al.}(2006{\natexlab{a}})\citenamefont {Flaten}, \citenamefont {Borden},
  \citenamefont {Lindensmith},\ and\ \citenamefont {{Zimmermann,
  Jr.}}}]{Flaten:06b}%
  \BibitemOpen
  \bibfield  {author} {\bibinfo {author} {\bibnamefont {Flaten}, \bibfnamefont
  {J.}}, \bibinfo {author} {\bibfnamefont {C.~T.}\ \bibnamefont {Borden}},
  \bibinfo {author} {\bibfnamefont {C.}~\bibnamefont {Lindensmith}}, \ and\
  \bibinfo {author} {\bibfnamefont {W.}~\bibnamefont {{Zimmermann, Jr.}}}}
  (\bibinfo {year} {2006}{\natexlab{a}}),\ \href@noop {} {\bibfield  {journal}
  {\bibinfo  {journal} {J. Low Temp. Phys.}\ }\textbf {\bibinfo {volume}
  {142}},\ \bibinfo {pages} {753}}\BibitemShut {NoStop}%
\bibitem [{\citenamefont {Flaten}\ \emph
  {et~al.}(2006{\natexlab{b}})\citenamefont {Flaten}, \citenamefont
  {Lindensmith},\ and\ \citenamefont {{Zimmermann, Jr.}}}]{Flaten:06a}%
  \BibitemOpen
  \bibfield  {author} {\bibinfo {author} {\bibnamefont {Flaten}, \bibfnamefont
  {J.}}, \bibinfo {author} {\bibfnamefont {C.}~\bibnamefont {Lindensmith}}, \
  and\ \bibinfo {author} {\bibfnamefont {W.}~\bibnamefont {{Zimmermann, Jr.}}}}
  (\bibinfo {year} {2006}{\natexlab{b}}),\ \href@noop {} {\bibfield  {journal}
  {\bibinfo  {journal} {J. Low Temp. Phys.}\ }\textbf {\bibinfo {volume}
  {142}},\ \bibinfo {pages} {725}}\BibitemShut {NoStop}%
\bibitem [{\citenamefont {Frisch}\ \emph {et~al.}(1992)\citenamefont {Frisch},
  \citenamefont {Pomeau},\ and\ \citenamefont {Rica}}]{Frisch:92}%
  \BibitemOpen
  \bibfield  {author} {\bibinfo {author} {\bibnamefont {Frisch}, \bibfnamefont
  {T.}}, \bibinfo {author} {\bibfnamefont {Y.}~\bibnamefont {Pomeau}}, \ and\
  \bibinfo {author} {\bibfnamefont {S.}~\bibnamefont {Rica}}} (\bibinfo {year}
  {1992}),\ \href@noop {} {\bibfield  {journal} {\bibinfo  {journal} {Phys.
  Rev. Lett.}\ }\textbf {\bibinfo {volume} {69}},\ \bibinfo {pages}
  {1644}}\BibitemShut {NoStop}%
\bibitem [{\citenamefont {Gammel}\ \emph {et~al.}(1985)\citenamefont {Gammel},
  \citenamefont {Ho},\ and\ \citenamefont {Reppy}}]{Gammel:85}%
  \BibitemOpen
  \bibfield  {author} {\bibinfo {author} {\bibnamefont {Gammel}, \bibfnamefont
  {P.~L.}}, \bibinfo {author} {\bibfnamefont {T.-L.}\ \bibnamefont {Ho}}, \
  and\ \bibinfo {author} {\bibfnamefont {J.~D.}\ \bibnamefont {Reppy}}}
  (\bibinfo {year} {1985}),\ \href@noop {} {\bibfield  {journal} {\bibinfo
  {journal} {Phys. Rev. Lett.}\ }\textbf {\bibinfo {volume} {55}},\ \bibinfo
  {pages} {2708}}\BibitemShut {NoStop}%
\bibitem [{\citenamefont {Gamota}(1973)}]{Gamota:73}%
  \BibitemOpen
  \bibfield  {author} {\bibinfo {author} {\bibnamefont {Gamota}, \bibfnamefont
  {G.}}} (\bibinfo {year} {1973}),\ \href@noop {} {\bibfield  {journal}
  {\bibinfo  {journal} {Phys. Rev. Lett.}\ }\textbf {\bibinfo {volume} {31}},\
  \bibinfo {pages} {517}}\BibitemShut {NoStop}%
\bibitem [{\citenamefont {Gamota}(1974)}]{Gamota:74}%
  \BibitemOpen
  \bibfield  {author} {\bibinfo {author} {\bibnamefont {Gamota}, \bibfnamefont
  {G.}}} (\bibinfo {year} {1974}),\ \href@noop {} {\bibfield  {journal}
  {\bibinfo  {journal} {Phys. Rev. Lett.}\ }\textbf {\bibinfo {volume} {33}},\
  \bibinfo {pages} {1428}}\BibitemShut {NoStop}%
\bibitem [{\citenamefont {Glaberson}\ and\ \citenamefont
  {Donnelly}(1966)}]{Glaberson:66}%
  \BibitemOpen
  \bibfield  {author} {\bibinfo {author} {\bibnamefont {Glaberson},
  \bibfnamefont {W.}}, \ and\ \bibinfo {author} {\bibfnamefont
  {R.}~\bibnamefont {Donnelly}}} (\bibinfo {year} {1966}),\ \href@noop {}
  {\bibfield  {journal} {\bibinfo  {journal} {Phys. Rev.}\ }\textbf {\bibinfo
  {volume} {141}},\ \bibinfo {pages} {208}}\BibitemShut {NoStop}%
\bibitem [{\citenamefont {Glaberson}\ and\ \citenamefont
  {Donnelly}(1986)}]{Glaberson:86}%
  \BibitemOpen
  \bibfield  {author} {\bibinfo {author} {\bibnamefont {Glaberson},
  \bibfnamefont {W.}}, \ and\ \bibinfo {author} {\bibfnamefont
  {R.}~\bibnamefont {Donnelly}}} (\bibinfo {year} {1986}),\ in\ \href@noop {}
  {\emph {\bibinfo {booktitle} {Prog. Low Temp. Phys.}}},\ Vol.~\bibinfo
  {volume} {IX},\ Chap.~\bibinfo {chapter} {1}\ (\bibinfo  {publisher} {ed. D.
  F. Brewer, Elsevier Science - Amsterdam})\ \bibinfo {note} {see also
  \citet{Donnelly:91}}\BibitemShut {NoStop}%
\bibitem [{\citenamefont {Glyde}(1993)}]{Glyde:93}%
  \BibitemOpen
  \bibfield  {author} {\bibinfo {author} {\bibnamefont {Glyde}, \bibfnamefont
  {H.}}} (\bibinfo {year} {1993}),\ \href@noop {} {\bibfield  {journal}
  {\bibinfo  {journal} {J. Low Temp. Phys.}\ }\textbf {\bibinfo {volume}
  {93}},\ \bibinfo {pages} {861}}\BibitemShut {NoStop}%
\bibitem [{\citenamefont {Glyde}(2013)}]{Glyde:13}%
  \BibitemOpen
  \bibfield  {author} {\bibinfo {author} {\bibnamefont {Glyde}, \bibfnamefont
  {H.}}} (\bibinfo {year} {2013}),\ \href@noop {} {\bibfield  {journal}
  {\bibinfo  {journal} {J. Low Temp. Phys.}\ }\textbf {\bibinfo {volume}
  {172}},\ \bibinfo {pages} {364}}\BibitemShut {NoStop}%
\bibitem [{\citenamefont {Golubov}\ \emph {et~al.}(2004)\citenamefont
  {Golubov}, \citenamefont {Kupriyanov},\ and\ \citenamefont
  {Il'ichev}}]{Golubov:04}%
  \BibitemOpen
  \bibfield  {author} {\bibinfo {author} {\bibnamefont {Golubov}, \bibfnamefont
  {A.~A.}}, \bibinfo {author} {\bibfnamefont {M.~Y.}\ \bibnamefont
  {Kupriyanov}}, \ and\ \bibinfo {author} {\bibfnamefont {E.}~\bibnamefont
  {Il'ichev}}} (\bibinfo {year} {2004}),\ \href@noop {} {\bibfield  {journal}
  {\bibinfo  {journal} {Rev. Mod. Phys.}\ }\textbf {\bibinfo {volume} {76}},\
  \bibinfo {pages} {411}}\BibitemShut {NoStop}%
\bibitem [{\citenamefont {Grabert}(1988)}]{Grabert:88}%
  \BibitemOpen
  \bibfield  {author} {\bibinfo {author} {\bibnamefont {Grabert}, \bibfnamefont
  {H.}}} (\bibinfo {year} {1988}),\ \href@noop {} {\bibfield  {journal}
  {\bibinfo  {journal} {Phys. Rev. Lett.}\ }\textbf {\bibinfo {volume} {61}},\
  \bibinfo {pages} {1683}}\BibitemShut {NoStop}%
\bibitem [{\citenamefont {Grabert}\ \emph {et~al.}(1987)\citenamefont
  {Grabert}, \citenamefont {Olschowski},\ and\ \citenamefont
  {Weiss}}]{Grabert:87}%
  \BibitemOpen
  \bibfield  {author} {\bibinfo {author} {\bibnamefont {Grabert}, \bibfnamefont
  {H.}}, \bibinfo {author} {\bibfnamefont {P.}~\bibnamefont {Olschowski}}, \
  and\ \bibinfo {author} {\bibfnamefont {U.}~\bibnamefont {Weiss}}} (\bibinfo
  {year} {1987}),\ \href@noop {} {\bibfield  {journal} {\bibinfo  {journal}
  {Phys. Rev.}\ }\textbf {\bibinfo {volume} {B36}},\ \bibinfo {pages}
  {1931}}\BibitemShut {NoStop}%
\bibitem [{\citenamefont {Greenberger}(1983)}]{Greenberger:83}%
  \BibitemOpen
  \bibfield  {author} {\bibinfo {author} {\bibnamefont {Greenberger},
  \bibfnamefont {D.}}} (\bibinfo {year} {1983}),\ \href@noop {} {\bibfield
  {journal} {\bibinfo  {journal} {Rev. Mod. Phys.}\ }\textbf {\bibinfo {volume}
  {55}},\ \bibinfo {pages} {875}},\ \bibinfo {note} {in Sec. IX}\BibitemShut
  {NoStop}%
\bibitem [{\citenamefont {Gregory}(1972)}]{Gregory:72}%
  \BibitemOpen
  \bibfield  {author} {\bibinfo {author} {\bibnamefont {Gregory}, \bibfnamefont
  {D.}}} (\bibinfo {year} {1972}),\ \href@noop {} {Ph.D. thesis}\ (\bibinfo
  {school} {University of California - San Diego}),\ \bibinfo {note}
  {unpublished}\BibitemShut {NoStop}%
\bibitem [{\citenamefont {Greiter}(2005)}]{Greiter:05}%
  \BibitemOpen
  \bibfield  {author} {\bibinfo {author} {\bibnamefont {Greiter}, \bibfnamefont
  {M.}}} (\bibinfo {year} {2005}),\ \href@noop {} {\bibfield  {journal}
  {\bibinfo  {journal} {Ann. Phys.}\ }\textbf {\bibinfo {volume} {319}},\
  \bibinfo {pages} {217}}\BibitemShut {NoStop}%
\bibitem [{\citenamefont {Griffin}(1987)}]{Griffin:87}%
  \BibitemOpen
  \bibfield  {author} {\bibinfo {author} {\bibnamefont {Griffin}, \bibfnamefont
  {A.}}} (\bibinfo {year} {1987}),\ \href@noop {} {\bibfield  {journal}
  {\bibinfo  {journal} {Can. J. Phys.}\ }\textbf {\bibinfo {volume} {65}},\
  \bibinfo {pages} {1368}}\BibitemShut {NoStop}%
\bibitem [{\citenamefont {Griffin}(1993)}]{Griffin:93}%
  \BibitemOpen
  \bibfield  {author} {\bibinfo {author} {\bibnamefont {Griffin}, \bibfnamefont
  {A.}}} (\bibinfo {year} {1993}),\ \href@noop {} {\emph {\bibinfo {title}
  {Excitations in a Bose-Condensed Liquid}}}\ (\bibinfo  {publisher} {Cambridge
  University Press})\BibitemShut {NoStop}%
\bibitem [{\citenamefont {Griffin}(1999)}]{Griffin:99}%
  \BibitemOpen
  \bibfield  {author} {\bibinfo {author} {\bibnamefont {Griffin}, \bibfnamefont
  {A.}}} (\bibinfo {year} {1999}),\ \enquote {\bibinfo {title} {Bose-{E}instein
  condensation in atomic gases},}\ \ (\bibinfo  {publisher} {Societ\`a Italiana
  di Fisica})\ p.~\bibinfo {pages} {1},\ \bibinfo {note} {also on
  arXiv:cond-mat/9901123}\BibitemShut {NoStop}%
\bibitem [{\citenamefont {Gross}(1961)}]{Gross:61}%
  \BibitemOpen
  \bibfield  {author} {\bibinfo {author} {\bibnamefont {Gross}, \bibfnamefont
  {E.}}} (\bibinfo {year} {1961}),\ \href@noop {} {\bibfield  {journal}
  {\bibinfo  {journal} {Nuovo Cimento}\ }\textbf {\bibinfo {volume} {20}},\
  \bibinfo {pages} {454}}\BibitemShut {NoStop}%
\bibitem [{\citenamefont {Guernsey}(1971)}]{Guernsey:71}%
  \BibitemOpen
  \bibfield  {author} {\bibinfo {author} {\bibnamefont {Guernsey},
  \bibfnamefont {R.}}} (\bibinfo {year} {1971}),\ in\ \href@noop {} {\emph
  {\bibinfo {booktitle} {Proc. 12th International Conference on Low Temperature
  Physics}}},\ \bibinfo {editor} {edited by\ \bibinfo {editor} {\bibfnamefont
  {E.}~\bibnamefont {Kanda}}}\ (\bibinfo  {publisher} {Keigahn - Tokyo})\
  p.~\bibinfo {pages} {79}\BibitemShut {NoStop}%
\bibitem [{\citenamefont {Hafele}\ and\ \citenamefont
  {Keating}(1972)}]{Hafele:72}%
  \BibitemOpen
  \bibfield  {author} {\bibinfo {author} {\bibnamefont {Hafele}, \bibfnamefont
  {J.~C.}}, \ and\ \bibinfo {author} {\bibfnamefont {R.~E.}\ \bibnamefont
  {Keating}}} (\bibinfo {year} {1972}),\ \href@noop {} {\bibfield  {journal}
  {\bibinfo  {journal} {Science}\ }\textbf {\bibinfo {volume} {177}},\ \bibinfo
  {pages} {168}}\BibitemShut {NoStop}%
\bibitem [{\citenamefont {Hakonen}\ \emph {et~al.}(1998)\citenamefont
  {Hakonen}, \citenamefont {Avenel},\ and\ \citenamefont
  {Varoquaux}}]{Hakonen:98b}%
  \BibitemOpen
  \bibfield  {author} {\bibinfo {author} {\bibnamefont {Hakonen}, \bibfnamefont
  {P.}}, \bibinfo {author} {\bibfnamefont {O.}~\bibnamefont {Avenel}}, \ and\
  \bibinfo {author} {\bibfnamefont {E.}~\bibnamefont {Varoquaux}}} (\bibinfo
  {year} {1998}),\ \href@noop {} {\bibfield  {journal} {\bibinfo  {journal}
  {Phys. Rev. Lett.}\ }\textbf {\bibinfo {volume} {81}},\ \bibinfo {pages}
  {3451}}\BibitemShut {NoStop}%
\bibitem [{\citenamefont {Hakonen}\ \emph {et~al.}(1987)\citenamefont
  {Hakonen}, \citenamefont {Nummila}, \citenamefont {Simola}, \citenamefont
  {Skrbek},\ and\ \citenamefont {Mamniashvili}}]{Hakonen:87}%
  \BibitemOpen
  \bibfield  {author} {\bibinfo {author} {\bibnamefont {Hakonen}, \bibfnamefont
  {P.}}, \bibinfo {author} {\bibfnamefont {K.}~\bibnamefont {Nummila}},
  \bibinfo {author} {\bibfnamefont {J.}~\bibnamefont {Simola}}, \bibinfo
  {author} {\bibfnamefont {L.}~\bibnamefont {Skrbek}}, \ and\ \bibinfo {author}
  {\bibfnamefont {G.}~\bibnamefont {Mamniashvili}}} (\bibinfo {year} {1987}),\
  \href@noop {} {\bibfield  {journal} {\bibinfo  {journal} {Phys. Rev. Lett.}\
  }\textbf {\bibinfo {volume} {58}},\ \bibinfo {pages} {678}}\BibitemShut
  {NoStop}%
\bibitem [{\citenamefont {Hall}(1960)}]{Hall:60}%
  \BibitemOpen
  \bibfield  {author} {\bibinfo {author} {\bibnamefont {Hall}, \bibfnamefont
  {H.}}} (\bibinfo {year} {1960}),\ \href@noop {} {\bibfield  {journal}
  {\bibinfo  {journal} {Adv. Phys.}\ }\textbf {\bibinfo {volume} {9}},\
  \bibinfo {pages} {89}}\BibitemShut {NoStop}%
\bibitem [{\citenamefont {Hall}\ and\ \citenamefont {Hook}(1986)}]{Hall:86}%
  \BibitemOpen
  \bibfield  {author} {\bibinfo {author} {\bibnamefont {Hall}, \bibfnamefont
  {H.}}, \ and\ \bibinfo {author} {\bibfnamefont {J.}~\bibnamefont {Hook}}}
  (\bibinfo {year} {1986}),\ in\ \href@noop {} {\emph {\bibinfo {booktitle}
  {Prog. Low Temp. Physics}}},\ Vol.~\bibinfo {volume} {IX},\ \bibinfo {editor}
  {edited by\ \bibinfo {editor} {\bibfnamefont {D.}~\bibnamefont {Brewer}}}\
  (\bibinfo  {publisher} {North-Holland - Amsterdam})\ p.\ \bibinfo {pages}
  {143},\ \bibinfo {note} {also, M. C Cross, in ``{\it Quantum Fluids and
  Solids -- 1983}'', ed. E. D. Adams, G. G. Ihas American Institute of Physics,
  N. Y., 1983}\BibitemShut {NoStop}%
\bibitem [{\citenamefont {Halperin}\ \emph {et~al.}(1976)\citenamefont
  {Halperin}, \citenamefont {Hohenberg},\ and\ \citenamefont
  {Siggia}}]{Halperin:76}%
  \BibitemOpen
  \bibfield  {author} {\bibinfo {author} {\bibnamefont {Halperin},
  \bibfnamefont {B.~I.}}, \bibinfo {author} {\bibfnamefont {P.~C.}\
  \bibnamefont {Hohenberg}}, \ and\ \bibinfo {author} {\bibfnamefont {E.~D.}\
  \bibnamefont {Siggia}}} (\bibinfo {year} {1976}),\ \href@noop {} {\bibfield
  {journal} {\bibinfo  {journal} {Phys. Rev. B}\ }\textbf {\bibinfo {volume}
  {13}},\ \bibinfo {pages} {1299}}\BibitemShut {NoStop}%
\bibitem [{\citenamefont {H\"anggi}\ \emph {et~al.}(1990)\citenamefont
  {H\"anggi}, \citenamefont {Talkner},\ and\ \citenamefont
  {Borkovec}}]{Hanggi:90}%
  \BibitemOpen
  \bibfield  {author} {\bibinfo {author} {\bibnamefont {H\"anggi},
  \bibfnamefont {P.}}, \bibinfo {author} {\bibfnamefont {P.}~\bibnamefont
  {Talkner}}, \ and\ \bibinfo {author} {\bibfnamefont {M.}~\bibnamefont
  {Borkovec}}} (\bibinfo {year} {1990}),\ \href@noop {} {\bibfield  {journal}
  {\bibinfo  {journal} {Rev. Mod. Phys.}\ }\textbf {\bibinfo {volume} {62}},\
  \bibinfo {pages} {251}}\BibitemShut {NoStop}%
\bibitem [{\citenamefont {Hasselbach}\ and\ \citenamefont
  {Niklaus}(1993)}]{Hasselbach:93}%
  \BibitemOpen
  \bibfield  {author} {\bibinfo {author} {\bibnamefont {Hasselbach},
  \bibfnamefont {F.}}, \ and\ \bibinfo {author} {\bibfnamefont
  {M.}~\bibnamefont {Niklaus}}} (\bibinfo {year} {1993}),\ \href@noop {}
  {\bibfield  {journal} {\bibinfo  {journal} {Phys. Rev. A}\ }\textbf {\bibinfo
  {volume} {48}},\ \bibinfo {pages} {143}}\BibitemShut {NoStop}%
\bibitem [{\citenamefont {Heitler}(1954)}]{Heitler:54}%
  \BibitemOpen
  \bibfield  {author} {\bibinfo {author} {\bibnamefont {Heitler}, \bibfnamefont
  {W.}}} (\bibinfo {year} {1954}),\ \href@noop {} {\emph {\bibinfo {title} {The
  Quantum Theory of Radiation}}}\ (\bibinfo  {publisher} {Oxford University
  Press - Oxford})\ \bibinfo {note} {chap. II,\S 7}\BibitemShut {NoStop}%
\bibitem [{\citenamefont {Hendry}\ \emph {et~al.}(1988)\citenamefont {Hendry},
  \citenamefont {Lawson}, \citenamefont {McClintock}, \citenamefont
  {Williams},\ and\ \citenamefont {Bowley}}]{Hendry:88}%
  \BibitemOpen
  \bibfield  {author} {\bibinfo {author} {\bibnamefont {Hendry}, \bibfnamefont
  {P.}}, \bibinfo {author} {\bibfnamefont {N.}~\bibnamefont {Lawson}}, \bibinfo
  {author} {\bibfnamefont {P.}~\bibnamefont {McClintock}}, \bibinfo {author}
  {\bibfnamefont {C.}~\bibnamefont {Williams}}, \ and\ \bibinfo {author}
  {\bibfnamefont {R.}~\bibnamefont {Bowley}}} (\bibinfo {year} {1988}),\
  \href@noop {} {\bibfield  {journal} {\bibinfo  {journal} {Phys. Rev. Lett.}\
  }\textbf {\bibinfo {volume} {60}},\ \bibinfo {pages} {604}}\BibitemShut
  {NoStop}%
\bibitem [{\citenamefont {Hess}(1977)}]{Hess:77}%
  \BibitemOpen
  \bibfield  {author} {\bibinfo {author} {\bibnamefont {Hess}, \bibfnamefont
  {G.}}} (\bibinfo {year} {1977}),\ \href@noop {} {\bibfield  {journal}
  {\bibinfo  {journal} {Phys. Rev. B}\ }\textbf {\bibinfo {volume} {15}},\
  \bibinfo {pages} {5204}}\BibitemShut {NoStop}%
\bibitem [{\citenamefont {Hess}(1992)}]{Hess:92}%
  \BibitemOpen
  \bibfield  {author} {\bibinfo {author} {\bibnamefont {Hess}, \bibfnamefont
  {G.}}} (\bibinfo {year} {1992}),\ \href@noop {} {\bibfield  {journal}
  {\bibinfo  {journal} {Nature}\ }\textbf {\bibinfo {volume} {359}},\ \bibinfo
  {pages} {192}}\BibitemShut {NoStop}%
\bibitem [{\citenamefont {Ho}(1978)}]{Ho:78}%
  \BibitemOpen
  \bibfield  {author} {\bibinfo {author} {\bibnamefont {Ho}, \bibfnamefont
  {T.-L.}}} (\bibinfo {year} {1978}),\ \href@noop {} {\bibfield  {journal}
  {\bibinfo  {journal} {Phys. Rev. B}\ }\textbf {\bibinfo {volume} {18}},\
  \bibinfo {pages} {1144}}\BibitemShut {NoStop}%
\bibitem [{\citenamefont {Ho}\ and\ \citenamefont
  {Mermin}(1980{\natexlab{a}})}]{Ho:80b}%
  \BibitemOpen
  \bibfield  {author} {\bibinfo {author} {\bibnamefont {Ho}, \bibfnamefont
  {T.-L.}}, \ and\ \bibinfo {author} {\bibfnamefont {N.~D.}\ \bibnamefont
  {Mermin}}} (\bibinfo {year} {1980}{\natexlab{a}}),\ \href@noop {} {\bibfield
  {journal} {\bibinfo  {journal} {Phys. Rev. B}\ }\textbf {\bibinfo {volume}
  {21}},\ \bibinfo {pages} {5190}}\BibitemShut {NoStop}%
\bibitem [{\citenamefont {Ho}\ and\ \citenamefont
  {Mermin}(1980{\natexlab{b}})}]{Ho:80a}%
  \BibitemOpen
  \bibfield  {author} {\bibinfo {author} {\bibnamefont {Ho}, \bibfnamefont
  {T.~L.}}, \ and\ \bibinfo {author} {\bibfnamefont {N.~D.}\ \bibnamefont
  {Mermin}}} (\bibinfo {year} {1980}{\natexlab{b}}),\ \href@noop {} {\bibfield
  {journal} {\bibinfo  {journal} {Phys. Rev. Lett.}\ }\textbf {\bibinfo
  {volume} {44}},\ \bibinfo {pages} {330}}\BibitemShut {NoStop}%
\bibitem [{\citenamefont {Hohenberg}\ and\ \citenamefont
  {Halperin}(1977)}]{Hohenberg:77}%
  \BibitemOpen
  \bibfield  {author} {\bibinfo {author} {\bibnamefont {Hohenberg},
  \bibfnamefont {P.~C.}}, \ and\ \bibinfo {author} {\bibfnamefont {B.~I.}\
  \bibnamefont {Halperin}}} (\bibinfo {year} {1977}),\ \href@noop {} {\bibfield
   {journal} {\bibinfo  {journal} {Rev. Mod. Phys.}\ }\textbf {\bibinfo
  {volume} {49}},\ \bibinfo {pages} {436}}\BibitemShut {NoStop}%
\bibitem [{\citenamefont {Hook}(1987)}]{Hook:87}%
  \BibitemOpen
  \bibfield  {author} {\bibinfo {author} {\bibnamefont {Hook}, \bibfnamefont
  {J.~R.}}} (\bibinfo {year} {1987}),\ \href@noop {} {\bibfield  {journal}
  {\bibinfo  {journal} {Jpn. J. Appl. Phys.}\ }\textbf {\bibinfo {volume}
  {26}},\ \bibinfo {pages} {159}}\BibitemShut {NoStop}%
\bibitem [{\citenamefont {Horak}\ and\ \citenamefont
  {Barnett}(1999)}]{Horak:99}%
  \BibitemOpen
  \bibfield  {author} {\bibinfo {author} {\bibnamefont {Horak}, \bibfnamefont
  {P.}}, \ and\ \bibinfo {author} {\bibfnamefont {S.~M.}\ \bibnamefont
  {Barnett}}} (\bibinfo {year} {1999}),\ \href@noop {} {\bibfield  {journal}
  {\bibinfo  {journal} {J. Phys. B: At. Mol. Opt. Phys.}\ }\textbf {\bibinfo
  {volume} {32}},\ \bibinfo {pages} {3421}}\BibitemShut {NoStop}%
\bibitem [{\citenamefont {Hoskinson}\ \emph
  {et~al.}(2006{\natexlab{a}})\citenamefont {Hoskinson}, \citenamefont {Sato},
  \citenamefont {Hahn},\ and\ \citenamefont {Packard}}]{Hoskinson:06-Nature}%
  \BibitemOpen
  \bibfield  {author} {\bibinfo {author} {\bibnamefont {Hoskinson},
  \bibfnamefont {E.}}, \bibinfo {author} {\bibfnamefont {Y.}~\bibnamefont
  {Sato}}, \bibinfo {author} {\bibfnamefont {I.}~\bibnamefont {Hahn}}, \ and\
  \bibinfo {author} {\bibfnamefont {R.~E.}\ \bibnamefont {Packard}}} (\bibinfo
  {year} {2006}{\natexlab{a}}),\ \href@noop {} {\bibfield  {journal} {\bibinfo
  {journal} {Nature Physics}\ }\textbf {\bibinfo {volume} {2}},\ \bibinfo
  {pages} {23}}\BibitemShut {NoStop}%
\bibitem [{\citenamefont {Hoskinson}\ \emph
  {et~al.}(2006{\natexlab{b}})\citenamefont {Hoskinson}, \citenamefont {Sato},\
  and\ \citenamefont {Packard}}]{Hoskinson:06}%
  \BibitemOpen
  \bibfield  {author} {\bibinfo {author} {\bibnamefont {Hoskinson},
  \bibfnamefont {E.}}, \bibinfo {author} {\bibfnamefont {Y.}~\bibnamefont
  {Sato}}, \ and\ \bibinfo {author} {\bibfnamefont {R.}~\bibnamefont
  {Packard}}} (\bibinfo {year} {2006}{\natexlab{b}}),\ \href@noop {} {\bibfield
   {journal} {\bibinfo  {journal} {Phys. Rev. B}\ }\textbf {\bibinfo {volume}
  {74}},\ \bibinfo {pages} {100509}}\BibitemShut {NoStop}%
\bibitem [{\citenamefont {Huggins}(1970)}]{Huggins:70}%
  \BibitemOpen
  \bibfield  {author} {\bibinfo {author} {\bibnamefont {Huggins}, \bibfnamefont
  {E.~R.}}} (\bibinfo {year} {1970}),\ \href@noop {} {\bibfield  {journal}
  {\bibinfo  {journal} {Phys. Rev. A}\ }\textbf {\bibinfo {volume} {1}},\
  \bibinfo {pages} {332}}\BibitemShut {NoStop}%
\bibitem [{\citenamefont {Hulin}\ \emph {et~al.}(1974)\citenamefont {Hulin},
  \citenamefont {D'Humi\`eres}, \citenamefont {Perrin},\ and\ \citenamefont
  {Libchaber}}]{Hulin:74}%
  \BibitemOpen
  \bibfield  {author} {\bibinfo {author} {\bibnamefont {Hulin}, \bibfnamefont
  {J.}}, \bibinfo {author} {\bibfnamefont {D.}~\bibnamefont {D'Humi\`eres}},
  \bibinfo {author} {\bibfnamefont {B.}~\bibnamefont {Perrin}}, \ and\ \bibinfo
  {author} {\bibfnamefont {A.}~\bibnamefont {Libchaber}}} (\bibinfo {year}
  {1974}),\ \href@noop {} {\bibfield  {journal} {\bibinfo  {journal} {Phys.
  Rev.}\ }\textbf {\bibinfo {volume} {A9}},\ \bibinfo {pages}
  {885}}\BibitemShut {NoStop}%
\bibitem [{\citenamefont {Hulin}\ \emph {et~al.}(1972)\citenamefont {Hulin},
  \citenamefont {Laroche}, \citenamefont {Libchaber},\ and\ \citenamefont
  {Perrin}}]{Hulin:72}%
  \BibitemOpen
  \bibfield  {author} {\bibinfo {author} {\bibnamefont {Hulin}, \bibfnamefont
  {J.}}, \bibinfo {author} {\bibfnamefont {C.}~\bibnamefont {Laroche}},
  \bibinfo {author} {\bibfnamefont {A.}~\bibnamefont {Libchaber}}, \ and\
  \bibinfo {author} {\bibfnamefont {B.}~\bibnamefont {Perrin}}} (\bibinfo
  {year} {1972}),\ \href@noop {} {\bibfield  {journal} {\bibinfo  {journal}
  {Phys. Rev.}\ }\textbf {\bibinfo {volume} {A5}},\ \bibinfo {pages}
  {1830}}\BibitemShut {NoStop}%
\bibitem [{\citenamefont {Hulin}\ \emph {et~al.}(1971)\citenamefont {Hulin},
  \citenamefont {Perrin}, \citenamefont {Laroche},\ and\ \citenamefont
  {Libchaber}}]{Hulin:71}%
  \BibitemOpen
  \bibfield  {author} {\bibinfo {author} {\bibnamefont {Hulin}, \bibfnamefont
  {J.}}, \bibinfo {author} {\bibfnamefont {B.}~\bibnamefont {Perrin}}, \bibinfo
  {author} {\bibfnamefont {C.}~\bibnamefont {Laroche}}, \ and\ \bibinfo
  {author} {\bibfnamefont {A.}~\bibnamefont {Libchaber}}} (\bibinfo {year}
  {1971}),\ in\ \href@noop {} {\emph {\bibinfo {booktitle} {Proc. 12th
  International Conference on Low Temperature Physics}}},\ \bibinfo {editor}
  {edited by\ \bibinfo {editor} {\bibfnamefont {E.}~\bibnamefont {Kanda}}}\
  (\bibinfo  {publisher} {Keigahn - Tokyo})\ p.~\bibinfo {pages}
  {83}\BibitemShut {NoStop}%
\bibitem [{\citenamefont {Idowu}\ \emph
  {et~al.}(2000{\natexlab{a}})\citenamefont {Idowu}, \citenamefont {Kivotides},
  \citenamefont {Barenghi},\ and\ \citenamefont {Samuels}}]{Idowu:00a}%
  \BibitemOpen
  \bibfield  {author} {\bibinfo {author} {\bibnamefont {Idowu}, \bibfnamefont
  {O.~C.}}, \bibinfo {author} {\bibfnamefont {D.}~\bibnamefont {Kivotides}},
  \bibinfo {author} {\bibfnamefont {C.~F.}\ \bibnamefont {Barenghi}}, \ and\
  \bibinfo {author} {\bibfnamefont {D.~C.}\ \bibnamefont {Samuels}}} (\bibinfo
  {year} {2000}{\natexlab{a}}),\ \href@noop {} {\bibfield  {journal} {\bibinfo
  {journal} {J. Low Temp. Phys.}\ }\textbf {\bibinfo {volume} {120}},\ \bibinfo
  {pages} {269}}\BibitemShut {NoStop}%
\bibitem [{\citenamefont {Idowu}\ \emph
  {et~al.}(2000{\natexlab{b}})\citenamefont {Idowu}, \citenamefont {Willis},
  \citenamefont {Barenghi},\ and\ \citenamefont {Samuels}}]{Idowu:00b}%
  \BibitemOpen
  \bibfield  {author} {\bibinfo {author} {\bibnamefont {Idowu}, \bibfnamefont
  {O.~C.}}, \bibinfo {author} {\bibfnamefont {A.}~\bibnamefont {Willis}},
  \bibinfo {author} {\bibfnamefont {C.~F.}\ \bibnamefont {Barenghi}}, \ and\
  \bibinfo {author} {\bibfnamefont {D.~C.}\ \bibnamefont {Samuels}}} (\bibinfo
  {year} {2000}{\natexlab{b}}),\ \href@noop {} {\bibfield  {journal} {\bibinfo
  {journal} {Phys. Rev. B}\ }\textbf {\bibinfo {volume} {62}},\ \bibinfo
  {pages} {3409}}\BibitemShut {NoStop}%
\bibitem [{\citenamefont {Ihas}\ \emph {et~al.}(1992)\citenamefont {Ihas},
  \citenamefont {Avenel}, \citenamefont {Aarts}, \citenamefont {Salmelin},\
  and\ \citenamefont {Varoquaux}}]{Ihas:92}%
  \BibitemOpen
  \bibfield  {author} {\bibinfo {author} {\bibnamefont {Ihas}, \bibfnamefont
  {G.}}, \bibinfo {author} {\bibfnamefont {O.}~\bibnamefont {Avenel}}, \bibinfo
  {author} {\bibfnamefont {R.}~\bibnamefont {Aarts}}, \bibinfo {author}
  {\bibfnamefont {R.}~\bibnamefont {Salmelin}}, \ and\ \bibinfo {author}
  {\bibfnamefont {E.}~\bibnamefont {Varoquaux}}} (\bibinfo {year} {1992}),\
  \href@noop {} {\bibfield  {journal} {\bibinfo  {journal} {Phys. Rev. Lett.}\
  }\textbf {\bibinfo {volume} {69}},\ \bibinfo {pages} {327}}\BibitemShut
  {NoStop}%
\bibitem [{\citenamefont {Javanainen}\ and\ \citenamefont
  {Yoo}(1996)}]{Javanainen:96}%
  \BibitemOpen
  \bibfield  {author} {\bibinfo {author} {\bibnamefont {Javanainen},
  \bibfnamefont {J.}}, \ and\ \bibinfo {author} {\bibfnamefont {S.~M.}\
  \bibnamefont {Yoo}}} (\bibinfo {year} {1996}),\ \href@noop {} {\bibfield
  {journal} {\bibinfo  {journal} {Phys. Rev. Lett.}\ }\textbf {\bibinfo
  {volume} {76}},\ \bibinfo {pages} {161}}\BibitemShut {NoStop}%
\bibitem [{\citenamefont {Josephson}(1962)}]{Josephson:62}%
  \BibitemOpen
  \bibfield  {author} {\bibinfo {author} {\bibnamefont {Josephson},
  \bibfnamefont {B.}}} (\bibinfo {year} {1962}),\ \href@noop {} {\bibfield
  {journal} {\bibinfo  {journal} {Phys. Lett.}\ }\textbf {\bibinfo {volume}
  {1}},\ \bibinfo {pages} {251}}\BibitemShut {NoStop}%
\bibitem [{\citenamefont {Josephson}(1965)}]{Josephson:65}%
  \BibitemOpen
  \bibfield  {author} {\bibinfo {author} {\bibnamefont {Josephson},
  \bibfnamefont {B.}}} (\bibinfo {year} {1965}),\ \href@noop {} {\bibfield
  {journal} {\bibinfo  {journal} {Adv. Phys.}\ }\textbf {\bibinfo {volume}
  {14}},\ \bibinfo {pages} {419}}\BibitemShut {NoStop}%
\bibitem [{\citenamefont {Josephson}(1964)}]{Josephson:64}%
  \BibitemOpen
  \bibfield  {author} {\bibinfo {author} {\bibnamefont {Josephson},
  \bibfnamefont {B.~D.}}} (\bibinfo {year} {1964}),\ \href@noop {} {\bibfield
  {journal} {\bibinfo  {journal} {Rev. Mod. Phys.}\ }\textbf {\bibinfo {volume}
  {36}},\ \bibinfo {pages} {216}}\BibitemShut {NoStop}%
\bibitem [{\citenamefont {Josephson}(1966)}]{Josephson:66}%
  \BibitemOpen
  \bibfield  {author} {\bibinfo {author} {\bibnamefont {Josephson},
  \bibfnamefont {B.~D.}}} (\bibinfo {year} {1966}),\ \href@noop {} {\bibfield
  {journal} {\bibinfo  {journal} {Phys. Lett.}\ }\textbf {\bibinfo {volume}
  {21}},\ \bibinfo {pages} {608}}\BibitemShut {NoStop}%
\bibitem [{\citenamefont {Josserand}\ and\ \citenamefont
  {Pomeau}(1995)}]{Josserand:95}%
  \BibitemOpen
  \bibfield  {author} {\bibinfo {author} {\bibnamefont {Josserand},
  \bibfnamefont {C.}}, \ and\ \bibinfo {author} {\bibfnamefont
  {Y.}~\bibnamefont {Pomeau}}} (\bibinfo {year} {1995}),\ \href@noop {}
  {\bibfield  {journal} {\bibinfo  {journal} {Europhys. Lett.}\ }\textbf
  {\bibinfo {volume} {30}},\ \bibinfo {pages} {43}}\BibitemShut {NoStop}%
\bibitem [{\citenamefont {Josserand}\ \emph {et~al.}(1995)\citenamefont
  {Josserand}, \citenamefont {Pomeau},\ and\ \citenamefont
  {Rica}}]{Josserand:95b}%
  \BibitemOpen
  \bibfield  {author} {\bibinfo {author} {\bibnamefont {Josserand},
  \bibfnamefont {C.}}, \bibinfo {author} {\bibfnamefont {Y.}~\bibnamefont
  {Pomeau}}, \ and\ \bibinfo {author} {\bibfnamefont {S.}~\bibnamefont {Rica}}}
  (\bibinfo {year} {1995}),\ \href@noop {} {\bibfield  {journal} {\bibinfo
  {journal} {Phys. Rev. Lett.}\ }\textbf {\bibinfo {volume} {75}},\ \bibinfo
  {pages} {3150}}\BibitemShut {NoStop}%
\bibitem [{\citenamefont {Kadanoff}(2013)}]{Kadanoff:13}%
  \BibitemOpen
  \bibfield  {author} {\bibinfo {author} {\bibnamefont {Kadanoff},
  \bibfnamefont {L.~P.}}} (\bibinfo {year} {2013}),\ \href@noop {} {\bibfield
  {journal} {\bibinfo  {journal} {J. Stat. Phys.}\ }\textbf {\bibinfo {volume}
  {152}},\ \bibinfo {pages} {805}}\BibitemShut {NoStop}%
\bibitem [{\citenamefont {Karn}\ \emph {et~al.}(1980)\citenamefont {Karn},
  \citenamefont {Starks},\ and\ \citenamefont {W.~Zimmermann}}]{Karn:80}%
  \BibitemOpen
  \bibfield  {author} {\bibinfo {author} {\bibnamefont {Karn}, \bibfnamefont
  {P.~W.}}, \bibinfo {author} {\bibfnamefont {D.~R.}\ \bibnamefont {Starks}}, \
  and\ \bibinfo {author} {\bibfnamefont {J.}~\bibnamefont {W.~Zimmermann}}}
  (\bibinfo {year} {1980}),\ \href@noop {} {\bibfield  {journal} {\bibinfo
  {journal} {Phys. Rev. B}\ }\textbf {\bibinfo {volume} {21}},\ \bibinfo
  {pages} {1797}}\BibitemShut {NoStop}%
\bibitem [{\citenamefont {Khalatnikov}(1965)}]{Khalatnikov:65}%
  \BibitemOpen
  \bibfield  {author} {\bibinfo {author} {\bibnamefont {Khalatnikov},
  \bibfnamefont {I.}}} (\bibinfo {year} {1965}),\ \href@noop {} {\emph
  {\bibinfo {title} {An Introduction to the Theory of Superfluidity}}}\
  (\bibinfo  {publisher} {W.A. Benjamin, N.Y.})\BibitemShut {NoStop}%
\bibitem [{\citenamefont {Khorana}(1969)}]{Khorana:69}%
  \BibitemOpen
  \bibfield  {author} {\bibinfo {author} {\bibnamefont {Khorana}, \bibfnamefont
  {B.~M.}}} (\bibinfo {year} {1969}),\ \href@noop {} {\bibfield  {journal}
  {\bibinfo  {journal} {Phys. Rev.}\ }\textbf {\bibinfo {volume} {185}},\
  \bibinfo {pages} {299}}\BibitemShut {NoStop}%
\bibitem [{\citenamefont {Khorana}\ and\ \citenamefont
  {Chandrasekhar}(1967)}]{Khorana:67}%
  \BibitemOpen
  \bibfield  {author} {\bibinfo {author} {\bibnamefont {Khorana}, \bibfnamefont
  {B.~M.}}, \ and\ \bibinfo {author} {\bibfnamefont {B.~S.}\ \bibnamefont
  {Chandrasekhar}}} (\bibinfo {year} {1967}),\ \href@noop {} {\bibfield
  {journal} {\bibinfo  {journal} {Phys. Rev. Lett.}\ }\textbf {\bibinfo
  {volume} {18}},\ \bibinfo {pages} {230}}\BibitemShut {NoStop}%
\bibitem [{\citenamefont {Kopnin}(1978)}]{Kopnin:78}%
  \BibitemOpen
  \bibfield  {author} {\bibinfo {author} {\bibnamefont {Kopnin}, \bibfnamefont
  {N.}}} (\bibinfo {year} {1978}),\ \href@noop {} {\bibfield  {journal}
  {\bibinfo  {journal} {Pis'ma Zh. Eksp. Teor. Fiz..}\ }\textbf {\bibinfo
  {volume} {27}},\ \bibinfo {pages} {417}},\ \bibinfo {note} {{J}ETP Lett.,
  {\bf 27}, 390 (1978)}\BibitemShut {NoStop}%
\bibitem [{\citenamefont {Kopnin}(1986)}]{Kopnin:86}%
  \BibitemOpen
  \bibfield  {author} {\bibinfo {author} {\bibnamefont {Kopnin}, \bibfnamefont
  {N.}}} (\bibinfo {year} {1986}),\ \href@noop {} {\bibfield  {journal}
  {\bibinfo  {journal} {Pis'ma Zh. Eksp. Teor. Fiz..}\ }\textbf {\bibinfo
  {volume} {43}},\ \bibinfo {pages} {541}},\ \bibinfo {note} {{J}ETP Lett.,
  {\bf 43}, 700 (1986)}\BibitemShut {NoStop}%
\bibitem [{\citenamefont {Kopnin}(1995)}]{Kopnin:95}%
  \BibitemOpen
  \bibfield  {author} {\bibinfo {author} {\bibnamefont {Kopnin}, \bibfnamefont
  {N.}}} (\bibinfo {year} {1995}),\ \href@noop {} {\bibfield  {journal}
  {\bibinfo  {journal} {Physica B}\ }\textbf {\bibinfo {volume} {210}},\
  \bibinfo {pages} {267}}\BibitemShut {NoStop}%
\bibitem [{\citenamefont {Kopnin}\ and\ \citenamefont
  {Salomaa}(1990)}]{Kopnin:90}%
  \BibitemOpen
  \bibfield  {author} {\bibinfo {author} {\bibnamefont {Kopnin}, \bibfnamefont
  {N.}}, \ and\ \bibinfo {author} {\bibfnamefont {M.}~\bibnamefont {Salomaa}}}
  (\bibinfo {year} {1990}),\ \href@noop {} {\bibfield  {journal} {\bibinfo
  {journal} {Phys. Rev.}\ }\textbf {\bibinfo {volume} {B41}},\ \bibinfo {pages}
  {2601}}\BibitemShut {NoStop}%
\bibitem [{\citenamefont {Kopnin}\ \emph {et~al.}(1991)\citenamefont {Kopnin},
  \citenamefont {Soininen},\ and\ \citenamefont {Salomaa}}]{Kopnin:91}%
  \BibitemOpen
  \bibfield  {author} {\bibinfo {author} {\bibnamefont {Kopnin}, \bibfnamefont
  {N.}}, \bibinfo {author} {\bibfnamefont {P.}~\bibnamefont {Soininen}}, \ and\
  \bibinfo {author} {\bibfnamefont {M.}~\bibnamefont {Salomaa}}} (\bibinfo
  {year} {1991}),\ \href@noop {} {\bibfield  {journal} {\bibinfo  {journal}
  {Physica B}\ }\textbf {\bibinfo {volume} {169}},\ \bibinfo {pages}
  {535}}\BibitemShut {NoStop}%
\bibitem [{\citenamefont {Kopnin}\ \emph {et~al.}(1992)\citenamefont {Kopnin},
  \citenamefont {Soininen},\ and\ \citenamefont {Salomaa}}]{Kopnin:92}%
  \BibitemOpen
  \bibfield  {author} {\bibinfo {author} {\bibnamefont {Kopnin}, \bibfnamefont
  {N.}}, \bibinfo {author} {\bibfnamefont {P.}~\bibnamefont {Soininen}}, \ and\
  \bibinfo {author} {\bibfnamefont {M.}~\bibnamefont {Salomaa}}} (\bibinfo
  {year} {1992}),\ \href@noop {} {\bibfield  {journal} {\bibinfo  {journal}
  {Phys. Rev. B}\ }\textbf {\bibinfo {volume} {45}},\ \bibinfo {pages}
  {5491}}\BibitemShut {NoStop}%
\bibitem [{\citenamefont {Kramers}(1940)}]{Kramers:40}%
  \BibitemOpen
  \bibfield  {author} {\bibinfo {author} {\bibnamefont {Kramers}, \bibfnamefont
  {H.}}} (\bibinfo {year} {1940}),\ \href@noop {} {\bibfield  {journal}
  {\bibinfo  {journal} {Physica}\ }\textbf {\bibinfo {volume} {7}},\ \bibinfo
  {pages} {284}}\BibitemShut {NoStop}%
\bibitem [{\citenamefont {Krusius}\ \emph {et~al.}(1993)\citenamefont
  {Krusius}, \citenamefont {Korhonen}, \citenamefont {Kondo},\ and\
  \citenamefont {Sonin}}]{Krusius:93}%
  \BibitemOpen
  \bibfield  {author} {\bibinfo {author} {\bibnamefont {Krusius}, \bibfnamefont
  {M.}}, \bibinfo {author} {\bibfnamefont {J.}~\bibnamefont {Korhonen}},
  \bibinfo {author} {\bibfnamefont {Y.}~\bibnamefont {Kondo}}, \ and\ \bibinfo
  {author} {\bibfnamefont {E.}~\bibnamefont {Sonin}}} (\bibinfo {year}
  {1993}),\ \href@noop {} {\bibfield  {journal} {\bibinfo  {journal} {Phys.
  Rev.}\ }\textbf {\bibinfo {volume} {B47}},\ \bibinfo {pages}
  {15113}}\BibitemShut {NoStop}%
\bibitem [{\citenamefont {Kurkij\"arvi}(1988)}]{Kurkijarvi:88}%
  \BibitemOpen
  \bibfield  {author} {\bibinfo {author} {\bibnamefont {Kurkij\"arvi},
  \bibfnamefont {J.}}} (\bibinfo {year} {1988}),\ \href@noop {} {\bibfield
  {journal} {\bibinfo  {journal} {Phys. Rev.}\ }\textbf {\bibinfo {volume}
  {38}},\ \bibinfo {pages} {11184}}\BibitemShut {NoStop}%
\bibitem [{\citenamefont {Lamb}(1945)}]{Lamb:45}%
  \BibitemOpen
  \bibfield  {author} {\bibinfo {author} {\bibnamefont {Lamb}, \bibfnamefont
  {S.~H.}}} (\bibinfo {year} {1945}),\ \href@noop {} {\emph {\bibinfo {title}
  {Hydrodynamics}}}\ (\bibinfo  {publisher} {Cambridge University
  Press})\BibitemShut {NoStop}%
\bibitem [{\citenamefont {Landau}(1941)}]{Landau:41}%
  \BibitemOpen
  \bibfield  {author} {\bibinfo {author} {\bibnamefont {Landau}, \bibfnamefont
  {L.}}} (\bibinfo {year} {1941}),\ \href@noop {} {\bibfield  {journal}
  {\bibinfo  {journal} {J. Phys. USSR}\ }\textbf {\bibinfo {volume} {V}},\
  \bibinfo {pages} {71}},\ \bibinfo {note} {reprinted in
  \citet{Khalatnikov:65}}\BibitemShut {NoStop}%
\bibitem [{\citenamefont {Landau}(1947)}]{Landau:47}%
  \BibitemOpen
  \bibfield  {author} {\bibinfo {author} {\bibnamefont {Landau}, \bibfnamefont
  {L.}}} (\bibinfo {year} {1947}),\ \href@noop {} {\bibfield  {journal}
  {\bibinfo  {journal} {J. Phys. USSR}\ }\textbf {\bibinfo {volume} {XI}},\
  \bibinfo {pages} {91}},\ \bibinfo {note} {reprinted in
  \citet{Khalatnikov:65}}\BibitemShut {NoStop}%
\bibitem [{\citenamefont {Landau}\ and\ \citenamefont
  {Lifshitz}(1958)}]{Landau:Quantum}%
  \BibitemOpen
  \bibfield  {author} {\bibinfo {author} {\bibnamefont {Landau}, \bibfnamefont
  {L.}}, \ and\ \bibinfo {author} {\bibfnamefont {E.}~\bibnamefont {Lifshitz}}}
  (\bibinfo {year} {1958}),\ \href@noop {} {\emph {\bibinfo {title} {Quantum
  Mechanics}}}\ (\bibinfo  {publisher} {Pergamon Press - Oxford})\BibitemShut
  {NoStop}%
\bibitem [{\citenamefont {Landau}\ and\ \citenamefont
  {Lifshitz}(1959)}]{Landau:Hydro}%
  \BibitemOpen
  \bibfield  {author} {\bibinfo {author} {\bibnamefont {Landau}, \bibfnamefont
  {L.}}, \ and\ \bibinfo {author} {\bibfnamefont {E.}~\bibnamefont {Lifshitz}}}
  (\bibinfo {year} {1959}),\ \href@noop {} {\emph {\bibinfo {title} {Fluid
  Mechanics}}}\ (\bibinfo  {publisher} {Pergamon Press - Oxford})\BibitemShut
  {NoStop}%
\bibitem [{\citenamefont {Landau}\ and\ \citenamefont
  {Lifshitz}(1971)}]{Landau:ClassicalFields}%
  \BibitemOpen
  \bibfield  {author} {\bibinfo {author} {\bibnamefont {Landau}, \bibfnamefont
  {L.}}, \ and\ \bibinfo {author} {\bibfnamefont {E.}~\bibnamefont {Lifshitz}}}
  (\bibinfo {year} {1971}),\ \href@noop {} {\emph {\bibinfo {title} {Classical
  Theory of Fields}}}\ (\bibinfo  {publisher} {Pergamon Press - Oxford})\
  \bibinfo {note} {\S 89}\BibitemShut {NoStop}%
\bibitem [{\citenamefont {Langer}(1968)}]{Langer:68}%
  \BibitemOpen
  \bibfield  {author} {\bibinfo {author} {\bibnamefont {Langer}, \bibfnamefont
  {J.}}} (\bibinfo {year} {1968}),\ \href@noop {} {\bibfield  {journal}
  {\bibinfo  {journal} {Phys. Rev.}\ }\textbf {\bibinfo {volume} {167}},\
  \bibinfo {pages} {183}}\BibitemShut {NoStop}%
\bibitem [{\citenamefont {Langer}\ and\ \citenamefont
  {Fisher}(1967)}]{Langer:67b}%
  \BibitemOpen
  \bibfield  {author} {\bibinfo {author} {\bibnamefont {Langer}, \bibfnamefont
  {J.}}, \ and\ \bibinfo {author} {\bibfnamefont {M.}~\bibnamefont {Fisher}}}
  (\bibinfo {year} {1967}),\ \href@noop {} {\bibfield  {journal} {\bibinfo
  {journal} {Phys. Rev. Lett.}\ }\textbf {\bibinfo {volume} {19}},\ \bibinfo
  {pages} {560}}\BibitemShut {NoStop}%
\bibitem [{\citenamefont {Langer}\ and\ \citenamefont
  {Reppy}(1970)}]{Langer:70}%
  \BibitemOpen
  \bibfield  {author} {\bibinfo {author} {\bibnamefont {Langer}, \bibfnamefont
  {J.}}, \ and\ \bibinfo {author} {\bibfnamefont {J.}~\bibnamefont {Reppy}}}
  (\bibinfo {year} {1970}),\ in\ \href@noop {} {\emph {\bibinfo {booktitle}
  {Prog. Low Temp. Phys.}}},\ Vol.~\bibinfo {volume} {6},\ Chap.~\bibinfo
  {chapter} {1}\ (\bibinfo  {publisher} {ed. C. J. Gorter, North-Holland -
  Amsterdam})\ p.~\bibinfo {pages} {1}\BibitemShut {NoStop}%
\bibitem [{\citenamefont {Langevin}(1921)}]{Langevin:21}%
  \BibitemOpen
  \bibfield  {author} {\bibinfo {author} {\bibnamefont {Langevin},
  \bibfnamefont {P.}}} (\bibinfo {year} {1921}),\ \href@noop {} {\bibfield
  {journal} {\bibinfo  {journal} {C.R. Acad. Sc. (Paris)}\ }\textbf {\bibinfo
  {volume} {173}},\ \bibinfo {pages} {831}},\ \bibinfo {note} {{\bf 200}, 48
  (1935), {\bf 205}, 304 (1937); http://gallica.bnf.fr}\BibitemShut {NoStop}%
\bibitem [{\citenamefont {Langlois}(2000)}]{Langlois:00}%
  \BibitemOpen
  \bibfield  {author} {\bibinfo {author} {\bibnamefont {Langlois},
  \bibfnamefont {D.}}} (\bibinfo {year} {2000}),\ \href@noop {} {\enquote
  {\bibinfo {title} {Superfluidity in relativistic neutron starts},}\ }\bibinfo
  {howpublished} {astro-ph/0008161}\BibitemShut {NoStop}%
\bibitem [{\citenamefont {Larkin}\ \emph {et~al.}(1984)\citenamefont {Larkin},
  \citenamefont {Likharev},\ and\ \citenamefont {Ovchinnikov}}]{Larkin:84}%
  \BibitemOpen
  \bibfield  {author} {\bibinfo {author} {\bibnamefont {Larkin}, \bibfnamefont
  {A.}}, \bibinfo {author} {\bibfnamefont {K.}~\bibnamefont {Likharev}}, \ and\
  \bibinfo {author} {\bibfnamefont {Y.}~\bibnamefont {Ovchinnikov}}} (\bibinfo
  {year} {1984}),\ \href@noop {} {\bibfield  {journal} {\bibinfo  {journal}
  {Physica B}\ }\textbf {\bibinfo {volume} {126}},\ \bibinfo {pages}
  {414}}\BibitemShut {NoStop}%
\bibitem [{\citenamefont {Lee}\ and\ \citenamefont
  {Richardson}(1978)}]{Lee:78}%
  \BibitemOpen
  \bibfield  {author} {\bibinfo {author} {\bibnamefont {Lee}, \bibfnamefont
  {D.~M.}}, \ and\ \bibinfo {author} {\bibfnamefont {R.~C.}\ \bibnamefont
  {Richardson}}} (\bibinfo {year} {1978}),\ in\ \href@noop {} {\emph {\bibinfo
  {booktitle} {Physics of Liquid and Solid Helium}}},\ \bibinfo {editor}
  {edited by\ \bibinfo {editor} {\bibfnamefont {K.~H.}\ \bibnamefont
  {Bennemann}}\ and\ \bibinfo {editor} {\bibfnamefont {J.~B.}\ \bibnamefont
  {Ketterson}}},\ Chap.\ \bibinfo {chapter} {Part II, p. 287}\ (\bibinfo
  {publisher} {Wiley, New York})\BibitemShut {NoStop}%
\bibitem [{\citenamefont {Leggett}(1975)}]{Leggett:75}%
  \BibitemOpen
  \bibfield  {author} {\bibinfo {author} {\bibnamefont {Leggett}, \bibfnamefont
  {A.~J.}}} (\bibinfo {year} {1975}),\ \href@noop {} {\bibfield  {journal}
  {\bibinfo  {journal} {Rev. Mod. Phys.}\ }\textbf {\bibinfo {volume} {47}},\
  \bibinfo {pages} {331}},\ \bibinfo {note} {erratum, {\bf 48}, 357
  (1976)}\BibitemShut {NoStop}%
\bibitem [{\citenamefont {Leggett}(1980)}]{Leggett:80}%
  \BibitemOpen
  \bibfield  {author} {\bibinfo {author} {\bibnamefont {Leggett}, \bibfnamefont
  {A.~J.}}} (\bibinfo {year} {1980}),\ \href@noop {} {\bibfield  {journal}
  {\bibinfo  {journal} {Prog. Theor. Phys. Suppl.}\ }\textbf {\bibinfo {volume}
  {69}},\ \bibinfo {pages} {80}}\BibitemShut {NoStop}%
\bibitem [{\citenamefont {Leggett}(1995)}]{Leggett:95}%
  \BibitemOpen
  \bibfield  {author} {\bibinfo {author} {\bibnamefont {Leggett}, \bibfnamefont
  {A.~J.}}} (\bibinfo {year} {1995}),\ in\ \href@noop {} {\emph {\bibinfo
  {booktitle} {Bose Einstein Condensation}}},\ \bibinfo {editor} {edited by\
  \bibinfo {editor} {\bibfnamefont {A.}~\bibnamefont {Griffin}}, \bibinfo
  {editor} {\bibfnamefont {D.~W.}\ \bibnamefont {Snoke}}, \ and\ \bibinfo
  {editor} {\bibfnamefont {S.}~\bibnamefont {Stringari}}},\ Chap.~\bibinfo
  {chapter} {19}\ (\bibinfo  {publisher} {Cambridge University
  Press})\BibitemShut {NoStop}%
\bibitem [{\citenamefont {Leggett}(2002)}]{Leggett:02}%
  \BibitemOpen
  \bibfield  {author} {\bibinfo {author} {\bibnamefont {Leggett}, \bibfnamefont
  {A.~J.}}} (\bibinfo {year} {2002}),\ \href@noop {} {\bibfield  {journal}
  {\bibinfo  {journal} {J. Phys.: Condens. Matter}\ }\textbf {\bibinfo {volume}
  {14}},\ \bibinfo {pages} {R415}}\BibitemShut {NoStop}%
\bibitem [{\citenamefont {Leggett}\ and\ \citenamefont
  {Sols}(1991)}]{Leggett:91}%
  \BibitemOpen
  \bibfield  {author} {\bibinfo {author} {\bibnamefont {Leggett}, \bibfnamefont
  {A.~J.}}, \ and\ \bibinfo {author} {\bibfnamefont {F.}~\bibnamefont {Sols}}}
  (\bibinfo {year} {1991}),\ \href@noop {} {\bibfield  {journal} {\bibinfo
  {journal} {Found. Phys.}\ }\textbf {\bibinfo {volume} {21}},\ \bibinfo
  {pages} {353}}\BibitemShut {NoStop}%
\bibitem [{\citenamefont {Leiderer}\ and\ \citenamefont
  {Pobell}(1973)}]{Leiderer:73}%
  \BibitemOpen
  \bibfield  {author} {\bibinfo {author} {\bibnamefont {Leiderer},
  \bibfnamefont {P.}}, \ and\ \bibinfo {author} {\bibfnamefont
  {F.}~\bibnamefont {Pobell}}} (\bibinfo {year} {1973}),\ \href@noop {}
  {\bibfield  {journal} {\bibinfo  {journal} {Phys. Rev.}\ }\textbf {\bibinfo
  {volume} {A7}},\ \bibinfo {pages} {1130}}\BibitemShut {NoStop}%
\bibitem [{\citenamefont {Li}\ and\ \citenamefont {Ho}(1988)}]{Li:88}%
  \BibitemOpen
  \bibfield  {author} {\bibinfo {author} {\bibnamefont {Li}, \bibfnamefont
  {Y.-H.}}, \ and\ \bibinfo {author} {\bibfnamefont {T.-L.}\ \bibnamefont
  {Ho}}} (\bibinfo {year} {1988}),\ \href@noop {} {\bibfield  {journal}
  {\bibinfo  {journal} {Phys. Rev. B}\ }\textbf {\bibinfo {volume} {38}},\
  \bibinfo {pages} {2362}}\BibitemShut {NoStop}%
\bibitem [{\citenamefont {Lifshitz}\ and\ \citenamefont
  {Pitaevskii}(1980)}]{Lifshitz:80}%
  \BibitemOpen
  \bibfield  {author} {\bibinfo {author} {\bibnamefont {Lifshitz},
  \bibfnamefont {E.~M.}}, \ and\ \bibinfo {author} {\bibfnamefont {L.~P.}\
  \bibnamefont {Pitaevskii}}} (\bibinfo {year} {1980}),\ \href@noop {} {\emph
  {\bibinfo {title} {Statistical Physics Part 2}}}\ (\bibinfo  {publisher}
  {Pergamon, Oxford})\ \bibinfo {note} {\S 24, p.93}\BibitemShut {NoStop}%
\bibitem [{\citenamefont {Likharev}(1979)}]{Likharev:79}%
  \BibitemOpen
  \bibfield  {author} {\bibinfo {author} {\bibnamefont {Likharev},
  \bibfnamefont {K.~K.}}} (\bibinfo {year} {1979}),\ \href@noop {} {\bibfield
  {journal} {\bibinfo  {journal} {Rev. Mod. Phys.}\ }\textbf {\bibinfo {volume}
  {51}},\ \bibinfo {pages} {101}},\ \bibinfo {note} {{\it Dynamics of Josephson
  Junctions and Circuits} (Gordon and Breach, New York (1986)}\BibitemShut
  {NoStop}%
\bibitem [{\citenamefont {London}(1938)}]{London:38}%
  \BibitemOpen
  \bibfield  {author} {\bibinfo {author} {\bibnamefont {London}, \bibfnamefont
  {F.}}} (\bibinfo {year} {1938}),\ \href@noop {} {\bibfield  {journal}
  {\bibinfo  {journal} {Nature}\ }\textbf {\bibinfo {volume} {141}},\ \bibinfo
  {pages} {643}}\BibitemShut {NoStop}%
\bibitem [{\citenamefont {London}(1954)}]{London:54}%
  \BibitemOpen
  \bibfield  {author} {\bibinfo {author} {\bibnamefont {London}, \bibfnamefont
  {F.}}} (\bibinfo {year} {1954}),\ \href@noop {} {\emph {\bibinfo {title}
  {Superfluids}}},\ Vol.~\bibinfo {volume} {II}\ (\bibinfo  {publisher} {John
  Wiley and sons, reprinted by Dover Publications -- New York
  (1964)})\BibitemShut {NoStop}%
\bibitem [{\citenamefont {Lounasmaa}\ \emph {et~al.}(1983)\citenamefont
  {Lounasmaa}, \citenamefont {Manninen}, \citenamefont {Nenonen}, \citenamefont
  {Pekola}, \citenamefont {Sharma},\ and\ \citenamefont
  {Tagirov}}]{Lounasmaa:83}%
  \BibitemOpen
  \bibfield  {author} {\bibinfo {author} {\bibnamefont {Lounasmaa},
  \bibfnamefont {O.~V.}}, \bibinfo {author} {\bibfnamefont {M.~T.}\
  \bibnamefont {Manninen}}, \bibinfo {author} {\bibfnamefont {S.~A.}\
  \bibnamefont {Nenonen}}, \bibinfo {author} {\bibfnamefont {J.~P.}\
  \bibnamefont {Pekola}}, \bibinfo {author} {\bibfnamefont {R.~G.}\
  \bibnamefont {Sharma}}, \ and\ \bibinfo {author} {\bibfnamefont {M.~S.}\
  \bibnamefont {Tagirov}}} (\bibinfo {year} {1983}),\ \href@noop {} {\bibfield
  {journal} {\bibinfo  {journal} {Phys. Rev. B}\ }\textbf {\bibinfo {volume}
  {28}},\ \bibinfo {pages} {6536}}\BibitemShut {NoStop}%
\bibitem [{\citenamefont {Madison}\ \emph {et~al.}(2001)\citenamefont
  {Madison}, \citenamefont {Chevy}, \citenamefont {Bretin},\ and\ \citenamefont
  {Dalibard}}]{Madison:01}%
  \BibitemOpen
  \bibfield  {author} {\bibinfo {author} {\bibnamefont {Madison}, \bibfnamefont
  {K.~W.}}, \bibinfo {author} {\bibfnamefont {F.}~\bibnamefont {Chevy}},
  \bibinfo {author} {\bibfnamefont {V.}~\bibnamefont {Bretin}}, \ and\ \bibinfo
  {author} {\bibfnamefont {J.}~\bibnamefont {Dalibard}}} (\bibinfo {year}
  {2001}),\ \href@noop {} {\bibfield  {journal} {\bibinfo  {journal} {Phys.
  Rev. Lett.}\ }\textbf {\bibinfo {volume} {86}},\ \bibinfo {pages}
  {4443}}\BibitemShut {NoStop}%
\bibitem [{\citenamefont {Manninen}\ and\ \citenamefont
  {Pekola}(1983)}]{Manninen:83}%
  \BibitemOpen
  \bibfield  {author} {\bibinfo {author} {\bibnamefont {Manninen},
  \bibfnamefont {M.}}, \ and\ \bibinfo {author} {\bibfnamefont
  {J.}~\bibnamefont {Pekola}}} (\bibinfo {year} {1983}),\ \href@noop {}
  {\bibfield  {journal} {\bibinfo  {journal} {J. Low Temp. Phys.}\ }\textbf
  {\bibinfo {volume} {52}},\ \bibinfo {pages} {497}}\BibitemShut {NoStop}%
\bibitem [{\citenamefont {Marchenkov}\ \emph {et~al.}(1999)\citenamefont
  {Marchenkov}, \citenamefont {Simmonds}, \citenamefont {Backhaus},
  \citenamefont {Loshak}, \citenamefont {Davis},\ and\ \citenamefont
  {Packard}}]{Marchenkov:99}%
  \BibitemOpen
  \bibfield  {author} {\bibinfo {author} {\bibnamefont {Marchenkov},
  \bibfnamefont {A.}}, \bibinfo {author} {\bibfnamefont {R.~W.}\ \bibnamefont
  {Simmonds}}, \bibinfo {author} {\bibfnamefont {S.}~\bibnamefont {Backhaus}},
  \bibinfo {author} {\bibfnamefont {A.}~\bibnamefont {Loshak}}, \bibinfo
  {author} {\bibfnamefont {J.~C.}\ \bibnamefont {Davis}}, \ and\ \bibinfo
  {author} {\bibfnamefont {R.~E.}\ \bibnamefont {Packard}}} (\bibinfo {year}
  {1999}),\ \href@noop {} {\bibfield  {journal} {\bibinfo  {journal} {Phys.
  Rev. Lett.}\ }\textbf {\bibinfo {volume} {83}},\ \bibinfo {pages}
  {3860}}\BibitemShut {NoStop}%
\bibitem [{\citenamefont {Martinis}\ and\ \citenamefont
  {Grabert}(1988)}]{Martinis:88}%
  \BibitemOpen
  \bibfield  {author} {\bibinfo {author} {\bibnamefont {Martinis},
  \bibfnamefont {J.}}, \ and\ \bibinfo {author} {\bibfnamefont
  {H.}~\bibnamefont {Grabert}}} (\bibinfo {year} {1988}),\ \href@noop {}
  {\bibfield  {journal} {\bibinfo  {journal} {Phys. Rev. B}\ }\textbf {\bibinfo
  {volume} {38}},\ \bibinfo {pages} {2371}}\BibitemShut {NoStop}%
\bibitem [{\citenamefont {Martinis}\ \emph {et~al.}(1987)\citenamefont
  {Martinis}, \citenamefont {Devoret},\ and\ \citenamefont
  {Clarke}}]{Martinis:87}%
  \BibitemOpen
  \bibfield  {author} {\bibinfo {author} {\bibnamefont {Martinis},
  \bibfnamefont {J.~M.}}, \bibinfo {author} {\bibfnamefont {M.~H.}\
  \bibnamefont {Devoret}}, \ and\ \bibinfo {author} {\bibfnamefont
  {J.}~\bibnamefont {Clarke}}} (\bibinfo {year} {1987}),\ \href@noop {}
  {\bibfield  {journal} {\bibinfo  {journal} {Phys. Rev. B}\ }\textbf {\bibinfo
  {volume} {35}},\ \bibinfo {pages} {4682}}\BibitemShut {NoStop}%
\bibitem [{\citenamefont {McClintock}\ and\ \citenamefont
  {Bowley}(1991)}]{McClintock:91}%
  \BibitemOpen
  \bibfield  {author} {\bibinfo {author} {\bibnamefont {McClintock},
  \bibfnamefont {P.}}, \ and\ \bibinfo {author} {\bibfnamefont
  {R.}~\bibnamefont {Bowley}}} (\bibinfo {year} {1991}),\ \bibinfo {note} {{\it
  loc. cit.} \citet{Wyatt:91}, p. 567}\BibitemShut {NoStop}%
\bibitem [{\citenamefont {McClintock}\ and\ \citenamefont
  {Bowley}(1995)}]{McClintock:95}%
  \BibitemOpen
  \bibfield  {author} {\bibinfo {author} {\bibnamefont {McClintock},
  \bibfnamefont {P.}}, \ and\ \bibinfo {author} {\bibfnamefont
  {R.}~\bibnamefont {Bowley}}} (\bibinfo {year} {1995}),\ in\ \href@noop {}
  {\emph {\bibinfo {booktitle} {Prog. Low Temp. Physics}}},\ Vol.\ \bibinfo
  {volume} {XIV},\ \bibinfo {editor} {edited by\ \bibinfo {editor}
  {\bibfnamefont {W.}~\bibnamefont {Halperin}}}\ (\bibinfo  {publisher}
  {Elsevier})\ p.~\bibinfo {pages} {1}\BibitemShut {NoStop}%
\bibitem [{\citenamefont {Mel'nikov}(1991)}]{Melnikov:91}%
  \BibitemOpen
  \bibfield  {author} {\bibinfo {author} {\bibnamefont {Mel'nikov},
  \bibfnamefont {V.}}} (\bibinfo {year} {1991}),\ \href@noop {} {\bibfield
  {journal} {\bibinfo  {journal} {Phys. Reports}\ }\textbf {\bibinfo {volume}
  {209}},\ \bibinfo {pages} {1}}\BibitemShut {NoStop}%
\bibitem [{\citenamefont {Monien}\ and\ \citenamefont
  {Tewordt}(1986)}]{Monien:86}%
  \BibitemOpen
  \bibfield  {author} {\bibinfo {author} {\bibnamefont {Monien}, \bibfnamefont
  {H.}}, \ and\ \bibinfo {author} {\bibfnamefont {L.}~\bibnamefont {Tewordt}}}
  (\bibinfo {year} {1986}),\ \href@noop {} {\bibfield  {journal} {\bibinfo
  {journal} {J. Low Temp. Phys.}\ }\textbf {\bibinfo {volume} {62}},\ \bibinfo
  {pages} {277}},\ \bibinfo {note} {also Can. J. Phys. {\bf 65}, 1388
  (1987)}\BibitemShut {NoStop}%
\bibitem [{\citenamefont {Muirhead}\ \emph {et~al.}(1984)\citenamefont
  {Muirhead}, \citenamefont {Vinen},\ and\ \citenamefont
  {Donnelly}}]{Muirhead:84}%
  \BibitemOpen
  \bibfield  {author} {\bibinfo {author} {\bibnamefont {Muirhead},
  \bibfnamefont {C.}}, \bibinfo {author} {\bibfnamefont {W.}~\bibnamefont
  {Vinen}}, \ and\ \bibinfo {author} {\bibfnamefont {R.}~\bibnamefont
  {Donnelly}}} (\bibinfo {year} {1984}),\ \href@noop {} {\bibfield  {journal}
  {\bibinfo  {journal} {Phil. Trans. Roy. Soc. A}\ }\textbf {\bibinfo {volume}
  {311}},\ \bibinfo {pages} {433}}\BibitemShut {NoStop}%
\bibitem [{\citenamefont {Muirhead}\ \emph {et~al.}(1985)\citenamefont
  {Muirhead}, \citenamefont {Vinen},\ and\ \citenamefont
  {Donnelly}}]{Muirhead:85}%
  \BibitemOpen
  \bibfield  {author} {\bibinfo {author} {\bibnamefont {Muirhead},
  \bibfnamefont {C.}}, \bibinfo {author} {\bibfnamefont {W.}~\bibnamefont
  {Vinen}}, \ and\ \bibinfo {author} {\bibfnamefont {R.}~\bibnamefont
  {Donnelly}}} (\bibinfo {year} {1985}),\ \href@noop {} {\bibfield  {journal}
  {\bibinfo  {journal} {Proc. R. Soc. London A}\ }\textbf {\bibinfo {volume}
  {402}},\ \bibinfo {pages} {225}}\BibitemShut {NoStop}%
\bibitem [{\citenamefont {Mukharsky}(2004)}]{Mukharsky:04-JLTP}%
  \BibitemOpen
  \bibfield  {author} {\bibinfo {author} {\bibnamefont {Mukharsky},
  \bibfnamefont {Y.}}} (\bibinfo {year} {2004}),\ \href@noop {} {\bibfield
  {journal} {\bibinfo  {journal} {J. Low Temp. Phys.}\ }\textbf {\bibinfo
  {volume} {134}},\ \bibinfo {pages} {731}}\BibitemShut {NoStop}%
\bibitem [{\citenamefont {Mukharsky}\ \emph {et~al.}(2004)\citenamefont
  {Mukharsky}, \citenamefont {Avenel},\ and\ \citenamefont
  {Varoquaux}}]{Mukharsky:04}%
  \BibitemOpen
  \bibfield  {author} {\bibinfo {author} {\bibnamefont {Mukharsky},
  \bibfnamefont {Y.}}, \bibinfo {author} {\bibfnamefont {O.}~\bibnamefont
  {Avenel}}, \ and\ \bibinfo {author} {\bibfnamefont {E.}~\bibnamefont
  {Varoquaux}}} (\bibinfo {year} {2004}),\ \href@noop {} {\bibfield  {journal}
  {\bibinfo  {journal} {Phys. Rev. Lett.}\ }\textbf {\bibinfo {volume} {92}},\
  \bibinfo {pages} {210402}}\BibitemShut {NoStop}%
\bibitem [{\citenamefont {Musinski}(1973)}]{Musinski:73}%
  \BibitemOpen
  \bibfield  {author} {\bibinfo {author} {\bibnamefont {Musinski},
  \bibfnamefont {D.}}} (\bibinfo {year} {1973}),\ \href@noop {} {\bibfield
  {journal} {\bibinfo  {journal} {J. Low Temp. Phys.}\ }\textbf {\bibinfo
  {volume} {13}},\ \bibinfo {pages} {287}}\BibitemShut {NoStop}%
\bibitem [{\citenamefont {Musinski}\ and\ \citenamefont
  {Douglass}(1972)}]{Musinski:72}%
  \BibitemOpen
  \bibfield  {author} {\bibinfo {author} {\bibnamefont {Musinski},
  \bibfnamefont {D.~L.}}, \ and\ \bibinfo {author} {\bibfnamefont {D.~H.}\
  \bibnamefont {Douglass}}} (\bibinfo {year} {1972}),\ \href@noop {} {\bibfield
   {journal} {\bibinfo  {journal} {Phys. Rev. Lett.}\ }\textbf {\bibinfo
  {volume} {29}},\ \bibinfo {pages} {1541}}\BibitemShut {NoStop}%
\bibitem [{\citenamefont {Narayana}\ and\ \citenamefont
  {Sato}(2010)}]{Narayana:10}%
  \BibitemOpen
  \bibfield  {author} {\bibinfo {author} {\bibnamefont {Narayana},
  \bibfnamefont {S.}}, \ and\ \bibinfo {author} {\bibfnamefont
  {Y.}~\bibnamefont {Sato}}} (\bibinfo {year} {2010}),\ \href@noop {}
  {\bibfield  {journal} {\bibinfo  {journal} {Phys. Rev. Lett.}\ }\textbf
  {\bibinfo {volume} {105}},\ \bibinfo {pages} {205302}}\BibitemShut {NoStop}%
\bibitem [{\citenamefont {Narayana}\ and\ \citenamefont
  {Sato}(2011)}]{Narayana:11}%
  \BibitemOpen
  \bibfield  {author} {\bibinfo {author} {\bibnamefont {Narayana},
  \bibfnamefont {S.}}, \ and\ \bibinfo {author} {\bibfnamefont
  {Y.}~\bibnamefont {Sato}}} (\bibinfo {year} {2011}),\ \href@noop {}
  {\bibfield  {journal} {\bibinfo  {journal} {Phys. Rev. Lett.}\ }\textbf
  {\bibinfo {volume} {106}},\ \bibinfo {pages} {055302}}\BibitemShut {NoStop}%
\bibitem [{\citenamefont {Neumann}\ and\ \citenamefont
  {Zieve}(2014)}]{Neumann:14}%
  \BibitemOpen
  \bibfield  {author} {\bibinfo {author} {\bibnamefont {Neumann}, \bibfnamefont
  {I.~H.}}, \ and\ \bibinfo {author} {\bibfnamefont {R.~J.}\ \bibnamefont
  {Zieve}}} (\bibinfo {year} {2014}),\ \href@noop {} {\bibfield  {journal}
  {\bibinfo  {journal} {Phys. Rev. B}\ }\textbf {\bibinfo {volume} {89}},\
  \bibinfo {pages} {104521}}\BibitemShut {NoStop}%
\bibitem [{\citenamefont {Neutze}\ and\ \citenamefont
  {Hasselbach}(1998)}]{Neutze:98}%
  \BibitemOpen
  \bibfield  {author} {\bibinfo {author} {\bibnamefont {Neutze}, \bibfnamefont
  {R.}}, \ and\ \bibinfo {author} {\bibfnamefont {F.}~\bibnamefont
  {Hasselbach}}} (\bibinfo {year} {1998}),\ \href@noop {} {\bibfield  {journal}
  {\bibinfo  {journal} {Phys. Rev. A}\ }\textbf {\bibinfo {volume} {48}},\
  \bibinfo {pages} {143}}\BibitemShut {NoStop}%
\bibitem [{\citenamefont {Nienhuis}(2001)}]{Nienhuis:01}%
  \BibitemOpen
  \bibfield  {author} {\bibinfo {author} {\bibnamefont {Nienhuis},
  \bibfnamefont {G.}}} (\bibinfo {year} {2001}),\ \enquote {\bibinfo {title}
  {Macroscopic entanglement and relative phase},}\ \ (\bibinfo  {publisher}
  {Lecture Notes in Physics 575, Springer-Verlag})\ p.~\bibinfo {pages}
  {95}\BibitemShut {NoStop}%
\bibitem [{\citenamefont {Nishida}\ \emph {et~al.}(2002)\citenamefont
  {Nishida}, \citenamefont {Hatakenaka},\ and\ \citenamefont
  {Kurihara}}]{Nishida:02}%
  \BibitemOpen
  \bibfield  {author} {\bibinfo {author} {\bibnamefont {Nishida}, \bibfnamefont
  {M.}}, \bibinfo {author} {\bibfnamefont {N.}~\bibnamefont {Hatakenaka}}, \
  and\ \bibinfo {author} {\bibfnamefont {S.}~\bibnamefont {Kurihara}}}
  (\bibinfo {year} {2002}),\ \href@noop {} {\bibfield  {journal} {\bibinfo
  {journal} {Phys. Rev. Lett.}\ }\textbf {\bibinfo {volume} {88}},\ \bibinfo
  {pages} {145302}}\BibitemShut {NoStop}%
\bibitem [{\citenamefont {Nore}\ \emph {et~al.}(2000)\citenamefont {Nore},
  \citenamefont {Huepe},\ and\ \citenamefont {Brachet}}]{Nore:00}%
  \BibitemOpen
  \bibfield  {author} {\bibinfo {author} {\bibnamefont {Nore}, \bibfnamefont
  {C.}}, \bibinfo {author} {\bibfnamefont {C.}~\bibnamefont {Huepe}}, \ and\
  \bibinfo {author} {\bibfnamefont {M.}~\bibnamefont {Brachet}}} (\bibinfo
  {year} {2000}),\ \href@noop {} {\bibfield  {journal} {\bibinfo  {journal}
  {Phys. Rev. Lett.}\ }\textbf {\bibinfo {volume} {84}},\ \bibinfo {pages}
  {2191}}\BibitemShut {NoStop}%
\bibitem [{\citenamefont {Nozi\`eres}\ and\ \citenamefont
  {Pines}(1990)}]{Nozieres:90}%
  \BibitemOpen
  \bibfield  {author} {\bibinfo {author} {\bibnamefont {Nozi\`eres},
  \bibfnamefont {P.}}, \ and\ \bibinfo {author} {\bibfnamefont
  {D.}~\bibnamefont {Pines}}} (\bibinfo {year} {1990}),\ \href@noop {} {\emph
  {\bibinfo {title} {The Theory of Quantum Liquids}}},\ Vol.~\bibinfo {volume}
  {II}\ (\bibinfo  {publisher} {Addison-Wesley - Redwood City (CA)})\
  Chap.~\bibinfo {chapter} {5}\BibitemShut {NoStop}%
\bibitem [{\citenamefont {Onsager}(1949)}]{Onsager:49}%
  \BibitemOpen
  \bibfield  {author} {\bibinfo {author} {\bibnamefont {Onsager}, \bibfnamefont
  {L.}}} (\bibinfo {year} {1949}),\ \href@noop {} {\bibfield  {journal}
  {\bibinfo  {journal} {Nuovo Cimento Suppl.}\ }\textbf {\bibinfo {volume}
  {6}},\ \bibinfo {pages} {249}},\ \bibinfo {note} {available at
  http://bookos.org/book/653790 and reprinted in {\it The collected Works of
  Lars Onsager}, Eds. P. C. Hemmer, H. Holden and S. Kjelstrup Ratkje, World
  Scientific, Singapore 1996}\BibitemShut {NoStop}%
\bibitem [{\citenamefont {Osheroff}\ \emph {et~al.}(1972)\citenamefont
  {Osheroff}, \citenamefont {Gully}, \citenamefont {Richardson},\ and\
  \citenamefont {Lee}}]{Osheroff:72}%
  \BibitemOpen
  \bibfield  {author} {\bibinfo {author} {\bibnamefont {Osheroff},
  \bibfnamefont {D.~D.}}, \bibinfo {author} {\bibfnamefont {W.~J.}\
  \bibnamefont {Gully}}, \bibinfo {author} {\bibfnamefont {R.~C.}\ \bibnamefont
  {Richardson}}, \ and\ \bibinfo {author} {\bibfnamefont {D.~M.}\ \bibnamefont
  {Lee}}} (\bibinfo {year} {1972}),\ \href@noop {} {\bibfield  {journal}
  {\bibinfo  {journal} {Phys. Rev. Lett}\ }\textbf {\bibinfo {volume} {29}},\
  \bibinfo {pages} {920}}\BibitemShut {NoStop}%
\bibitem [{\citenamefont {Packard}(1972)}]{Packard:72}%
  \BibitemOpen
  \bibfield  {author} {\bibinfo {author} {\bibnamefont {Packard}, \bibfnamefont
  {R.~E.}}} (\bibinfo {year} {1972}),\ \href@noop {} {\bibfield  {journal}
  {\bibinfo  {journal} {Phys. Rev. Lett.}\ }\textbf {\bibinfo {volume} {28}},\
  \bibinfo {pages} {1080}}\BibitemShut {NoStop}%
\bibitem [{\citenamefont {Packard}(1998)}]{Packard:98}%
  \BibitemOpen
  \bibfield  {author} {\bibinfo {author} {\bibnamefont {Packard}, \bibfnamefont
  {R.~E.}}} (\bibinfo {year} {1998}),\ \href@noop {} {\bibfield  {journal}
  {\bibinfo  {journal} {Rev. Mod. Phys.}\ }\textbf {\bibinfo {volume} {70}},\
  \bibinfo {pages} {641}}\BibitemShut {NoStop}%
\bibitem [{\citenamefont {Packard}(2004)}]{Packard:04}%
  \BibitemOpen
  \bibfield  {author} {\bibinfo {author} {\bibnamefont {Packard}, \bibfnamefont
  {R.~E.}}} (\bibinfo {year} {2004}),\ \href@noop {} {\bibfield  {journal}
  {\bibinfo  {journal} {J. Low Temp. Phys.}\ }\textbf {\bibinfo {volume}
  {135}},\ \bibinfo {pages} {471}}\BibitemShut {NoStop}%
\bibitem [{\citenamefont {Packard}\ and\ \citenamefont
  {Vitale}(1992)}]{Packard:92b}%
  \BibitemOpen
  \bibfield  {author} {\bibinfo {author} {\bibnamefont {Packard}, \bibfnamefont
  {R.~E.}}, \ and\ \bibinfo {author} {\bibfnamefont {S.}~\bibnamefont
  {Vitale}}} (\bibinfo {year} {1992}),\ \href@noop {} {\bibfield  {journal}
  {\bibinfo  {journal} {Phys. Rev.}\ }\textbf {\bibinfo {volume} {B46}},\
  \bibinfo {pages} {3540}}\BibitemShut {NoStop}%
\bibitem [{\citenamefont {Parts}\ \emph {et~al.}(1994)\citenamefont {Parts},
  \citenamefont {Thuneberg}, \citenamefont {Volovik}, \citenamefont
  {Koivuniemi}, \citenamefont {Ruutu}, \citenamefont {Heinil{\"a}},
  \citenamefont {Karim{\"a}ki},\ and\ \citenamefont {Krusius}}]{Parts:94}%
  \BibitemOpen
  \bibfield  {author} {\bibinfo {author} {\bibnamefont {Parts}, \bibfnamefont
  {{\"U}.}}, \bibinfo {author} {\bibfnamefont {E.~V.}\ \bibnamefont
  {Thuneberg}}, \bibinfo {author} {\bibfnamefont {G.~E.}\ \bibnamefont
  {Volovik}}, \bibinfo {author} {\bibfnamefont {J.~H.}\ \bibnamefont
  {Koivuniemi}}, \bibinfo {author} {\bibfnamefont {V.~M.~H.}\ \bibnamefont
  {Ruutu}}, \bibinfo {author} {\bibfnamefont {M.}~\bibnamefont {Heinil{\"a}}},
  \bibinfo {author} {\bibfnamefont {J.~M.}\ \bibnamefont {Karim{\"a}ki}}, \
  and\ \bibinfo {author} {\bibfnamefont {M.}~\bibnamefont {Krusius}}} (\bibinfo
  {year} {1994}),\ \href@noop {} {\bibfield  {journal} {\bibinfo  {journal}
  {Phys. Rev. Lett.}\ }\textbf {\bibinfo {volume} {72}},\ \bibinfo {pages}
  {3839}}\BibitemShut {NoStop}%
\bibitem [{\citenamefont {Pekker}\ \emph {et~al.}(2007)\citenamefont {Pekker},
  \citenamefont {Barankov},\ and\ \citenamefont {Goldbart}}]{Pekker:07}%
  \BibitemOpen
  \bibfield  {author} {\bibinfo {author} {\bibnamefont {Pekker}, \bibfnamefont
  {D.}}, \bibinfo {author} {\bibfnamefont {R.}~\bibnamefont {Barankov}}, \ and\
  \bibinfo {author} {\bibfnamefont {P.~M.}\ \bibnamefont {Goldbart}}} (\bibinfo
  {year} {2007}),\ \href@noop {} {\bibfield  {journal} {\bibinfo  {journal}
  {Phys. Rev. Lett.}\ }\textbf {\bibinfo {volume} {98}},\ \bibinfo {pages}
  {175301}}\BibitemShut {NoStop}%
\bibitem [{\citenamefont {Penrose}\ and\ \citenamefont
  {Onsager}(1956)}]{Penrose:56}%
  \BibitemOpen
  \bibfield  {author} {\bibinfo {author} {\bibnamefont {Penrose}, \bibfnamefont
  {O.}}, \ and\ \bibinfo {author} {\bibfnamefont {L.}~\bibnamefont {Onsager}}}
  (\bibinfo {year} {1956}),\ \href@noop {} {\bibfield  {journal} {\bibinfo
  {journal} {Phys. Rev.}\ }\textbf {\bibinfo {volume} {104}},\ \bibinfo {pages}
  {576}},\ \bibinfo {note} {see also O. Penrose, Phil. Mag. {\bf 42}, 1373
  (1951), reprinted in\citet{Anderson:84b}}\BibitemShut {NoStop}%
\bibitem [{\citenamefont {Pereverzev}\ and\ \citenamefont
  {Eska}(2001)}]{Pereverzev:01}%
  \BibitemOpen
  \bibfield  {author} {\bibinfo {author} {\bibnamefont {Pereverzev},
  \bibfnamefont {S.~V.}}, \ and\ \bibinfo {author} {\bibfnamefont
  {G.}~\bibnamefont {Eska}}} (\bibinfo {year} {2001}),\ \href@noop {}
  {\bibfield  {journal} {\bibinfo  {journal} {J. Low Temp. Phys.}\ }\textbf
  {\bibinfo {volume} {124}},\ \bibinfo {pages} {383}}\BibitemShut {NoStop}%
\bibitem [{\citenamefont {Perron}\ \emph {et~al.}(2013)\citenamefont {Perron},
  \citenamefont {Kimball}, \citenamefont {Mooney},\ and\ \citenamefont
  {Gasparini}}]{Perron:13}%
  \BibitemOpen
  \bibfield  {author} {\bibinfo {author} {\bibnamefont {Perron}, \bibfnamefont
  {J.~K.}}, \bibinfo {author} {\bibfnamefont {M.~O.}\ \bibnamefont {Kimball}},
  \bibinfo {author} {\bibfnamefont {K.~P.}\ \bibnamefont {Mooney}}, \ and\
  \bibinfo {author} {\bibfnamefont {F.~M.}\ \bibnamefont {Gasparini}}}
  (\bibinfo {year} {2013}),\ \href@noop {} {\bibfield  {journal} {\bibinfo
  {journal} {Phys.Rev. B}\ }\textbf {\bibinfo {volume} {87}},\ \bibinfo {pages}
  {094507}}\BibitemShut {NoStop}%
\bibitem [{\citenamefont {Pines}(1962)}]{Pines:62}%
  \BibitemOpen
  \bibfield  {author} {\bibinfo {author} {\bibnamefont {Pines}, \bibfnamefont
  {D.}}} (\bibinfo {year} {1962}),\ \href@noop {} {\emph {\bibinfo {title} {The
  Many-Body Problem}}}\ (\bibinfo  {publisher} {W. A. Benjamin, inc, New
  York})\BibitemShut {NoStop}%
\bibitem [{\citenamefont {Pitaevskii}(1961)}]{Pitaevskii:61}%
  \BibitemOpen
  \bibfield  {author} {\bibinfo {author} {\bibnamefont {Pitaevskii},
  \bibfnamefont {L.}}} (\bibinfo {year} {1961}),\ \href@noop {} {\bibfield
  {journal} {\bibinfo  {journal} {{J}. Exp. Theor. Phys. USSR}\ }\textbf
  {\bibinfo {volume} {40}},\ \bibinfo {pages} {646}},\ \bibinfo {note} {{S}ov.
  Phys. JETP, {\bf 13}, 451 (1961)}\BibitemShut {NoStop}%
\bibitem [{\citenamefont {Pomeau}\ and\ \citenamefont
  {Rica}(1993)}]{Pomeau:93}%
  \BibitemOpen
  \bibfield  {author} {\bibinfo {author} {\bibnamefont {Pomeau}, \bibfnamefont
  {Y.}}, \ and\ \bibinfo {author} {\bibfnamefont {S.}~\bibnamefont {Rica}}}
  (\bibinfo {year} {1993}),\ \href@noop {} {\bibfield  {journal} {\bibinfo
  {journal} {Phys. Rev. Lett.}\ }\textbf {\bibinfo {volume} {71}},\ \bibinfo
  {pages} {247}}\BibitemShut {NoStop}%
\bibitem [{\citenamefont {Putterman}(1974)}]{Putterman:74}%
  \BibitemOpen
  \bibfield  {author} {\bibinfo {author} {\bibnamefont {Putterman},
  \bibfnamefont {S.}}} (\bibinfo {year} {1974}),\ \href@noop {} {\emph
  {\bibinfo {title} {Superfluid Hydrodynamics}}}\ (\bibinfo  {publisher}
  {North-Holland, Amsterdam})\ \bibinfo {note} {in the Preface}\BibitemShut
  {NoStop}%
\bibitem [{\citenamefont {Putterman}\ and\ \citenamefont
  {Rudnick}(1971)}]{Putterman:71}%
  \BibitemOpen
  \bibfield  {author} {\bibinfo {author} {\bibnamefont {Putterman},
  \bibfnamefont {S.}}, \ and\ \bibinfo {author} {\bibfnamefont
  {I.}~\bibnamefont {Rudnick}}} (\bibinfo {year} {1971}),\ \href@noop {}
  {\bibinfo  {journal} {Physics Today}\ ,\ \bibinfo {pages} {39}}\BibitemShut
  {NoStop}%
\bibitem [{\citenamefont {Rainer}\ and\ \citenamefont {Lee}(1987)}]{Rainer:87}%
  \BibitemOpen
\bibfield  {journal} {  }\bibfield  {author} {\bibinfo {author} {\bibnamefont
  {Rainer}, \bibfnamefont {D.}}, \ and\ \bibinfo {author} {\bibfnamefont
  {P.~A.}\ \bibnamefont {Lee}}} (\bibinfo {year} {1987}),\ \href@noop {}
  {\bibfield  {journal} {\bibinfo  {journal} {Phys. Rev. B}\ }\textbf {\bibinfo
  {volume} {35}},\ \bibinfo {pages} {1987}}\BibitemShut {NoStop}%
\bibitem [{\citenamefont {Rayfield}\ and\ \citenamefont
  {Reif}(1964)}]{Rayfield:64}%
  \BibitemOpen
  \bibfield  {author} {\bibinfo {author} {\bibnamefont {Rayfield},
  \bibfnamefont {G.~W.}}, \ and\ \bibinfo {author} {\bibfnamefont
  {F.}~\bibnamefont {Reif}}} (\bibinfo {year} {1964}),\ \href@noop {}
  {\bibfield  {journal} {\bibinfo  {journal} {Phys. Rev.}\ }\textbf {\bibinfo
  {volume} {136}},\ \bibinfo {pages} {A1194}}\BibitemShut {NoStop}%
\bibitem [{\citenamefont {Reppy}\ and\ \citenamefont {Lane}(1965)}]{Reppy:65}%
  \BibitemOpen
  \bibfield  {author} {\bibinfo {author} {\bibnamefont {Reppy}, \bibfnamefont
  {J.}}, \ and\ \bibinfo {author} {\bibfnamefont {C.}~\bibnamefont {Lane}}}
  (\bibinfo {year} {1965}),\ \href@noop {} {\bibfield  {journal} {\bibinfo
  {journal} {Phys. Rev.}\ }\textbf {\bibinfo {volume} {140}},\ \bibinfo {pages}
  {A 106}}\BibitemShut {NoStop}%
\bibitem [{\citenamefont {Rica}(2001)}]{Rica:01}%
  \BibitemOpen
  \bibfield  {author} {\bibinfo {author} {\bibnamefont {Rica}, \bibfnamefont
  {S.}}} (\bibinfo {year} {2001}),\ \bibinfo {note} {{\it loc.cit.}
  \citet{Barenghi:01}, p. 258}\BibitemShut {NoStop}%
\bibitem [{\citenamefont {Richards}(1970)}]{Richards:70}%
  \BibitemOpen
  \bibfield  {author} {\bibinfo {author} {\bibnamefont {Richards},
  \bibfnamefont {P.~L.}}} (\bibinfo {year} {1970}),\ \href@noop {} {\bibfield
  {journal} {\bibinfo  {journal} {Phys. Rev.}\ }\textbf {\bibinfo {volume} {A
  2}},\ \bibinfo {pages} {1532}}\BibitemShut {NoStop}%
\bibitem [{\citenamefont {Richards}\ and\ \citenamefont
  {Anderson}(1965)}]{Richards:65}%
  \BibitemOpen
  \bibfield  {author} {\bibinfo {author} {\bibnamefont {Richards},
  \bibfnamefont {P.~L.}}, \ and\ \bibinfo {author} {\bibfnamefont {P.~W.}\
  \bibnamefont {Anderson}}} (\bibinfo {year} {1965}),\ \href@noop {} {\bibfield
   {journal} {\bibinfo  {journal} {Phys. Rev. Lett.}\ }\textbf {\bibinfo
  {volume} {14}},\ \bibinfo {pages} {540}}\BibitemShut {NoStop}%
\bibitem [{\citenamefont {Rips}\ and\ \citenamefont {Pollak}(1989)}]{Rips:89}%
  \BibitemOpen
  \bibfield  {author} {\bibinfo {author} {\bibnamefont {Rips}, \bibfnamefont
  {I.}}, \ and\ \bibinfo {author} {\bibfnamefont {E.}~\bibnamefont {Pollak}}}
  (\bibinfo {year} {1989}),\ \href@noop {} {\bibfield  {journal} {\bibinfo
  {journal} {Phys. Rev. A}\ }\textbf {\bibinfo {volume} {41}},\ \bibinfo
  {pages} {5366}}\BibitemShut {NoStop}%
\bibitem [{\citenamefont {Rizzi}\ and\ \citenamefont
  {Ruggiero}(2004)}]{Rizzi:04}%
  \BibitemOpen
  \bibfield  {author} {\bibinfo {author} {\bibnamefont {Rizzi}, \bibfnamefont
  {G.}}, \ and\ \bibinfo {author} {\bibfnamefont {M.~L.}\ \bibnamefont
  {Ruggiero}}} (\bibinfo {year} {2004}),\ \enquote {\bibinfo {title}
  {Relativity in rotating frames},}\ \ (\bibinfo  {publisher} {Springer Verlag,
  Heidelberg})\ p.\ \bibinfo {pages} {179},\ \bibinfo {note}
  {http://digilander.libero.it/solciclos/; also Gen. Rel. Grav. {\bf 35}, 1745,
  2129 (2003)}\BibitemShut {NoStop}%
\bibitem [{\citenamefont {Roberts}\ and\ \citenamefont
  {Donnelly}(1970)}]{Roberts:70}%
  \BibitemOpen
  \bibfield  {author} {\bibinfo {author} {\bibnamefont {Roberts}, \bibfnamefont
  {P.~H.}}, \ and\ \bibinfo {author} {\bibfnamefont {R.~J.}\ \bibnamefont
  {Donnelly}}} (\bibinfo {year} {1970}),\ \href@noop {} {\bibfield  {journal}
  {\bibinfo  {journal} {Phys. Lett.}\ }\textbf {\bibinfo {volume} {31}},\
  \bibinfo {pages} {137}}\BibitemShut {NoStop}%
\bibitem [{\citenamefont {Roberts}\ and\ \citenamefont
  {Grant}(1971)}]{Roberts:71}%
  \BibitemOpen
  \bibfield  {author} {\bibinfo {author} {\bibnamefont {Roberts}, \bibfnamefont
  {P.~H.}}, \ and\ \bibinfo {author} {\bibfnamefont {J.}~\bibnamefont {Grant}}}
  (\bibinfo {year} {1971}),\ \href@noop {} {\bibfield  {journal} {\bibinfo
  {journal} {J. Phys. A: Gen. Phys.}\ }\textbf {\bibinfo {volume} {4}},\
  \bibinfo {pages} {55}}\BibitemShut {NoStop}%
\bibitem [{\citenamefont {Rudnick}(1973)}]{Rudnick:73}%
  \BibitemOpen
  \bibfield  {author} {\bibinfo {author} {\bibnamefont {Rudnick}, \bibfnamefont
  {I.}}} (\bibinfo {year} {1973}),\ \href@noop {} {\bibfield  {journal}
  {\bibinfo  {journal} {Phys. Rev.}\ }\textbf {\bibinfo {volume} {A8}},\
  \bibinfo {pages} {1969}}\BibitemShut {NoStop}%
\bibitem [{\citenamefont {Saba}\ \emph {et~al.}(2005)\citenamefont {Saba},
  \citenamefont {Pasquini}, \citenamefont {Sanner}, \citenamefont {Shin},
  \citenamefont {Ketterle},\ and\ \citenamefont {Pritchard}}]{Saba:05}%
  \BibitemOpen
  \bibfield  {author} {\bibinfo {author} {\bibnamefont {Saba}, \bibfnamefont
  {M.}}, \bibinfo {author} {\bibfnamefont {T.~A.}\ \bibnamefont {Pasquini}},
  \bibinfo {author} {\bibfnamefont {C.}~\bibnamefont {Sanner}}, \bibinfo
  {author} {\bibfnamefont {Y.}~\bibnamefont {Shin}}, \bibinfo {author}
  {\bibfnamefont {W.}~\bibnamefont {Ketterle}}, \ and\ \bibinfo {author}
  {\bibfnamefont {D.~E.}\ \bibnamefont {Pritchard}}} (\bibinfo {year} {2005}),\
  \href@noop {} {\bibfield  {journal} {\bibinfo  {journal} {Science}\ }\textbf
  {\bibinfo {volume} {307}},\ \bibinfo {pages} {1945}}\BibitemShut {NoStop}%
\bibitem [{\citenamefont {Saffman}(1992)}]{Saffman:92}%
  \BibitemOpen
  \bibfield  {author} {\bibinfo {author} {\bibnamefont {Saffman}, \bibfnamefont
  {P.~G.}}} (\bibinfo {year} {1992}),\ \href@noop {} {\emph {\bibinfo {title}
  {Vortex Dynamics}}}\ (\bibinfo  {publisher} {Cambridge University Press -
  Cambridge})\BibitemShut {NoStop}%
\bibitem [{\citenamefont {Salomaa}\ and\ \citenamefont
  {Volovik}(1987)}]{Salomaa:87}%
  \BibitemOpen
  \bibfield  {author} {\bibinfo {author} {\bibnamefont {Salomaa}, \bibfnamefont
  {M.~M.}}, \ and\ \bibinfo {author} {\bibfnamefont {G.~E.}\ \bibnamefont
  {Volovik}}} (\bibinfo {year} {1987}),\ \href@noop {} {\bibfield  {journal}
  {\bibinfo  {journal} {Rev. Mod. Phys.}\ }\textbf {\bibinfo {volume} {59}},\
  \bibinfo {pages} {533}}\BibitemShut {NoStop}%
\bibitem [{\citenamefont {Salomaa}\ and\ \citenamefont
  {Volovik}(1988)}]{Salomaa:88}%
  \BibitemOpen
  \bibfield  {author} {\bibinfo {author} {\bibnamefont {Salomaa}, \bibfnamefont
  {M.~M.}}, \ and\ \bibinfo {author} {\bibfnamefont {G.~E.}\ \bibnamefont
  {Volovik}}} (\bibinfo {year} {1988}),\ \href@noop {} {\bibfield  {journal}
  {\bibinfo  {journal} {Phys. Rev. B}\ }\textbf {\bibinfo {volume} {37}},\
  \bibinfo {pages} {9298}}\BibitemShut {NoStop}%
\bibitem [{\citenamefont {Sato}(2014)}]{Sato:14}%
  \BibitemOpen
  \bibfield  {author} {\bibinfo {author} {\bibnamefont {Sato}, \bibfnamefont
  {Y.}}} (\bibinfo {year} {2014}),\ \href@noop {} {\bibfield  {journal}
  {\bibinfo  {journal} {C.R. Physique}\ }\textbf {\bibinfo {volume} {15}},\
  \bibinfo {pages} {898}}\BibitemShut {NoStop}%
\bibitem [{\citenamefont {Sato}\ \emph {et~al.}(2006)\citenamefont {Sato},
  \citenamefont {Hoskinson},\ and\ \citenamefont {Packard}}]{Sato:06}%
  \BibitemOpen
  \bibfield  {author} {\bibinfo {author} {\bibnamefont {Sato}, \bibfnamefont
  {Y.}}, \bibinfo {author} {\bibfnamefont {E.}~\bibnamefont {Hoskinson}}, \
  and\ \bibinfo {author} {\bibfnamefont {R.~E.}\ \bibnamefont {Packard}}}
  (\bibinfo {year} {2006}),\ \href@noop {} {\bibfield  {journal} {\bibinfo
  {journal} {Phys. Rev. B}\ }\textbf {\bibinfo {volume} {74}},\ \bibinfo
  {pages} {144502}}\BibitemShut {NoStop}%
\bibitem [{\citenamefont {Sato}\ \emph {et~al.}(2007)\citenamefont {Sato},
  \citenamefont {Hoskinson},\ and\ \citenamefont {Packard}}]{Sato:07}%
  \BibitemOpen
  \bibfield  {author} {\bibinfo {author} {\bibnamefont {Sato}, \bibfnamefont
  {Y.}}, \bibinfo {author} {\bibfnamefont {E.}~\bibnamefont {Hoskinson}}, \
  and\ \bibinfo {author} {\bibfnamefont {R.~E.}\ \bibnamefont {Packard}}}
  (\bibinfo {year} {2007}),\ \href@noop {} {\bibfield  {journal} {\bibinfo
  {journal} {J. Low Temp. Phys.}\ }\textbf {\bibinfo {volume} {149}},\ \bibinfo
  {pages} {222}}\BibitemShut {NoStop}%
\bibitem [{\citenamefont {Sato}\ \emph {et~al.}(2008)\citenamefont {Sato},
  \citenamefont {Joshi},\ and\ \citenamefont {Packard}}]{Sato:08}%
  \BibitemOpen
  \bibfield  {author} {\bibinfo {author} {\bibnamefont {Sato}, \bibfnamefont
  {Y.}}, \bibinfo {author} {\bibfnamefont {A.}~\bibnamefont {Joshi}}, \ and\
  \bibinfo {author} {\bibfnamefont {R.~E.}\ \bibnamefont {Packard}}} (\bibinfo
  {year} {2008}),\ \href@noop {} {\bibfield  {journal} {\bibinfo  {journal}
  {Phys. Rev. Lett.}\ }\textbf {\bibinfo {volume} {101}},\ \bibinfo {pages}
  {085302}}\BibitemShut {NoStop}%
\bibitem [{\citenamefont {Sato}\ and\ \citenamefont {Packard}(2012)}]{Sato:12}%
  \BibitemOpen
  \bibfield  {author} {\bibinfo {author} {\bibnamefont {Sato}, \bibfnamefont
  {Y.}}, \ and\ \bibinfo {author} {\bibfnamefont {R.~E.}\ \bibnamefont
  {Packard}}} (\bibinfo {year} {2012}),\ \href@noop {} {\bibfield  {journal}
  {\bibinfo  {journal} {Rep. Prog. Phys.}\ }\textbf {\bibinfo {volume} {75}},\
  \bibinfo {pages} {016401}}\BibitemShut {NoStop}%
\bibitem [{\citenamefont {{Schofield, Jr.}}(1971)}]{Schofield:71}%
  \BibitemOpen
  \bibfield  {author} {\bibinfo {author} {\bibnamefont {{Schofield, Jr.}},
  \bibfnamefont {G.}}} (\bibinfo {year} {1971}),\ \href@noop {} {Ph.D. thesis}\
  (\bibinfo  {school} {University of Michigan - Ann Arbor}),\ \bibinfo {note}
  {unpublished and private communication of T.M. Sanders.}\BibitemShut {Stop}%
\bibitem [{\citenamefont {Schwab}\ \emph {et~al.}(1997)\citenamefont {Schwab},
  \citenamefont {Bruckner},\ and\ \citenamefont {Packard}}]{Schwab:97}%
  \BibitemOpen
  \bibfield  {author} {\bibinfo {author} {\bibnamefont {Schwab}, \bibfnamefont
  {K.}}, \bibinfo {author} {\bibfnamefont {N.}~\bibnamefont {Bruckner}}, \ and\
  \bibinfo {author} {\bibfnamefont {R.~E.}\ \bibnamefont {Packard}}} (\bibinfo
  {year} {1997}),\ \href@noop {} {\bibfield  {journal} {\bibinfo  {journal}
  {Nature}\ }\textbf {\bibinfo {volume} {386}},\ \bibinfo {pages}
  {585}}\BibitemShut {NoStop}%
\bibitem [{\citenamefont {Schwab}\ \emph {et~al.}(1998)\citenamefont {Schwab},
  \citenamefont {Bruckner},\ and\ \citenamefont {Packard}}]{Schwab:98}%
  \BibitemOpen
  \bibfield  {author} {\bibinfo {author} {\bibnamefont {Schwab}, \bibfnamefont
  {K.}}, \bibinfo {author} {\bibfnamefont {N.}~\bibnamefont {Bruckner}}, \ and\
  \bibinfo {author} {\bibfnamefont {R.~E.}\ \bibnamefont {Packard}}} (\bibinfo
  {year} {1998}),\ \href@noop {} {\bibfield  {journal} {\bibinfo  {journal} {J.
  Low Temp. Phys.}\ }\textbf {\bibinfo {volume} {110}},\ \bibinfo {pages}
  {1043}}\BibitemShut {NoStop}%
\bibitem [{\citenamefont {Schwab}\ \emph
  {et~al.}(1996{\natexlab{a}})\citenamefont {Schwab}, \citenamefont {Davis},\
  and\ \citenamefont {Packard}}]{Schwab:96a}%
  \BibitemOpen
  \bibfield  {author} {\bibinfo {author} {\bibnamefont {Schwab}, \bibfnamefont
  {K.}}, \bibinfo {author} {\bibfnamefont {J.~C.}\ \bibnamefont {Davis}}, \
  and\ \bibinfo {author} {\bibfnamefont {R.~E.}\ \bibnamefont {Packard}}}
  (\bibinfo {year} {1996}{\natexlab{a}}),\ \href@noop {} {\bibfield  {journal}
  {\bibinfo  {journal} {Czech. J. Phys.}\ }\textbf {\bibinfo {volume}
  {46-S5}},\ \bibinfo {pages} {2739}}\BibitemShut {NoStop}%
\bibitem [{\citenamefont {Schwab}\ \emph
  {et~al.}(1996{\natexlab{b}})\citenamefont {Schwab}, \citenamefont
  {Steinhauer}, \citenamefont {Davis},\ and\ \citenamefont
  {Packard}}]{Schwab:96b}%
  \BibitemOpen
  \bibfield  {author} {\bibinfo {author} {\bibnamefont {Schwab}, \bibfnamefont
  {K.}}, \bibinfo {author} {\bibfnamefont {J.}~\bibnamefont {Steinhauer}},
  \bibinfo {author} {\bibfnamefont {J.~C.}\ \bibnamefont {Davis}}, \ and\
  \bibinfo {author} {\bibfnamefont {R.~E.}\ \bibnamefont {Packard}}} (\bibinfo
  {year} {1996}{\natexlab{b}}),\ \href@noop {} {\bibfield  {journal} {\bibinfo
  {journal} {J. Micromechanical Systems}\ }\textbf {\bibinfo {volume} {5}},\
  \bibinfo {pages} {180}}\BibitemShut {NoStop}%
\bibitem [{\citenamefont {Schwarz}(1978)}]{Schwarz:78}%
  \BibitemOpen
  \bibfield  {author} {\bibinfo {author} {\bibnamefont {Schwarz}, \bibfnamefont
  {K.~W.}}} (\bibinfo {year} {1978}),\ \href@noop {} {\bibfield  {journal}
  {\bibinfo  {journal} {Phys. Rev. B}\ }\textbf {\bibinfo {volume} {18}},\
  \bibinfo {pages} {245}}\BibitemShut {NoStop}%
\bibitem [{\citenamefont {Schwarz}(1981)}]{Schwarz:81}%
  \BibitemOpen
  \bibfield  {author} {\bibinfo {author} {\bibnamefont {Schwarz}, \bibfnamefont
  {K.~W.}}} (\bibinfo {year} {1981}),\ \href@noop {} {\bibfield  {journal}
  {\bibinfo  {journal} {Phys. Rev. Lett.}\ }\textbf {\bibinfo {volume} {47}},\
  \bibinfo {pages} {251}}\BibitemShut {NoStop}%
\bibitem [{\citenamefont {Schwarz}(1983)}]{Schwarz:83}%
  \BibitemOpen
  \bibfield  {author} {\bibinfo {author} {\bibnamefont {Schwarz}, \bibfnamefont
  {K.~W.}}} (\bibinfo {year} {1983}),\ \href@noop {} {\bibfield  {journal}
  {\bibinfo  {journal} {Phys. Rev. Lett.}\ }\textbf {\bibinfo {volume} {50}},\
  \bibinfo {pages} {364}}\BibitemShut {NoStop}%
\bibitem [{\citenamefont {Schwarz}(1985)}]{Schwarz:85}%
  \BibitemOpen
  \bibfield  {author} {\bibinfo {author} {\bibnamefont {Schwarz}, \bibfnamefont
  {K.~W.}}} (\bibinfo {year} {1985}),\ \href@noop {} {\bibfield  {journal}
  {\bibinfo  {journal} {Phys. Rev. B}\ }\textbf {\bibinfo {volume} {31}},\
  \bibinfo {pages} {5782}}\BibitemShut {NoStop}%
\bibitem [{\citenamefont {Schwarz}(1990)}]{Schwarz:90}%
  \BibitemOpen
  \bibfield  {author} {\bibinfo {author} {\bibnamefont {Schwarz}, \bibfnamefont
  {K.~W.}}} (\bibinfo {year} {1990}),\ \href@noop {} {\bibfield  {journal}
  {\bibinfo  {journal} {Phys. Rev. Lett.}\ }\textbf {\bibinfo {volume} {64}},\
  \bibinfo {pages} {1130}}\BibitemShut {NoStop}%
\bibitem [{\citenamefont {Schwarz}(1992)}]{Schwarz:92}%
  \BibitemOpen
  \bibfield  {author} {\bibinfo {author} {\bibnamefont {Schwarz}, \bibfnamefont
  {K.~W.}}} (\bibinfo {year} {1992}),\ \href@noop {} {\bibfield  {journal}
  {\bibinfo  {journal} {Phys. Rev. Lett.}\ }\textbf {\bibinfo {volume} {69}},\
  \bibinfo {pages} {3342}}\BibitemShut {NoStop}%
\bibitem [{\citenamefont {Schwarz}(1993{\natexlab{a}})}]{Schwarz:93c}%
  \BibitemOpen
  \bibfield  {author} {\bibinfo {author} {\bibnamefont {Schwarz}, \bibfnamefont
  {K.~W.}}} (\bibinfo {year} {1993}{\natexlab{a}}),\ \href@noop {} {\bibfield
  {journal} {\bibinfo  {journal} {J. Low Temp. Phys.}\ }\textbf {\bibinfo
  {volume} {93}},\ \bibinfo {pages} {1019}}\BibitemShut {NoStop}%
\bibitem [{\citenamefont {Schwarz}(1993{\natexlab{b}})}]{Schwarz:93}%
  \BibitemOpen
  \bibfield  {author} {\bibinfo {author} {\bibnamefont {Schwarz}, \bibfnamefont
  {K.~W.}}} (\bibinfo {year} {1993}{\natexlab{b}}),\ \href@noop {} {\bibfield
  {journal} {\bibinfo  {journal} {Phys. Rev. Lett.}\ }\textbf {\bibinfo
  {volume} {71}},\ \bibinfo {pages} {259}}\BibitemShut {NoStop}%
\bibitem [{\citenamefont {Schwarz}\ and\ \citenamefont
  {Rozen}(1991)}]{Schwarz:91}%
  \BibitemOpen
  \bibfield  {author} {\bibinfo {author} {\bibnamefont {Schwarz}, \bibfnamefont
  {K.~W.}}, \ and\ \bibinfo {author} {\bibfnamefont {J.~R.}\ \bibnamefont
  {Rozen}}} (\bibinfo {year} {1991}),\ \href@noop {} {\bibfield  {journal}
  {\bibinfo  {journal} {Phys. Rev. B}\ }\textbf {\bibinfo {volume} {44}},\
  \bibinfo {pages} {7563}}\BibitemShut {NoStop}%
\bibitem [{\citenamefont {Shapiro}(1963)}]{Shapiro:63}%
  \BibitemOpen
  \bibfield  {author} {\bibinfo {author} {\bibnamefont {Shapiro}, \bibfnamefont
  {S.}}} (\bibinfo {year} {1963}),\ \href@noop {} {\bibfield  {journal}
  {\bibinfo  {journal} {Phys. Rev. Lett.}\ }\textbf {\bibinfo {volume} {11}},\
  \bibinfo {pages} {80}}\BibitemShut {NoStop}%
\bibitem [{\citenamefont {Shifflett}\ and\ \citenamefont
  {Hess}(1995)}]{Shifflett:95}%
  \BibitemOpen
  \bibfield  {author} {\bibinfo {author} {\bibnamefont {Shifflett},
  \bibfnamefont {G.}}, \ and\ \bibinfo {author} {\bibfnamefont
  {G.}~\bibnamefont {Hess}}} (\bibinfo {year} {1995}),\ \href@noop {}
  {\bibfield  {journal} {\bibinfo  {journal} {J. Low Temp. Phys.}\ }\textbf
  {\bibinfo {volume} {98}},\ \bibinfo {pages} {591}}\BibitemShut {NoStop}%
\bibitem [{\citenamefont {Silaev}(2012)}]{Silaev:12}%
  \BibitemOpen
  \bibfield  {author} {\bibinfo {author} {\bibnamefont {Silaev}, \bibfnamefont
  {M.~A.}}} (\bibinfo {year} {2012}),\ \href@noop {} {\bibfield  {journal}
  {\bibinfo  {journal} {Phys. Rev. Lett.}\ }\textbf {\bibinfo {volume} {108}},\
  \bibinfo {pages} {045303}}\BibitemShut {NoStop}%
\bibitem [{\citenamefont {Silveri}\ \emph {et~al.}(2014)\citenamefont
  {Silveri}, \citenamefont {Turunen},\ and\ \citenamefont
  {Thuneberg}}]{Silveri:14}%
  \BibitemOpen
  \bibfield  {author} {\bibinfo {author} {\bibnamefont {Silveri}, \bibfnamefont
  {M.}}, \bibinfo {author} {\bibfnamefont {T.}~\bibnamefont {Turunen}}, \ and\
  \bibinfo {author} {\bibfnamefont {E.}~\bibnamefont {Thuneberg}}} (\bibinfo
  {year} {2014}),\ \href@noop {} {\bibfield  {journal} {\bibinfo  {journal}
  {Phys. Rev. B}\ }\textbf {\bibinfo {volume} {90}},\ \bibinfo {pages}
  {184513}}\BibitemShut {NoStop}%
\bibitem [{\citenamefont {Simmonds}\ \emph {et~al.}(2000)\citenamefont
  {Simmonds}, \citenamefont {Marchenkov}, \citenamefont {Vitale}, \citenamefont
  {Davis},\ and\ \citenamefont {Packard}}]{Simmonds:00}%
  \BibitemOpen
  \bibfield  {author} {\bibinfo {author} {\bibnamefont {Simmonds},
  \bibfnamefont {R.~W.}}, \bibinfo {author} {\bibfnamefont {A.}~\bibnamefont
  {Marchenkov}}, \bibinfo {author} {\bibfnamefont {S.}~\bibnamefont {Vitale}},
  \bibinfo {author} {\bibfnamefont {J.~C.}\ \bibnamefont {Davis}}, \ and\
  \bibinfo {author} {\bibfnamefont {R.~E.}\ \bibnamefont {Packard}}} (\bibinfo
  {year} {2000}),\ \href@noop {} {\bibfield  {journal} {\bibinfo  {journal}
  {Phys. Rev. Lett.}\ }\textbf {\bibinfo {volume} {84}},\ \bibinfo {pages}
  {6062}}\BibitemShut {NoStop}%
\bibitem [{\citenamefont {Smerzi}\ \emph {et~al.}(2001)\citenamefont {Smerzi},
  \citenamefont {Raghavan}, \citenamefont {Fantoni},\ and\ \citenamefont
  {Shenoy}}]{Smerzi:01}%
  \BibitemOpen
  \bibfield  {author} {\bibinfo {author} {\bibnamefont {Smerzi}, \bibfnamefont
  {A.}}, \bibinfo {author} {\bibfnamefont {S.}~\bibnamefont {Raghavan}},
  \bibinfo {author} {\bibfnamefont {S.}~\bibnamefont {Fantoni}}, \ and\
  \bibinfo {author} {\bibfnamefont {S.}~\bibnamefont {Shenoy}}} (\bibinfo
  {year} {2001}),\ \href@noop {} {\bibfield  {journal} {\bibinfo  {journal}
  {Eur. Phys. J. B}\ }\textbf {\bibinfo {volume} {24}},\ \bibinfo {pages}
  {431}}\BibitemShut {NoStop}%
\bibitem [{\citenamefont {Soininen}\ \emph {et~al.}(1991)\citenamefont
  {Soininen}, \citenamefont {Kopnin},\ and\ \citenamefont
  {Salomaa}}]{Soininen:91}%
  \BibitemOpen
  \bibfield  {author} {\bibinfo {author} {\bibnamefont {Soininen},
  \bibfnamefont {P.~I.}}, \bibinfo {author} {\bibfnamefont {N.}~\bibnamefont
  {Kopnin}}, \ and\ \bibinfo {author} {\bibfnamefont {M.}~\bibnamefont
  {Salomaa}}} (\bibinfo {year} {1991}),\ \href@noop {} {\bibfield  {journal}
  {\bibinfo  {journal} {Europhys. Lett.}\ }\textbf {\bibinfo {volume} {14}},\
  \bibinfo {pages} {49}}\BibitemShut {NoStop}%
\bibitem [{\citenamefont {Soininen}\ \emph {et~al.}(1992)\citenamefont
  {Soininen}, \citenamefont {Kopnin},\ and\ \citenamefont
  {Salomaa}}]{Soininen:92}%
  \BibitemOpen
  \bibfield  {author} {\bibinfo {author} {\bibnamefont {Soininen},
  \bibfnamefont {P.~I.}}, \bibinfo {author} {\bibfnamefont {N.}~\bibnamefont
  {Kopnin}}, \ and\ \bibinfo {author} {\bibfnamefont {M.}~\bibnamefont
  {Salomaa}}} (\bibinfo {year} {1992}),\ \href@noop {} {\bibfield  {journal}
  {\bibinfo  {journal} {Europhys. Lett.}\ }\textbf {\bibinfo {volume} {17}},\
  \bibinfo {pages} {429}},\ \bibinfo {note} {{P}hysica B {\bf 178},
  318}\BibitemShut {NoStop}%
\bibitem [{\citenamefont {Sols}(1994)}]{Sols:94}%
  \BibitemOpen
  \bibfield  {author} {\bibinfo {author} {\bibnamefont {Sols}, \bibfnamefont
  {F.}}} (\bibinfo {year} {1994}),\ \href@noop {} {\bibfield  {journal}
  {\bibinfo  {journal} {Physica B}\ }\textbf {\bibinfo {volume} {194-196}},\
  \bibinfo {pages} {1389}}\BibitemShut {NoStop}%
\bibitem [{\citenamefont {Sonin}(1987)}]{Sonin:87}%
  \BibitemOpen
  \bibfield  {author} {\bibinfo {author} {\bibnamefont {Sonin}, \bibfnamefont
  {E.}}} (\bibinfo {year} {1987}),\ \href@noop {} {\bibfield  {journal}
  {\bibinfo  {journal} {Rev. Mod. Phys.}\ }\textbf {\bibinfo {volume} {59}},\
  \bibinfo {pages} {87}}\BibitemShut {NoStop}%
\bibitem [{\citenamefont {Sonin}(1994)}]{Sonin:94b}%
  \BibitemOpen
  \bibfield  {author} {\bibinfo {author} {\bibnamefont {Sonin}, \bibfnamefont
  {E.}}} (\bibinfo {year} {1994}),\ \href@noop {} {\bibfield  {journal}
  {\bibinfo  {journal} {J. Low Temp. Phys.}\ }\textbf {\bibinfo {volume}
  {97}},\ \bibinfo {pages} {145}}\BibitemShut {NoStop}%
\bibitem [{\citenamefont {Sonin}(1995)}]{Sonin:95}%
  \BibitemOpen
  \bibfield  {author} {\bibinfo {author} {\bibnamefont {Sonin}, \bibfnamefont
  {E.}}} (\bibinfo {year} {1995}),\ \href@noop {} {\bibfield  {journal}
  {\bibinfo  {journal} {Physica B}\ }\textbf {\bibinfo {volume} {210}},\
  \bibinfo {pages} {234}}\BibitemShut {NoStop}%
\bibitem [{\citenamefont {Sonin}(1997)}]{Sonin:97}%
  \BibitemOpen
  \bibfield  {author} {\bibinfo {author} {\bibnamefont {Sonin}, \bibfnamefont
  {E.}}} (\bibinfo {year} {1997}),\ \href@noop {} {\bibfield  {journal}
  {\bibinfo  {journal} {Phys. Rev. B}\ }\textbf {\bibinfo {volume} {55}},\
  \bibinfo {pages} {485}}\BibitemShut {NoStop}%
\bibitem [{\citenamefont {Sonin}\ \emph {et~al.}(1998)\citenamefont {Sonin},
  \citenamefont {Geshkenbein}, \citenamefont {van Otterlo},\ and\ \citenamefont
  {Blatter}}]{Sonin:98}%
  \BibitemOpen
  \bibfield  {author} {\bibinfo {author} {\bibnamefont {Sonin}, \bibfnamefont
  {E.}}, \bibinfo {author} {\bibfnamefont {V.}~\bibnamefont {Geshkenbein}},
  \bibinfo {author} {\bibfnamefont {A.}~\bibnamefont {van Otterlo}}, \ and\
  \bibinfo {author} {\bibfnamefont {G.}~\bibnamefont {Blatter}}} (\bibinfo
  {year} {1998}),\ \href@noop {} {\bibfield  {journal} {\bibinfo  {journal}
  {Phys. Rev. B}\ }\textbf {\bibinfo {volume} {57}},\ \bibinfo {pages}
  {575}}\BibitemShut {NoStop}%
\bibitem [{\citenamefont {Sonin}\ and\ \citenamefont
  {Krusius}(1994)}]{Sonin:94}%
  \BibitemOpen
  \bibfield  {author} {\bibinfo {author} {\bibnamefont {Sonin}, \bibfnamefont
  {E.}}, \ and\ \bibinfo {author} {\bibfnamefont {M.}~\bibnamefont {Krusius}}}
  (\bibinfo {year} {1994}),\ in\ \href@noop {} {\emph {\bibinfo {booktitle}
  {The Vortex State}}},\ \bibinfo {editor} {edited by\ \bibinfo {editor}
  {\bibnamefont {{N. Bontemps \it et al.}}}}\ (\bibinfo  {publisher} {Kluwer
  Academic Publishers - Dordrecht})\ p.\ \bibinfo {pages} {193}\BibitemShut
  {NoStop}%
\bibitem [{\citenamefont {Sonin}(2015)}]{Sonin:15}%
  \BibitemOpen
  \bibfield  {author} {\bibinfo {author} {\bibnamefont {Sonin}, \bibfnamefont
  {E.~B.}}} (\bibinfo {year} {2015}),\ \href@noop {} {\enquote {\bibinfo
  {title} {Dynamics of quantised vortices in superfluids},}\ }\bibinfo {note}
  {In preparation}\BibitemShut {NoStop}%
\bibitem [{\citenamefont {Stedman}(1997)}]{Stedman:97}%
  \BibitemOpen
  \bibfield  {author} {\bibinfo {author} {\bibnamefont {Stedman}, \bibfnamefont
  {G.~E.}}} (\bibinfo {year} {1997}),\ \href@noop {} {\bibfield  {journal}
  {\bibinfo  {journal} {Rep. Prog. Phys.}\ }\textbf {\bibinfo {volume} {60}},\
  \bibinfo {pages} {615}}\BibitemShut {NoStop}%
\bibitem [{\citenamefont {Steinhauer}\ \emph {et~al.}(1995)\citenamefont
  {Steinhauer}, \citenamefont {Schwab}, \citenamefont {Mukharsky},
  \citenamefont {Davis},\ and\ \citenamefont {Packard}}]{Steinhauer:95}%
  \BibitemOpen
  \bibfield  {author} {\bibinfo {author} {\bibnamefont {Steinhauer},
  \bibfnamefont {J.}}, \bibinfo {author} {\bibfnamefont {K.}~\bibnamefont
  {Schwab}}, \bibinfo {author} {\bibfnamefont {Y.}~\bibnamefont {Mukharsky}},
  \bibinfo {author} {\bibfnamefont {J.}~\bibnamefont {Davis}}, \ and\ \bibinfo
  {author} {\bibfnamefont {R.~E.}\ \bibnamefont {Packard}}} (\bibinfo {year}
  {1995}),\ \href@noop {} {\bibfield  {journal} {\bibinfo  {journal} {Phys.
  Rev. Lett.}\ }\textbf {\bibinfo {volume} {74}},\ \bibinfo {pages}
  {5056}}\BibitemShut {NoStop}%
\bibitem [{\citenamefont {Stirling}\ \emph {et~al.}(1976)\citenamefont
  {Stirling}, \citenamefont {Scherm}, \citenamefont {Hilton},\ and\
  \citenamefont {Cowley}}]{Stirling:76}%
  \BibitemOpen
  \bibfield  {author} {\bibinfo {author} {\bibnamefont {Stirling},
  \bibfnamefont {W.~G.}}, \bibinfo {author} {\bibfnamefont {R.}~\bibnamefont
  {Scherm}}, \bibinfo {author} {\bibfnamefont {P.~A.}\ \bibnamefont {Hilton}},
  \ and\ \bibinfo {author} {\bibfnamefont {R.~A.}\ \bibnamefont {Cowley}}}
  (\bibinfo {year} {1976}),\ \href@noop {} {\bibfield  {journal} {\bibinfo
  {journal} {J. Phys. C: Solid State Phys.}\ }\textbf {\bibinfo {volume} {9}},\
  \bibinfo {pages} {1643}}\BibitemShut {NoStop}%
\bibitem [{\citenamefont {Stringari}(2001)}]{Stringari:01}%
  \BibitemOpen
  \bibfield  {author} {\bibinfo {author} {\bibnamefont {Stringari},
  \bibfnamefont {S.}}} (\bibinfo {year} {2001}),\ \href@noop {} {\bibfield
  {journal} {\bibinfo  {journal} {Phys. Rev. Lett.}\ }\textbf {\bibinfo
  {volume} {21}},\ \bibinfo {pages} {4725}}\BibitemShut {NoStop}%
\bibitem [{\citenamefont {Sudraud}\ \emph {et~al.}(1987)\citenamefont
  {Sudraud}, \citenamefont {Ballongue}, \citenamefont {Varoquaux},\ and\
  \citenamefont {Avenel}}]{Sudraud:87}%
  \BibitemOpen
  \bibfield  {author} {\bibinfo {author} {\bibnamefont {Sudraud}, \bibfnamefont
  {P.}}, \bibinfo {author} {\bibfnamefont {P.}~\bibnamefont {Ballongue}},
  \bibinfo {author} {\bibfnamefont {E.}~\bibnamefont {Varoquaux}}, \ and\
  \bibinfo {author} {\bibfnamefont {O.}~\bibnamefont {Avenel}}} (\bibinfo
  {year} {1987}),\ \href@noop {} {\bibfield  {journal} {\bibinfo  {journal} {J.
  Appl. Phys.}\ }\textbf {\bibinfo {volume} {62}},\ \bibinfo {pages}
  {2163}}\BibitemShut {NoStop}%
\bibitem [{\citenamefont {Sukhatme}\ \emph {et~al.}(2001)\citenamefont
  {Sukhatme}, \citenamefont {Mukharsky}, \citenamefont {Chui},\ and\
  \citenamefont {Pearson}}]{Sukhatme:01}%
  \BibitemOpen
  \bibfield  {author} {\bibinfo {author} {\bibnamefont {Sukhatme},
  \bibfnamefont {K.}}, \bibinfo {author} {\bibfnamefont {Y.}~\bibnamefont
  {Mukharsky}}, \bibinfo {author} {\bibfnamefont {T.}~\bibnamefont {Chui}}, \
  and\ \bibinfo {author} {\bibfnamefont {D.}~\bibnamefont {Pearson}}} (\bibinfo
  {year} {2001}),\ \href@noop {} {\bibfield  {journal} {\bibinfo  {journal}
  {Nature}\ }\textbf {\bibinfo {volume} {411}},\ \bibinfo {pages}
  {280}}\BibitemShut {NoStop}%
\bibitem [{\citenamefont {Svistunov}(1995)}]{Svistunov:95}%
  \BibitemOpen
  \bibfield  {author} {\bibinfo {author} {\bibnamefont {Svistunov},
  \bibfnamefont {B.}}} (\bibinfo {year} {1995}),\ \href@noop {} {\bibfield
  {journal} {\bibinfo  {journal} {Phys. Rev. B}\ }\textbf {\bibinfo {volume}
  {52}},\ \bibinfo {pages} {3647}}\BibitemShut {NoStop}%
\bibitem [{\citenamefont {Thuneberg}(1988)}]{Thuneberg:88}%
  \BibitemOpen
  \bibfield  {author} {\bibinfo {author} {\bibnamefont {Thuneberg},
  \bibfnamefont {E.}}} (\bibinfo {year} {1988}),\ \href@noop {} {\bibfield
  {journal} {\bibinfo  {journal} {Europhys. Lett.}\ }\textbf {\bibinfo {volume}
  {7}},\ \bibinfo {pages} {441}}\BibitemShut {NoStop}%
\bibitem [{\citenamefont {Thuneberg}(2005)}]{Thuneberg:05}%
  \BibitemOpen
  \bibfield  {author} {\bibinfo {author} {\bibnamefont {Thuneberg},
  \bibfnamefont {E.}}} (\bibinfo {year} {2005}),\ \href@noop {} {\enquote
  {\bibinfo {title} {Theory of {J}osephson phenomena in superfluid 3{H}e},}\
  }\bibinfo {note} {ArXiv:cond-mat 0509504}\BibitemShut {NoStop}%
\bibitem [{\citenamefont {Thuneberg}\ \emph {et~al.}(1990)\citenamefont
  {Thuneberg}, \citenamefont {Kurkij\"arvi},\ and\ \citenamefont
  {Sauls}}]{Thuneberg:90}%
  \BibitemOpen
  \bibfield  {author} {\bibinfo {author} {\bibnamefont {Thuneberg},
  \bibfnamefont {E.}}, \bibinfo {author} {\bibfnamefont {J.}~\bibnamefont
  {Kurkij\"arvi}}, \ and\ \bibinfo {author} {\bibfnamefont {J.}~\bibnamefont
  {Sauls}}} (\bibinfo {year} {1990}),\ \href@noop {} {\bibfield  {journal}
  {\bibinfo  {journal} {Physica B}\ }\textbf {\bibinfo {volume} {165\&166}},\
  \bibinfo {pages} {755}}\BibitemShut {NoStop}%
\bibitem [{\citenamefont {Tilley}\ and\ \citenamefont
  {Tilley}(1990)}]{Tilley:90}%
  \BibitemOpen
  \bibfield  {author} {\bibinfo {author} {\bibnamefont {Tilley}, \bibfnamefont
  {D.}}, \ and\ \bibinfo {author} {\bibfnamefont {J.}~\bibnamefont {Tilley}}}
  (\bibinfo {year} {1990}),\ \href@noop {} {\emph {\bibinfo {title}
  {Superfluidity and Superconductivity}}},\ \bibinfo {edition} {third edition}\
  ed.\ (\bibinfo  {publisher} {IOP Publishing Ltd - Bristol})\BibitemShut
  {NoStop}%
\bibitem [{\citenamefont {Tisza}(1938)}]{Tisza:38}%
  \BibitemOpen
  \bibfield  {author} {\bibinfo {author} {\bibnamefont {Tisza}, \bibfnamefont
  {L.}}} (\bibinfo {year} {1938}),\ \href@noop {} {\bibfield  {journal}
  {\bibinfo  {journal} {Nature}\ }\textbf {\bibinfo {volume} {141}},\ \bibinfo
  {pages} {913}},\ \bibinfo {note} {\mbox{C}ompt. \mbox{R}end. \mbox{A}cad.
  \mbox{S}ciences (\mbox{P}aris), {\bf 207}, 1035, 1186 (1938)}\BibitemShut
  {NoStop}%
\bibitem [{\citenamefont {Trela}\ and\ \citenamefont
  {Fairbank}(1967)}]{Trela:67}%
  \BibitemOpen
  \bibfield  {author} {\bibinfo {author} {\bibnamefont {Trela}, \bibfnamefont
  {W.}}, \ and\ \bibinfo {author} {\bibfnamefont {W.}~\bibnamefont {Fairbank}}}
  (\bibinfo {year} {1967}),\ \href@noop {} {\bibfield  {journal} {\bibinfo
  {journal} {Phys. Rev. Lett.}\ }\textbf {\bibinfo {volume} {19}},\ \bibinfo
  {pages} {822}}\BibitemShut {NoStop}%
\bibitem [{\citenamefont {Tsubota}\ \emph {et~al.}(2000)\citenamefont
  {Tsubota}, \citenamefont {Araki},\ and\ \citenamefont
  {Nemirovskii}}]{Tsubota:00}%
  \BibitemOpen
  \bibfield  {author} {\bibinfo {author} {\bibnamefont {Tsubota}, \bibfnamefont
  {M.}}, \bibinfo {author} {\bibfnamefont {T.}~\bibnamefont {Araki}}, \ and\
  \bibinfo {author} {\bibfnamefont {S.~K.}\ \bibnamefont {Nemirovskii}}}
  (\bibinfo {year} {2000}),\ \href@noop {} {\bibfield  {journal} {\bibinfo
  {journal} {Phys. Rev. B}\ }\textbf {\bibinfo {volume} {62}},\ \bibinfo
  {pages} {11751}}\BibitemShut {NoStop}%
\bibitem [{\citenamefont {Tsubota}\ and\ \citenamefont
  {Kobayashi}(2009)}]{Tsubota:09}%
  \BibitemOpen
  \bibfield  {author} {\bibinfo {author} {\bibnamefont {Tsubota}, \bibfnamefont
  {M.}}, \ and\ \bibinfo {author} {\bibfnamefont {M.}~\bibnamefont
  {Kobayashi}}} (\bibinfo {year} {2009}),\ \enquote {\bibinfo {title} {{Prog.
  Low T. Phys., Vol. IX}},}\ Chap.~\bibinfo {chapter} {1}\ (\bibinfo
  {publisher} {Elsevier - Amsterdam})\ \bibinfo {note} {and references
  therein}\BibitemShut {NoStop}%
\bibitem [{\citenamefont {Tsubota}\ and\ \citenamefont
  {Maekawa}(1994)}]{Tsubota:94}%
  \BibitemOpen
  \bibfield  {author} {\bibinfo {author} {\bibnamefont {Tsubota}, \bibfnamefont
  {M.}}, \ and\ \bibinfo {author} {\bibfnamefont {S.}~\bibnamefont {Maekawa}}}
  (\bibinfo {year} {1994}),\ \href@noop {} {\bibfield  {journal} {\bibinfo
  {journal} {Physica B}\ }\textbf {\bibinfo {volume} {194-196}},\ \bibinfo
  {pages} {721}}\BibitemShut {NoStop}%
\bibitem [{\citenamefont {Ullah}\ and\ \citenamefont
  {Fetter}(1989)}]{Ullah:89}%
  \BibitemOpen
  \bibfield  {author} {\bibinfo {author} {\bibnamefont {Ullah}, \bibfnamefont
  {S.}}, \ and\ \bibinfo {author} {\bibfnamefont {A.}~\bibnamefont {Fetter}}}
  (\bibinfo {year} {1989}),\ \href@noop {} {\bibfield  {journal} {\bibinfo
  {journal} {Phys. Rev. B}\ }\textbf {\bibinfo {volume} {39}},\ \bibinfo
  {pages} {4186}}\BibitemShut {NoStop}%
\bibitem [{\citenamefont {Varoquaux}(2000)}]{Varoquaux:00a}%
  \BibitemOpen
  \bibfield  {author} {\bibinfo {author} {\bibnamefont {Varoquaux},
  \bibfnamefont {E.}}} (\bibinfo {year} {2000}),\ in\ \href@noop {} {\emph
  {\bibinfo {booktitle} {Topological Defects and the Non-Equilibrium Dynamics
  of Symmetry Breaking Phase Transitions}}},\ \bibinfo {editor} {edited by\
  \bibinfo {editor} {\bibfnamefont {Y.}~\bibnamefont {Bunkov}}\ and\ \bibinfo
  {editor} {\bibfnamefont {H.}~\bibnamefont {Godfrin}}}\ (\bibinfo  {publisher}
  {Kluwer Academic Publishers - Dordrecht})\ p.\ \bibinfo {pages}
  {303}\BibitemShut {NoStop}%
\bibitem [{\citenamefont {Varoquaux}(2001)}]{Varoquaux:01b}%
  \BibitemOpen
  \bibfield  {author} {\bibinfo {author} {\bibnamefont {Varoquaux},
  \bibfnamefont {E.}}} (\bibinfo {year} {2001}),\ in\ \href@noop {} {\emph
  {\bibinfo {booktitle} {Bose-Einstein condensates and atom lasers}}},\
  \bibinfo {editor} {edited by\ \bibinfo {editor} {\bibfnamefont
  {A.}~\bibnamefont {Aspect}}\ and\ \bibinfo {editor} {\bibfnamefont
  {J.}~\bibnamefont {Dalibard}}}\ (\bibinfo  {publisher} {C.R. Acad. Sci.
  Paris, t.2, S\'erie IV})\ p.\ \bibinfo {pages} {531}\BibitemShut {NoStop}%
\bibitem [{\citenamefont {Varoquaux}(2006)}]{Varoquaux:06}%
  \BibitemOpen
  \bibfield  {author} {\bibinfo {author} {\bibnamefont {Varoquaux},
  \bibfnamefont {E.}}} (\bibinfo {year} {2006}),\ \href@noop {} {\bibfield
  {journal} {\bibinfo  {journal} {C. R. Physique}\ }\textbf {\bibinfo {volume}
  {7}},\ \bibinfo {pages} {1101}}\BibitemShut {NoStop}%
\bibitem [{\citenamefont {Varoquaux}\ and\ \citenamefont
  {Avenel}(1994)}]{Varoquaux:94}%
  \BibitemOpen
  \bibfield  {author} {\bibinfo {author} {\bibnamefont {Varoquaux},
  \bibfnamefont {E.}}, \ and\ \bibinfo {author} {\bibfnamefont
  {O.}~\bibnamefont {Avenel}}} (\bibinfo {year} {1994}),\ \href@noop {}
  {\bibfield  {journal} {\bibinfo  {journal} {Physica B}\ }\textbf {\bibinfo
  {volume} {197}},\ \bibinfo {pages} {306}}\BibitemShut {NoStop}%
\bibitem [{\citenamefont {Varoquaux}\ and\ \citenamefont
  {Avenel}(1996{\natexlab{a}})}]{Varoquaux:96}%
  \BibitemOpen
  \bibfield  {author} {\bibinfo {author} {\bibnamefont {Varoquaux},
  \bibfnamefont {E.}}, \ and\ \bibinfo {author} {\bibfnamefont
  {O.}~\bibnamefont {Avenel}}} (\bibinfo {year} {1996}{\natexlab{a}}),\
  \href@noop {} {\bibfield  {journal} {\bibinfo  {journal} {Phys. Rev. Lett.}\
  }\textbf {\bibinfo {volume} {76}},\ \bibinfo {pages} {1180}}\BibitemShut
  {NoStop}%
\bibitem [{\citenamefont {Varoquaux}\ and\ \citenamefont
  {Avenel}(1996{\natexlab{b}})}]{Varoquaux:96a}%
  \BibitemOpen
  \bibfield  {author} {\bibinfo {author} {\bibnamefont {Varoquaux},
  \bibfnamefont {E.}}, \ and\ \bibinfo {author} {\bibfnamefont
  {O.}~\bibnamefont {Avenel}}} (\bibinfo {year} {1996}{\natexlab{b}}),\
  \href@noop {} {\bibfield  {journal} {\bibinfo  {journal} {Czech. J. Phys.}\
  }\textbf {\bibinfo {volume} {46, Suppl. S1}},\ \bibinfo {pages}
  {41}}\BibitemShut {NoStop}%
\bibitem [{\citenamefont {Varoquaux}\ and\ \citenamefont
  {Avenel}(2003)}]{Varoquaux:03}%
  \BibitemOpen
  \bibfield  {author} {\bibinfo {author} {\bibnamefont {Varoquaux},
  \bibfnamefont {E.}}, \ and\ \bibinfo {author} {\bibfnamefont
  {O.}~\bibnamefont {Avenel}}} (\bibinfo {year} {2003}),\ \href@noop {}
  {\bibfield  {journal} {\bibinfo  {journal} {Phys. Rev. B}\ }\textbf {\bibinfo
  {volume} {68}},\ \bibinfo {pages} {054515}}\BibitemShut {NoStop}%
\bibitem [{\citenamefont {Varoquaux}\ \emph {et~al.}(1995)\citenamefont
  {Varoquaux}, \citenamefont {Avenel}, \citenamefont {Bernard},\ and\
  \citenamefont {Burkhart}}]{Varoquaux:95}%
  \BibitemOpen
  \bibfield  {author} {\bibinfo {author} {\bibnamefont {Varoquaux},
  \bibfnamefont {E.}}, \bibinfo {author} {\bibfnamefont {O.}~\bibnamefont
  {Avenel}}, \bibinfo {author} {\bibfnamefont {M.}~\bibnamefont {Bernard}}, \
  and\ \bibinfo {author} {\bibfnamefont {S.}~\bibnamefont {Burkhart}}}
  (\bibinfo {year} {1995}),\ \href@noop {} {\bibfield  {journal} {\bibinfo
  {journal} {J. Low Temp. Phys.}\ }\textbf {\bibinfo {volume} {101}},\ \bibinfo
  {pages} {821}}\BibitemShut {NoStop}%
\bibitem [{\citenamefont {Varoquaux}\ \emph {et~al.}(1998)\citenamefont
  {Varoquaux}, \citenamefont {Avenel}, \citenamefont {Hakonen},\ and\
  \citenamefont {Mukharsky}}]{Varoquaux:98}%
  \BibitemOpen
  \bibfield  {author} {\bibinfo {author} {\bibnamefont {Varoquaux},
  \bibfnamefont {E.}}, \bibinfo {author} {\bibfnamefont {O.}~\bibnamefont
  {Avenel}}, \bibinfo {author} {\bibfnamefont {P.}~\bibnamefont {Hakonen}}, \
  and\ \bibinfo {author} {\bibfnamefont {Y.}~\bibnamefont {Mukharsky}}}
  (\bibinfo {year} {1998}),\ \href@noop {} {\bibfield  {journal} {\bibinfo
  {journal} {Physica B}\ }\textbf {\bibinfo {volume} {255}},\ \bibinfo {pages}
  {55}}\BibitemShut {NoStop}%
\bibitem [{\citenamefont {Varoquaux}\ \emph {et~al.}(1999)\citenamefont
  {Varoquaux}, \citenamefont {Avenel}, \citenamefont {Hakonen},\ and\
  \citenamefont {Mukharsky}}]{Varoquaux:99}%
  \BibitemOpen
  \bibfield  {author} {\bibinfo {author} {\bibnamefont {Varoquaux},
  \bibfnamefont {E.}}, \bibinfo {author} {\bibfnamefont {O.}~\bibnamefont
  {Avenel}}, \bibinfo {author} {\bibfnamefont {P.}~\bibnamefont {Hakonen}}, \
  and\ \bibinfo {author} {\bibfnamefont {Y.}~\bibnamefont {Mukharsky}}}
  (\bibinfo {year} {1999}),\ in\ \href@noop {} {\emph {\bibinfo {booktitle}
  {Quantum Coherence and Decoherence - ISQM - Tokyo '98}}},\ \bibinfo {editor}
  {edited by\ \bibinfo {editor} {\bibfnamefont {Y.}~\bibnamefont {Ono}}\ and\
  \bibinfo {editor} {\bibfnamefont {K.}~\bibnamefont {Fujikawa}}}\ (\bibinfo
  {publisher} {Elsevier Science B.V.})\ p.\ \bibinfo {pages} {287}\BibitemShut
  {NoStop}%
\bibitem [{\citenamefont {Varoquaux}\ \emph {et~al.}(2000)\citenamefont
  {Varoquaux}, \citenamefont {Avenel}, \citenamefont {Hakonen},\ and\
  \citenamefont {Mukharsky}}]{Varoquaux:00b}%
  \BibitemOpen
  \bibfield  {author} {\bibinfo {author} {\bibnamefont {Varoquaux},
  \bibfnamefont {E.}}, \bibinfo {author} {\bibfnamefont {O.}~\bibnamefont
  {Avenel}}, \bibinfo {author} {\bibfnamefont {P.}~\bibnamefont {Hakonen}}, \
  and\ \bibinfo {author} {\bibfnamefont {Y.}~\bibnamefont {Mukharsky}}}
  (\bibinfo {year} {2000}),\ \href@noop {} {\bibfield  {journal} {\bibinfo
  {journal} {Physica B}\ }\textbf {\bibinfo {volume} {87-88}},\ \bibinfo
  {pages} {284}}\BibitemShut {NoStop}%
\bibitem [{\citenamefont {Varoquaux}\ \emph {et~al.}(1992)\citenamefont
  {Varoquaux}, \citenamefont {Avenel}, \citenamefont {Ihas},\ and\
  \citenamefont {Salmelin}}]{Varoquaux:92}%
  \BibitemOpen
  \bibfield  {author} {\bibinfo {author} {\bibnamefont {Varoquaux},
  \bibfnamefont {E.}}, \bibinfo {author} {\bibfnamefont {O.}~\bibnamefont
  {Avenel}}, \bibinfo {author} {\bibfnamefont {G.}~\bibnamefont {Ihas}}, \ and\
  \bibinfo {author} {\bibfnamefont {R.}~\bibnamefont {Salmelin}}} (\bibinfo
  {year} {1992}),\ \href@noop {} {\bibfield  {journal} {\bibinfo  {journal}
  {Physica B}\ }\textbf {\bibinfo {volume} {178}},\ \bibinfo {pages}
  {309}}\BibitemShut {NoStop}%
\bibitem [{\citenamefont {Varoquaux}\ \emph {et~al.}(1987)\citenamefont
  {Varoquaux}, \citenamefont {Avenel},\ and\ \citenamefont
  {Meisel}}]{Varoquaux:87}%
  \BibitemOpen
  \bibfield  {author} {\bibinfo {author} {\bibnamefont {Varoquaux},
  \bibfnamefont {E.}}, \bibinfo {author} {\bibfnamefont {O.}~\bibnamefont
  {Avenel}}, \ and\ \bibinfo {author} {\bibfnamefont {M.}~\bibnamefont
  {Meisel}}} (\bibinfo {year} {1987}),\ \href@noop {} {\bibfield  {journal}
  {\bibinfo  {journal} {Can. J. Phys.}\ }\textbf {\bibinfo {volume} {65}},\
  \bibinfo {pages} {1377}}\BibitemShut {NoStop}%
\bibitem [{\citenamefont {Varoquaux}\ \emph {et~al.}(2001)\citenamefont
  {Varoquaux}, \citenamefont {Avenel}, \citenamefont {Mukharsky},\ and\
  \citenamefont {Hakonen}}]{Varoquaux:01a}%
  \BibitemOpen
  \bibfield  {author} {\bibinfo {author} {\bibnamefont {Varoquaux},
  \bibfnamefont {E.}}, \bibinfo {author} {\bibfnamefont {O.}~\bibnamefont
  {Avenel}}, \bibinfo {author} {\bibfnamefont {Y.}~\bibnamefont {Mukharsky}}, \
  and\ \bibinfo {author} {\bibfnamefont {P.}~\bibnamefont {Hakonen}}} (\bibinfo
  {year} {2001}),\ \bibinfo {note} {{\it loc. cit.} \citet{Barenghi:01},
  p.36}\BibitemShut {NoStop}%
\bibitem [{\citenamefont {Varoquaux}\ \emph {et~al.}(1993)\citenamefont
  {Varoquaux}, \citenamefont {Ihas}, \citenamefont {Avenel},\ and\
  \citenamefont {Aarts}}]{Varoquaux:93}%
  \BibitemOpen
  \bibfield  {author} {\bibinfo {author} {\bibnamefont {Varoquaux},
  \bibfnamefont {E.}}, \bibinfo {author} {\bibfnamefont {G.}~\bibnamefont
  {Ihas}}, \bibinfo {author} {\bibfnamefont {O.}~\bibnamefont {Avenel}}, \ and\
  \bibinfo {author} {\bibfnamefont {R.}~\bibnamefont {Aarts}}} (\bibinfo {year}
  {1993}),\ \href@noop {} {\bibfield  {journal} {\bibinfo  {journal} {Phys.
  Rev. Lett.}\ }\textbf {\bibinfo {volume} {70}},\ \bibinfo {pages}
  {2114}}\BibitemShut {NoStop}%
\bibitem [{\citenamefont {Varoquaux}\ \emph {et~al.}(1986)\citenamefont
  {Varoquaux}, \citenamefont {Meisel},\ and\ \citenamefont
  {Avenel}}]{Varoquaux:86}%
  \BibitemOpen
  \bibfield  {author} {\bibinfo {author} {\bibnamefont {Varoquaux},
  \bibfnamefont {E.}}, \bibinfo {author} {\bibfnamefont {M.}~\bibnamefont
  {Meisel}}, \ and\ \bibinfo {author} {\bibfnamefont {O.}~\bibnamefont
  {Avenel}}} (\bibinfo {year} {1986}),\ \href@noop {} {\bibfield  {journal}
  {\bibinfo  {journal} {Phys. Rev. Lett.}\ }\textbf {\bibinfo {volume} {57}},\
  \bibinfo {pages} {2291}}\BibitemShut {NoStop}%
\bibitem [{\citenamefont {Varoquaux}\ and\ \citenamefont
  {Varoquaux}(2008)}]{Varoquaux:08}%
  \BibitemOpen
  \bibfield  {author} {\bibinfo {author} {\bibnamefont {Varoquaux},
  \bibfnamefont {E.}}, \ and\ \bibinfo {author} {\bibfnamefont
  {G.}~\bibnamefont {Varoquaux}}} (\bibinfo {year} {2008}),\ \href@noop {}
  {\bibfield  {journal} {\bibinfo  {journal} {{U}sp. Fiz. Nauk}\ }\textbf
  {\bibinfo {volume} {178}},\ \bibinfo {pages} {217}},\ \bibinfo {note}
  {[{P}hysics-{U}spekhi {\bf{51}}, 205 (2008)]}\BibitemShut {NoStop}%
\bibitem [{\citenamefont {Varoquaux}\ \emph {et~al.}(1991)\citenamefont
  {Varoquaux}, \citenamefont {{Zimmermann, Jr.}},\ and\ \citenamefont
  {Avenel}}]{Varoquaux:91}%
  \BibitemOpen
  \bibfield  {author} {\bibinfo {author} {\bibnamefont {Varoquaux},
  \bibfnamefont {E.}}, \bibinfo {author} {\bibfnamefont {W.}~\bibnamefont
  {{Zimmermann, Jr.}}}, \ and\ \bibinfo {author} {\bibfnamefont
  {O.}~\bibnamefont {Avenel}}} (\bibinfo {year} {1991}),\ \bibinfo {note} {{\it
  loc. cit.}\citet{Wyatt:91}, p.343}\BibitemShut {NoStop}%
\bibitem [{\citenamefont {Verbeek}\ \emph {et~al.}(1974)\citenamefont
  {Verbeek}, \citenamefont {Spronsen}, \citenamefont {Mars}, \citenamefont
  {Beelen}, \citenamefont {Ouboter},\ and\ \citenamefont
  {Taconise}}]{Verbeek:74}%
  \BibitemOpen
  \bibfield  {author} {\bibinfo {author} {\bibnamefont {Verbeek}, \bibfnamefont
  {H.}}, \bibinfo {author} {\bibfnamefont {E.~V.}\ \bibnamefont {Spronsen}},
  \bibinfo {author} {\bibfnamefont {H.}~\bibnamefont {Mars}}, \bibinfo {author}
  {\bibfnamefont {H.~V.}\ \bibnamefont {Beelen}}, \bibinfo {author}
  {\bibfnamefont {R.~D.~B.}\ \bibnamefont {Ouboter}}, \ and\ \bibinfo {author}
  {\bibfnamefont {K.~W.}\ \bibnamefont {Taconise}}} (\bibinfo {year} {1974}),\
  \href@noop {} {\bibfield  {journal} {\bibinfo  {journal} {Physica}\ }\textbf
  {\bibinfo {volume} {73}},\ \bibinfo {pages} {621}}\BibitemShut {NoStop}%
\bibitem [{\citenamefont {Viljas}\ and\ \citenamefont
  {Thuneberg}(1999)}]{Viljas:99}%
  \BibitemOpen
  \bibfield  {author} {\bibinfo {author} {\bibnamefont {Viljas}, \bibfnamefont
  {J.}}, \ and\ \bibinfo {author} {\bibfnamefont {E.}~\bibnamefont
  {Thuneberg}}} (\bibinfo {year} {1999}),\ \href@noop {} {\bibfield  {journal}
  {\bibinfo  {journal} {Phys. Rev. Lett.}\ }\textbf {\bibinfo {volume} {83}},\
  \bibinfo {pages} {3868}}\BibitemShut {NoStop}%
\bibitem [{\citenamefont {Viljas}\ and\ \citenamefont
  {Thuneberg}(2002{\natexlab{a}})}]{Viljas:02-JLTP}%
  \BibitemOpen
  \bibfield  {author} {\bibinfo {author} {\bibnamefont {Viljas}, \bibfnamefont
  {J.}}, \ and\ \bibinfo {author} {\bibfnamefont {E.}~\bibnamefont
  {Thuneberg}}} (\bibinfo {year} {2002}{\natexlab{a}}),\ \href@noop {}
  {\bibfield  {journal} {\bibinfo  {journal} {J. Low Temp. Phys.}\ }\textbf
  {\bibinfo {volume} {129}},\ \bibinfo {pages} {423}}\BibitemShut {NoStop}%
\bibitem [{\citenamefont {Viljas}\ and\ \citenamefont
  {Thuneberg}(2002{\natexlab{b}})}]{Viljas:02-PRB}%
  \BibitemOpen
  \bibfield  {author} {\bibinfo {author} {\bibnamefont {Viljas}, \bibfnamefont
  {J.}}, \ and\ \bibinfo {author} {\bibfnamefont {E.}~\bibnamefont
  {Thuneberg}}} (\bibinfo {year} {2002}{\natexlab{b}}),\ \href@noop {}
  {\bibfield  {journal} {\bibinfo  {journal} {Phys. Rev. B}\ }\textbf {\bibinfo
  {volume} {65}},\ \bibinfo {pages} {064530}}\BibitemShut {NoStop}%
\bibitem [{\citenamefont {Viljas}(2005)}]{Viljas:05}%
  \BibitemOpen
  \bibfield  {author} {\bibinfo {author} {\bibnamefont {Viljas}, \bibfnamefont
  {J.~K.}}} (\bibinfo {year} {2005}),\ \href@noop {} {\bibfield  {journal}
  {\bibinfo  {journal} {Phys. Rev. B}\ }\textbf {\bibinfo {volume} {71}},\
  \bibinfo {pages} {064509}}\BibitemShut {NoStop}%
\bibitem [{\citenamefont {Viljas}\ and\ \citenamefont
  {Thuneberg}(2004{\natexlab{a}})}]{Viljas:04-PRL}%
  \BibitemOpen
  \bibfield  {author} {\bibinfo {author} {\bibnamefont {Viljas}, \bibfnamefont
  {J.~K.}}, \ and\ \bibinfo {author} {\bibfnamefont {E.~V.}\ \bibnamefont
  {Thuneberg}}} (\bibinfo {year} {2004}{\natexlab{a}}),\ \href@noop {}
  {\bibfield  {journal} {\bibinfo  {journal} {Phys. Rev. lett.}\ }\textbf
  {\bibinfo {volume} {93}},\ \bibinfo {pages} {205301}}\BibitemShut {NoStop}%
\bibitem [{\citenamefont {Viljas}\ and\ \citenamefont
  {Thuneberg}(2004{\natexlab{b}})}]{Viljas:04-JLTP}%
  \BibitemOpen
  \bibfield  {author} {\bibinfo {author} {\bibnamefont {Viljas}, \bibfnamefont
  {J.~K.}}, \ and\ \bibinfo {author} {\bibfnamefont {E.~V.}\ \bibnamefont
  {Thuneberg}}} (\bibinfo {year} {2004}{\natexlab{b}}),\ \href@noop {}
  {\bibfield  {journal} {\bibinfo  {journal} {J. Low Temp. Phys.}\ }\textbf
  {\bibinfo {volume} {136}},\ \bibinfo {pages} {329}}\BibitemShut {NoStop}%
\bibitem [{\citenamefont {Vinen}(1961)}]{Vinen:61}%
  \BibitemOpen
  \bibfield  {author} {\bibinfo {author} {\bibnamefont {Vinen}, \bibfnamefont
  {W.}}} (\bibinfo {year} {1961}),\ \href@noop {} {\bibfield  {journal}
  {\bibinfo  {journal} {Proc. Roy. Soc. London}\ }\textbf {\bibinfo {volume}
  {260}},\ \bibinfo {pages} {218}}\BibitemShut {NoStop}%
\bibitem [{\citenamefont {Vinen}(1963)}]{Vinen:63}%
  \BibitemOpen
  \bibfield  {author} {\bibinfo {author} {\bibnamefont {Vinen}, \bibfnamefont
  {W.}}} (\bibinfo {year} {1963}),\ in\ \href@noop {} {\emph {\bibinfo
  {booktitle} {Liquid Helium}}}\ (\bibinfo  {publisher} {ed. G. Careri,
  Academic Press - New-York})\ p.\ \bibinfo {pages} {336}\BibitemShut {NoStop}%
\bibitem [{\citenamefont {Vinen}(1966)}]{Vinen:66}%
  \BibitemOpen
  \bibfield  {author} {\bibinfo {author} {\bibnamefont {Vinen}, \bibfnamefont
  {W.}}} (\bibinfo {year} {1966}),\ in\ \href@noop {} {\emph {\bibinfo
  {booktitle} {Quantum Fluids}}}\ (\bibinfo  {publisher} {ed. D. F. Brewer,
  North-Holland - Amsterdam})\ p.~\bibinfo {pages} {74}\BibitemShut {NoStop}%
\bibitem [{\citenamefont {Vinen}(1968)}]{Vinen:68}%
  \BibitemOpen
  \bibfield  {author} {\bibinfo {author} {\bibnamefont {Vinen}, \bibfnamefont
  {W.}}} (\bibinfo {year} {1968}),\ \href@noop {} {\bibfield  {journal}
  {\bibinfo  {journal} {Rep. Prog. Phys.}\ }\textbf {\bibinfo {volume}
  {XXXI}},\ \bibinfo {pages} {61}}\BibitemShut {NoStop}%
\bibitem [{\citenamefont {Vollhardt}\ and\ \citenamefont
  {W\"olfle}(1990)}]{Vollhardt:90}%
  \BibitemOpen
  \bibfield  {author} {\bibinfo {author} {\bibnamefont {Vollhardt},
  \bibfnamefont {D.}}, \ and\ \bibinfo {author} {\bibfnamefont
  {P.}~\bibnamefont {W\"olfle}}} (\bibinfo {year} {1990}),\ \href@noop {}
  {\emph {\bibinfo {title} {The Superfluid Phases of Helium 3}}}\ (\bibinfo
  {publisher} {Taylor\&Francis - London})\BibitemShut {NoStop}%
\bibitem [{\citenamefont {Volovik}(1972)}]{Volovik:72}%
  \BibitemOpen
  \bibfield  {author} {\bibinfo {author} {\bibnamefont {Volovik}, \bibfnamefont
  {G.}}} (\bibinfo {year} {1972}),\ \href@noop {} {\bibfield  {journal}
  {\bibinfo  {journal} {Sov. Phys. JETP Lett.}\ }\textbf {\bibinfo {volume}
  {15}},\ \bibinfo {pages} {81}}\BibitemShut {NoStop}%
\bibitem [{\citenamefont {Volovik}(1997)}]{Volovik:97}%
  \BibitemOpen
  \bibfield  {author} {\bibinfo {author} {\bibnamefont {Volovik}, \bibfnamefont
  {G.}}} (\bibinfo {year} {1997}),\ \href@noop {} {\bibfield  {journal}
  {\bibinfo  {journal} {JETP Lett.}\ }\textbf {\bibinfo {volume} {65}},\
  \bibinfo {pages} {217}}\BibitemShut {NoStop}%
\bibitem [{\citenamefont {Volovik}(2003)}]{Volovik:03}%
  \BibitemOpen
  \bibfield  {author} {\bibinfo {author} {\bibnamefont {Volovik}, \bibfnamefont
  {G.~E.}}} (\bibinfo {year} {2003}),\ \href@noop {} {\emph {\bibinfo {title}
  {The Universe in a Helium Droplet}}}\ (\bibinfo  {publisher} {Oxford
  University Press - Oxford})\ \bibinfo {note} {ch. 31}\BibitemShut {NoStop}%
\bibitem [{\citenamefont {Waxman}\ and\ \citenamefont
  {Leggett}(1985)}]{Waxman:85}%
  \BibitemOpen
  \bibfield  {author} {\bibinfo {author} {\bibnamefont {Waxman}, \bibfnamefont
  {D.}}, \ and\ \bibinfo {author} {\bibfnamefont {A.}~\bibnamefont {Leggett}}}
  (\bibinfo {year} {1985}),\ \href@noop {} {\bibfield  {journal} {\bibinfo
  {journal} {Phys. Rev. B}\ }\textbf {\bibinfo {volume} {32}},\ \bibinfo
  {pages} {4450}}\BibitemShut {NoStop}%
\bibitem [{\citenamefont {Wheatley}(1975{\natexlab{a}})}]{Wheatley:75a}%
  \BibitemOpen
  \bibfield  {author} {\bibinfo {author} {\bibnamefont {Wheatley},
  \bibfnamefont {J.}}} (\bibinfo {year} {1975}{\natexlab{a}}),\ \href@noop {}
  {\bibfield  {journal} {\bibinfo  {journal} {Rev. Mod. Phys.}\ }\textbf
  {\bibinfo {volume} {47}},\ \bibinfo {pages} {415}}\BibitemShut {NoStop}%
\bibitem [{\citenamefont {Wheatley}(1975{\natexlab{b}})}]{Wheatley:75b}%
  \BibitemOpen
  \bibfield  {author} {\bibinfo {author} {\bibnamefont {Wheatley},
  \bibfnamefont {J.}}} (\bibinfo {year} {1975}{\natexlab{b}}),\ in\ \href@noop
  {} {\emph {\bibinfo {booktitle} {The Helium Liquids}}},\ \bibinfo {editor}
  {edited by\ \bibinfo {editor} {\bibfnamefont {J.}~\bibnamefont {Armitage}}\
  and\ \bibinfo {editor} {\bibfnamefont {I.}~\bibnamefont {Farquhar}}}\
  (\bibinfo  {publisher} {Academic Press - London})\ p.\ \bibinfo {pages}
  {241}\BibitemShut {NoStop}%
\bibitem [{\citenamefont {Whitmore}\ and\ \citenamefont {{Zimmermann,
  Jr.}}(1968)}]{Whitmore:68}%
  \BibitemOpen
  \bibfield  {author} {\bibinfo {author} {\bibnamefont {Whitmore},
  \bibfnamefont {S.}}, \ and\ \bibinfo {author} {\bibfnamefont
  {W.}~\bibnamefont {{Zimmermann, Jr.}}}} (\bibinfo {year} {1968}),\ \href@noop
  {} {\bibfield  {journal} {\bibinfo  {journal} {Phys. Rev.}\ }\textbf
  {\bibinfo {volume} {66}},\ \bibinfo {pages} {181}}\BibitemShut {NoStop}%
\bibitem [{\citenamefont {Wilks}(1967)}]{Wilks:67}%
  \BibitemOpen
  \bibfield  {author} {\bibinfo {author} {\bibnamefont {Wilks}, \bibfnamefont
  {J.}}} (\bibinfo {year} {1967}),\ \href@noop {} {\emph {\bibinfo {title} {The
  Properties of Liquid and Solid Helium}}}\ (\bibinfo  {publisher} {Clarendon
  Press - Oxford})\BibitemShut {NoStop}%
\bibitem [{\citenamefont {Wirth}\ and\ \citenamefont {{Zimmermann,
  Jr.}}(1981)}]{Wirth:81}%
  \BibitemOpen
  \bibfield  {author} {\bibinfo {author} {\bibnamefont {Wirth}, \bibfnamefont
  {F.~H.}}, \ and\ \bibinfo {author} {\bibfnamefont {W.}~\bibnamefont
  {{Zimmermann, Jr.}}}} (\bibinfo {year} {1981}),\ \href@noop {} {\bibfield
  {journal} {\bibinfo  {journal} {Physica B+C}\ }\textbf {\bibinfo {volume}
  {107}},\ \bibinfo {pages} {579}}\BibitemShut {NoStop}%
\bibitem [{\citenamefont {Wyatt}\ and\ \citenamefont
  {Lauter}(1991)}]{Wyatt:91}%
  \BibitemOpen
  \bibinfo {editor} {\bibnamefont {Wyatt}, \bibfnamefont {A.~F.}}, \ and\
  \bibinfo {editor} {\bibfnamefont {H.~J.}\ \bibnamefont {Lauter}},\ Eds.
  (\bibinfo {year} {1991}),\ \href@noop {} {\emph {\bibinfo {title}
  {Excitations in Two-Dimensional and Three-Dimensional Quantum Fluids}}}\
  (\bibinfo  {publisher} {Plenum press, New-York})\BibitemShut {NoStop}%
\bibitem [{\citenamefont {Yang}(1962)}]{Yang:62}%
  \BibitemOpen
  \bibfield  {author} {\bibinfo {author} {\bibnamefont {Yang}, \bibfnamefont
  {C.~N.}}} (\bibinfo {year} {1962}),\ \href@noop {} {\bibfield  {journal}
  {\bibinfo  {journal} {Rev. Mod. Phys.}\ }\textbf {\bibinfo {volume} {34}},\
  \bibinfo {pages} {694}}\BibitemShut {NoStop}%
\bibitem [{\citenamefont {Yang}(2003)}]{Yang:03}%
  \BibitemOpen
  \bibfield  {author} {\bibinfo {author} {\bibnamefont {Yang}, \bibfnamefont
  {C.~N.}}} (\bibinfo {year} {2003}),\ \href@noop {} {\bibfield  {journal}
  {\bibinfo  {journal} {Int. J. Mod. Phys. A}\ }\textbf {\bibinfo {volume}
  {18}},\ \bibinfo {pages} {1}}\BibitemShut {NoStop}%
\bibitem [{\citenamefont {Yip}(1999)}]{Yip:99}%
  \BibitemOpen
  \bibfield  {author} {\bibinfo {author} {\bibnamefont {Yip}, \bibfnamefont
  {S.-K.}}} (\bibinfo {year} {1999}),\ \href@noop {} {\bibfield  {journal}
  {\bibinfo  {journal} {Phys. Rev. Lett.}\ }\textbf {\bibinfo {volume} {83}},\
  \bibinfo {pages} {3864}}\BibitemShut {NoStop}%
\bibitem [{\citenamefont {Zapata}\ \emph {et~al.}(2003)\citenamefont {Zapata},
  \citenamefont {Sols},\ and\ \citenamefont {Leggett}}]{Zapata:03}%
  \BibitemOpen
  \bibfield  {author} {\bibinfo {author} {\bibnamefont {Zapata}, \bibfnamefont
  {I.}}, \bibinfo {author} {\bibfnamefont {F.}~\bibnamefont {Sols}}, \ and\
  \bibinfo {author} {\bibfnamefont {A.~J.}\ \bibnamefont {Leggett}}} (\bibinfo
  {year} {2003}),\ \href@noop {} {\bibfield  {journal} {\bibinfo  {journal}
  {Phys. Rev. A}\ }\textbf {\bibinfo {volume} {67}},\ \bibinfo {pages}
  {021603}}\BibitemShut {NoStop}%
\bibitem [{\citenamefont {Zhang}\ and\ \citenamefont {Wang}(2001)}]{Zhang:01}%
  \BibitemOpen
  \bibfield  {author} {\bibinfo {author} {\bibnamefont {Zhang}, \bibfnamefont
  {W.}}, \ and\ \bibinfo {author} {\bibfnamefont {Z.~D.}\ \bibnamefont {Wang}}}
  (\bibinfo {year} {2001}),\ \href@noop {} {\bibfield  {journal} {\bibinfo
  {journal} {Phys. Rev. B}\ }\textbf {\bibinfo {volume} {64}},\ \bibinfo
  {pages} {214501}}\BibitemShut {NoStop}%
\bibitem [{\citenamefont {Zieve}\ \emph {et~al.}(1992)\citenamefont {Zieve},
  \citenamefont {Mukharsky}, \citenamefont {Close}, \citenamefont {Davis},\
  and\ \citenamefont {Packard}}]{Zieve:92}%
  \BibitemOpen
  \bibfield  {author} {\bibinfo {author} {\bibnamefont {Zieve}, \bibfnamefont
  {R.~J.}}, \bibinfo {author} {\bibfnamefont {Y.}~\bibnamefont {Mukharsky}},
  \bibinfo {author} {\bibfnamefont {J.~D.}\ \bibnamefont {Close}}, \bibinfo
  {author} {\bibfnamefont {J.~C.}\ \bibnamefont {Davis}}, \ and\ \bibinfo
  {author} {\bibfnamefont {R.~E.}\ \bibnamefont {Packard}}} (\bibinfo {year}
  {1992}),\ \href@noop {} {\bibfield  {journal} {\bibinfo  {journal} {Phys.
  Rev. Lett.}\ }\textbf {\bibinfo {volume} {68}},\ \bibinfo {pages}
  {1327}}\BibitemShut {NoStop}%
\bibitem [{\citenamefont {Ziff}\ \emph {et~al.}(1977)\citenamefont {Ziff},
  \citenamefont {Uhlenbeck},\ and\ \citenamefont {Kac}}]{Ziff:77}%
  \BibitemOpen
  \bibfield  {author} {\bibinfo {author} {\bibnamefont {Ziff}, \bibfnamefont
  {R.~M.}}, \bibinfo {author} {\bibfnamefont {G.~E.}\ \bibnamefont
  {Uhlenbeck}}, \ and\ \bibinfo {author} {\bibfnamefont {M.}~\bibnamefont
  {Kac}}} (\bibinfo {year} {1977}),\ \href@noop {} {\bibfield  {journal}
  {\bibinfo  {journal} {Phys. Lett. C}\ }\textbf {\bibinfo {volume} {32}},\
  \bibinfo {pages} {169}}\BibitemShut {NoStop}%
\bibitem [{\citenamefont {{Zimmermann,
  Jr.}}(1993{\natexlab{a}})}]{Zimmermann:93}%
  \BibitemOpen
  \bibfield  {author} {\bibinfo {author} {\bibnamefont {{Zimmermann, Jr.}},
  \bibfnamefont {W.}}} (\bibinfo {year} {1993}{\natexlab{a}}),\ \href@noop {}
  {\bibfield  {journal} {\bibinfo  {journal} {J. Low Temp. Phys.}\ }\textbf
  {\bibinfo {volume} {91}},\ \bibinfo {pages} {219}}\BibitemShut {NoStop}%
\bibitem [{\citenamefont {{Zimmermann,
  Jr.}}(1993{\natexlab{b}})}]{Zimmermann:93b}%
  \BibitemOpen
  \bibfield  {author} {\bibinfo {author} {\bibnamefont {{Zimmermann, Jr.}},
  \bibfnamefont {W.}}} (\bibinfo {year} {1993}{\natexlab{b}}),\ \href@noop {}
  {\bibfield  {journal} {\bibinfo  {journal} {J. Low Temp. Phys.}\ }\textbf
  {\bibinfo {volume} {93}},\ \bibinfo {pages} {1003}}\BibitemShut {NoStop}%
\bibitem [{\citenamefont {{Zimmermann, Jr.}}(1994)}]{Zimmermann:94}%
  \BibitemOpen
  \bibfield  {author} {\bibinfo {author} {\bibnamefont {{Zimmermann, Jr.}},
  \bibfnamefont {W.}}} (\bibinfo {year} {1994}),\ \href@noop {} {\bibfield
  {journal} {\bibinfo  {journal} {Physica B}\ }\textbf {\bibinfo {volume}
  {194-196}},\ \bibinfo {pages} {585}}\BibitemShut {NoStop}%
\bibitem [{\citenamefont {{Zimmermann, Jr.}}(1996)}]{Zimmermann:96}%
  \BibitemOpen
  \bibfield  {author} {\bibinfo {author} {\bibnamefont {{Zimmermann, Jr.}},
  \bibfnamefont {W.}}} (\bibinfo {year} {1996}),\ \href@noop {} {\bibfield
  {journal} {\bibinfo  {journal} {Contemp. Phys.}\ }\textbf {\bibinfo {volume}
  {37}},\ \bibinfo {pages} {219}}\BibitemShut {NoStop}%
\bibitem [{\citenamefont {{Zimmermann, Jr.}}\ \emph {et~al.}(1990)\citenamefont
  {{Zimmermann, Jr.}}, \citenamefont {Avenel},\ and\ \citenamefont
  {Varoquaux}}]{Zimmermann:90}%
  \BibitemOpen
  \bibfield  {author} {\bibinfo {author} {\bibnamefont {{Zimmermann, Jr.}},
  \bibfnamefont {W.}}, \bibinfo {author} {\bibfnamefont {O.}~\bibnamefont
  {Avenel}}, \ and\ \bibinfo {author} {\bibfnamefont {E.}~\bibnamefont
  {Varoquaux}}} (\bibinfo {year} {1990}),\ \href@noop {} {\bibfield  {journal}
  {\bibinfo  {journal} {Physica B}\ }\textbf {\bibinfo {volume} {165\&166}},\
  \bibinfo {pages} {749}}\BibitemShut {NoStop}%
\bibitem [{\citenamefont {{Zimmermann, Jr.}}\ \emph {et~al.}(1998)\citenamefont
  {{Zimmermann, Jr.}}, \citenamefont {Lindensmith},\ and\ \citenamefont
  {Flaten}}]{Zimmermann:98}%
  \BibitemOpen
  \bibfield  {author} {\bibinfo {author} {\bibnamefont {{Zimmermann, Jr.}},
  \bibfnamefont {W.}}, \bibinfo {author} {\bibfnamefont {C.~A.}\ \bibnamefont
  {Lindensmith}}, \ and\ \bibinfo {author} {\bibfnamefont {J.~A.}\ \bibnamefont
  {Flaten}}} (\bibinfo {year} {1998}),\ \href@noop {} {\bibfield  {journal}
  {\bibinfo  {journal} {J. Low Temp. Phys.}\ }\textbf {\bibinfo {volume}
  {110}},\ \bibinfo {pages} {497}}\BibitemShut {NoStop}%
\end{thebibliography}

\newcommand{\noopsort}[1]{}

\end{document}